\documentclass{nature}

\usepackage{lineno}

\usepackage{setspace}
\usepackage{enumitem}
\usepackage[margin=0.7in]{geometry}
\usepackage{caption}

\usepackage{amsfonts}
\usepackage{amssymb}
\usepackage{amsmath}

\usepackage{placeins}
\usepackage{graphicx}
\usepackage[T1]{fontenc}

\usepackage[export]{adjustbox}

\usepackage[labelformat=empty]{caption}

\usepackage{xspace}
\usepackage{subfig}
\usepackage{ulem}
\usepackage{longtable}
\usepackage{multirow}
\usepackage{adjustbox}
\usepackage{booktabs}
\usepackage{array}
\usepackage{float}

\newcommand{\apjl}{The Astrophysical Journal Letters}
\newcommand{\apjs}{The Astrophysical Journal Supplement Series}

\newcommand{\aap}{Astronomy and Astrophysics}

\newcommand{\arcsecs}{{$^{\prime\prime}$}\xspace}

\newcommand{\ks}[1]{{\color{black} #1}}
\newcommand{\sw}[1]{\texttt{#1}}

\newcommand{\arcsec}{{$^{\prime\prime}$}}

\newcommand{\frbmark}{20220319D\xspace}
\newcommand{\frbmizu}{20231120A\xspace}
\newcommand{\frbzach}{20220207C\xspace}
\newcommand{\frbjackie}{20220509G\xspace}
\newcommand{\frbleonidas}{20230124\xspace}
\newcommand{\frbelektra}{20220914A\xspace}
\newcommand{\frbtildis}{20230628A\xspace}
\newcommand{\frbetienne}{20220920A\xspace}
\newcommand{\frbnina}{20221101B\xspace}
\newcommand{\frbansel}{20220825A\xspace}
\newcommand{\frbalex}{20220307B\xspace}
\newcommand{\frbisha}{20221113A\xspace}
\newcommand{\frbsquanto}{20231123B\xspace}
\newcommand{\frbphineas}{20230307A\xspace}
\newcommand{\frbcharlotte}{20221116A\xspace}
\newcommand{\frbjuan}{20221012A\xspace}
\newcommand{\frboran}{20220506D\xspace}
\newcommand{\frbmikayla}{20230501A\xspace}
\newcommand{\frbfatima}{20230626A\xspace}
\newcommand{\frbishita}{20220208A\xspace}
\newcommand{\frbgertrude}{20220726A\xspace}
\newcommand{\frberdos}{20220330D\xspace}
\newcommand{\frbfen}{20220204A\xspace}
\newcommand{\frbpingu}{20230712A\xspace}
\newcommand{\frbwhitney}{20220310F\xspace}
\newcommand{\frbferb}{20230216A\xspace}
\newcommand{\frbkoyaanisqatsi}{20221027A\xspace}
\newcommand{\frbnihari}{20221219A\xspace}
\newcommand{\frbquincy}{20220418A\xspace}
\newcommand{\frbmifanshan}{20221029A\xspace}

\newcolumntype{C}{>{\footnotesize\raggedright\arraybackslash}c}

\newcolumntype{L}{>{\footnotesize\raggedright\arraybackslash}l}

\newcounter{mybodyfigure}
\newcounter{myedfigure}
\newcounter{mysdfigure}

\newcommand{\beginbodyfigures}{\renewcommand{figure}{\let\caption\NAT@figcaption}{{\themybodyfigure}}}
\newcommand{\beginedfigures}{\renewcommand{figure}{\let\caption\NAT@figcaption}{{\themyedfigure}}}
\newcommand{\beginsdfigures}{\renewcommand{figure}{\let\caption\NAT@figcaption}{{\themysdfigure}}}

\newcommand{\stepbodyfigure}{\refstepcounter{mybodyfigure}}
\newcommand{\stepedfigure}{\refstepcounter{myedfigure}}
\newcommand{\stepsdfigure}{\refstepcounter{mysdfigure}}

\newcommand{\bodyfigurelabel}[1]{\bf{Fig.\bodyfigure{#1}:}}
\newcommand{\edfigurelabel}[1]{\bf{Extended Data Fig.\edfigure{#1}:}}
\newcommand{\supfigurelabel}[1]{\bf{Supplementary Fig.\sdfigure{#1}:}}

\DeclareRobustCommand{\bodyfigure}[1]{\stepbodyfigure\label{#1}{\themybodyfigure}}
\DeclareRobustCommand{\edfigure}[1]{\stepedfigure\label{#1}{\themyedfigure}}
\DeclareRobustCommand{\sdfigure}[1]{\stepsdfigure\label{#1}{\themysdfigure}}

\makeatletter
\g@addto@macro\caption@prepareslc{%
  \renewcommand{\stepbodyfigure}{\caption@l@stepcounter{mybodyfigure}}}
\makeatother

\makeatletter
\g@addto@macro\caption@prepareslc{%
  \renewcommand{\stepedfigure}{\caption@l@stepcounter{myedfigure}}}
\makeatother

\makeatletter
\g@addto@macro\caption@prepareslc{%
  \renewcommand{\stepsdfigure}{\caption@l@stepcounter{mysdfigure}}}
\makeatother

\DeclareCaptionLabelFormat{supplementarytablelabel}{\bf{Supplementary Table #2}} 

\DeclareCaptionLabelFormat{tablelabel}{\bf{Extended Data Table #2}} 

\usepackage[super,comma,sort&compress]{natbib}

\usepackage{hyperref}
\usepackage{verbatim}

\usepackage{ulem} 
\usepackage{rotating} 
\usepackage{soul}

\usepackage{color} 

\usepackage[colorinlistoftodos]{todonotes}

\makeatletter
\newsavebox\myboxA
\newsavebox\myboxB
\newlength\mylenA

\newcommand*\xoverline[2][0.75]{%
    \sbox{\myboxA}{$\m@th#2$}%
    \setbox\myboxB\null
    \ht\myboxB=\ht\myboxA%
    \dp\myboxB=\dp\myboxA%
    \wd\myboxB=#1\wd\myboxA
    \sbox\myboxB{$\m@th\overline{\copy\myboxB}$}
    \setlength\mylenA{\the\wd\myboxA}
    \addtolength\mylenA{-\the\wd\myboxB}%
    \ifdim\wd\myboxB<\wd\myboxA%
       \rlap{\hskip 0.5\mylenA\usebox\myboxB}{\usebox\myboxA}%
    \else
        \hskip -0.5\mylenA\rlap{\usebox\myboxA}{\hskip 0.5\mylenA\usebox\myboxB}%
    \fi}
\makeatother

\usepackage{multibib}
\newcites{main}{\noindent{\bfseries \LARGE References}}
\newcites{methods}{\noindent{\bfseries \LARGE Additional References}}

\title{Preferential Occurrence of Fast Radio Bursts in Massive Star-Forming Galaxies}

\author{\large 
Kritti Sharma$^{1,^\ast}$, 
Vikram Ravi$^{1,2}$, 
Liam Connor$^{1}$, 
Casey Law$^{1,2}$, 
Stella Koch Ocker$^{1,3}$, 
Myles Sherman$^{1}$, 
Nikita Kosogorov$^{1,2}$, 
Jakob Faber$^{1,}$, 
Gregg Hallinan$^{1,2}$, 
Charlie Harnach$^{1,2}$, 
Greg Hellbourg$^{1,2}$, 
Rick Hobbs$^{1,2}$, 
David Hodge$^{1,2}$, 
Mark Hodges$^{1,2}$, 
James Lamb$^{1,2}$, 
Paul Rasmussen$^{1,2}$, 
Jean Somalwar$^{1}$, 
Sander Weinreb$^{1}$, 
David Woody$^{1,2}$,
Joel Leja$^{4,5,6}$, 
Shreya Anand$^{1}$, 
Kaustav Kashyap Das$^{1}$, 
Yu-Jing Qin$^{1}$, 
Sam Rose$^{1}$, 
Dillon Z. Dong$^{1, 7}$, 
Jessie Miller$^{1}$,
Yuhan Yao$^{1, 8}$
}

\begin{document}

\maketitle

\begin{affiliations}
 \item Cahill Center for Astronomy and Astrophysics, MC 249-17 California Institute of Technology, Pasadena, CA 91125, USA.
 \item Owens Valley Radio Observatory, California Institute of Technology, Big Pine, CA 93513, USA.
 \item The Observatories of the Carnegie Institution for Science, Pasadena, CA 91101, USA.
 \item Department of Astronomy \& Astrophysics, The Pennsylvania State University, University Park, PA 16802, USA.
 \item Institute for Computational \& Data Sciences, The Pennsylvania State University, University Park, PA, USA.
 \item Institute for Gravitation \& the Cosmos, The Pennsylvania State University, University Park, PA 16802, USA.
 \item National Radio Astronomy Observatory, 1003 Lopezville Road, Socorro, NM, 87801, USA.
 \item Department of Astronomy, University of California, Berkeley, CA 94720-3411, USA.
 \item[] $^{\ast}$Correspondence email: kritti@caltech.edu
\end{affiliations}

\begin{abstract}

Fast Radio Bursts (FRBs) are millisecond-duration events detected from beyond the Milky Way. FRB emission characteristics favor highly magnetized neutron stars, known as magnetars, as the sources~\citep{2023RvMP...95c5005Z}. This is supported by the detection of FRB-like bursts from a magnetar in the Milky Way~\citep{2020Natur.587...54C,2020Natur.587...59B}\ks{, and the typically star-forming nature of FRB host galaxies~\citep{2023arXiv230205465G, 2023arXiv231010018B}}. However, the specific stellar-evolutionary pathways by which FRB sources, and indeed all magnetars, are formed remain unknown. Although several lines of evidence link Galactic magnetars to core-collapse supernovae \ks{(CCSNe)}~\citep{2017ARA&A..55..261K}, the factors that determine which supernovae result in magnetars is unclear~\citep{2023IAUS..363...61P}. The galactic environments of FRB sources harbor a wealth of information that can be harnessed to probe their progenitors. Here, we present the stellar population properties of a sample of 30 FRB host galaxies discovered by the Deep Synoptic Array (DSA-110). We find that \ks{there is a significant deficit of low-mass FRB hosts relative to the occurrence of star-formation in background galaxies. This implies that unlike the most prevalent class of CCSNe (Type II), which accurately trace star-formation in the universe,} FRBs are a biased tracer of star-formation\ks{, preferentially selecting for massive star-forming galaxies}. We suggest that this preference for massive star-forming galaxies is driven by the underlying galaxy metallicity, which is known to be positively correlated with galaxy stellar mass~\citep{2005MNRAS.362...41G}. \ks{CCSNe} of stellar merger remnants are more likely in metal-rich environments~\citep{2017A&A...601A..29Z, 2019A&A...631A...5Z}\ks{, as higher metallicity stars are less compact and are more likely to fill their Roche lobes in binaries, leading to unstable mass transfer and mergers that potentially producing magnetar progenitors. Furthermore, massive stars do not have convective interiors and may not be able to generate strong magnetic fields by dynamo~\citep{2024Sci...384..214F}, whereas stellar-merger remnants are thought to have the requisite internal magnetic-field strengths to result in magnetars~\citep{2019Natur.574..211S, 2024Sci...384..214F}. The preferential occurrence of FRBs in massive star-forming galaxies suggests that CCSN of stellar merger remnants preferentially form magnetars.} 
\end{abstract}

The Deep Synoptic Array (DSA-110), situated at the Owens Valley Radio Observatory (OVRO) near Bishop, California, is a radio interferometer built for simultaneous FRB discovery and arcsecond -scale localization. The DSA-110 underwent science commissioning and performed observations between February 2022 and March 2024 with a coherent core of 48 4.65~m antennas used for FRB searching combined with 15 outrigger antennas (maximum baseline of 2.5\,km) used for localization. Each antenna is equipped with a dual-polarization ambient-temperature 1.28--1.53\,GHz receiver. A custom low-noise amplifier design delivering 7\,K noise temperature~\citep{2021ITMTT..69.2345W} was central to achieving sensitivity to 1.9\,Jy\,ms FRBs (for millisecond-duration events). A real-time search for FRBs with 0.262\,ms sampling and a dispersion-measure (DM) range up to 1500\,pc\,cm$^{-3}$ was conducted. Localization accuracies of better than $\pm2$\,arcsecond (90\% confidence) were achieved by comparison with coeval observations of standard astrometric reference sources (see Methods and Supplementary Fig.~\ref{fig:offsets_test},~\ref{fig:repeater_localization}). During these observations, 60 FRBs were successfully localized.

In this work, we limit our analysis to FRBs discovered up to November 2023 which have redshifts for all hosts detectable down to $r = 23.5$~mag, to ensure a uniform sample selection. The follow-up of a subset of FRBs discovered post November 2023 is presented in our companion paper (Connor et al.). Among the 42 FRBs localized by DSA-110 up to November 2023, 30 had a potential host-galaxy candidate in the vicinity of the FRB localization (within 10\arcsecs), detectable at $\leq 23.5$~mag in archival $r$-band data from PanSTARRS1 (PS1)~\citep{2016arXiv161205560C} or the Beijing-Arizona Sky Survey (BASS) from the Dark Energy Survey~\citep{2017PASP..129f4101Z}. We complement these archival data with deeper ground-based optical or near-infrared imaging observations with the Wafer-Scale Imager for Prime focus (WaSP)~\citep{2017JATIS...3c6002N} and the Wide Field Infrared Camera (WIRC)~\citep{2003SPIE.4841..451W} instruments, mounted on the 200-inch Hale Telescope at the Palomar Observatory in our follow-up campaigns (see Methods). We use the Bayesian Probabilistic Association of Transients to their Hosts (PATH) formalism~\citep{2021ApJ...911...95A} on the deepest available imaging data to estimate the association probability (P$_{\mathrm{host}}$) of the most likely host galaxy (see Methods). The PATH analysis finds secure host associations for 26 FRBs with P$_{\mathrm{host}} \geq 90$\% (see Extended Data Fig.~\ref{fig:path_analysis}). Of the remaining four events, FRBs \frbkoyaanisqatsi and \frberdos have two possible hosts, one of which is favored by both the localizations and the DMs (see Methods). FRB \frbferb is found at a large offset from the preferred host, which lowers the association probability according to the chosen PATH setup, and the localization of FRB\,\frbishita is confused by the presence of a faint (23.4~mag in $J$-band data, spectroscopic redshift not available) alternative host. We further validate our host associations in Methods and Supplementary Fig.~\ref{fig:association_capabilities_demonstration},~\ref{fig:association_capabilities_demonstration_highz}. We also discuss the hostless FRBs in Methods and Supplementary Fig.~\ref{fig:hostDMdistribution}. The imaging mosaic of 30 FRB hosts included in our sample is displayed in Fig.~\ref{fig:host_cutouts}, and the discovery properties of the host galaxies are tabulated in Extended Data Table~\ref{table:basic_frb_properties}. For all quantitative arguments in our work, we only consider secure host associations with P$_{\mathrm{host}} \geq 90\%$.

Having identified the most probable host galaxies, we obtained optical spectroscopy with the Low Resolution Imaging Spectrometer (LRIS)~\citep{1995PASP..107..375O} on Keck-I, DEep Imaging Multi-Object Spectrograph (DEIMOS)~\citep{2003SPIE.4841.1657F} on Keck-II at W. M. Keck Observatory and the Double Spectrograph (DBSP)~\citep{1982PASP...94..586O} on the 200-inch Hale Telescope at Palomar Observatory (see Methods). The spectroscopic redshifts ($z$) and emission line fluxes are measured by jointly fitting the stellar continuum and nebular emission using the penalized PiXel-Fitting (\sw{pPXF}) software~\citep{2022arXiv220814974C} (see Supplementary Fig.~\ref{fig:host_spectra} and Table~\ref{table:host_spectroscopy_details}). Next, we model the spectral energy distributions (SEDs) of the FRB host galaxies using the \sw{Prospector} software~\citep{2021ApJS..254...22J}, where we jointly forward model the observed spectra, archival photometry from PS1, BASS, Mayall z-band Legacy Survey (MzLS)~\citep{2019AJ....157..168D}, Sloan Digital Sky Survey (SDSS)~\citep{2015ApJS..219...12A}, Two Micron All Sky Survey (2MASS)~\citep{2006AJ....131.1163S}, Wide-field Infrared Survey Explorer (WISE)~\citep{2014yCat.2328....0C}, and Galaxy Evolution Explorer (GALEX)~\citep{2005ApJ...619L...1M} surveys and photometry of data obtained with WaSP and WIRC instruments (see Supplementary Table~\ref{table:imaging_log}). We model the galaxies with a seven-component non-parametric star-formation history (SFH), a two-component dust attenuation model, a flexible dust attenuation curve, dust emission, and a self-consistent nebular emission model (see Methods and Supplementary Table~\ref{table:sed_params} for a summary of model parameters). Using standard empirical optical emission-line diagnostic diagrams~\citep{1981PASP...93....5B, 2006MNRAS.372..961K} (see Extended Data Fig.~\ref{fig:BPT_analysis}) and WISE color-color galaxy classifications~\citep{2010AJ....140.1868W} (see Extended Data Fig.~\ref{fig:wise_color_color}), we find that the dominant ionization mechanism in FRB host galaxies is consistent with the locus of star-forming galaxies (late-type spirals) and emission line galaxies with active galactic nuclei (AGN, either LINERs or Seyferts) (see Methods and Supplementary Table~\ref{table:BPT_analysis}). Therefore, we also include the emission from dust-enshrouded AGN in our SED modeling. The derived properties from our SED fits (see Supplementary Fig.~\ref{fig:host_seds}) and constrained SFHs (see Supplementary Fig.~\ref{fig:host_sfh}) for FRB host galaxies are tabulated in Extended Data Table~\ref{table:derived_galaxy_properties}, and their distributions are shown in Extended Data Fig.~\ref{fig:all_params_box_plot}.

To contextualize FRB host galaxies within the broader framework of star-formation and stellar mass in the universe, we compare them with the background galaxy population. In our comparison sample, alongside our 26 secure host associations, we include a complete literature sample of 26 FRB hosts~\citep{2023arXiv230205465G,2023arXiv231010018B} that adhere to our selection criteria of $r$-band magnitude $\leq 23.5$~mag and secure host association (P$_{\mathrm{host}} \geq 90$\%). To address incompleteness inherent in magnitude-limited galaxy surveys, we adopted a hybrid approach to simulate the complete background galaxy population. We sample the galaxy stellar masses, $\mathrm{M}_\ast$, from the stellar mass function, $\Phi(\mathrm{M}_\ast, z)$~\citep{2020ApJ...893..111L} and then compute the corresponding star-formation rate (SFR) using the star-forming main sequence, $\mathrm{SFR (M}_\ast, z)$~\citep{2022ApJ...936..165L} and the distribution of galaxies in $\log \mathrm{M}_\ast - \log \mathrm{SFR} - z$ space~\citep{2022ApJ...936..165L} (see Methods and Supplementary Fig.~\ref{fig:extreme_cases_sanity_check},~\ref{fig:additional_uncertainties_no_z_evolution}). We compare the stellar mass distribution of FRB hosts with the distributions of stellar mass of background galaxies selected by two methods -- weighted by SFR and weighted by stellar mass. We split the FRB comparison sample into three redshift bins to mitigate  biases from the evolution of the background galaxy population: $z \leq 0.2$ with 20 FRBs, $0.2 < z \leq 0.4$ with 24 FRBs, and $0.4 < z \leq 0.7$ with 7 FRBs. The lowest redshift bin edge was chosen based on our capability to confidently identify low mass galaxies, given the optical imaging depths (see Methods). Notably, FRB \frbmifanshan was excluded from this analysis due to its solitary occurrence at $z \sim 1$, rendering meaningful comparisons challenging at high redshifts owing to limited statistical power. We perform one-sample Kolmogorov-Smirnov (KS) tests between the sample of FRB stellar masses, and the background distributions corrected for optical selection effect of $r$-band magnitude $\leq 23.5$~mag (see Methods). The results are shown in Fig.~\ref{fig:frbs_trace_sfr}.

We find that the sample of FRB host-galaxy stellar masses is inconsistent with the stellar mass distribution in the universe, but broadly consistent with the distribution of galaxies selected according to SFR. In all three redshift bins, the KS-test $p$-value from the comparison between FRBs and galaxies selected according to stellar mass is $<0.001$ (i.e., $>3\sigma$ significance). Conversely, the comparison with the stellar mass distribution of galaxies selected according to SFR yields $p$-values greater than $0.01$ in all the three redshift bins. This similarity to galaxies selected by SFR is further emphasized by the close alignment of FRB host galaxies with the star-forming main sequence of galaxies~\citep{2023arXiv230205465G,law2023deep} (see Extended Data Fig.~\ref{fig:gal_sfms_tm_ssfr}). However, for $z \leq 0.2$, despite our sensitivity to optically faint galaxies, we observe a notable scarcity of FRBs in the 
galaxies with $\log \mathrm{M}_\ast \lesssim 9$ 
(see Fig.~\ref{fig:frbs_trace_sfr}a). 
This is indicated by the low associated KS-test result of $p=0.030$; we note that the KS-test is not optimal to quantify the significance of this claim. Radio selection effects are not expected to contribute to this scarcity of low-mass FRB hosts at $z \leq 0.2$~\citep{2022ApJ...934...71O} (see Methods).

The dearth of $z \leq 0.2$ low-mass FRB host galaxies becomes even clearer when we compare them to host galaxies of the most prevalent class of CCSNe (Type II), which trace the occurrence of star-formation in the universe, with no dependence on other galaxy properties~\citep{2021ApJS..255...29S} (see Fig.~\ref{fig:frbs_are_biased_tracer_of_sf_in_universe}b). We show the distribution of stellar masses of Type~II CCSNe and FRB host galaxies in the $r$-band magnitude and redshift space in Fig.~\ref{fig:frbs_are_biased_tracer_of_sf_in_universe}a. FRB hosts trace the locus of $0.1-1$~L$_\ast$ background galaxies, and are more massive than typical Type~II CCSNe host galaxies. To contextualize the rarity of the occurrence of Type~II CCSNe in only massive galaxies on the scale of our $z\leq0.2$ FRB sample size (N$_{\mathrm{FRB}}$), we perform $1,000$ Monte-Carlo simulations where we sample N$_{\mathrm{FRB}}$ galaxy stellar masses from the Type~II CCSNe host distributions. We compute the fraction of these samples with all stellar masses above a particular stellar mass $\log \mathrm{M}_\ast$ (see Extended Data Fig.~\ref{fig:significance_of_dwarfs}). We find that for our complete local universe FRB sample of size N$_{\mathrm{FRB}} = 20$, the probability that all Type~II CCSNe occur in galaxies more massive than $10^9$~M$_\odot$ is $p=0.0014$ ($\sim 3.2\sigma$ significance). If FRBs were an unbiased tracer of star-formation in the universe, then this quantifies the significance of the deficit of low-mass FRB hosts.

Contrary to previous studies which suggest that FRBs trace the occurrence of star-formation in the universe~\citep{2023arXiv231010018B, 2022MNRAS.510L..18J}, we have shown that FRBs preferentially occur in massive star-forming galaxies and are a biased tracer of star-formation in the universe. This could point to an environment-dependent production efficiency of FRB sources. The primary driver of changes in stellar population properties with galaxy mass is the galaxy mass-metallicity relation~\citep{2005MNRAS.362...41G}. Increased metallicity 
affects the evolution of massive stars by line-driven stellar winds, where the mass-loss rate positively correlates with metallicity. Certain classes of supernova preferentially occur in low-metallicity environments~\citep{2021MNRAS.503.3931T}, such as those that produce long-duration gamma-ray bursts (lGRBs) and superluminous supernovae (SLSNe)~\citep{2021ApJS..255...29S}. We quantify the effect of metallicity on the selection of FRB host galaxies by constructing background stellar mass distributions weighted by SFR together with a metallicity-dependent FRB source formation efficiency $\rho = (1+(-\mathrm{M}/\mathrm{M}_c)^\beta)^{-1}$. Here, M$_c$ is a characteristic cut-off mass that regulates the production of FRB sources, ceasing their occurrence in lower stellar mass (and hence, lower metallicity) galaxies and $\beta$ regulates the strength of the metallicity cutoff. The best-fitting model suggests a strong cutoff with $\log \mathrm{M}_c = 9.02$ (see Fig.~\ref{fig:frbs_are_biased_tracer_of_sf_in_universe}b), thus implying that the formation efficiency of FRB sources is suppressed at oxygen abundances below $12 + \log \mathrm{O/H} \sim 8.08_{-0.47}^{+0.61}$, corresponding to a cutoff metallicity of $\log(Z/Z_\odot) = -0.61_{-0.47}^{+0.61}$. We determine this threshold metallicity by employing the galaxy mass-metallicity relation~\citep{2005MNRAS.362...41G}, which is incorporated as a prior in our SED modeling methodology (see Methods).

We have interpreted the preferential occurrence of FRBs in massive star-forming galaxies as a preference for high metallicity environments, as inferred from the positive correlation between galaxy stellar mass and metallicity~\citep{2005MNRAS.362...41G}. Magnetars are known to be potential FRB sources~\citep{2023RvMP...95c5005Z} and the preferential occurrence of FRBs in higher metallicity environments may be expected in the scenario~\citep{2023RvMP...95c5005Z} that FRBs are emitted by magnetars formed in a sub-population of CCSNe. First, for single-star progenitors, elevated metallicity would favor the formation of neutron star remnants over black holes due to increased mass-loss in higher metallicity stars~\citep{2003ApJ...591..288H}. Further, stellar mergers have been theoretically demonstrated as the origin of magnetic blue straggler stars, which undergo rejuvenation by burning the accreted fuel from their companions, and are believed to be potential progenitors of magnetars due to the amplified magnetic fields of the merger remnants~\citep{2019Natur.574..211S}. The increase in the metallicity of intermediate-mass progenitor stars evolving in such binaries, which eventually culminate in CCSNe, increases the proportion of CCSNe occurring through this delayed binary evolution channel~\citep{2017A&A...601A..29Z, 2019A&A...631A...5Z}. The heightened efficiency of CCSN formation through binary interactions in high-metallicity settings likely stems from the association between metallicity and stellar size~\citep{2019A&A...631A...5Z}. A star with higher metallicity is less compact as it evolves beyond the main sequence, thereby affecting the progression of mass transfer in binary systems~\citep{2020A&A...638A..55K}. At high metallicity, stars in binaries are more likely to evolve to fill their Roche lobes, leading to unstable mass transfer and stellar mergers that potentially produce magnetar progenitors. A stellar-merger formation channel for magnetar progenitors may indeed be observationally favored for the Galactic magnetar population~\citep{sherman2024searching}.

We broaden our understanding of FRB sources by comparing the distributions of host-normalized projected galactocentric offsets and host galaxy stellar mass with various classes of transients (see Fig.~\ref{fig:compare_with_other_transients_in_local_univ_bin} and Methods for a description of the literature samples used). We limit our comparisons to the local universe ($z \leq 0.2$) to potentially mitigate any unknown incompleteness that might be inherent to other transients at high redshifts. We also show the distribution for the entire redshift range in Extended Data Fig.~\ref{fig:offset_distribution_compare_with_other_transients} and \ref{fig:compare_with_other_transients_at_zero_redshift}. We correct the galaxy stellar mass distributions for the redshift evolution and perform two-sample KS-tests to quantify the potential similarities (see Methods and Supplementary Table~\ref{table:compare_with_other_transients}). In contrast to FRBs, the SLSNe and lGRBs predominantly manifest in the central star-forming regions of low-mass galaxies characterized by low metallicity and high specific SFR, thus underscoring the dissimilarities with FRBs. Although the offset distribution of ultra-luminous X-ray (ULX) sources is consistent with FRBs ($p_{\mathrm{KS}} = 0.09$), they demonstrate a preference for occurrence in massive galaxies and trace the background galaxy population selected by stellar mass, not star-formation. The stellar mass distribution of FRB host galaxies is comparable to those of other classes of transient that trace star-formation, including Type~II CCSNe, Type~Ia supernovae, and short-duration GRBs (sGRBs), but with the deficit of low-mass galaxies. 

Some differences are apparent in the offset distributions of FRBs and classes of transients that trace star formation. Although FRBs are systematically found at larger offsets than Type~II CCSNe and Type~Ia supernovae, but smaller offsets than sGRBs, the host-normalized offsets are consistent with these three transient classes, owing to massive FRB host galaxies and the positive galaxy stellar mass - radius correlation. The larger absolute offset values may be a consequence of the radio-observation bias, where bursts originating closer to the center of star-forming spiral galaxies are over-dispersed and exhibit higher scattering timescales~\citep{2022ApJ...934...71O}, thus preventing their detection. If FRBs were to trace the locations of star-formation within their host galaxies, this radio selection bias may shift the FRB offset distribution to lower offsets by up to $\sim 1$~kpc~\citep{2021arXiv211207639S}. On the other hand, the larger FRB offsets may be indicative of the long delays in CCSNe involving interacting binaries, which would imply that the CCSNe occur significantly displaced from the birth sites~\citep{2017A&A...601A..29Z}. For example, if the typical stellar motions at the birth site are $\sim 10$~km s$^{-1}$ and the delay-time is 75~Myr, then the system would have drifted by $750$~pc before the explosion. Alternatively, the larger offsets of FRBs may also arise from the contribution of non-CCSNe formation channels, such as the accretion- or merger-induced collapse (AIC/MIC) of massive white dwarfs and binary neutron star mergers, towards FRB sources. The existence of these FRB source formation channels is indicated by the globular cluster FRB source 20201120E~\citep{2021ApJ...910L..18B, 2021ApJ...917L..11K}, and early DSA-110 results~\citep{law2023deep}. To conclude, the larger offsets of FRBs may either be due to delayed pre-CCSNe stellar merger magnetar formation scenario, or due to contributions from non-CCSN formation channels. However, we note that the current data shows no evidence for the existence of multiple statistically different FRB host galaxy populations (see Methods).

Further insight into source formation channels may be gained through a detailed analysis of the distribution of FRB delay-times with respect to the formation of their stellar progenitors~\citep{2023ApJ...950..175S}. Non-CCSN channels (e.g., AIC/MIC of white dwarfs) are expected to have extended delay-time distributions of several Gyr~\citep{2013A&A...558A..39T}, whereas CCSNe of isolated stars occur on $\sim 3-50$~Myr stellar lifetimes, and the CCSNe of stellar-merger remnants are expected to occur promptly within $\sim 50-250$~Myr of the birth of binary components~\citep{2017A&A...601A..29Z}. The preferential occurrence of FRBs in massive star-forming galaxies is a constraint that applies to any model for FRB source formation. 
The influence of metallicity on the formation of FRB sources can be independently corroborated using forthcoming surveys. Given that star-formation in the early universe predominantly occurs within low-mass galaxies, and galaxies of the same stellar mass at higher redshifts are less chemically enriched~\citep{2016MNRAS.456.2140M}, the preference of FRBs for metal-rich environments implies a suppression of the proposed FRB source formation channel at high redshifts. However, scenarios proposed for the repeating FRB 121102~\citep{2018ApJ...868L...4M}, which is found in a low-metallicity dwarf star-forming galaxy, may become more common at high redshifts. If most FRBs are emitted by magnetars like those observed in the Milky Way, our results favor a scenario where magnetars are generally formed from the CCSN of stellar merger remnants in interacting binaries. 

\newpage

\begin{figure}[h]
\centering
\includegraphics[width=0.975\textwidth]{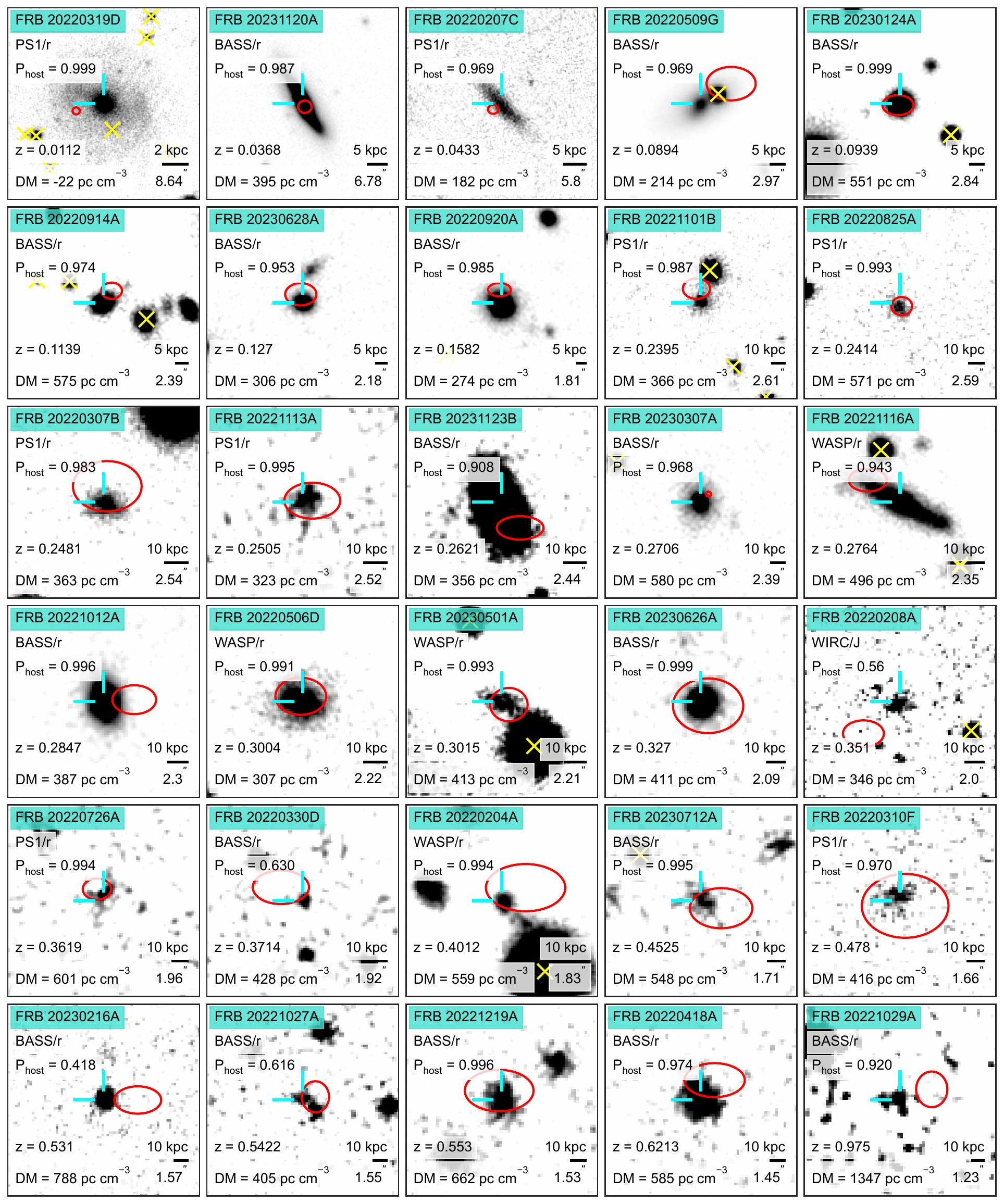}
\caption{{\bodyfigurelabel{fig:host_cutouts}} \textbf{Optical/IR imaging of the fields of DSA-110 discovered FRBs.} The images are centered on the PATH-identified host galaxies (cyan crosshairs), and panels are arranged in increasing order of redshifts (see Extended Data Table~\ref{table:basic_frb_properties}). The 90\% confidence FRB localization regions are marked as red ellipses and stars are marked as yellow crosses. These images reach 3$\sigma$ depths of $\gtrsim 23-24$~mag and are oriented with north up and east to the left. The imaging instrument, association probability, extragalactic DM, redshift, and physical scales are marked on the panels for reference. All images were smoothed with a Gaussian kernel of $\sigma = 0$\arcsec$.15$ to improve visibility.}
\end{figure}

\newpage
\clearpage
\newpage
\newpage

\begin{figure*}[h]
    \includegraphics[width=\textwidth]{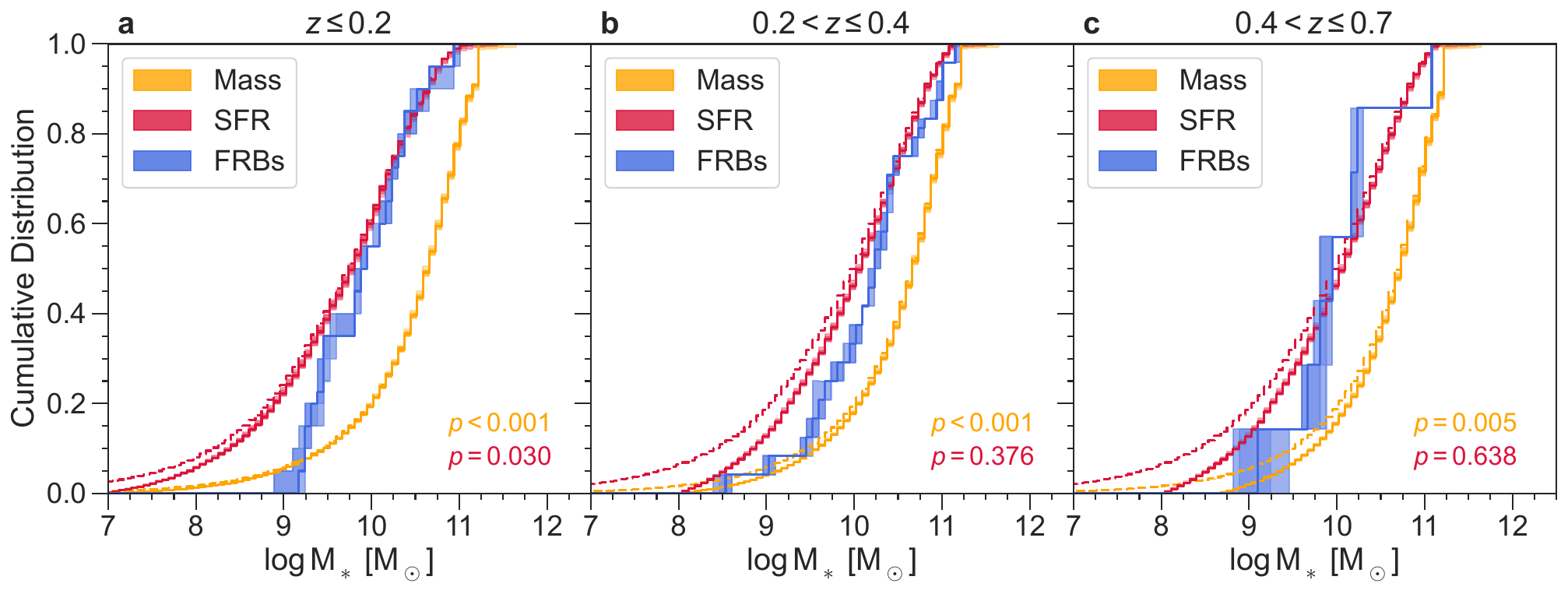}
    \caption{{\bodyfigurelabel{fig:frbs_trace_sfr}} \textbf{Comparison of FRB host galaxies with the distribution of galaxies in the Universe selected by stellar mass and star-formation.} We show cumulative distributions of galaxy stellar mass of samples selected in three ways: the occurrence of FRBs (blue), SFR (red), and stellar mass (orange). 
    We correct the background stellar mass distributions for optical selection effects by employing an $r$-band magnitude threshold of 23.5~mag (solid lines, see Methods). For reference, we also plot the distributions without this selection (dashed lines). 
    The shaded regions represent the $1,2,3\sigma$ bands. 
    Along with 26 secure host associations of DSA-110 FRBs from this work, we also include the Gordon et al.~\citep{2023arXiv230205465G} and Bhardwaj et al.~\citep{2023arXiv231010018B} sample of FRB host galaxies that follow our selection criterion. 
    The distribution of FRB-host stellar masses is inconsistent with the distribution of background galaxies selected by stellar mass in all redshift bins with $>3\sigma$ confidence. 
    The $p$-value computed using the KS-test for similarity with the distribution of background galaxies selected by SFR (red) is $\gtrsim 0.01$ in all redshift bins, indicating that the occurrence of FRBs is correlated with the occurrence of star-formation. 
    However, despite our sensitivity, there is a deficit of low-mass FRB hosts in $z \leq 0.2$ bin.}
\end{figure*}

\newpage
\clearpage
\newpage
\newpage

\begin{figure*}[h]
\centering
\includegraphics[width=\textwidth]{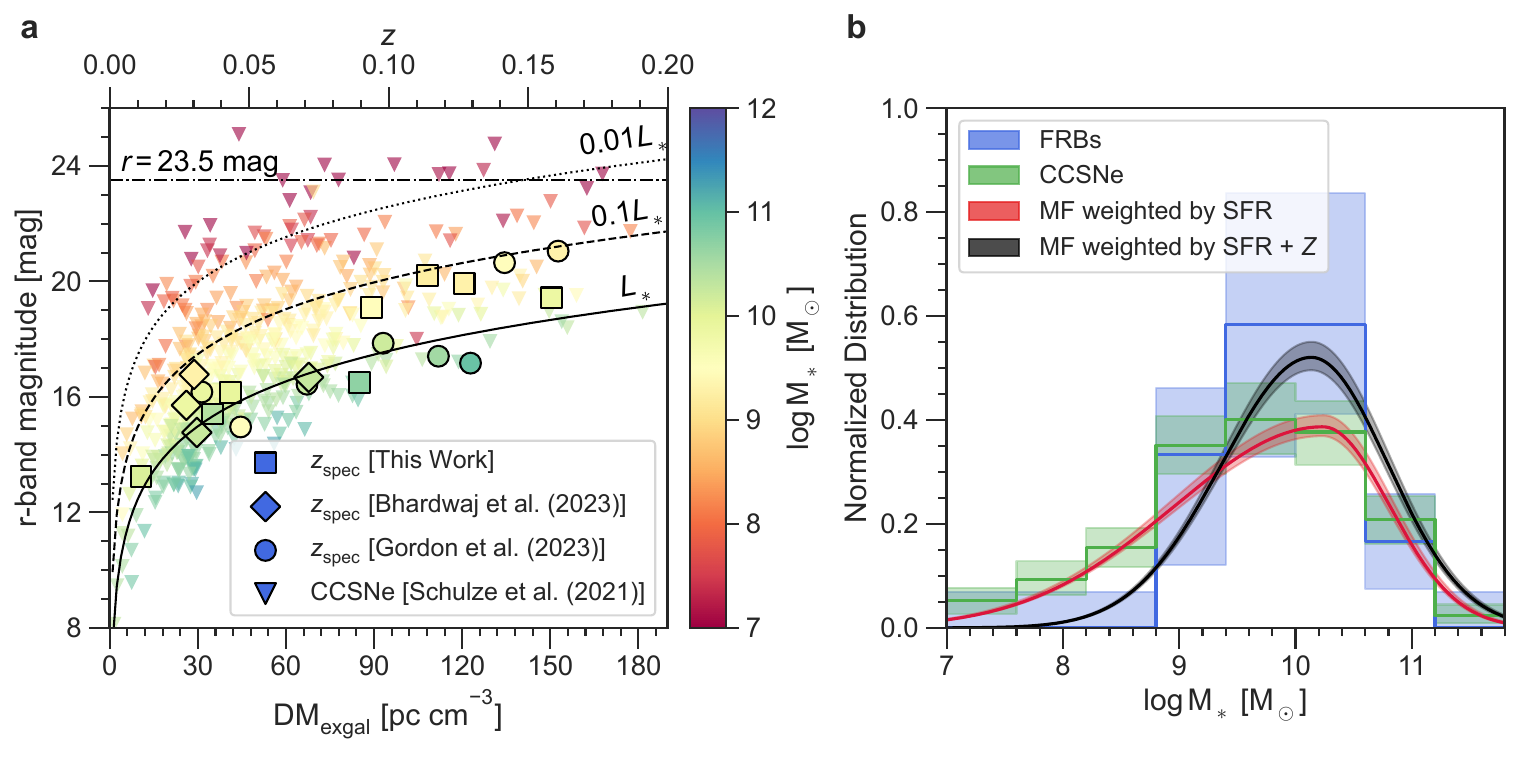}
\caption{{\bodyfigurelabel{fig:frbs_are_biased_tracer_of_sf_in_universe}} \textbf{Investigation of whether FRBs trace star-formation in the universe using the $z \leq 0.2$ sample.} 
The panel \textbf{a} shows the distribution of the $r$-band magnitude and redshift of FRB hosts published in this work (squares), alongside Gordon et al.~\citep{2023arXiv230205465G} (circles) and Bhardwaj et al.~\citep{2023arXiv231010018B} (diamonds) FRB host galaxies samples with $r$-band magnitude $\lesssim 23.5$~mag (dashdot line) and $z \leq 0.2$. 
On comparing with the redshift evolution of galaxies with characteristic luminosities $L_\ast$ (solid line), $0.1L_\ast$ (dashed line), and $0.01L_\ast$ (dotted line)
, we find that the FRB hosts trace $\sim 0.1-1L_\ast$ galaxies. 
A comparison with the host galaxies of Type~II CCSNe~\citep{2021ApJS..255...29S} (triangles) reveals that FRB host galaxies are relatively massive. 
This result is also evident in panel \textbf{b}, where we show the host galaxy mass distributions (solid lines) with Poisson errors (shaded regions). Since Type~II CCSNe (green) are unbiased tracers of star-formation in the universe, the SFR-weighted galaxy mass distribution (red) provides an adequate description of their host mass distribution. 
On the other hand, the host galaxies of FRBs (blue) show a clear dearth of low-mass galaxies. 
This absence can be accounted for by adding a metallicity-dependent FRB progenitor formation efficiency (black), 
which is stifled in environments with oxygen abundances, 12+log(O/H) $\leq 8.08_{-0.47}^{+0.61}$, 
corresponding to a cutoff metallicity of 
$\log(Z/Z_\odot) = -0.61_{-0.47}^{+0.61}$.
}
\end{figure*}

\newpage
\clearpage
\newpage
\newpage

\begin{figure*}[h]
\centering
\includegraphics[width=0.9\textwidth]{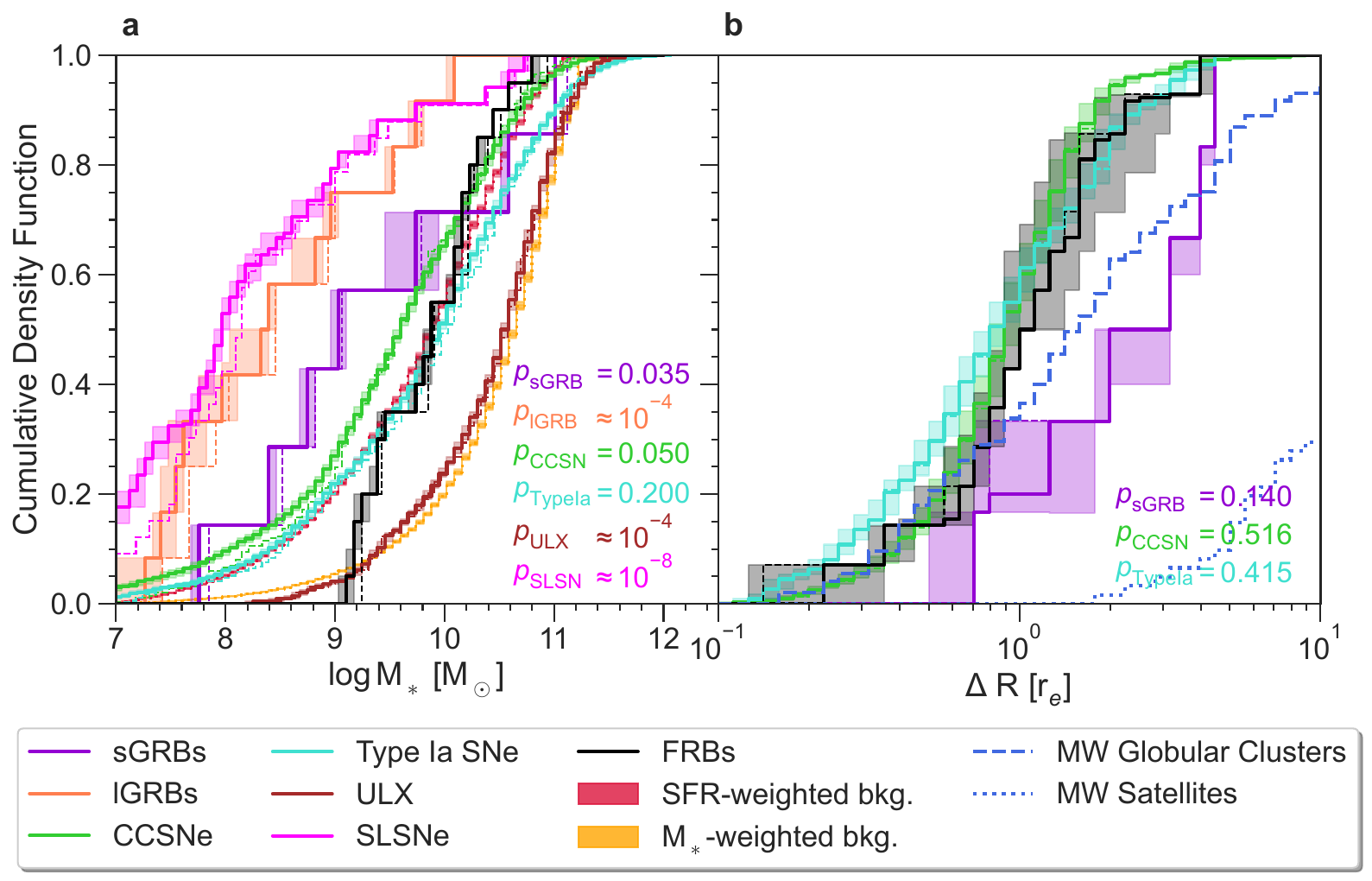}

\caption{{\bodyfigurelabel{fig:compare_with_other_transients_in_local_univ_bin}} \textbf{Comparison of FRB host galaxy properties with those of various transient classes at $z \leq 0.2$.} We compare host galaxy stellar mass (panel \textbf{a}) and host-normalized galactocentric offset (panel \textbf{b}) distributions of FRBs with Type Ia supernovae
, ultra-luminous X-ray sources (ULX)
, superluminous supernovae (SLSNe)
, core-collapse supernovae (CCSNe)
, short-duration gamma ray bursts (sGRBs)
and long-duration gamma ray bursts (lGRBs) (see Methods for a description of the comparison samples). For comparisons, we only use our secure FRB host associations, together with the literature sample of FRB host galaxies and offset measurements (see Methods). We correct stellar masses for redshift evolution~\citep{2021arXiv211207639S} (see Methods). The measured values (dashed lines), median (thick lines) and $1\sigma$ errors (shaded regions) computed using $1,000$ Monte Carlo samples of measurements reported in literature are plotted. For reference, we also plot the background population selected by stellar mass (orange) and SFR (red) in panel \textbf{a} (see Fig.~\ref{fig:frbs_trace_sfr}) and offsets of the satellites and globular clusters of Milky Way (MW) in panel \textbf{b}.}
\end{figure*}

\newpage
\clearpage
\newpage
\newpage

\vspace{0.5cm}  
\noindent{\bfseries \LARGE References}\setlength{\parskip}{12pt}
\setlength{\parskip}{17pt}%
\bibliographystyle{naturemag}
\bibliography{manuscript}

\begin{thebibliography}{10}
\urlstyle{rm}
\expandafter\ifx\csname url\endcsname\relax
  \def\url#1{\texttt{#1}}\fi
\expandafter\ifx\csname urlprefix\endcsname\relax\def\urlprefix{URL }\fi
\expandafter\ifx\csname doiprefix\endcsname\relax\def\doiprefix{DOI: }\fi
\providecommand{\bibinfo}[2]{#2}
\providecommand{\eprint}[2][]{\url{#2}}

\bibitem{2023RvMP...95c5005Z}
\bibinfo{author}{{Zhang}, B.}
\newblock \bibinfo{journal}{\bibinfo{title}{{The physics of fast radio bursts}}}.
\newblock {\emph{\JournalTitle{Reviews of Modern Physics}}} \textbf{\bibinfo{volume}{95}}, \bibinfo{pages}{035005}, \doiprefix\url{https://dx.doi.org/10.1103/RevModPhys.95.035005} (\bibinfo{year}{2023}).
\newblock \eprint{2212.03972}.

\bibitem{2020Natur.587...54C}
\bibinfo{author}{{CHIME/FRB Collaboration}} \emph{et~al.}
\newblock \bibinfo{journal}{\bibinfo{title}{{A bright millisecond-duration radio burst from a Galactic magnetar}}}.
\newblock {\emph{\JournalTitle{\nat}}} \textbf{\bibinfo{volume}{587}}, \bibinfo{pages}{54--58}, \doiprefix\url{https://dx.doi.org/10.1038/s41586-020-2863-y} (\bibinfo{year}{2020}).
\newblock \eprint{2005.10324}.

\bibitem{2020Natur.587...59B}
\bibinfo{author}{{Bochenek}, C.~D.} \emph{et~al.}
\newblock \bibinfo{journal}{\bibinfo{title}{{A fast radio burst associated with a Galactic magnetar}}}.
\newblock {\emph{\JournalTitle{\nat}}} \textbf{\bibinfo{volume}{587}}, \bibinfo{pages}{59--62}, \doiprefix\url{https://dx.doi.org/10.1038/s41586-020-2872-x} (\bibinfo{year}{2020}).
\newblock \eprint{2005.10828}.

\bibitem{2023arXiv230205465G}
\bibinfo{author}{{Gordon}, A.~C.} \emph{et~al.}
\newblock \bibinfo{journal}{\bibinfo{title}{{The Demographics, Stellar Populations, and Star Formation Histories of Fast Radio Burst Host Galaxies: Implications for the Progenitors}}}.
\newblock {\emph{\JournalTitle{\apj}}} \textbf{\bibinfo{volume}{954}}, \bibinfo{pages}{80}, \doiprefix\url{https://dx.doi.org/10.3847/1538-4357/ace5aa} (\bibinfo{year}{2023}).
\newblock \eprint{2302.05465}.

\bibitem{2023arXiv231010018B}
\bibinfo{author}{{Bhardwaj}, M.} \emph{et~al.}
\newblock \bibinfo{journal}{\bibinfo{title}{{Host Galaxies for Four Nearby CHIME/FRB Sources and the Local Universe FRB Host Galaxy Population}}}.
\newblock {\emph{\JournalTitle{\apjl}}} \textbf{\bibinfo{volume}{971}}, \bibinfo{pages}{L51}, \doiprefix\url{https://dx.doi.org/10.3847/2041-8213/ad64d1} (\bibinfo{year}{2024}).
\newblock \eprint{2310.10018}.

\bibitem{2017ARA&A..55..261K}
\bibinfo{author}{{Kaspi}, V.~M.} \& \bibinfo{author}{{Beloborodov}, A.~M.}
\newblock \bibinfo{journal}{\bibinfo{title}{{Magnetars}}}.
\newblock {\emph{\JournalTitle{\araa}}} \textbf{\bibinfo{volume}{55}}, \bibinfo{pages}{261--301}, \doiprefix\url{https://dx.doi.org/10.1146/annurev-astro-081915-023329} (\bibinfo{year}{2017}).
\newblock \eprint{1703.00068}.

\bibitem{2023IAUS..363...61P}
\bibinfo{author}{{Popov}, S.~B.}
\newblock \bibinfo{title}{{High magnetic field neutron stars and magnetars in binary systems}}.
\newblock In \bibinfo{editor}{{Troja}, E.} \& \bibinfo{editor}{{Baring}, M.~G.} (eds.) \emph{\bibinfo{booktitle}{Neutron Star Astrophysics at the Crossroads: Magnetars and the Multimessenger Revolution}}, vol. \bibinfo{volume}{363}, \bibinfo{pages}{61--71}, \doiprefix\url{https://dx.doi.org/10.1017/S1743921322000308} (\bibinfo{year}{2023}).
\newblock \eprint{2201.07507}.

\bibitem{2005MNRAS.362...41G}
\bibinfo{author}{{Gallazzi}, A.}, \bibinfo{author}{{Charlot}, S.}, \bibinfo{author}{{Brinchmann}, J.}, \bibinfo{author}{{White}, S. D.~M.} \& \bibinfo{author}{{Tremonti}, C.~A.}
\newblock \bibinfo{journal}{\bibinfo{title}{{The ages and metallicities of galaxies in the local universe}}}.
\newblock {\emph{\JournalTitle{\mnras}}} \textbf{\bibinfo{volume}{362}}, \bibinfo{pages}{41--58}, \doiprefix\url{https://dx.doi.org/10.1111/j.1365-2966.2005.09321.x} (\bibinfo{year}{2005}).
\newblock \eprint{astro-ph/0506539}.

\bibitem{2017A&A...601A..29Z}
\bibinfo{author}{{Zapartas}, E.} \emph{et~al.}
\newblock \bibinfo{journal}{\bibinfo{title}{{Delay-time distribution of core-collapse supernovae with late events resulting from binary interaction}}}.
\newblock {\emph{\JournalTitle{\aap}}} \textbf{\bibinfo{volume}{601}}, \bibinfo{pages}{A29}, \doiprefix\url{https://dx.doi.org/10.1051/0004-6361/201629685} (\bibinfo{year}{2017}).
\newblock \eprint{1701.07032}.

\bibitem{2019A&A...631A...5Z}
\bibinfo{author}{{Zapartas}, E.} \emph{et~al.}
\newblock \bibinfo{journal}{\bibinfo{title}{{The diverse lives of progenitors of hydrogen-rich core-collapse supernovae: the role of binary interaction}}}.
\newblock {\emph{\JournalTitle{\aap}}} \textbf{\bibinfo{volume}{631}}, \bibinfo{pages}{A5}, \doiprefix\url{https://dx.doi.org/10.1051/0004-6361/201935854} (\bibinfo{year}{2019}).
\newblock \eprint{1907.06687}.

\bibitem{2024Sci...384..214F}
\bibinfo{author}{{Frost}, A.~J.} \emph{et~al.}
\newblock \bibinfo{journal}{\bibinfo{title}{{A magnetic massive star has experienced a stellar merger}}}.
\newblock {\emph{\JournalTitle{Science}}} \textbf{\bibinfo{volume}{384}}, \bibinfo{pages}{214--217}, \doiprefix\url{https://dx.doi.org/10.1126/science.adg7700} (\bibinfo{year}{2024}).
\newblock \eprint{2404.10167}.

\bibitem{2019Natur.574..211S}
\bibinfo{author}{{Schneider}, F. R.~N.} \emph{et~al.}
\newblock \bibinfo{journal}{\bibinfo{title}{{Stellar mergers as the origin of magnetic massive stars}}}.
\newblock {\emph{\JournalTitle{\nat}}} \textbf{\bibinfo{volume}{574}}, \bibinfo{pages}{211--214}, \doiprefix\url{https://dx.doi.org/10.1038/s41586-019-1621-5} (\bibinfo{year}{2019}).
\newblock \eprint{1910.14058}.

\bibitem{2021ITMTT..69.2345W}
\bibinfo{author}{{Weinreb}, S.} \& \bibinfo{author}{{Shi}, J.}
\newblock \bibinfo{journal}{\bibinfo{title}{{Low Noise Amplifier With 7-K Noise at 1.4 GHz and 25 {\textdegree}C}}}.
\newblock {\emph{\JournalTitle{IEEE Transactions on Microwave Theory Techniques}}} \textbf{\bibinfo{volume}{69}}, \bibinfo{pages}{2345--2351}, \doiprefix\url{https://dx.doi.org/10.1109/TMTT.2021.3061459} (\bibinfo{year}{2021}).

\bibitem{2016arXiv161205560C}
\bibinfo{author}{{Chambers}, K.~C.} \emph{et~al.}
\newblock \bibinfo{journal}{\bibinfo{title}{{The Pan-STARRS1 Surveys}}}.
\newblock {\emph{\JournalTitle{arXiv e-prints}}} \bibinfo{pages}{arXiv:1612.05560} (\bibinfo{year}{2016}).
\newblock \eprint{1612.05560}.

\bibitem{2017PASP..129f4101Z}
\bibinfo{author}{{Zou}, H.} \emph{et~al.}
\newblock \bibinfo{journal}{\bibinfo{title}{{Project Overview of the Beijing-Arizona Sky Survey}}}.
\newblock {\emph{\JournalTitle{\pasp}}} \textbf{\bibinfo{volume}{129}}, \bibinfo{pages}{064101}, \doiprefix\url{https://dx.doi.org/10.1088/1538-3873/aa65ba} (\bibinfo{year}{2017}).
\newblock \eprint{1702.03653}.

\bibitem{2017JATIS...3c6002N}
\bibinfo{author}{{Nikzad}, S.} \emph{et~al.}
\newblock \bibinfo{journal}{\bibinfo{title}{{High-efficiency UV/optical/NIR detectors for large aperture telescopes and UV explorer missions: development of and field observations with delta-doped arrays}}}.
\newblock {\emph{\JournalTitle{Journal of Astronomical Telescopes, Instruments, and Systems}}} \textbf{\bibinfo{volume}{3}}, \bibinfo{pages}{036002}, \doiprefix\url{https://dx.doi.org/10.1117/1.JATIS.3.3.036002} (\bibinfo{year}{2017}).
\newblock \eprint{1612.04734}.

\bibitem{2003SPIE.4841..451W}
\bibinfo{author}{{Wilson}, J.~C.} \emph{et~al.}
\newblock \bibinfo{title}{{A Wide-Field Infrared Camera for the Palomar 200-inch Telescope}}.
\newblock In \bibinfo{editor}{{Iye}, M.} \& \bibinfo{editor}{{Moorwood}, A. F.~M.} (eds.) \emph{\bibinfo{booktitle}{Instrument Design and Performance for Optical/Infrared Ground-based Telescopes}}, vol. \bibinfo{volume}{4841} of \emph{\bibinfo{series}{Society of Photo-Optical Instrumentation Engineers (SPIE) Conference Series}}, \bibinfo{pages}{451--458}, \doiprefix\url{https://dx.doi.org/10.1117/12.460336} (\bibinfo{year}{2003}).

\bibitem{2021ApJ...911...95A}
\bibinfo{author}{{Aggarwal}, K.} \emph{et~al.}
\newblock \bibinfo{journal}{\bibinfo{title}{{Probabilistic Association of Transients to their Hosts (PATH)}}}.
\newblock {\emph{\JournalTitle{\apj}}} \textbf{\bibinfo{volume}{911}}, \bibinfo{pages}{95}, \doiprefix\url{https://dx.doi.org/10.3847/1538-4357/abe8d2} (\bibinfo{year}{2021}).
\newblock \eprint{2102.10627}.

\bibitem{1995PASP..107..375O}
\bibinfo{author}{{Oke}, J.~B.} \emph{et~al.}
\newblock \bibinfo{journal}{\bibinfo{title}{{The Keck Low-Resolution Imaging Spectrometer}}}.
\newblock {\emph{\JournalTitle{\pasp}}} \textbf{\bibinfo{volume}{107}}, \bibinfo{pages}{375}, \doiprefix\url{https://dx.doi.org/10.1086/133562} (\bibinfo{year}{1995}).

\bibitem{2003SPIE.4841.1657F}
\bibinfo{author}{{Faber}, S.~M.} \emph{et~al.}
\newblock \bibinfo{title}{{The DEIMOS spectrograph for the Keck II Telescope: integration and testing}}.
\newblock In \bibinfo{editor}{{Iye}, M.} \& \bibinfo{editor}{{Moorwood}, A. F.~M.} (eds.) \emph{\bibinfo{booktitle}{Instrument Design and Performance for Optical/Infrared Ground-based Telescopes}}, vol. \bibinfo{volume}{4841} of \emph{\bibinfo{series}{Society of Photo-Optical Instrumentation Engineers (SPIE) Conference Series}}, \bibinfo{pages}{1657--1669}, \doiprefix\url{https://dx.doi.org/10.1117/12.460346} (\bibinfo{year}{2003}).

\bibitem{1982PASP...94..586O}
\bibinfo{author}{{Oke}, J.~B.} \& \bibinfo{author}{{Gunn}, J.~E.}
\newblock \bibinfo{journal}{\bibinfo{title}{{An Efficient Low Resolution and Moderate Resolution Spectrograph for the Hale Telescope}}}.
\newblock {\emph{\JournalTitle{\pasp}}} \textbf{\bibinfo{volume}{94}}, \bibinfo{pages}{586}, \doiprefix\url{https://dx.doi.org/10.1086/131027} (\bibinfo{year}{1982}).

\bibitem{2022arXiv220814974C}
\bibinfo{author}{{Cappellari}, M.}
\newblock \bibinfo{journal}{\bibinfo{title}{{Full spectrum fitting with photometry in PPXF: stellar population versus dynamical masses, non-parametric star formation history and metallicity for 3200 LEGA-C galaxies at redshift z {\ensuremath{\approx}} 0.8}}}.
\newblock {\emph{\JournalTitle{\mnras}}} \textbf{\bibinfo{volume}{526}}, \bibinfo{pages}{3273--3300}, \doiprefix\url{https://dx.doi.org/10.1093/mnras/stad2597} (\bibinfo{year}{2023}).
\newblock \eprint{2208.14974}.

\bibitem{2021ApJS..254...22J}
\bibinfo{author}{{Johnson}, B.~D.}, \bibinfo{author}{{Leja}, J.}, \bibinfo{author}{{Conroy}, C.} \& \bibinfo{author}{{Speagle}, J.~S.}
\newblock \bibinfo{journal}{\bibinfo{title}{{Stellar Population Inference with Prospector}}}.
\newblock {\emph{\JournalTitle{\apjs}}} \textbf{\bibinfo{volume}{254}}, \bibinfo{pages}{22}, \doiprefix\url{https://dx.doi.org/10.3847/1538-4365/abef67} (\bibinfo{year}{2021}).
\newblock \eprint{2012.01426}.

\bibitem{2019AJ....157..168D}
\bibinfo{author}{{Dey}, A.} \emph{et~al.}
\newblock \bibinfo{journal}{\bibinfo{title}{{Overview of the DESI Legacy Imaging Surveys}}}.
\newblock {\emph{\JournalTitle{\aj}}} \textbf{\bibinfo{volume}{157}}, \bibinfo{pages}{168}, \doiprefix\url{https://dx.doi.org/10.3847/1538-3881/ab089d10.48550/arXiv.1804.08657} (\bibinfo{year}{2019}).
\newblock \eprint{1804.08657}.

\bibitem{2015ApJS..219...12A}
\bibinfo{author}{{Alam}, S.} \emph{et~al.}
\newblock \bibinfo{journal}{\bibinfo{title}{{The Eleventh and Twelfth Data Releases of the Sloan Digital Sky Survey: Final Data from SDSS-III}}}.
\newblock {\emph{\JournalTitle{\apjs}}} \textbf{\bibinfo{volume}{219}}, \bibinfo{pages}{12}, \doiprefix\url{https://dx.doi.org/10.1088/0067-0049/219/1/12} (\bibinfo{year}{2015}).
\newblock \eprint{1501.00963}.

\bibitem{2006AJ....131.1163S}
\bibinfo{author}{{Skrutskie}, M.~F.} \emph{et~al.}
\newblock \bibinfo{journal}{\bibinfo{title}{{The Two Micron All Sky Survey (2MASS)}}}.
\newblock {\emph{\JournalTitle{\aj}}} \textbf{\bibinfo{volume}{131}}, \bibinfo{pages}{1163--1183}, \doiprefix\url{https://dx.doi.org/10.1086/498708} (\bibinfo{year}{2006}).

\bibitem{2014yCat.2328....0C}
\bibinfo{author}{{Cutri}, R.~M.} \emph{et~al.}
\newblock \bibinfo{journal}{\bibinfo{title}{{VizieR Online Data Catalog: AllWISE Data Release (Cutri+ 2013)}}}.
\newblock {\emph{\JournalTitle{VizieR Online Data Catalog}}} \bibinfo{pages}{II/328} (\bibinfo{year}{2021}).

\bibitem{2005ApJ...619L...1M}
\bibinfo{author}{{Martin}, D.~C.} \emph{et~al.}
\newblock \bibinfo{journal}{\bibinfo{title}{{The Galaxy Evolution Explorer: A Space Ultraviolet Survey Mission}}}.
\newblock {\emph{\JournalTitle{\apjl}}} \textbf{\bibinfo{volume}{619}}, \bibinfo{pages}{L1--L6}, \doiprefix\url{https://dx.doi.org/10.1086/426387} (\bibinfo{year}{2005}).
\newblock \eprint{astro-ph/0411302}.

\bibitem{1981PASP...93....5B}
\bibinfo{author}{{Baldwin}, J.~A.}, \bibinfo{author}{{Phillips}, M.~M.} \& \bibinfo{author}{{Terlevich}, R.}
\newblock \bibinfo{journal}{\bibinfo{title}{{Classification parameters for the emission-line spectra of extragalactic objects.}}}
\newblock {\emph{\JournalTitle{\pasp}}} \textbf{\bibinfo{volume}{93}}, \bibinfo{pages}{5--19}, \doiprefix\url{https://dx.doi.org/10.1086/130766} (\bibinfo{year}{1981}).

\bibitem{2006MNRAS.372..961K}
\bibinfo{author}{{Kewley}, L.~J.}, \bibinfo{author}{{Groves}, B.}, \bibinfo{author}{{Kauffmann}, G.} \& \bibinfo{author}{{Heckman}, T.}
\newblock \bibinfo{journal}{\bibinfo{title}{{The host galaxies and classification of active galactic nuclei}}}.
\newblock {\emph{\JournalTitle{\mnras}}} \textbf{\bibinfo{volume}{372}}, \bibinfo{pages}{961--976}, \doiprefix\url{https://dx.doi.org/10.1111/j.1365-2966.2006.10859.x} (\bibinfo{year}{2006}).
\newblock \eprint{astro-ph/0605681}.

\bibitem{2010AJ....140.1868W}
\bibinfo{author}{{Wright}, E.~L.} \emph{et~al.}
\newblock \bibinfo{journal}{\bibinfo{title}{{The Wide-field Infrared Survey Explorer (WISE): Mission Description and Initial On-orbit Performance}}}.
\newblock {\emph{\JournalTitle{\aj}}} \textbf{\bibinfo{volume}{140}}, \bibinfo{pages}{1868--1881}, \doiprefix\url{https://dx.doi.org/10.1088/0004-6256/140/6/1868} (\bibinfo{year}{2010}).
\newblock \eprint{1008.0031}.

\bibitem{2020ApJ...893..111L}
\bibinfo{author}{{Leja}, J.} \emph{et~al.}
\newblock \bibinfo{journal}{\bibinfo{title}{{A New Census of the 0.2 < z < 3.0 Universe. I. The Stellar Mass Function}}}.
\newblock {\emph{\JournalTitle{\apj}}} \textbf{\bibinfo{volume}{893}}, \bibinfo{pages}{111}, \doiprefix\url{https://dx.doi.org/10.3847/1538-4357/ab7e27} (\bibinfo{year}{2020}).
\newblock \eprint{1910.04168}.

\bibitem{2022ApJ...936..165L}
\bibinfo{author}{{Leja}, J.} \emph{et~al.}
\newblock \bibinfo{journal}{\bibinfo{title}{{A New Census of the 0.2 < z < 3.0 Universe. II. The Star-forming Sequence}}}.
\newblock {\emph{\JournalTitle{\apj}}} \textbf{\bibinfo{volume}{936}}, \bibinfo{pages}{165}, \doiprefix\url{https://dx.doi.org/10.3847/1538-4357/ac887d} (\bibinfo{year}{2022}).
\newblock \eprint{2110.04314}.

\bibitem{law2023deep}
\bibinfo{author}{{Law}, C.~J.} \emph{et~al.}
\newblock \bibinfo{journal}{\bibinfo{title}{{Deep Synoptic Array Science: First FRB and Host Galaxy Catalog}}}.
\newblock {\emph{\JournalTitle{\apj}}} \textbf{\bibinfo{volume}{967}}, \bibinfo{pages}{29}, \doiprefix\url{https://dx.doi.org/10.3847/1538-4357/ad3736} (\bibinfo{year}{2024}).
\newblock \eprint{2307.03344}.

\bibitem{2022ApJ...934...71O}
\bibinfo{author}{{Ocker}, S.~K.}, \bibinfo{author}{{Cordes}, J.~M.}, \bibinfo{author}{{Chatterjee}, S.} \& \bibinfo{author}{{Gorsuch}, M.~R.}
\newblock \bibinfo{journal}{\bibinfo{title}{{Radio Scattering Horizons for Galactic and Extragalactic Transients}}}.
\newblock {\emph{\JournalTitle{\apj}}} \textbf{\bibinfo{volume}{934}}, \bibinfo{pages}{71}, \doiprefix\url{https://dx.doi.org/10.3847/1538-4357/ac75ba} (\bibinfo{year}{2022}).
\newblock \eprint{2203.16716}.

\bibitem{2021ApJS..255...29S}
\bibinfo{author}{{Schulze}, S.} \emph{et~al.}
\newblock \bibinfo{journal}{\bibinfo{title}{{The Palomar Transient Factory Core-collapse Supernova Host-galaxy Sample. I. Host-galaxy Distribution Functions and Environment Dependence of Core-collapse Supernovae}}}.
\newblock {\emph{\JournalTitle{\apjs}}} \textbf{\bibinfo{volume}{255}}, \bibinfo{pages}{29}, \doiprefix\url{https://dx.doi.org/10.3847/1538-4365/abff5e} (\bibinfo{year}{2021}).
\newblock \eprint{2008.05988}.

\bibitem{2022MNRAS.510L..18J}
\bibinfo{author}{{James}, C.~W.} \emph{et~al.}
\newblock \bibinfo{journal}{\bibinfo{title}{{The fast radio burst population evolves, consistent with the star formation rate}}}.
\newblock {\emph{\JournalTitle{\mnras}}} \textbf{\bibinfo{volume}{510}}, \bibinfo{pages}{L18--L23}, \doiprefix\url{https://dx.doi.org/10.1093/mnrasl/slab117} (\bibinfo{year}{2022}).
\newblock \eprint{2101.07998}.

\bibitem{2021MNRAS.503.3931T}
\bibinfo{author}{{Taggart}, K.} \& \bibinfo{author}{{Perley}, D.~A.}
\newblock \bibinfo{journal}{\bibinfo{title}{{Core-collapse, superluminous, and gamma-ray burst supernova host galaxy populations at low redshift: the importance of dwarf and starbursting galaxies}}}.
\newblock {\emph{\JournalTitle{\mnras}}} \textbf{\bibinfo{volume}{503}}, \bibinfo{pages}{3931--3952}, \doiprefix\url{https://dx.doi.org/10.1093/mnras/stab174} (\bibinfo{year}{2021}).
\newblock \eprint{1911.09112}.

\bibitem{2003ApJ...591..288H}
\bibinfo{author}{{Heger}, A.}, \bibinfo{author}{{Fryer}, C.~L.}, \bibinfo{author}{{Woosley}, S.~E.}, \bibinfo{author}{{Langer}, N.} \& \bibinfo{author}{{Hartmann}, D.~H.}
\newblock \bibinfo{journal}{\bibinfo{title}{{How Massive Single Stars End Their Life}}}.
\newblock {\emph{\JournalTitle{\apj}}} \textbf{\bibinfo{volume}{591}}, \bibinfo{pages}{288--300}, \doiprefix\url{https://dx.doi.org/10.1086/375341} (\bibinfo{year}{2003}).
\newblock \eprint{astro-ph/0212469}.

\bibitem{2020A&A...638A..55K}
\bibinfo{author}{{Klencki}, J.}, \bibinfo{author}{{Nelemans}, G.}, \bibinfo{author}{{Istrate}, A.~G.} \& \bibinfo{author}{{Pols}, O.}
\newblock \bibinfo{journal}{\bibinfo{title}{{Massive donors in interacting binaries: effect of metallicity}}}.
\newblock {\emph{\JournalTitle{\aap}}} \textbf{\bibinfo{volume}{638}}, \bibinfo{pages}{A55}, \doiprefix\url{https://dx.doi.org/10.1051/0004-6361/202037694} (\bibinfo{year}{2020}).
\newblock \eprint{2004.00628}.

\bibitem{sherman2024searching}
\bibinfo{author}{{Sherman}, M.~B.} \emph{et~al.}
\newblock \bibinfo{journal}{\bibinfo{title}{{Searching for magnetar binaries disrupted by core-collapse supernovae}}}.
\newblock {\emph{\JournalTitle{\mnras}}} \textbf{\bibinfo{volume}{531}}, \bibinfo{pages}{2379--2414}, \doiprefix\url{https://dx.doi.org/10.1093/mnras/stae1289} (\bibinfo{year}{2024}).
\newblock \eprint{2404.05135}.

\bibitem{2021arXiv211207639S}
\bibinfo{author}{{Seebeck}, J.} \emph{et~al.}
\newblock \bibinfo{journal}{\bibinfo{title}{{The Effects of Selection Biases on the Analysis of Localised Fast Radio Bursts}}}.
\newblock {\emph{\JournalTitle{arXiv e-prints}}} \bibinfo{pages}{arXiv:2112.07639}, \doiprefix\url{https://dx.doi.org/10.48550/arXiv.2112.07639} (\bibinfo{year}{2021}).
\newblock \eprint{2112.07639}.

\bibitem{2021ApJ...910L..18B}
\bibinfo{author}{{Bhardwaj}, M.} \emph{et~al.}
\newblock \bibinfo{journal}{\bibinfo{title}{{A Nearby Repeating Fast Radio Burst in the Direction of M81}}}.
\newblock {\emph{\JournalTitle{\apjl}}} \textbf{\bibinfo{volume}{910}}, \bibinfo{pages}{L18}, \doiprefix\url{https://dx.doi.org/10.3847/2041-8213/abeaa6} (\bibinfo{year}{2021}).
\newblock \eprint{2103.01295}.

\bibitem{2021ApJ...917L..11K}
\bibinfo{author}{{Kremer}, K.}, \bibinfo{author}{{Piro}, A.~L.} \& \bibinfo{author}{{Li}, D.}
\newblock \bibinfo{journal}{\bibinfo{title}{{Dynamical Formation Channels for Fast Radio Bursts in Globular Clusters}}}.
\newblock {\emph{\JournalTitle{\apjl}}} \textbf{\bibinfo{volume}{917}}, \bibinfo{pages}{L11}, \doiprefix\url{https://dx.doi.org/10.3847/2041-8213/ac13a010.48550/arXiv.2107.03394} (\bibinfo{year}{2021}).
\newblock \eprint{2107.03394}.

\bibitem{2023ApJ...950..175S}
\bibinfo{author}{{Sharma}, K.} \emph{et~al.}
\newblock \bibinfo{journal}{\bibinfo{title}{{Deep Synoptic Array Science: A Massive Elliptical Host Among Two Galaxy-cluster Fast Radio Bursts}}}.
\newblock {\emph{\JournalTitle{\apj}}} \textbf{\bibinfo{volume}{950}}, \bibinfo{pages}{175}, \doiprefix\url{https://dx.doi.org/10.3847/1538-4357/accf1d} (\bibinfo{year}{2023}).
\newblock \eprint{2302.14782}.

\bibitem{2013A&A...558A..39T}
\bibinfo{author}{{Tauris}, T.~M.}, \bibinfo{author}{{Sanyal}, D.}, \bibinfo{author}{{Yoon}, S.~C.} \& \bibinfo{author}{{Langer}, N.}
\newblock \bibinfo{journal}{\bibinfo{title}{{Evolution towards and beyond accretion-induced collapse of massive white dwarfs and formation of millisecond pulsars}}}.
\newblock {\emph{\JournalTitle{\aap}}} \textbf{\bibinfo{volume}{558}}, \bibinfo{pages}{A39}, \doiprefix\url{https://dx.doi.org/10.1051/0004-6361/201321662} (\bibinfo{year}{2013}).
\newblock \eprint{1308.4887}.

\bibitem{2016MNRAS.456.2140M}
\bibinfo{author}{{Ma}, X.} \emph{et~al.}
\newblock \bibinfo{journal}{\bibinfo{title}{{The origin and evolution of the galaxy mass-metallicity relation}}}.
\newblock {\emph{\JournalTitle{\mnras}}} \textbf{\bibinfo{volume}{456}}, \bibinfo{pages}{2140--2156}, \doiprefix\url{https://dx.doi.org/10.1093/mnras/stv2659} (\bibinfo{year}{2016}).
\newblock \eprint{1504.02097}.

\bibitem{2018ApJ...868L...4M}
\bibinfo{author}{{Margalit}, B.} \& \bibinfo{author}{{Metzger}, B.~D.}
\newblock \bibinfo{journal}{\bibinfo{title}{{A Concordance Picture of FRB 121102 as a Flaring Magnetar Embedded in a Magnetized Ion-Electron Wind Nebula}}}.
\newblock {\emph{\JournalTitle{\apjl}}} \textbf{\bibinfo{volume}{868}}, \bibinfo{pages}{L4}, \doiprefix\url{https://dx.doi.org/10.3847/2041-8213/aaedad} (\bibinfo{year}{2018}).
\newblock \eprint{1808.09969}.

\bibitem{2022ApJ...940...56F}
\bibinfo{author}{{Fong}, W.-f.} \emph{et~al.}
\newblock \bibinfo{journal}{\bibinfo{title}{{Short GRB Host Galaxies. I. Photometric and Spectroscopic Catalogs, Host Associations, and Galactocentric Offsets}}}.
\newblock {\emph{\JournalTitle{\apj}}} \textbf{\bibinfo{volume}{940}}, \bibinfo{pages}{56}, \doiprefix\url{https://dx.doi.org/10.3847/1538-4357/ac91d010.48550/arXiv.2206.01763} (\bibinfo{year}{2022}).
\newblock \eprint{2206.01763}.

\end{thebibliography}


\begin{thebibliography}{10}
\urlstyle{rm}
\expandafter\ifx\csname url\endcsname\relax
  \def\url#1{\texttt{#1}}\fi
\expandafter\ifx\csname urlprefix\endcsname\relax\def\urlprefix{URL }\fi
\expandafter\ifx\csname doiprefix\endcsname\relax\def\doiprefix{DOI: }\fi
\providecommand{\bibinfo}[2]{#2}
\providecommand{\eprint}[2][]{\url{#2}}

\bibitem{2023ApJ...949L...3R}
\bibinfo{author}{{Ravi}, V.} \emph{et~al.}
\newblock \bibinfo{journal}{\bibinfo{title}{{Deep Synoptic Array Science: Discovery of the Host Galaxy of FRB 20220912A}}}.
\newblock {\emph{\JournalTitle{\apjl}}} \textbf{\bibinfo{volume}{949}}, \bibinfo{pages}{L3}, \doiprefix\url{https://dx.doi.org/10.3847/2041-8213/acc4b6} (\bibinfo{year}{2023}).
\newblock \eprint{2211.09049}.

\bibitem{2012MNRAS.422..379B}
\bibinfo{author}{{Barsdell}, B.~R.}, \bibinfo{author}{{Bailes}, M.}, \bibinfo{author}{{Barnes}, D.~G.} \& \bibinfo{author}{{Fluke}, C.~J.}
\newblock \bibinfo{journal}{\bibinfo{title}{{Accelerating incoherent dedispersion}}}.
\newblock {\emph{\JournalTitle{\mnras}}} \textbf{\bibinfo{volume}{422}}, \bibinfo{pages}{379--392}, \doiprefix\url{https://dx.doi.org/10.1111/j.1365-2966.2012.20622.x} (\bibinfo{year}{2012}).
\newblock \eprint{1201.5380}.

\bibitem{2022PASP..134k4501C}
\bibinfo{author}{{CASA Team}} \emph{et~al.}
\newblock \bibinfo{journal}{\bibinfo{title}{{CASA, the Common Astronomy Software Applications for Radio Astronomy}}}.
\newblock {\emph{\JournalTitle{\pasp}}} \textbf{\bibinfo{volume}{134}}, \bibinfo{pages}{114501}, \doiprefix\url{https://dx.doi.org/10.1088/1538-3873/ac9642} (\bibinfo{year}{2022}).
\newblock \eprint{2210.02276}.

\bibitem{offringa-wsclean-2014}
\bibinfo{author}{Offringa, A.~R.}, \bibinfo{author}{McKinley, B.}, \bibinfo{author}{Hurley-Walker} \emph{et~al.}
\newblock \bibinfo{journal}{\bibinfo{title}{{WSClean: an implementation of a fast, generic wide-field imager for radio astronomy}}}.
\newblock {\emph{\JournalTitle{MNRAS}}} \textbf{\bibinfo{volume}{444}}, \bibinfo{pages}{606--619}, \doiprefix\url{https://dx.doi.org/10.1093/mnras/stu1368} (\bibinfo{year}{2014}).

\bibitem{1998AJ....115.1693C}
\bibinfo{author}{{Condon}, J.~J.} \emph{et~al.}
\newblock \bibinfo{journal}{\bibinfo{title}{{The NRAO VLA Sky Survey}}}.
\newblock {\emph{\JournalTitle{\aj}}} \textbf{\bibinfo{volume}{115}}, \bibinfo{pages}{1693--1716}, \doiprefix\url{https://dx.doi.org/10.1086/300337} (\bibinfo{year}{1998}).

\bibitem{2024MNRAS.529.1814H}
\bibinfo{author}{{Hewitt}, D.~M.} \emph{et~al.}
\newblock \bibinfo{journal}{\bibinfo{title}{{Milliarcsecond localization of the hyperactive repeating FRB 20220912A}}}.
\newblock {\emph{\JournalTitle{\mnras}}} \textbf{\bibinfo{volume}{529}}, \bibinfo{pages}{1814--1826}, \doiprefix\url{https://dx.doi.org/10.1093/mnras/stae632} (\bibinfo{year}{2024}).
\newblock \eprint{2312.14490}.

\bibitem{2021ApJ...906...49Z}
\bibinfo{author}{{Zhang}, Z.~J.}, \bibinfo{author}{{Yan}, K.}, \bibinfo{author}{{Li}, C.~M.}, \bibinfo{author}{{Zhang}, G.~Q.} \& \bibinfo{author}{{Wang}, F.~Y.}
\newblock \bibinfo{journal}{\bibinfo{title}{{Intergalactic Medium Dispersion Measures of Fast Radio Bursts Estimated from IllustrisTNG Simulation and Their Cosmological Applications}}}.
\newblock {\emph{\JournalTitle{\apj}}} \textbf{\bibinfo{volume}{906}}, \bibinfo{pages}{49}, \doiprefix\url{https://dx.doi.org/10.3847/1538-4357/abceb9} (\bibinfo{year}{2021}).
\newblock \eprint{2011.14494}.

\bibitem{2018ApJS..239...18A}
\bibinfo{author}{{Abbott}, T.~M.~C.} \emph{et~al.}
\newblock \bibinfo{journal}{\bibinfo{title}{{The Dark Energy Survey: Data Release 1}}}.
\newblock {\emph{\JournalTitle{\apjs}}} \textbf{\bibinfo{volume}{239}}, \bibinfo{pages}{18}, \doiprefix\url{https://dx.doi.org/10.3847/1538-4365/aae9f0} (\bibinfo{year}{2018}).
\newblock \eprint{1801.03181}.

\bibitem{2000A&AS..143...23O}
\bibinfo{author}{{Ochsenbein}, F.}, \bibinfo{author}{{Bauer}, P.} \& \bibinfo{author}{{Marcout}, J.}
\newblock \bibinfo{journal}{\bibinfo{title}{{The VizieR database of astronomical catalogues}}}.
\newblock {\emph{\JournalTitle{\aaps}}} \textbf{\bibinfo{volume}{143}}, \bibinfo{pages}{23--32}, \doiprefix\url{https://dx.doi.org/10.1051/aas:2000169} (\bibinfo{year}{2000}).
\newblock \eprint{astro-ph/0002122}.

\bibitem{2013A&A...558A..33A}
\bibinfo{author}{{Astropy Collaboration}} \emph{et~al.}
\newblock \bibinfo{journal}{\bibinfo{title}{{Astropy: A community Python package for astronomy}}}.
\newblock {\emph{\JournalTitle{\aap}}} \textbf{\bibinfo{volume}{558}}, \bibinfo{pages}{A33}, \doiprefix\url{https://dx.doi.org/10.1051/0004-6361/201322068} (\bibinfo{year}{2013}).
\newblock \eprint{1307.6212}.

\bibitem{2018AJ....156..123A}
\bibinfo{author}{{Astropy Collaboration}} \emph{et~al.}
\newblock \bibinfo{journal}{\bibinfo{title}{{The Astropy Project: Building an Open-science Project and Status of the v2.0 Core Package}}}.
\newblock {\emph{\JournalTitle{\aj}}} \textbf{\bibinfo{volume}{156}}, \bibinfo{pages}{123}, \doiprefix\url{https://dx.doi.org/10.3847/1538-3881/aabc4f} (\bibinfo{year}{2018}).
\newblock \eprint{1801.02634}.

\bibitem{2018PASP..130l8001T}
\bibinfo{author}{{Tachibana}, Y.} \& \bibinfo{author}{{Miller}, A.~A.}
\newblock \bibinfo{journal}{\bibinfo{title}{{A Morphological Classification Model to Identify Unresolved PanSTARRS1 Sources: Application in the ZTF Real-time Pipeline}}}.
\newblock {\emph{\JournalTitle{\pasp}}} \textbf{\bibinfo{volume}{130}}, \bibinfo{pages}{128001}, \doiprefix\url{https://dx.doi.org/10.1088/1538-3873/aae3d9} (\bibinfo{year}{2018}).
\newblock \eprint{1902.01935}.

\bibitem{2005cs........2072O}
\bibinfo{author}{{OMullane}, W.} \emph{et~al.}
\newblock \bibinfo{journal}{\bibinfo{title}{{Batch is back: CasJobs, serving multi-TB data on the Web}}}.
\newblock {\emph{\JournalTitle{arXiv e-prints}}} \bibinfo{pages}{cs/0502072} (\bibinfo{year}{2005}).
\newblock \eprint{cs/0502072}.

\bibitem{2021AAS...23723503S}
\bibinfo{author}{{Schlegel}, D.} \emph{et~al.}
\newblock \bibinfo{title}{{DESI Legacy Imaging Surveys Data Release 9}}.
\newblock In \emph{\bibinfo{booktitle}{American Astronomical Society Meeting Abstracts}}, vol.~\bibinfo{volume}{53} of \emph{\bibinfo{series}{American Astronomical Society Meeting Abstracts}}, \bibinfo{pages}{235.03} (\bibinfo{year}{2021}).

\bibitem{Lang_2010}
\bibinfo{author}{Lang, D.}, \bibinfo{author}{Hogg, D.~W.}, \bibinfo{author}{Mierle, K.}, \bibinfo{author}{Blanton, M.} \& \bibinfo{author}{Roweis, S.}
\newblock \bibinfo{journal}{\bibinfo{title}{{ASTROMETRY}.{NET}: {BLIND} {ASTROMETRIC} {CALIBRATION} {OF} {ARBITRARY} {ASTRONOMICAL} {IMAGES}}}.
\newblock {\emph{\JournalTitle{The Astronomical Journal}}} \textbf{\bibinfo{volume}{139}}, \bibinfo{pages}{1782--1800}, \doiprefix\url{https://dx.doi.org/10.1088/0004-6256/139/5/1782} (\bibinfo{year}{2010}).

\bibitem{bertin11}
\bibinfo{author}{{Bertin}, E.}
\newblock \bibinfo{title}{{Automated Morphometry with SExtractor and PSFEx}}.
\newblock In \bibinfo{editor}{{Evans}, I.~N.}, \bibinfo{editor}{{Accomazzi}, A.}, \bibinfo{editor}{{Mink}, D.~J.} \& \bibinfo{editor}{{Rots}, A.~H.} (eds.) \emph{\bibinfo{booktitle}{Astronomical Data Analysis Software and Systems XX}}, vol. \bibinfo{volume}{442} of \emph{\bibinfo{series}{Astronomical Society of the Pacific Conference Series}}, \bibinfo{pages}{435} (\bibinfo{year}{2011}).

\bibitem{2018AAS...23143601F}
\bibinfo{author}{{Flewelling}, H.}
\newblock \bibinfo{title}{{Pan-STARRS Data Release 2}}.
\newblock In \emph{\bibinfo{booktitle}{American Astronomical Society Meeting Abstracts \#231}}, vol. \bibinfo{volume}{231} of \emph{\bibinfo{series}{American Astronomical Society Meeting Abstracts}}, \bibinfo{pages}{436.01} (\bibinfo{year}{2018}).

\bibitem{larry_bradley_2022_6825092}
\bibinfo{author}{Bradley, L.} \emph{et~al.}
\newblock \bibinfo{title}{astropy/photutils: v1.5.0 zenodo release}, \doiprefix\url{https://dx.doi.org/10.5281/zenodo.6825092} (\bibinfo{year}{2022}).

\bibitem{2017ApJ...849..162E}
\bibinfo{author}{{Eftekhari}, T.} \& \bibinfo{author}{{Berger}, E.}
\newblock \bibinfo{journal}{\bibinfo{title}{{Associating Fast Radio Bursts with Their Host Galaxies}}}.
\newblock {\emph{\JournalTitle{\apj}}} \textbf{\bibinfo{volume}{849}}, \bibinfo{pages}{162}, \doiprefix\url{https://dx.doi.org/10.3847/1538-4357/aa90b9} (\bibinfo{year}{2017}).
\newblock \eprint{1705.02998}.

\bibitem{2014ApJ...788...28V}
\bibinfo{author}{{van der Wel}, A.} \emph{et~al.}
\newblock \bibinfo{journal}{\bibinfo{title}{{3D-HST+CANDELS: The Evolution of the Galaxy Size-Mass Distribution since z = 3}}}.
\newblock {\emph{\JournalTitle{\apj}}} \textbf{\bibinfo{volume}{788}}, \bibinfo{pages}{28}, \doiprefix\url{https://dx.doi.org/10.1088/0004-637X/788/1/28} (\bibinfo{year}{2014}).
\newblock \eprint{1404.2844}.

\bibitem{2016ApJ...827..108D}
\bibinfo{author}{{Driver}, S.~P.} \emph{et~al.}
\newblock \bibinfo{journal}{\bibinfo{title}{{Measurements of Extragalactic Background Light from the Far UV to the Far IR from Deep Ground- and Space-based Galaxy Counts}}}.
\newblock {\emph{\JournalTitle{\apj}}} \textbf{\bibinfo{volume}{827}}, \bibinfo{pages}{108}, \doiprefix\url{https://dx.doi.org/10.3847/0004-637X/827/2/108} (\bibinfo{year}{2016}).
\newblock \eprint{1605.01523}.

\bibitem{1980lssu.book.....P}
\bibinfo{author}{{Peebles}, P.~J.~E.}
\newblock \emph{\bibinfo{title}{{The large-scale structure of the universe}}} (\bibinfo{publisher}{Princeton University Press}, \bibinfo{year}{1980}).

\bibitem{curtis_mccully_2018_1482019}
\bibinfo{author}{McCully, C.} \emph{et~al.}
\newblock \bibinfo{title}{astropy/astroscrappy: v1.0.5 zenodo release}, \doiprefix\url{https://dx.doi.org/10.5281/zenodo.1482019} (\bibinfo{year}{2018}).

\bibitem{2018A&A...616A...1G}
\bibinfo{author}{{Gaia Collaboration}} \emph{et~al.}
\newblock \bibinfo{journal}{\bibinfo{title}{{Gaia Data Release 2. Summary of the contents and survey properties}}}.
\newblock {\emph{\JournalTitle{\aap}}} \textbf{\bibinfo{volume}{616}}, \bibinfo{pages}{A1}, \doiprefix\url{https://dx.doi.org/10.1051/0004-6361/201833051} (\bibinfo{year}{2018}).
\newblock \eprint{1804.09365}.

\bibitem{2002ASPC..281..228B}
\bibinfo{author}{{Bertin}, E.} \emph{et~al.}
\newblock \bibinfo{title}{{The TERAPIX Pipeline}}.
\newblock In \bibinfo{editor}{{Bohlender}, D.~A.}, \bibinfo{editor}{{Durand}, D.} \& \bibinfo{editor}{{Handley}, T.~H.} (eds.) \emph{\bibinfo{booktitle}{Astronomical Data Analysis Software and Systems XI}}, vol. \bibinfo{volume}{281} of \emph{\bibinfo{series}{Astronomical Society of the Pacific Conference Series}}, \bibinfo{pages}{228} (\bibinfo{year}{2002}).

\bibitem{2018JOSS....3..695M}
\bibinfo{author}{{Green}, G.}
\newblock \bibinfo{journal}{\bibinfo{title}{{dustmaps: A Python interface for maps of interstellar dust}}}.
\newblock {\emph{\JournalTitle{The Journal of Open Source Software}}} \textbf{\bibinfo{volume}{3}}, \bibinfo{pages}{695}, \doiprefix\url{https://dx.doi.org/10.21105/joss.00695} (\bibinfo{year}{2018}).

\bibitem{1999PASP..111...63F}
\bibinfo{author}{{Fitzpatrick}, E.~L.}
\newblock \bibinfo{journal}{\bibinfo{title}{{Correcting for the Effects of Interstellar Extinction}}}.
\newblock {\emph{\JournalTitle{\pasp}}} \textbf{\bibinfo{volume}{111}}, \bibinfo{pages}{63--75}, \doiprefix\url{https://dx.doi.org/10.1086/316293} (\bibinfo{year}{1999}).
\newblock \eprint{astro-ph/9809387}.

\bibitem{2019PASP..131h4503P}
\bibinfo{author}{{Perley}, D.~A.}
\newblock \bibinfo{journal}{\bibinfo{title}{{Fully Automated Reduction of Longslit Spectroscopy with the Low Resolution Imaging Spectrometer at the Keck Observatory}}}.
\newblock {\emph{\JournalTitle{\pasp}}} \textbf{\bibinfo{volume}{131}}, \bibinfo{pages}{084503}, \doiprefix\url{https://dx.doi.org/10.1088/1538-3873/ab215d} (\bibinfo{year}{2019}).
\newblock \eprint{1903.07629}.

\bibitem{pypeit:zenodov_v1_6}
\bibinfo{author}{Prochaska, J.~X.} \emph{et~al.}
\newblock \bibinfo{title}{pypeit/pypeit: Version 1.6.0}, \doiprefix\url{https://dx.doi.org/10.5281/zenodo.5548381} (\bibinfo{year}{2021}).

\bibitem{pypeit:joss_pub}
\bibinfo{author}{Prochaska, J.~X.} \emph{et~al.}
\newblock \bibinfo{journal}{\bibinfo{title}{Pypeit: The python spectroscopic data reduction pipeline}}.
\newblock {\emph{\JournalTitle{Journal of Open Source Software}}} \textbf{\bibinfo{volume}{5}}, \bibinfo{pages}{2308}, \doiprefix\url{https://dx.doi.org/10.21105/joss.02308} (\bibinfo{year}{2020}).

\bibitem{dbsp_drp:joss}
\bibinfo{author}{Mandigo-Stoba, M.~S.}, \bibinfo{author}{Fremling, C.} \& \bibinfo{author}{Kasliwal, M.~M.}
\newblock \bibinfo{journal}{\bibinfo{title}{Dbsp\_drp: A python package for automated spectroscopic data reduction of dbsp data}}.
\newblock {\emph{\JournalTitle{Journal of Open Source Software}}} \textbf{\bibinfo{volume}{7}}, \bibinfo{pages}{3612}, \doiprefix\url{https://dx.doi.org/10.21105/joss.03612} (\bibinfo{year}{2022}).

\bibitem{2017MNRAS.466..798C}
\bibinfo{author}{{Cappellari}, M.}
\newblock \bibinfo{journal}{\bibinfo{title}{{Improving the full spectrum fitting method: accurate convolution with Gauss-Hermite functions}}}.
\newblock {\emph{\JournalTitle{\mnras}}} \textbf{\bibinfo{volume}{466}}, \bibinfo{pages}{798--811}, \doiprefix\url{https://dx.doi.org/10.1093/mnras/stw3020} (\bibinfo{year}{2017}).
\newblock \eprint{1607.08538}.

\bibitem{2006MNRAS.371..703S}
\bibinfo{author}{{S{\'a}nchez-Bl{\'a}zquez}, P.} \emph{et~al.}
\newblock \bibinfo{journal}{\bibinfo{title}{{Medium-resolution Isaac Newton Telescope library of empirical spectra}}}.
\newblock {\emph{\JournalTitle{\mnras}}} \textbf{\bibinfo{volume}{371}}, \bibinfo{pages}{703--718}, \doiprefix\url{https://dx.doi.org/10.1111/j.1365-2966.2006.10699.x} (\bibinfo{year}{2006}).
\newblock \eprint{astro-ph/0607009}.

\bibitem{2009ApJ...699..486C}
\bibinfo{author}{{Conroy}, C.}, \bibinfo{author}{{Gunn}, J.~E.} \& \bibinfo{author}{{White}, M.}
\newblock \bibinfo{journal}{\bibinfo{title}{{The Propagation of Uncertainties in Stellar Population Synthesis Modeling. I. The Relevance of Uncertain Aspects of Stellar Evolution and the Initial Mass Function to the Derived Physical Properties of Galaxies}}}.
\newblock {\emph{\JournalTitle{\apj}}} \textbf{\bibinfo{volume}{699}}, \bibinfo{pages}{486--506}, \doiprefix\url{https://dx.doi.org/10.1088/0004-637X/699/1/486} (\bibinfo{year}{2009}).
\newblock \eprint{0809.4261}.

\bibitem{2010ascl.soft10043C}
\bibinfo{author}{{Conroy}, C.} \& \bibinfo{author}{{Gunn}, J.~E.}
\newblock \bibinfo{title}{{FSPS: Flexible Stellar Population Synthesis}}.
\newblock \bibinfo{howpublished}{Astrophysics Source Code Library, record ascl:1010.043} (\bibinfo{year}{2010}).
\newblock \eprint{1010.043}.

\bibitem{Speagle_2020}
\bibinfo{author}{Speagle, J.~S.}
\newblock \bibinfo{journal}{\bibinfo{title}{dynesty: a dynamic nested sampling package for estimating bayesian posteriors and evidences}}.
\newblock {\emph{\JournalTitle{Monthly Notices of the Royal Astronomical Society}}} \textbf{\bibinfo{volume}{493}}, \bibinfo{pages}{3132–3158}, \doiprefix\url{https://dx.doi.org/10.1093/mnras/staa278} (\bibinfo{year}{2020}).

\bibitem{2001MNRAS.322..231K}
\bibinfo{author}{{Kroupa}, P.}
\newblock \bibinfo{journal}{\bibinfo{title}{{On the variation of the initial mass function}}}.
\newblock {\emph{\JournalTitle{\mnras}}} \textbf{\bibinfo{volume}{322}}, \bibinfo{pages}{231--246}, \doiprefix\url{https://dx.doi.org/10.1046/j.1365-8711.2001.04022.x} (\bibinfo{year}{2001}).
\newblock \eprint{astro-ph/0009005}.

\bibitem{2013ApJ...775L..16K}
\bibinfo{author}{{Kriek}, M.} \& \bibinfo{author}{{Conroy}, C.}
\newblock \bibinfo{journal}{\bibinfo{title}{{The Dust Attenuation Law in Distant Galaxies: Evidence for Variation with Spectral Type}}}.
\newblock {\emph{\JournalTitle{\apjl}}} \textbf{\bibinfo{volume}{775}}, \bibinfo{pages}{L16}, \doiprefix\url{https://dx.doi.org/10.1088/2041-8205/775/1/L16} (\bibinfo{year}{2013}).
\newblock \eprint{1308.1099}.

\bibitem{2000ApJ...533..682C}
\bibinfo{author}{{Calzetti}, D.} \emph{et~al.}
\newblock \bibinfo{journal}{\bibinfo{title}{{The Dust Content and Opacity of Actively Star-forming Galaxies}}}.
\newblock {\emph{\JournalTitle{\apj}}} \textbf{\bibinfo{volume}{533}}, \bibinfo{pages}{682--695}, \doiprefix\url{https://dx.doi.org/10.1086/308692} (\bibinfo{year}{2000}).
\newblock \eprint{astro-ph/9911459}.

\bibitem{2007ApJ...657..810D}
\bibinfo{author}{{Draine}, B.~T.} \& \bibinfo{author}{{Li}, A.}
\newblock \bibinfo{journal}{\bibinfo{title}{{Infrared Emission from Interstellar Dust. IV. The Silicate-Graphite-PAH Model in the Post-Spitzer Era}}}.
\newblock {\emph{\JournalTitle{\apj}}} \textbf{\bibinfo{volume}{657}}, \bibinfo{pages}{810--837}, \doiprefix\url{https://dx.doi.org/10.1086/511055} (\bibinfo{year}{2007}).
\newblock \eprint{astro-ph/0608003}.

\bibitem{2008ApJ...685..160N}
\bibinfo{author}{{Nenkova}, M.}, \bibinfo{author}{{Sirocky}, M.~M.}, \bibinfo{author}{{Nikutta}, R.}, \bibinfo{author}{{Ivezi{\'c}}, {\v{Z}}.} \& \bibinfo{author}{{Elitzur}, M.}
\newblock \bibinfo{journal}{\bibinfo{title}{{AGN Dusty Tori. II. Observational Implications of Clumpiness}}}.
\newblock {\emph{\JournalTitle{\apj}}} \textbf{\bibinfo{volume}{685}}, \bibinfo{pages}{160--180}, \doiprefix\url{https://dx.doi.org/10.1086/590483} (\bibinfo{year}{2008}).
\newblock \eprint{0806.0512}.

\bibitem{2018ApJ...854...62L}
\bibinfo{author}{{Leja}, J.}, \bibinfo{author}{{Johnson}, B.~D.}, \bibinfo{author}{{Conroy}, C.} \& \bibinfo{author}{{van Dokkum}, P.}
\newblock \bibinfo{journal}{\bibinfo{title}{{Hot Dust in Panchromatic SED Fitting: Identification of Active Galactic Nuclei and Improved Galaxy Properties}}}.
\newblock {\emph{\JournalTitle{\apj}}} \textbf{\bibinfo{volume}{854}}, \bibinfo{pages}{62}, \doiprefix\url{https://dx.doi.org/10.3847/1538-4357/aaa8db} (\bibinfo{year}{2018}).
\newblock \eprint{1709.04469}.

\bibitem{2016ApJS..224...24L}
\bibinfo{author}{{Laigle}, C.} \emph{et~al.}
\newblock \bibinfo{journal}{\bibinfo{title}{{The COSMOS2015 Catalog: Exploring the 1 < z < 6 Universe with Half a Million Galaxies}}}.
\newblock {\emph{\JournalTitle{\apjs}}} \textbf{\bibinfo{volume}{224}}, \bibinfo{pages}{24}, \doiprefix\url{https://dx.doi.org/10.3847/0067-0049/224/2/24} (\bibinfo{year}{2016}).
\newblock \eprint{1604.02350}.

\bibitem{2014ApJS..214...24S}
\bibinfo{author}{{Skelton}, R.~E.} \emph{et~al.}
\newblock \bibinfo{journal}{\bibinfo{title}{{3D-HST WFC3-selected Photometric Catalogs in the Five CANDELS/3D-HST Fields: Photometry, Photometric Redshifts, and Stellar Masses}}}.
\newblock {\emph{\JournalTitle{\apjs}}} \textbf{\bibinfo{volume}{214}}, \bibinfo{pages}{24}, \doiprefix\url{https://dx.doi.org/10.1088/0067-0049/214/2/24} (\bibinfo{year}{2014}).
\newblock \eprint{1403.3689}.

\bibitem{2019ApJ...877..140L}
\bibinfo{author}{{Leja}, J.} \emph{et~al.}
\newblock \bibinfo{journal}{\bibinfo{title}{{An Older, More Quiescent Universe from Panchromatic SED Fitting of the 3D-HST Survey}}}.
\newblock {\emph{\JournalTitle{\apj}}} \textbf{\bibinfo{volume}{877}}, \bibinfo{pages}{140}, \doiprefix\url{https://dx.doi.org/10.3847/1538-4357/ab1d5a} (\bibinfo{year}{2019}).
\newblock \eprint{1812.05608}.

\bibitem{2022AJ....163...69B}
\bibinfo{author}{{Bhandari}, S.} \emph{et~al.}
\newblock \bibinfo{journal}{\bibinfo{title}{{Characterizing the Fast Radio Burst Host Galaxy Population and its Connection to Transients in the Local and Extragalactic Universe}}}.
\newblock {\emph{\JournalTitle{\aj}}} \textbf{\bibinfo{volume}{163}}, \bibinfo{pages}{69}, \doiprefix\url{https://dx.doi.org/10.3847/1538-3881/ac3aec} (\bibinfo{year}{2022}).
\newblock \eprint{2108.01282}.

\bibitem{2020ApJ...895L..37B}
\bibinfo{author}{{Bhandari}, S.} \emph{et~al.}
\newblock \bibinfo{journal}{\bibinfo{title}{{The Host Galaxies and Progenitors of Fast Radio Bursts Localized with the Australian Square Kilometre Array Pathfinder}}}.
\newblock {\emph{\JournalTitle{\apjl}}} \textbf{\bibinfo{volume}{895}}, \bibinfo{pages}{L37}, \doiprefix\url{https://dx.doi.org/10.3847/2041-8213/ab672e} (\bibinfo{year}{2020}).
\newblock \eprint{2005.13160}.

\bibitem{2020ApJ...903..152H}
\bibinfo{author}{{Heintz}, K.~E.} \emph{et~al.}
\newblock \bibinfo{journal}{\bibinfo{title}{{Host Galaxy Properties and Offset Distributions of Fast Radio Bursts: Implications for Their Progenitors}}}.
\newblock {\emph{\JournalTitle{\apj}}} \textbf{\bibinfo{volume}{903}}, \bibinfo{pages}{152}, \doiprefix\url{https://dx.doi.org/10.3847/1538-4357/abb6fb} (\bibinfo{year}{2020}).
\newblock \eprint{2009.10747}.

\bibitem{2010PASA...27..272M}
\bibinfo{author}{{Macquart}, J.-P.} \emph{et~al.}
\newblock \bibinfo{journal}{\bibinfo{title}{{The Commensal Real-Time ASKAP Fast-Transients (CRAFT) Survey}}}.
\newblock {\emph{\JournalTitle{\pasa}}} \textbf{\bibinfo{volume}{27}}, \bibinfo{pages}{272--282}, \doiprefix\url{https://dx.doi.org/10.1071/AS09082} (\bibinfo{year}{2010}).
\newblock \eprint{1001.2958}.

\bibitem{2016PASA...33...42M}
\bibinfo{author}{{McConnell}, D.} \emph{et~al.}
\newblock \bibinfo{journal}{\bibinfo{title}{{The Australian Square Kilometre Array Pathfinder: Performance of the Boolardy Engineering Test Array}}}.
\newblock {\emph{\JournalTitle{\pasa}}} \textbf{\bibinfo{volume}{33}}, \bibinfo{pages}{e042}, \doiprefix\url{https://dx.doi.org/10.1017/pasa.2016.37} (\bibinfo{year}{2016}).
\newblock \eprint{1608.00750}.

\bibitem{2022MNRAS.514.1961R}
\bibinfo{author}{{Rajwade}, K.~M.} \emph{et~al.}
\newblock \bibinfo{journal}{\bibinfo{title}{{First discoveries and localizations of Fast Radio Bursts with MeerTRAP: real-time, commensal MeerKAT survey}}}.
\newblock {\emph{\JournalTitle{\mnras}}} \textbf{\bibinfo{volume}{514}}, \bibinfo{pages}{1961--1974}, \doiprefix\url{https://dx.doi.org/10.1093/mnras/stac1450} (\bibinfo{year}{2022}).
\newblock \eprint{2205.14600}.

\bibitem{2016mks..confE...1J}
\bibinfo{author}{{Jonas}, J.} \& \bibinfo{author}{{MeerKAT Team}}.
\newblock \bibinfo{title}{{The MeerKAT Radio Telescope}}.
\newblock In \emph{\bibinfo{booktitle}{MeerKAT Science: On the Pathway to the SKA}}, \bibinfo{pages}{1}, \doiprefix\url{https://dx.doi.org/10.22323/1.277.0001} (\bibinfo{year}{2016}).

\bibitem{2019ApJ...885L..24C}
\bibinfo{author}{{CHIME/FRB Collaboration}} \emph{et~al.}
\newblock \bibinfo{journal}{\bibinfo{title}{{CHIME/FRB Discovery of Eight New Repeating Fast Radio Burst Sources}}}.
\newblock {\emph{\JournalTitle{\apjl}}} \textbf{\bibinfo{volume}{885}}, \bibinfo{pages}{L24}, \doiprefix\url{https://dx.doi.org/10.3847/2041-8213/ab4a80} (\bibinfo{year}{2019}).
\newblock \eprint{1908.03507}.

\bibitem{2014ApJ...790..101S}
\bibinfo{author}{{Spitler}, L.~G.} \emph{et~al.}
\newblock \bibinfo{journal}{\bibinfo{title}{{Fast Radio Burst Discovered in the Arecibo Pulsar ALFA Survey}}}.
\newblock {\emph{\JournalTitle{\apj}}} \textbf{\bibinfo{volume}{790}}, \bibinfo{pages}{101}, \doiprefix\url{https://dx.doi.org/10.1088/0004-637X/790/2/101} (\bibinfo{year}{2014}).
\newblock \eprint{1404.2934}.

\bibitem{2016ApJ...833..177S}
\bibinfo{author}{{Scholz}, P.} \emph{et~al.}
\newblock \bibinfo{journal}{\bibinfo{title}{{The Repeating Fast Radio Burst FRB 121102: Multi-wavelength Observations and Additional Bursts}}}.
\newblock {\emph{\JournalTitle{\apj}}} \textbf{\bibinfo{volume}{833}}, \bibinfo{pages}{177}, \doiprefix\url{https://dx.doi.org/10.3847/1538-4357/833/2/177} (\bibinfo{year}{2016}).
\newblock \eprint{1603.08880}.

\bibitem{2019MNRAS.486.3636P}
\bibinfo{author}{{Price}, D.~C.} \emph{et~al.}
\newblock \bibinfo{journal}{\bibinfo{title}{{A fast radio burst with frequency-dependent polarization detected during Breakthrough Listen observations}}}.
\newblock {\emph{\JournalTitle{\mnras}}} \textbf{\bibinfo{volume}{486}}, \bibinfo{pages}{3636--3646}, \doiprefix\url{https://dx.doi.org/10.1093/mnras/stz958} (\bibinfo{year}{2019}).
\newblock \eprint{1901.07412}.

\bibitem{2021ApJ...917...75M}
\bibinfo{author}{{Mannings}, A.~G.} \emph{et~al.}
\newblock \bibinfo{journal}{\bibinfo{title}{{A High-resolution View of Fast Radio Burst Host Environments}}}.
\newblock {\emph{\JournalTitle{\apj}}} \textbf{\bibinfo{volume}{917}}, \bibinfo{pages}{75}, \doiprefix\url{https://dx.doi.org/10.3847/1538-4357/abff56} (\bibinfo{year}{2021}).
\newblock \eprint{2012.11617}.

\bibitem{2023arXiv231201578W}
\bibinfo{author}{{Woodland}, M.~N.} \emph{et~al.}
\newblock \bibinfo{journal}{\bibinfo{title}{{The Environments of Fast Radio Bursts Viewed Using Adaptive Optics}}}.
\newblock {\emph{\JournalTitle{arXiv e-prints}}} \bibinfo{pages}{arXiv:2312.01578}, \doiprefix\url{https://dx.doi.org/10.48550/arXiv.2312.01578} (\bibinfo{year}{2023}).
\newblock \eprint{2312.01578}.

\bibitem{2008ApJS..178..247K}
\bibinfo{author}{{Kennicutt}, J., Robert~C.} \emph{et~al.}
\newblock \bibinfo{journal}{\bibinfo{title}{{An H{\ensuremath{\alpha}} Imaging Survey of Galaxies in the Local 11 Mpc Volume}}}.
\newblock {\emph{\JournalTitle{\apjs}}} \textbf{\bibinfo{volume}{178}}, \bibinfo{pages}{247--279}, \doiprefix\url{https://dx.doi.org/10.1086/590058} (\bibinfo{year}{2008}).
\newblock \eprint{0807.2035}.

\bibitem{2023ApJ...948...67B}
\bibinfo{author}{{Bhandari}, S.} \emph{et~al.}
\newblock \bibinfo{journal}{\bibinfo{title}{{A Nonrepeating Fast Radio Burst in a Dwarf Host Galaxy}}}.
\newblock {\emph{\JournalTitle{\apj}}} \textbf{\bibinfo{volume}{948}}, \bibinfo{pages}{67}, \doiprefix\url{https://dx.doi.org/10.3847/1538-4357/acc178} (\bibinfo{year}{2023}).
\newblock \eprint{2211.16790}.

\bibitem{2022Natur.606..873N}
\bibinfo{author}{{Niu}, C.~H.} \emph{et~al.}
\newblock \bibinfo{journal}{\bibinfo{title}{{A repeating fast radio burst associated with a persistent radio source}}}.
\newblock {\emph{\JournalTitle{\nat}}} \textbf{\bibinfo{volume}{606}}, \bibinfo{pages}{873--877}, \doiprefix\url{https://dx.doi.org/10.1038/s41586-022-04755-5} (\bibinfo{year}{2022}).
\newblock \eprint{2110.07418}.

\bibitem{2017ApJ...834L...7T}
\bibinfo{author}{{Tendulkar}, S.~P.} \emph{et~al.}
\newblock \bibinfo{journal}{\bibinfo{title}{{The Host Galaxy and Redshift of the Repeating Fast Radio Burst FRB 121102}}}.
\newblock {\emph{\JournalTitle{\apjl}}} \textbf{\bibinfo{volume}{834}}, \bibinfo{pages}{L7}, \doiprefix\url{https://dx.doi.org/10.3847/2041-8213/834/2/L7} (\bibinfo{year}{2017}).
\newblock \eprint{1701.01100}.

\bibitem{2012ApJ...759..107K}
\bibinfo{author}{{Kelly}, P.~L.} \& \bibinfo{author}{{Kirshner}, R.~P.}
\newblock \bibinfo{journal}{\bibinfo{title}{{Core-collapse Supernovae and Host Galaxy Stellar Populations}}}.
\newblock {\emph{\JournalTitle{\apj}}} \textbf{\bibinfo{volume}{759}}, \bibinfo{pages}{107}, \doiprefix\url{https://dx.doi.org/10.1088/0004-637X/759/2/107} (\bibinfo{year}{2012}).
\newblock \eprint{1110.1377}.

\bibitem{2015ApJ...804...90L}
\bibinfo{author}{{Lunnan}, R.} \emph{et~al.}
\newblock \bibinfo{journal}{\bibinfo{title}{{Zooming In on the Progenitors of Superluminous Supernovae With the HST}}}.
\newblock {\emph{\JournalTitle{\apj}}} \textbf{\bibinfo{volume}{804}}, \bibinfo{pages}{90}, \doiprefix\url{https://dx.doi.org/10.1088/0004-637X/804/2/90} (\bibinfo{year}{2015}).
\newblock \eprint{1411.1060}.

\bibitem{2015A&A...581A.102V}
\bibinfo{author}{{Vergani}, S.~D.} \emph{et~al.}
\newblock \bibinfo{journal}{\bibinfo{title}{{Are long gamma-ray bursts biased tracers of star formation? Clues from the host galaxies of the Swift/BAT6 complete sample of LGRBs . I. Stellar mass at z < 1}}}.
\newblock {\emph{\JournalTitle{\aap}}} \textbf{\bibinfo{volume}{581}}, \bibinfo{pages}{A102}, \doiprefix\url{https://dx.doi.org/10.1051/0004-6361/20142501310.48550/arXiv.1409.7064} (\bibinfo{year}{2015}).
\newblock \eprint{1409.7064}.

\bibitem{2016ApJ...817..144B}
\bibinfo{author}{{Blanchard}, P.~K.}, \bibinfo{author}{{Berger}, E.} \& \bibinfo{author}{{Fong}, W.-f.}
\newblock \bibinfo{journal}{\bibinfo{title}{{The Offset and Host Light Distributions of Long Gamma-Ray Bursts: A New View From HST Observations of Swift Bursts}}}.
\newblock {\emph{\JournalTitle{\apj}}} \textbf{\bibinfo{volume}{817}}, \bibinfo{pages}{144}, \doiprefix\url{https://dx.doi.org/10.3847/0004-637X/817/2/144} (\bibinfo{year}{2016}).
\newblock \eprint{1509.07866}.

\bibitem{2013ApJ...770..108C}
\bibinfo{author}{{Childress}, M.} \emph{et~al.}
\newblock \bibinfo{journal}{\bibinfo{title}{{Host Galaxy Properties and Hubble Residuals of Type Ia Supernovae from the Nearby Supernova Factory}}}.
\newblock {\emph{\JournalTitle{\apj}}} \textbf{\bibinfo{volume}{770}}, \bibinfo{pages}{108}, \doiprefix\url{https://dx.doi.org/10.1088/0004-637X/770/2/108} (\bibinfo{year}{2013}).
\newblock \eprint{1304.4720}.

\bibitem{2020ApJ...901..143U}
\bibinfo{author}{{Uddin}, S.~A.} \emph{et~al.}
\newblock \bibinfo{journal}{\bibinfo{title}{{The Carnegie Supernova Project-I: Correlation between Type Ia Supernovae and Their Host Galaxies from Optical to Near-infrared Bands}}}.
\newblock {\emph{\JournalTitle{\apj}}} \textbf{\bibinfo{volume}{901}}, \bibinfo{pages}{143}, \doiprefix\url{https://dx.doi.org/10.3847/1538-4357/abafb7} (\bibinfo{year}{2020}).
\newblock \eprint{2006.15164}.

\bibitem{2022ApJ...940...57N}
\bibinfo{author}{{Nugent}, A.~E.} \emph{et~al.}
\newblock \bibinfo{journal}{\bibinfo{title}{{Short GRB Host Galaxies. II. A Legacy Sample of Redshifts, Stellar Population Properties, and Implications for Their Neutron Star Merger Origins}}}.
\newblock {\emph{\JournalTitle{\apj}}} \textbf{\bibinfo{volume}{940}}, \bibinfo{pages}{57}, \doiprefix\url{https://dx.doi.org/10.3847/1538-4357/ac91d1} (\bibinfo{year}{2022}).
\newblock \eprint{2206.01764}.

\bibitem{2020MNRAS.498.4790K}
\bibinfo{author}{{Kovlakas}, K.} \emph{et~al.}
\newblock \bibinfo{journal}{\bibinfo{title}{{A census of ultraluminous X-ray sources in the local Universe}}}.
\newblock {\emph{\JournalTitle{\mnras}}} \textbf{\bibinfo{volume}{498}}, \bibinfo{pages}{4790--4810}, \doiprefix\url{https://dx.doi.org/10.1093/mnras/staa2481} (\bibinfo{year}{2020}).
\newblock \eprint{2008.10572}.

\bibitem{2021ApJ...907L..31B}
\bibinfo{author}{{Bochenek}, C.~D.}, \bibinfo{author}{{Ravi}, V.} \& \bibinfo{author}{{Dong}, D.}
\newblock \bibinfo{journal}{\bibinfo{title}{{Localized Fast Radio Bursts Are Consistent with Magnetar Progenitors Formed in Core-collapse Supernovae}}}.
\newblock {\emph{\JournalTitle{\apjl}}} \textbf{\bibinfo{volume}{907}}, \bibinfo{pages}{L31}, \doiprefix\url{https://dx.doi.org/10.3847/2041-8213/abd634} (\bibinfo{year}{2021}).
\newblock \eprint{2009.13030}.

\bibitem{2014ApJS..214...15S}
\bibinfo{author}{{Speagle}, J.~S.}, \bibinfo{author}{{Steinhardt}, C.~L.}, \bibinfo{author}{{Capak}, P.~L.} \& \bibinfo{author}{{Silverman}, J.~D.}
\newblock \bibinfo{journal}{\bibinfo{title}{{A Highly Consistent Framework for the Evolution of the Star-Forming ``Main Sequence'' from z \raisebox{-0.5ex}\textasciitilde 0-6}}}.
\newblock {\emph{\JournalTitle{\apjs}}} \textbf{\bibinfo{volume}{214}}, \bibinfo{pages}{15}, \doiprefix\url{https://dx.doi.org/10.1088/0067-0049/214/2/15} (\bibinfo{year}{2014}).
\newblock \eprint{1405.2041}.

\bibitem{2001ApJ...556..121K}
\bibinfo{author}{{Kewley}, L.~J.}, \bibinfo{author}{{Dopita}, M.~A.}, \bibinfo{author}{{Sutherland}, R.~S.}, \bibinfo{author}{{Heisler}, C.~A.} \& \bibinfo{author}{{Trevena}, J.}
\newblock \bibinfo{journal}{\bibinfo{title}{{Theoretical Modeling of Starburst Galaxies}}}.
\newblock {\emph{\JournalTitle{\apj}}} \textbf{\bibinfo{volume}{556}}, \bibinfo{pages}{121--140}, \doiprefix\url{https://dx.doi.org/10.1086/321545} (\bibinfo{year}{2001}).
\newblock \eprint{astro-ph/0106324}.

\bibitem{2003MNRAS.346.1055K}
\bibinfo{author}{{Kauffmann}, G.} \emph{et~al.}
\newblock \bibinfo{journal}{\bibinfo{title}{{The host galaxies of active galactic nuclei}}}.
\newblock {\emph{\JournalTitle{\mnras}}} \textbf{\bibinfo{volume}{346}}, \bibinfo{pages}{1055--1077}, \doiprefix\url{https://dx.doi.org/10.1111/j.1365-2966.2003.07154.x} (\bibinfo{year}{2003}).
\newblock \eprint{astro-ph/0304239}.

\bibitem{2020ARA&A..58..257G}
\bibinfo{author}{{Greene}, J.~E.}, \bibinfo{author}{{Strader}, J.} \& \bibinfo{author}{{Ho}, L.~C.}
\newblock \bibinfo{journal}{\bibinfo{title}{{Intermediate-Mass Black Holes}}}.
\newblock {\emph{\JournalTitle{\araa}}} \textbf{\bibinfo{volume}{58}}, \bibinfo{pages}{257--312}, \doiprefix\url{https://dx.doi.org/10.1146/annurev-astro-032620-021835} (\bibinfo{year}{2020}).
\newblock \eprint{1911.09678}.

\bibitem{2022MNRAS.511.6126W}
\bibinfo{author}{{Walters}, D.}, \bibinfo{author}{{Woo}, J.} \& \bibinfo{author}{{Ellison}, S.~L.}
\newblock \bibinfo{journal}{\bibinfo{title}{{Quenching time-scales in the IllustrisTNG simulation}}}.
\newblock {\emph{\JournalTitle{\mnras}}} \textbf{\bibinfo{volume}{511}}, \bibinfo{pages}{6126--6142}, \doiprefix\url{https://dx.doi.org/10.1093/mnras/stac283} (\bibinfo{year}{2022}).
\newblock \eprint{2202.00015}.

\bibitem{2023ApJ...958...66E}
\bibinfo{author}{{Eftekhari}, T.} \emph{et~al.}
\newblock \bibinfo{journal}{\bibinfo{title}{{An X-Ray Census of Fast Radio Burst Host Galaxies: Constraints on Active Galactic Nuclei and X-Ray Counterparts}}}.
\newblock {\emph{\JournalTitle{\apj}}} \textbf{\bibinfo{volume}{958}}, \bibinfo{pages}{66}, \doiprefix\url{https://dx.doi.org/10.3847/1538-4357/acf843} (\bibinfo{year}{2023}).
\newblock \eprint{2307.03766}.

\bibitem{2022ApJ...926..134T}
\bibinfo{author}{{Tacchella}, S.} \emph{et~al.}
\newblock \bibinfo{journal}{\bibinfo{title}{{Fast, Slow, Early, Late: Quenching Massive Galaxies at z {\ensuremath{\sim}} 0.8}}}.
\newblock {\emph{\JournalTitle{\apj}}} \textbf{\bibinfo{volume}{926}}, \bibinfo{pages}{134}, \doiprefix\url{https://dx.doi.org/10.3847/1538-4357/ac449b} (\bibinfo{year}{2022}).
\newblock \eprint{2102.12494}.

\bibitem{2020ApJ...896..142B}
\bibinfo{author}{{Beloborodov}, A.~M.}
\newblock \bibinfo{journal}{\bibinfo{title}{{Blast Waves from Magnetar Flares and Fast Radio Bursts}}}.
\newblock {\emph{\JournalTitle{\apj}}} \textbf{\bibinfo{volume}{896}}, \bibinfo{pages}{142}, \doiprefix\url{https://dx.doi.org/10.3847/1538-4357/ab83eb} (\bibinfo{year}{2020}).
\newblock \eprint{1908.07743}.

\bibitem{2019ApJ...886..110M}
\bibinfo{author}{{Margalit}, B.}, \bibinfo{author}{{Berger}, E.} \& \bibinfo{author}{{Metzger}, B.~D.}
\newblock \bibinfo{journal}{\bibinfo{title}{{Fast Radio Bursts from Magnetars Born in Binary Neutron Star Mergers and Accretion Induced Collapse}}}.
\newblock {\emph{\JournalTitle{\apj}}} \textbf{\bibinfo{volume}{886}}, \bibinfo{pages}{110}, \doiprefix\url{https://dx.doi.org/10.3847/1538-4357/ab4c31} (\bibinfo{year}{2019}).
\newblock \eprint{1907.00016}.

\bibitem{2017ApJ...843...84N}
\bibinfo{author}{{Nicholl}, M.} \emph{et~al.}
\newblock \bibinfo{journal}{\bibinfo{title}{{Empirical Constraints on the Origin of Fast Radio Bursts: Volumetric Rates and Host Galaxy Demographics as a Test of Millisecond Magnetar Connection}}}.
\newblock {\emph{\JournalTitle{\apj}}} \textbf{\bibinfo{volume}{843}}, \bibinfo{pages}{84}, \doiprefix\url{https://dx.doi.org/10.3847/1538-4357/aa794d} (\bibinfo{year}{2017}).
\newblock \eprint{1704.00022}.

\bibitem{2023arXiv230101000R}
\bibinfo{author}{{Ravi}, V.} \emph{et~al.}
\newblock \bibinfo{journal}{\bibinfo{title}{{Deep Synoptic Array science: a 50 Mpc fast radio burst constrains the mass of the Milky Way circumgalactic medium}}}.
\newblock {\emph{\JournalTitle{arXiv e-prints}}} \bibinfo{pages}{arXiv:2301.01000}, \doiprefix\url{https://dx.doi.org/10.48550/arXiv.2301.01000} (\bibinfo{year}{2023}).
\newblock \eprint{2301.01000}.

\bibitem{2020ApJ...893...47D}
\bibinfo{author}{{Drlica-Wagner}, A.} \emph{et~al.}
\newblock \bibinfo{journal}{\bibinfo{title}{{Milky Way Satellite Census. I. The Observational Selection Function for Milky Way Satellites in DES Y3 and Pan-STARRS DR1}}}.
\newblock {\emph{\JournalTitle{\apj}}} \textbf{\bibinfo{volume}{893}}, \bibinfo{pages}{47}, \doiprefix\url{https://dx.doi.org/10.3847/1538-4357/ab7eb9} (\bibinfo{year}{2020}).
\newblock \eprint{1912.03302}.

\bibitem{1996AJ....112.1487H}
\bibinfo{author}{{Harris}, W.~E.}
\newblock \bibinfo{journal}{\bibinfo{title}{{A Catalog of Parameters for Globular Clusters in the Milky Way}}}.
\newblock {\emph{\JournalTitle{\aj}}} \textbf{\bibinfo{volume}{112}}, \bibinfo{pages}{1487}, \doiprefix\url{https://dx.doi.org/10.1086/118116} (\bibinfo{year}{1996}).

\bibitem{2010ApJ...722..566L}
\bibinfo{author}{{Lampeitl}, H.} \emph{et~al.}
\newblock \bibinfo{journal}{\bibinfo{title}{{The Effect of Host Galaxies on Type Ia Supernovae in the SDSS-II Supernova Survey}}}.
\newblock {\emph{\JournalTitle{\apj}}} \textbf{\bibinfo{volume}{722}}, \bibinfo{pages}{566--576}, \doiprefix\url{https://dx.doi.org/10.1088/0004-637X/722/1/56610.48550/arXiv.1005.4687} (\bibinfo{year}{2010}).
\newblock \eprint{1005.4687}.

\end{thebibliography}
\vspace{-0.5cm}  

\begin{addendum}
\item [Acknowledgments] The authors thank staff members of the Owens Valley Radio Observatory and the Caltech radio group, including Kristen Bernasconi, Stephanie Cha-Ramos, Sarah Harnach, Tom Klinefelter, Lori McGraw, Corey Posner, Andres Rizo, Michael Virgin, Scott White, and Thomas Zentmyer. Their tireless efforts were instrumental to the success of the DSA-110. The DSA-110 is supported by the National Science Foundation Mid-Scale Innovations Program in Astronomical Sciences (MSIP) under grant AST-1836018. We acknowledge use of the VLA calibrator manual and the radio fundamental catalog. Some of the data presented herein were obtained at the W. M. Keck Observatory, which is operated as a scientific partnership among the California Institute of Technology, the University of California and the National Aeronautics and Space Administration. The Observatory was made possible by the generous financial support of the W. M. Keck Foundation.

K.S. thanks Alexa Gordon for guidance on SED modeling. K.S. thanks Anya Nugent and Harsh Kumar for valuable insights into the parallels with short-duration GRB host population. K.S. thanks Xavier Prochaska, Keith Bannister and Andrei Beloborodov for engaging discussions on the distribution of star-formation in the universe and core-collapse supernova formation channel of FRB sources. K.S. thanks Navin Sridhar for insightful discussions. K.S. thanks David Cook and Jenny Greene for discussions on star-formation in the local universe. K.S. thanks Yashvardhan Tomar for guidance in constructing mass functions and simulating background galaxy population. 

\item[Author Contributions] V.R. and G.H. led the development of the DSA-110. D.H., M.H., J.L., P.R., S.W., and D.W. contributed to the construction of the DSA-110. K.S. and V.R. led the writing of the manuscript. K.S. developed the statistical framework to analyze the FRB sample and undertook the majority of the optical/IR host galaxy data analysis and interpretation. K.S., V.R., L.C., C.L., J.S., J.F., N.K., M.S., S.A., K.D., Y.Q., S.R., D.D., J.M., Y.Y conducted the optical/IR follow-up observations presented in this work. V.R., C.L., L.C., G.H., and R.H. developed the software pipeline for detecting FRBs with DSA-110. J.L. and J.S. provided guidance for SED analysis.

\item[Competing Interests] The authors declare no competing interests.

\item[Correspondence] Correspondence and requests for materials should be addressed to K.S. and V.R.

\ks{\item[Code Availability Statement] We have created 
a reproduction package for our work that includes all 
code used for our data analysis. We have placed this code on GitHub at https://github.com/krittisharma/frb\_host\_sharma2024.}

\ks{\item[Data Availability Statement] The FRB data 
presented here is publicly available in a CSV file in the same GitHub repository.}

\end{addendum}

\newpage
\clearpage
\newpage
\newpage

\noindent{\bfseries \LARGE Methods}\setlength{\parskip}{12pt}

\setlength{\parskip}{12pt}

\noindent{\bfseries \large The DSA-110 Instrument and FRB Localizations}
\setlength{\parskip}{3pt}

All DSA-110 FRBs presented herein were detected between February 2022 and November 2023 with a limited array of 48 core antennas and 15 outriggers~\protect\citemethods{2023ApJ...949L...3R}. Here we summarize details of the FRB detection system and localization procedures. 

The DSA-110 is a radio interferometer operating between 1311--1499\,MHz with 4.65-m dish antennas. The antennas have an automatic elevation drive enabling repointing along the meridian, and are equipped with dual linearly polarized receivers and ambient temperature low noise amplifiers delivering a system temperature of 25~K~\citep{2021ITMTT..69.2345W}. A coherent real-time search for FRBs over approximately 14\,deg$^{2}$ is implemented using the core antennas, which are evenly spaced along a 400~m east-west infrastructure. During the observations presented here, the array was primarily pointed at a declination of 71$^\circ$.6. We use a modified version of \sw{heimdall}~\protect\citemethods{2012MNRAS.422..379B} software to perform the search on total intensity data in 256 fan-shaped beams spaced by 1\,arcmin perpendicularly to the meridian. The data are integrated over 262.144\,$\mu$s, and channelized into 768 244.14~kHz channels. We search for FRBs with boxcar time-domain filters between 1 and 32 times the minimum time-resolution, and between DMs of 75\% of the Galactic expectation from the NE2001 model, and 1500\,pc\,cm$^{-3}$. Triggered FRB events prompt the storage of 4-bit voltage data from the core and outrigger antennas (maximum baseline of 2.5~km) for 61440 samples at 32.768~$\mu$s time resolution in 6144$\times$30.518~kHz channels. Additionally, we store 4 hours of visibility data near the trigger time. 

\ks{We derive arcsecond-scale localizations using the triggered voltage-data dumps, combined with slower-cadence visibilities recorded over 4 hours surrounding each FRB, and 5 minutes of visibility data on the bandpass-calibrator source 3C309.1 obtained within 12 hours of each FRB. For each FRB, we form the following data sets as  Common Astronomy Software Applications (\sw{CASA}) software~\protect\citemethods{2022PASP..134k4501C} (version 5.4.1) measurement sets (MSs).

\vspace{-3mm} \begin{itemize}[itemsep=-1mm]

\item A 5-minute bandpass-calibrator MS with 3C309.1 at the phase center. This MS is formed using data obtained when 3C309.1 transits through the center of the primary beam. We assume a flux density of 7.6\,Jy for 3C309.1, based on the VLA Calibrator Manual. This assumption is not critical to the analysis, but is useful in some verification steps described below. 

\item 5-minute MSs recorded during the transits of between 8 and 18 Very Long Baseline Interferometry (VLBI) calibrator sources observed within $\pm2$\,hr of the FRB. VLBI calibrators are selected from the 2022b Radio Fundamental Catalog as sources with flux densities $>50$\,mJy, located within $\pm1.3$\,deg of the pointing declination. 

\item A 10-minute MS containing data in the FRB field, formed using data temporally centered on the FRB detection. These data are phased to the pointing center at the time of the FRB detection. 

\item A 1.1-second MS formed by correlating the triggered voltage dump, phased to the pointing center at the time of the FRB detection. 

\item An MS with a single integration formed by dedispersing and correlating the triggered voltage dump, and integrating only the data containing the FRB. The optimal FRB DM is used for dedispersion.

\end{itemize}

Our calibration and imaging procedure is then as follows. 

\vspace{-3mm} \begin{itemize}[itemsep=-1mm]

\item We excise frequency channels affected by radio-frequency interference (RFI) in all data via visual inspection. We also excise baselines to any malfunctioning antennas. 

\item We then use the \sw{CASA} task \sw{bandpass} to derive an antenna-based complex bandpass from the 3C309.1 data. In all analyses we exclude baselines shorter than 45\,m, because these are affected by spurious correlations due to cross-talk and low-level RFI. This solution is applied to every other data set. Any ``delay'' terms are absorbed into the complex bandpass. 

\item We use the \sw{wsclean} software~\protect\citemethods{offringa-wsclean-2014} to generate a 3.3\,deg image from the 10-minute FRB-field MS. Standard $w$-projection is applied. We verify that $>95\%$ of compact ($<20$\,arcsec major axes), bright ($>20$\,mJy) sources in the NRAO VLA Sky Survey (NVSS)~\protect\citemethods{1998AJ....115.1693C} catalog are detected with approximately the correct flux densities in this image. This is occasionally not the case, which we attribute to poor calibration, and we then perform a phase-only gain calibration using a sky model derived from NVSS. The model includes all sources within the primary beam modeled as elliptical Gaussians. This second calibration is applied to all data if necessary. 

\item We then use \sw{wsclean} to make 5.5\,deg images of the FRB MS and the 1.1-second voltage MS, and small 0.17\,deg images of each VLBI-calibrator MS. The small images in particular have pixel sizes of 0.3\,arcsec to enable accurate image-plane astrometry. 

\item We verify that extremely bright ($>400$\,mJy) continuum sources from NVSS are detected in the image made from the 2-second voltage MS, at approximately the correct positions. This is a basic check for valid voltage data and associated correlation products. 

\item We manually identify the approximate FRB position to within a few arc-seconds in the large image of the FRB MS, using the location of the detection beam as a guide and checking that it is undetected in the voltage-MS image. We then use the \sw{CASA} task \sw{tclean} to make a small image (0.17\,deg, 0.3\,arcsec pixels) with the approximate FRB position as the phase center, and fit the FRB position in the image plane. We measure the signal-to-noise (S/N) ratio, $\sigma_{\rm FRB}$, of the FRB in the image plane, and using the full-width half-maxima of the synthesized beam in right ascension and declination ($s_{\theta}$ and $s_{\phi}$ respectively), we derive the statistical localization uncertainty as $0.45 \times s_{\theta, \phi} / \sigma_{\rm FRB}$. 

\item We fit the derived positions of each VLBI calibrator using their associated images, and perform a weighted (according to the image S/N) linear-model fit to the right ascension and declination position offsets with time. We correct the FRB position by the predicted image offsets at the burst detection time, and add the formal uncertainty in this prediction in quadrature to the statistical localization uncertainty. Typical corrections are at the sub-arcsecond level, with maximum observed corrections of $\sim2$\,arcsec. This method is physically motivated by the fact that the DSA-110 array essentially fits entirely within a physical correlation length in the case of ionospheric path-length errors, such that the errors can be modeled as bulk astrometric shifts. 

\end{itemize}

Given the reliance of the present work on accurate FRB localizations, we performed and here present an end-to-end test of the procedure. We analyzed 35 archived voltage dumps obtained in a variety of observing conditions during the year 2023, including on FRBs and pulsars. In these data sets, we identified 69 continuum sources selected from the NVSS catalog with flux densities $>400$\,mJy, major axes $<35$\,arcsec, and locations within 2.1\,deg of the primary-beam center. For such bright sources, the NVSS catalog positions have uncertainties of approximately 0.5\,arcsec in right ascension and 0.6\,arcsec in declination~\protect\citemethods{1998AJ....115.1693C}. Using exactly the same procedures as for the FRBs, we determined the position of each source in the DSA-110 data using 1.1\,s of voltage data, and a random dispersion measure between 80 and 1500~pc~cm$^{-3}$. The sources were detected with S/N ratios between 8 and 20. 

In Supplementary Fig.~\ref{fig:offsets_test}, we show the offsets of the derived continuum-source positions from the NVSS-catalog positions, and the uncertainties in the derived positions. Just as for the FRBs, the localization uncertainties are the quadrature sums of the theoretical statistical uncertainties given the image S/N, and the uncertainties in the final astrometric corrections from the adjacent VLBI calibrator sources. We find that only 4/69 of the 90\% confidence error ellipses do not contain the true source positions, consistent with expectations. We also do not find any significant systematic biases in the derived positions. 

Finally, we address an issue with the published DSA-110 localization of FRB\,20220912A~\protect\citemethods{2023ApJ...949L...3R}, identified by Hewitt et al. (2024)~\protect\citemethods{2024MNRAS.529.1814H}. The DSA-110 90\% confidence localization ellipse was found to be approximately 0.6\,arcsec offset from the correct VLBI position. We attribute this error to poorly converged in-field calibration solutions, and the fact that we had not yet incorporated astrometric corrections derived from adjacent VLBI calibrators into the localization procedure. We reprocessed the DSA-110 data on the 2022 October 18 burst from FRB\,20220912A using the methods described above, and derived a revised burst position of (J2000) 23h09m04.83s $+$48d42m23.6s, with uncertainties ($1\sigma$) of $\pm1.3$\,arcsec in right ascension and $\pm0.9$\,arcsec in declination (see Supplementary Fig.~\ref{fig:repeater_localization}). This position is consistent with the results of Hewitt et al. (2024)~\protect\citemethods{2024MNRAS.529.1814H}.}

\ks{Next we discuss the possibility of any selection biases from the DM cut of 1500~pc~cm$^{-3}$ used in our FRB search pipeline. Essentially, a DM cut of 1500~pc~cm$^{-3}$ excludes FRB hosts under two scenarios: either the FRBs are very distant, or the FRBs have excess local DM. In the first scenario, assuming a median host DM contribution of DM$_\mathrm{host} = 120$~pc~cm$^{-3}$ (Connor et al.) and attributing the remaining DM to the diffuse baryons in the IGM, the expected host galaxy redshift for DM$_\mathrm{IGM}$ of 1380~pc~cm$^{-3}$, based on the DM$_\mathrm{IGM}$ - $z$ relation~\protect\citemethods{2021ApJ...906...49Z}, is $z = 1.51_{-0.38}^{+0.17}$. Since our analysis includes only FRBs, background galaxies, and other transients at $z \leq 1$, this suggests that limiting our sample to a DM of 1500~pc~cm$^{-3}$ does not introduce significant biases. In the second scenario, assuming a log-normal host galaxy DM distribution with $\mu = 4.8$ and $\sigma = 0.5$ (Connor et al.), we evaluate the likelihood of an FRB with DM $> 1500$~pc~cm$^{-3}$ occurring in host galaxies at lower redshifts. For a DM$_\mathrm{exgal} = 1500$~pc~cm$^{-3}$, the probabilities that the FRB has excess local DM and is associated with a host galaxy at various redshifts are: $P (\mathrm{DM}_\mathrm{host} > \mathrm{DM}_\mathrm{exgal} - \langle \mathrm{DM}_\mathrm{IGM}(z = 0.2) \rangle ) = 9.5 \times 10^{-7}$ at $z \leq 0.2$, $P (\mathrm{DM}_\mathrm{host} > \mathrm{DM}_\mathrm{exgal} - \langle \mathrm{DM}_\mathrm{IGM}(z = 0.4) \rangle ) = 3.7 \times 10^{-6}$ at $z \leq 0.4$, $P (\mathrm{DM}_\mathrm{host} > \mathrm{DM}_\mathrm{exgal} - \langle \mathrm{DM}_\mathrm{IGM}(z = 0.7) \rangle ) = 4 \times 10^{-5}$ at $z \leq 0.7$, and $P (\mathrm{DM}_\mathrm{host} > \mathrm{DM}_\mathrm{exgal} - \langle \mathrm{DM}_\mathrm{IGM}(z = 1) \rangle ) = 6.8 \times 10^{-4}$ at $z \leq 1$. These probabilities indicate that the occurrence of an FRB with DM $> 1500$~pc~cm$^{-3}$ in host galaxies at lower redshifts is highly unlikely. Thus, it is evident that our sample is not impacted by selection biases due to the DM cut of 1500~pc~cm$^{-3}$.}

\setlength{\parskip}{12pt}

\noindent{\bfseries \large FRB Host Association and Sample Selection}
\setlength{\parskip}{3pt}

The FRBs published in this work were selected such that there is a plausible candidate host galaxy in the vicinity of the localization region detected in either PanSTARRS1 (PS1)~\citep{2016arXiv161205560C} or Beijing-Arizona Sky Survey (BASS)~\citep{2017PASP..129f4101Z} from the Dark Energy Survey~\protect\citemethods{2018ApJS..239...18A} $r$-band data. Out of 42 FRBs discovered by DSA-110 during its science commissioning phase from February 2022 to November 2023, 30 FRBs had a plausible candidate host galaxy. In cases of marginal optical detections, we obtain deeper imaging in our follow-up campaigns. We then use the Bayesian Probabilistic Association of Transients to their Hosts (PATH) formalism to identify the host galaxy and estimate its probability of association~\citep{2021ApJ...911...95A}. This formalism invokes Bayes' rule to calculate the posterior probability $P(O_i|x)$, where $O_i$ is the case that the FRB is from galaxy $i$ and $x$ represents the observables such as FRB localization, galaxy coordinates, magnitude, and angular size.

We use Galactic-extinction corrected $r$-band magnitudes for our PATH analysis, unless deeper imaging in another band is available (see Extended Data Table~\ref{table:basic_frb_properties} for a summary of optical/IR imaging used for PATH analysis). When the deepest available imaging is PS1 or BASS $r$-band, we use their respective published source catalogs to identify the set of candidate host galaxies within 30\arcsecs of the localization. For PS1, we query \sw{VizieR}~\protect\citemethods{2000A&AS..143...23O} using \sw{Astropy}~\protect\citemethods{2013A&A...558A..33A, 2018AJ....156..123A} and remove point sources using the PS1-Point Source Catalog \protect\citemethods{2018PASP..130l8001T} queried using \sw{CasJobs}~\protect\citemethods{2005cs........2072O}. For BASS, we extract the DR9 source catalog from Legacy Survey Browser~\protect\citemethods{2021AAS...23723503S} and identify galaxy candidates using the \sw{Tractor} morphological classifications. We use the cataloged half-light radii and photometry in our PATH analysis. For FRB fields with deeper imaging from our follow-up campaigns, we obtain the astrometry solution using \sw{astrometry.net} \protect\citemethods{Lang_2010}. We use \sw{SExtractor}~\protect\citemethods{bertin11} to extract all $\geq2\sigma$ sources in the field and remove point sources using the \sw{CLASS\_STAR} classifier with an $85$\% threshold. The source magnitudes are calibrated by correcting for zero points computed by cross-matching \sw{SExtractor}-identified sources with the \#~II/349/ps1 catalogue~\protect\citemethods{2018AAS...23143601F}. The half light radii are measured by iteratively fitting elliptical isophotes using standard procedures defined in \sw{photutils}~\protect\citemethods{larry_bradley_2022_6825092}.

The key observables in our analysis include the candidate host magnitude ($m$), host-normalized offsets ($\theta/\phi$), and the probability of non-detection of the host galaxy due to limited imaging depth $P(U)$. We assume that the brighter candidate galaxies have a higher prior probability, $P(O_i) \propto \frac{1}{\Sigma (m_i)}$, where $\Sigma (m_i)$ is the angular surface density of galaxies on the sky with magnitude $m \leq m_i$. Since the true angular distribution of FRBs is unknown, we assume an exponential host-normalized offset distribution ($\theta/\phi$) and assert a maximum offset of $\theta_{\mathrm{max}} = 6\phi$. Although quantifying $P(U)$ is tricky, incorporating the probability of host invisibility in optically-limited data has crucial implications on (but not limited to) offsets and host galaxy properties distributions. We adopt $P(U) = 0.2$ for PS1 and $P(U) = 0.1$ for deeper BASS, WIRC, and WaSP imaging based on tests with simulated FRB populations~\citep{2021arXiv211207639S}.

We perform PATH analysis on the deepest available images of FRB fields using the procedure outlined above (see Fig.~\ref{fig:host_cutouts} for images). The PATH analysis confidently associates 26 out of 30 FRBs to a host galaxy with $P(O_i|x) \geq 90$\%. The distribution of PATH association probabilities in the $m - \theta/\phi$ space of observables in Extended Data Fig.~\ref{fig:path_analysis} indicates that the 4 FRBs with insecure host associations are \ks{the ones with} typically fainter \ks{hosts} and at higher host-normalized offsets, assuming the most likely host. We investigate the cause of lower host association probabilities for these four FRBs further. FRB \frbferb has $m = 21.4$~mag host candidate at host-normalized offsets of $\theta/\phi = 5.4$, respectively. Since this angular offset is very close to the maximum offset permitted by our PATH priors $\theta_{\mathrm{max}} = 6\phi$, the association probability of this candidate host is low ($P(O_i|x) = 0.42$). FRB \frbishita had an $m = 20.8$~mag secure host galaxy in PS1 $r$-band image with $P(O_i|x) = 0.98$. However, deeper imaging revealed another faint candidate host galaxy with $m = 23.4$~mag in WIRC $J$-band image, which lowered the association probability of the PS1 host candidate to $P(O_i|x) = 0.56$. FRB \frberdos had an $m = 22.5$~mag secure host galaxy in BASS $r$-band image with $P(O_i|x) = 0.94$. We noticed another faint source with $m = 23.4$~mag in the vicinity of the localization region, which \sw{Tractor} morphological classifications identified as a star. We obtained a spectrum of this source and identified it as a mis-classification in Legacy catalogs. Incorporating this galaxy in our PATH analysis identified it as the most probable host galaxy with $P(O_i|x) = 0.63$. The host association probability of FRB \frbkoyaanisqatsi is low ($P(O_i|x) = 0.62$) because it has two candidate host galaxies in BASS with $m = 22.1$~mag and $m = 22.8$~mag respectively. We investigate the cases of FRB \frberdos and FRB \frbkoyaanisqatsi further by obtaining the redshift of the second-likely host ($z_2$) and comparing the $P(z | \mathrm{DM}_{\mathrm{exgal}})$ for the two candidates computed using a standard DM$_{\mathrm{IGM}} - z$ relation~\protect\citemethods{2021ApJ...906...49Z}. For FRB \frbkoyaanisqatsi, $P(z_1 = 0.542 | \mathrm{DM}_{\mathrm{exgal}}) = 0.758$, $P(z_2 = 0.229 | \mathrm{DM}_{\mathrm{exgal}}) = 0.242$ and for FRB \frberdos, $P(z_1 = 0.371 | \mathrm{DM}_{\mathrm{exgal}}) = 0.634 $, $P(z_2 = 0.671 | \mathrm{DM}_{\mathrm{exgal}}) = 0.366$, \ks{where the subscript ``1'' denotes the PATH-identified host}. Since in both cases, the $\mathrm{DM}_{\mathrm{exgal}}$ considerations also prefer the host identified by PATH analysis, we declare them as the most likely host. Since the other candidate host of FRB \frbishita is very faint ($m = 23.4$~mag in $J$-band), we refrain from obtaining its redshift due to limited observational resources.

\setlength{\parskip}{12pt}

\noindent{\bfseries \large \ks{Validating Our Host Associations}}
\setlength{\parskip}{3pt}

\ks{We validate the DSA-110 FRB host associations using the following three step methodology. First, we show that with the DSA's localization capabilities and optical imaging depths, we confidently associate FRBs with low-mass galaxies at low redshifts ($z \lesssim 0.2$), strengthening our result of the deficit of low-mass galaxies. Second, for higher redshifts, we quantify the rarity of a luminous galaxy's presence along the FRB sight-line using the probability of chance coincidence to justify these associations. Third, we reject the hypothesis of another low-luminosity host galaxy along the FRB sight-line by quantifying the improbability of more than one galaxies existing within the cosmic volume defined by the localization region and the FRB's extragalactic DM. Below, we discuss these points in detail:

\textbf{Robust host galaxy associations at low redshifts, considering the DSA-110's localization capabilities and the optical imaging depth of the utilized surveys:} The robustness of FRB host galaxy associations is a function of FRB localization area from the radio interferometer and the limiting magnitude of the optical imaging surveys~\protect\citemethods{2017ApJ...849..162E}. To assess the extent of our proficiency in confidently associating an FRB to a low mass galaxy, we perform the following simulation. For r-band magnitudes in the range $m \in [12, 26]$~mag and redshifts in the range $z \in [0, 1.2]$, we compute the galaxy stellar mass ($M_\ast$) assuming mass-to-light ratio, $M_\ast/L_\ast = 1$. We then compute the representative host galaxy half-light radius ($R_h$) for a galaxy of stellar mass $M_\ast$ using the mass-radius correlation of late-type galaxies~\protect\citemethods{2014ApJ...788...28V}. We compute the probability of chance coincidence ($P_{cc}$) using galaxy number counts for apparent r-band magnitudes. We follow the methodology of Eftekhari and Berger (2017)~\protect\citemethods{2017ApJ...849..162E} and fit the Driver et al. (2016)~\protect\citemethods{2016ApJ...827..108D} r-band galaxy number counts to compute the projected areal number density of galaxies brighter than r-band magnitude, $\sigma(\leq m)$. We then compute the $P_{cc}$ of occurring within a radius $R$, assuming a Poisson distribution of galaxies across the sky as 

\begin{equation}
\label{eqn:Pcc}
    P_{cc} = 1 - e^{-\pi R^2 \sigma(\leq m)},
\end{equation}

where $R$ is parameterized as $R = \max \left(2R_{\mathrm{FRB}},~\sqrt{R_0^2 + 4R_h^2}\right)$, $R_{\mathrm{FRB}}$ is the equivalent radius that corresponds to $1\sigma$ of the localization area, $R_0$ is the radial angular separation between the FRB position and a presumed host and $R_h$ is the half-light radius. Typical $1\sigma$ localization uncertainty of DSA-110 is $\lesssim 2$\arcsecs. For the purpose of testing our host association capabilities and to put upper limits on the $P_{cc}$, we use $R_0 \approx 2R_h$ and $R_\mathrm{FRB} = 3$\arcsecs in our simulation. The results of this simulation are shown in Supplementary Fig.~\ref{fig:association_capabilities_demonstration}a. We observe that in the absence of any optical observation biases, the DSA-110 localizations are sufficient to confidently associate FRBs to their true host galaxies. The FRBs can be associated to their true low-mass $0.01-0.1~L_\ast$ galaxies with $P_{cc} \lesssim 0.1$. However, the optical observation bias of r-band magnitude $\lesssim 23.5$~mag limit our ability to associate FRBs to low-mass galaxies at high redshifts. But, at low redshifts ($\lesssim 0.2$), the optical observation biases barely impact our host associations and we are capable of confidently associating FRBs to low-mass $0.001-0.1~L_\ast$ galaxies with $P_{cc} \lesssim 0.1$. This implies that our host associations at $z \lesssim 0.2$ are robust.

\textbf{Rarity of existence of a massive galaxy within the small DSA-110 localization regions:} The probability of chance coincidence also quantifies the rarity of existence of a massive galaxy within the localization uncertainty of the FRB by virtue of the fact that massive galaxies in the universe are rare as compared to the low-mass galaxies and the probability of chance existence of a massive galaxy within the small DSA-110 localization area is small. Motivated by this, we compute the $P_{cc}$ for all DSA-110 FRBs and plot them in Supplementary Fig.~\ref{fig:association_capabilities_demonstration}b in the space of their characteristic radius (a function of $R_\mathrm{FRB},~R_0,~R_h$) and r-band magnitude. We observe that the probability of chance occurrence of our FRBs along a sightline close to these massive galaxies is smaller than $0.1$, thus strengthening our associations.

\textbf{Rarity of existence of multiple candidate galaxies within the cosmic volume permitted by DSA-110 localization uncertainties and FRB dispersion measure:} We begin by computing the average number density of galaxies in the universe~\citep{2020ApJ...893..111L} ($n$) as a function of redshift ($z$), while also accounting for galaxy clustering~\protect\citemethods{1980lssu.book.....P}. First, we compute the galaxies stellar mass function~\citep{2020ApJ...893..111L} $dn(z)/d\log M$. To compute the total number density of galaxies, we integrate this mass function in the range $\log M \in [5, 12]$, where the lower limit is set based on the lowest mass CCSNe host galaxy~\citep{2021ApJS..255...29S}. Since the distribution of galaxies in the low-redshift universe is non-homogeneous, we account for galaxy clustering~\protect\citemethods{1980lssu.book.....P} to compute the effective galaxy number density ($n_\mathrm{eff}$) in the redshift slice $z$ to $z+dz$ as

\begin{equation}
    n_{\mathrm{eff}}(z) = n(z) \frac{\int_0^\Delta \left( 1 + \xi (r) \right) r^2 dr}{\int_0^\Delta r^2 dr}, 
\end{equation}

where $\Delta$ is the thickness of the comoving shell between redshifts $z$ and $z+dz$, and $\xi(r)$ is the two-point correlation function which describes the excess probability, compared to a random distribution, of finding a pair of galaxies separated by a distance $r$,

\begin{equation}
    \xi(r) = \left( \frac{r}{r_0} \right)^{-\gamma},
\end{equation}

where $\gamma \approx 1.8$ is the power-law index~\protect\citemethods{1980lssu.book.....P} and $r_0$ is the characteristic correlation length~\protect\citemethods{1980lssu.book.....P}, $r_0 = 5.4~h^{-1}$~Mpc. Having computed the effective galaxy number density ($n_\mathrm{eff}$), we compute the expected number of galaxies as $N_{\mathrm{eff}} = n_{\mathrm{eff}} \times V_\mathrm{loc,~z}$, where $V_\mathrm{loc,~z}$ is the conical shell volume defined by the localization uncertainties and redshift slice $z$ to $z + dz$, computed as

\begin{equation}
    V_\mathrm{loc,~z} = \frac{\Omega}{4\pi} \times \frac{4\pi}{3} \left( D_c(z+dz)^3 - D_c(z)^3 \right),
\end{equation}

where $\Omega$ is the solid angle of the conical region computed as $\Omega = \pi \theta^2 D_A(z)^2/D_\mathrm{c}(z)^2$, where we use double of the maximum localization uncertainties, $\theta = 6$\arcsecs to present a conservative estimate on galaxy counts, $D_A(z)$ is the angular diameter distance and $D_\mathrm{c}(z)$ is the comoving distance at redshift $z$. We plot $N_\mathrm{eff}(z)$ in Supplementary Fig.~\ref{fig:association_capabilities_demonstration_highz}a for reference. We also include uncertainties on $N_\mathrm{eff}(z)$, which primarily stem from the uncertainties on the mass function.

Next, we compute the average number of galaxies within the localization region ($\bar{N}(\mathrm{DM})$), given the FRB dispersion measure (DM). To this end, for FRB extragalactic DM$_\mathrm{exgal} \in [0, 1000]$~pc~cm$^{-3}$, we first compute $P(z | \mathrm{DM}_\mathrm{exgal})$, by assuming that the extragalactic DM is attributed to diffuse baryons in the intergalactic medium (IGM)~\protect\citemethods{2021ApJ...906...49Z}. We then compute $\bar{N}(\mathrm{DM})$ as follows,

\begin{equation}
    \bar{N}(\mathrm{DM}_\mathrm{exgal}) = \int_0^2 N_\mathrm{eff}(z) \cdot P(z | \mathrm{DM}_\mathrm{exgal})~dz.
\end{equation}

Finally, we run the following simulation to compute $P (\text{galaxy~counts}~>1 | \mathrm{DM}_\mathrm{exgal})$. For each $\mathrm{DM}_\mathrm{exgal}$, 
we sample 10,000 galaxy counts from a Poisson distribution with the mean $\bar{N}(\mathrm{DM}_\mathrm{exgal})$
and then compute $P (\text{galaxy~counts}~>1 | \mathrm{DM}_\mathrm{exgal})$ as the fraction of these samples where the counts are greater than 1. We run this in a Monte Carlo simulation to also compute the uncertainties on the probabilities. We plot our results in Supplementary Fig.~\ref{fig:association_capabilities_demonstration_highz}b.

We find that for $\mathrm{DM}_\mathrm{exgal} \lesssim 200$~pc~cm$^{-3}$, the probability of existence of more than one galaxies within 6\arcsecs of the localization is $\lesssim 0.20_{-0.04}^{+0.04}$\%. Majority of DSA-110 FRBs have DM$_\mathrm{exgal} \lesssim 600$~pc~cm$^{-3}$, implying a probability of existence of more than one galaxies $\lesssim 2.00_{-0.14}^{+0.14}$\%. For increasing DM$_\mathrm{exgal}$, this probability increases to a maximum value of $2.93_{-0.17}^{+0.17}$\%, primarily due to the larger cosmic volume probed by the localization region. Therefore, the maximum probability of existence of more than one galaxies within 6\arcsecs of DSA-110 FRB localizations is $\approx 3$\%. This implies that out of 30 DSA-110 FRBs, we expect a confusion between more than one candidate host galaxies for at most $30 \times 0.03 \approx 1$ case, thus quantifying the rarity of this scenario.

For completeness, we also quantify the rarity of existence of more than one galaxies within 6\arcsecs of the localization by following another methodology. For a given FRB extragalactic DM, we estimate redshift upper limits, $z_\mathrm{max}$, assuming that $\mathrm{DM}_\mathrm{exgal}$ is attributed to diffuse baryons in the IGM and using the DM$_\mathrm{IGM} - z$ relation~\protect\citemethods{2021ApJ...906...49Z}. We estimate these redshift upper limits ($z_\mathrm{max}$) by computing the redshift corresponding to the 95th percentile of the probability distribution, $P(z | \mathrm{DM}_\mathrm{exgal})$. We then compute the mean number of galaxies within the comoving volume out to redshift $z_\mathrm{max}$ as

\begin{equation}
    \bar{N}(\mathrm{DM}_\mathrm{exgal}) = \int_0^{z_\mathrm{max}} N_\mathrm{eff}(z)~dz,
\end{equation}

and repeat the aforementioned simulation to compute $P$(galaxy counts $>1 | \mathrm{DM}_\mathrm{exgal}$). The results from this simulation are shown in Supplementary Fig.~\ref{fig:association_capabilities_demonstration_highz}c. We find that the probability of existence of more than one galaxies within 6\arcsecs of the localization is $\lesssim 3.48_{-0.19}^{+0.18}$\%, again quantifying the rarity of this scenario.}

\setlength{\parskip}{12pt}

\noindent{\bfseries \large \ks{Hostless DSA FRBs}}
\setlength{\parskip}{3pt}

\ks{We quantify the likelihood of the possibility that some of the 12 hostless DSA FRBs, which were not associated with hosts down to an $r$-band magnitude of $\lesssim 23.5$ mag, could actually have hosts within the local universe, specifically in the redshift bin $z \leq 0.2$. To investigate this, we use the extragalactic DMs of these FRBs to quantify how rare the excess local $\mathrm{DM}_\mathrm{host}$ would need to be for a sufficient number of $z \leq 0.2$ hosts to exist in this sample of hostless FRBs to break our low-mass host deficit statistics.}

\ks{Following our companion paper by Connor et al., we assume a log-normal distribution for the host contribution to the DM with parameters $\mu_\mathrm{host} = 4.8$ and $\sigma_\mathrm{host} = 0.5$ (shown in Fig.~\ref{fig:hostDMdistribution}). We compute the $p$-values for each FRB as the probability $P(\mathrm{DM}_\mathrm{host} > \mathrm{DM}_\mathrm{exgal} - \langle\mathrm{DM}_\mathrm{IGM}(z=0.2)\rangle )$. We subtract the median DM contribution from the IGM at $z = 0.2$ of $\mathrm{DM}_\mathrm{IGM} = 180$~pc~cm$^{-3}$. Under the hypothesis that some of these 12 FRBs may exist at $z \leq 0.2$, these $\mathrm{DM}_\mathrm{host}$ values would represent the extreme scenario. This is because if these FRBs were at $z < 0.2$, then $\mathrm{DM}_\mathrm{IGM}$ would be lower, and the corresponding $\mathrm{DM}_\mathrm{host}$ would be larger, thus pushing to the tail of the $\mathrm{DM}_\mathrm{host}$ distribution, making the scenario of the FRB existing at lower redshifts even rarer.}

\ks{Since $\sim$20\% of the star formation in the universe at $z \leq 0.2$ occurs in low-mass galaxies ($\log M_\ast \leq 9$), and the FRBs sample has a total of 20 FRBs with $\log M_\ast \geq 9$ at $z \leq 0.2$, we require 5 low-mass, low-redshift hosts in the hostless sample for FRBs to perfectly trace star formation in the universe and to break our low-mass hosts deficit statistics. We therefore compute the probability that 5 of 12 hostless FRBs are low-mass, low-redshift hosts as the sum of the product of $p$-values of all possible combinations of 5 FRBs out of the 12 hostless candidates. We find that the probability of this scenario is $\approx 10^{-7}$ ($5.3\sigma$ significance), implying that the probability of 5 of these 12 hostless FRBs being part of our local-universe sample is low. We note that this analysis assumed nothing about the host association procedures. Therefore, our analysis supports the conclusion that the deficit of low-mass hosts among the DSA-discovered FRBs is not due to observational or association limitations but reflects a genuine lack of such hosts in our sample.}

\setlength{\parskip}{12pt}
\newpage

\noindent{\bfseries \large Optical/IR Imaging and Spectroscopy}
\setlength{\parskip}{3pt}

Given the outlined selection criteria, we leverage a wealth of high-quality archival optical to near-infrared (NIR) imaging data for a majority of host galaxies in our sample, obtained from surveys such as PS1, BASS, Mayall z-band Legacy Survey (MzLS)~\citep{2019AJ....157..168D}, Sloan Digital Sky Survey (SDSS)~\citep{2015ApJS..219...12A}, Two Micron All Sky Survey (2MASS)~\citep{2006AJ....131.1163S}, and Wide-field Infrared Survey Explorer (WISE)~\citep{2014yCat.2328....0C}. Additionally, we incorporate near-ultraviolet (NUV) and far-ultraviolet (FUV) photometry from the Galaxy Evolution Explorer (GALEX)~\citep{2005ApJ...619L...1M} survey whenever available. For most of our FRB host galaxies, we utilize archival photometry from the respective published catalogs of these surveys, except for BASS, MzLS and SDSS. Specifically for WISE data, we use the aperture 2 instrumental photometry to accurately capture the flux solely from the host galaxy, thereby avoiding contamination from other sources and accounting for the change in point spread function with wavelength~\citep{2023arXiv230205465G}. We convert the photometry of 2MASS and WISE Catalogs from Vega to AB magnitudes. We execute photometry on SDSS, BASS and MzLS imaging by iteratively fitting elliptical isophotes to the BASS $r$-band images of the galaxy using standard procedures defined in \sw{photutils}~\protect\citemethods{larry_bradley_2022_6825092} to identify the isophote that captures $\gtrsim$95\% of the light from the galaxy and then convolving this aperture with the point spread function of all images to measure the instrumental magnitudes in all bands. These instrumental magnitudes were subsequently calibrated using the standard zero-point of 22.5 for SDSS, BASS and MzLS. For FRB host galaxies exhibiting extended spiral/disk-like features, we perform manual photometry on all optical imaging by employing isophotal analysis and consistent apertures across all bands, as described above. For PS1, we use zero-points from the headers. The data for FRB \frbjackie is contaminated by the presence of a star in its vicinity. We use the procedures from its discovery paper to remove the star~\citep{2023ApJ...950..175S} before performing photometry.

To enhance the marginal detections in PS1, BASS or MzLS surveys, we acquire deeper optical imaging in the SDSS $g'$, $r'$, $i'$, and $z'$-bands using the Wafer-Scale Imager for Prime focus (WaSP)~\citep{2017JATIS...3c6002N} instrument mounted on the Palomar 200-inch Hale Telescope. We complement optical data with NIR data in $J$, $H$, and $K_s$ bands obtained with the Wide Field Infrared Camera (WIRC)~\citep{2003SPIE.4841..451W} instrument, also mounted on the Palomar 200-inch Hale Telescope. The WaSP data are obtained as a set of six 300~s exposures in all bands. For WIRC data acquisition, we use 9-point dither patterns with appropriate exposure times and co-addition strategies: 45~s exposures with 1-coadd for $J$-band, 6~s exposures with 5-coadds for $H$-band, and 3~s exposures with 10-coadds for $K_s$-band. We calibrate the images by applying bias correction and flat-fielding, followed by cosmic rays removal using \sw{Astro-SCRAPPY} package~\protect\citemethods{curtis_mccully_2018_1482019}. We extract all point sources in the image using \sw{SExtractor} and obtain an astrometry solution by a cross-match with Gaia Data Release 2 catalog~\protect\citemethods{2018A&A...616A...1G}. We then resample all images on the same astrometric grid to align their position angles using \sw{SWarp}~\protect\citemethods{2002ASPC..281..228B} and median-combine them to obtain stacked images. We use \sw{SExtractor} to extract all $\geq2\sigma$ sources in the stacked images and calibrate the source magnitudes by correcting for zero points computed by cross-matching \sw{SExtractor}-identified sources with the \#~II/349/ps1 catalog for WaSP and 2MASS catalog for WIRC. Additionally, to ensure consistency across different photometric systems, we converted the Vega magnitudes of WIRC imaging to the AB magnitude system. Finally, we correct for interstellar dust reddening and galactic extinction along the line-of-sight~\protect\citemethods{2018JOSS....3..695M, 1999PASP..111...63F}. All the imaging and photometry used in our work are cataloged in Supplementary Table~\ref{table:imaging_log}.

In addition to the fundamental insights garnered through photometry alone, the spectrum of the host galaxy serves as an indispensable reservoir of information crucial for constraining its star-formation history (SFH). To this end, we obtain optical spectroscopy of FRB host galaxies. The observations were taken with Low Resolution Imaging Spectrometer (LRIS)~\citep{1995PASP..107..375O} on Keck-I, DEep Imaging Multi-Object Spectrograph (DEIMOS)~\citep{2003SPIE.4841.1657F} on Keck-II at W. M. Keck Observatory, and Double Spectrograph (DBSP)~\citep{1982PASP...94..586O} on the 200-inch Hale Telescope at Palomar Observatory. The facilities, instruments, configurations, and observation details for each FRB host are listed in Supplementary Table~\ref{table:host_spectroscopy_details}. The LRIS, DEIMOS, and DBSP data were reduced using the \sw{LPipe}~\protect\citemethods{2019PASP..131h4503P}, the Python Spectroscopic Data Reduction Pipeline (\sw{PypeIt})~\protect\citemethods{pypeit:zenodov_v1_6, pypeit:joss_pub}, and the \sw{DBSP\_DRP}~\protect\citemethods{dbsp_drp:joss} software. All the spectra are flux calibrated using observations of standard stars. We measure the spectroscopic redshift and emission line fluxes of the host galaxies using the Penalized PiXel-Fitting (\sw{pPXF}) software~\citep{2022arXiv220814974C}\protect\citemethods{2017MNRAS.466..798C}. We jointly fit the stellar continuum and nebular emission using the MILES stellar library~\protect\citemethods{2006MNRAS.371..703S}. The reduced spectra with \sw{pPXF} fits in the rest-frame of hosts are displayed in Supplementary Fig.~\ref{fig:host_spectra}. We fit the stellar-continuum subtracted emission lines with Gaussian profiles to measure line fluxes.

\setlength{\parskip}{12pt}

\noindent{\bfseries \large Stellar Population Modeling}
\setlength{\parskip}{3pt}

We forward model our photometry and spectroscopy data to derive the posterior parameter distribution of our host galaxies under the statistical inference framework offered by the \sw{Prospector} software~\citep{2021ApJS..254...22J}. The model parameters describe the stellar, nebular, active galactic nucleus (AGN), and dust components of the galaxy along with instrument parameters, and we specify the noise model and priors. For Bayesian forward-modeling, \sw{Prospector} uses the Flexible Stellar Population Synthesis code (\sw{PYTHON-FSPS})~\protect\citemethods{2009ApJ...699..486C, 2010ascl.soft10043C} to compute the emergent galaxy spectrum by combining simple stellar populations, given a set of stellar population parameters. We sample the posteriors using the dynamic nested sampling routine \sw{dynesty}~\protect\citemethods{Speagle_2020}.

We use a non-parametric model which consists of a piece-wise constant SFH where the ratio of the average star-formation rate (SFR) in each lookback time bin is used to parameterize the stellar mass formed in that bin. This approach offers more flexibility in modeling the unusual shapes of SFHs that are not captured by parametric forms, thus reducing any biases that may exist from the strong prior of a particular parametric form of the SFH. Non-parametric models have been shown to be better suited to recover the shape of recent and older SFHs~\citep{2021ApJS..254...22J}. This is important in inferring the stellar masses, mass-weighted stellar ages, recent SFRs, and delays from star-formation events. Hence, we choose to use non-parametric models in our work. We employ a continuity non-parametric SFH with seven bins and use the recommended StudentT prior on the SFR ratios. The SFRs in the two most recent bins (0-30 Myr and 30-100 Myr) are expected to be constrained by the nebular emission line features in the spectrum, the ultraviolet photometry and the infrared photometry (via dust emission), if available. The remaining five bins are spaced logarithmically uniform in time up to the age of the universe at the redshift of the galaxy.

For spectral energy distribution (SED) modeling, we assume the Kroupa initial mass function~\protect\citemethods{2001MNRAS.322..231K}. Since a reliable measurement of stellar metallicity in the absence of high-quality absorption features in the spectrum can be tricky, we enforce the mass-metallicity relation of galaxies~\citep{2005MNRAS.362...41G} in our model. This prior also helps in breaking the age-metallicity degeneracy. We include the effects of dust attenuation as a dust screen affecting stars of all ages, where the normalization of the wavelength dependence of the optical depth~\protect\citemethods{2013ApJ...775L..16K} at 5500~\AA, the extra attenuation towards young stars and the offset in slope from the dust attenuation curve~\protect\citemethods{2000ApJ...533..682C} are the free parameters. We also use the three-component dust emission model~\protect\citemethods{2007ApJ...657..810D} provided in \sw{FSPS} where only the component describing the grain size distribution through the fraction of grain mass in polycyclic aromatic hydrocarbons (PAHs) is left as a free parameter. Since this grain model produces dust emission above 1~$\mu$m, we include it in our models only when data at these rest-frame wavelengths are available. Due to substantial mid-infrared emission and important implications of dust-obscured AGNs on the stellar ages and SFRs, we also include an AGN dust torus emission component~\protect\citemethods{2008ApJ...685..160N} in our models whenever mid-infrared data are available. The inclusion of AGN templates smoothes out the derived SFH by removing the mis-identified dust emission around AGB stars in the 0.1-1 Gyr lookback time bins~\protect\citemethods{2018ApJ...854...62L}. Since we are jointly fitting photometry and spectroscopy with abundant emission line features, we also include a nebular emission model. We tie the gas-phase metallicity to the stellar metallicity and float the nebular ionization parameter. The nebular emission model in \sw{Prospector} assumes that all the emission lines are powered by young stars, which may not be true in the presence of hard ionizing fields of AGNs or shock-heated emission. Hence, we marginalize over the amplitude of emission lines in our observed spectrum. We initialize the redshift to the value obtained from \sw{pPXF} fits with a uniform prior width of 1\% to allow for a better fit to emission and absorption lines.

Due to systematics in measuring photometry and uncertainties in the underlying stellar and photoionization models, we assume additional 10\% photometric errors as done in previous works~\citep{2021ApJS..254...22J}. To avoid corrupting normalization-sensitive parameters by gross aperture corrections, we scale the spectrum to match the integrated photometry probed by the population and fit for a 12th-order Chebyshev polynomial as a multiplicative calibration function. We include spectral smoothing to account for the line-of-sight velocity distribution of stars and the instrumental resolution to better fit the emission lines and incorporate a pixel outlier model to marginalize over poorly modeled noise. We apply additional masking in the noisy parts of the spectrum.

All the fixed and free parameters in our SED models with their priors are summarized in Supplementary Table~\ref{table:sed_params}. The Supplementary Fig.~\ref{fig:host_seds} shows the SED fits for all FRBs included in our sample, and the key derived properties from our analysis are listed in Extended Data Table~\ref{table:derived_galaxy_properties}, where we report the median and 68\% credible intervals. By default, Prospector computes the formed stellar mass, and we compute the current stellar mass by multiplying it with the surviving mass fraction, which takes into account mass loss during evolution from stellar winds and supernovae. We convert the optical depth towards young and old stellar light to V-band attenuation in magnitudes by multiplying with 1.086. We compute the average SFR over the recent 20~Myr and 100~Myr using our constrained non-parametric SFH. We compute the rest-frame $r$-band absolute magnitude, $(g-r)^0$ color, and $(u-r)^0$ color using posterior distributions of modeled SEDs.

\setlength{\parskip}{12pt}

\noindent{\bfseries \large Comparison Samples and Ensemble Statistics}
\setlength{\parskip}{3pt}

We construct multiple comparison sets to place FRB host galaxies within the broader context of the background galaxy population and to compare them with various transient classes. In this section, we summarize all comparison sets and the observational biases inherent in their selection. Further, we also discuss the statistical techniques used to compare the distributions of their properties.

\textbf{Background Galaxy Population:} \ks{With the ultimate goal of avoiding any systematics from the differences in stellar population modeling techniques for our FRBs and the background galaxy population, we use galaxy population data products presented in Leja et al. (2020~\citep{2020ApJ...893..111L} \& 2022~\citep{2022ApJ...936..165L}) to simulate a complete population of background galaxies. These data products were derived using the properties of galaxies in the COSMOS-2015~\protect\citemethods{2016ApJS..224...24L} and 3D-HST~\protect\citemethods{2014ApJS..214...24S} galaxy catalogs, inferred using similar \sw{Prospector}-based procedures as ours,} including a seven-component SFH, a two-component flexible dust attenuation model, nebular emission, and AGN-heated dust-torus emission~\protect\citemethods{2019ApJ...877..140L}\citep{2020ApJ...893..111L, 2022ApJ...936..165L}. The 3D-HST photometric catalog comprises of $\sim23,000$ galaxies from $\sim900$~arcmin$^2$ field of view at redshifts $z \in [0.5, 1]$ with rest-frame photometry in the range 0.3-8~$\mu$m and spectroscopic/grism redshifts for 30\% of the objects. The COSMOS-2015 photometric catalog comprises of $\sim48,500$ galaxies from 2~deg$^2$ COSMOS field at redshifts $z \in [0.005, 0.8]$ with photometry covering rest-frame 0.2-24~$\mu$m and photometric redshifts for a majority of objects.

\ks{Here, we first summarize the reasons for not directly using the COSMOS-2015 and 3D-HST galaxy catalogs as our comparison sample.} Firstly, since the massive galaxies in the universe are rare and the field surveyed is order a degree-squared, the volume probed at low redshifts is insufficient to find enough massive galaxies. This would lead to limited-number statistics at high mass end of the background galaxy population, thus preventing any meaningful comparisons. Furthermore, both of these surveys are luminosity-limited. Consequently, the 3D-HST survey is complete down to $\sim 10^{8.7}$~M$_\odot$ at a redshift of $z=0.65$, and COSMOS-2015 survey is complete down to $\sim 10^{8.6}$~M$_\odot$ at a redshift of $z=0.175$, and $\sim 10^{9.1}$~M$_\odot$ at a redshift of $z=0.5$~\citep{2020ApJ...893..111L}. This mass-incompleteness for low-mass galaxies would imply that they are substantially under-represented in these galaxy catalogs, thus again preventing any meaningful comparisons.

To address these limitations, we take steps to compensate for incompleteness in data by simulating a representative sample of the background galaxy population. This involves using established models for the stellar mass function~\citep{2020ApJ...893..111L}, star-forming main sequence (SFMS)~\citep{2022ApJ...936..165L}, and star-formation density~\citep{2022ApJ...936..165L} within ranges where existing catalogs are complete in stellar mass and SFR. \ks{Since the models presented in Leja et al. (2020~\citep{2020ApJ...893..111L} \& 2022~\citep{2022ApJ...936..165L}) are for redshifts $z > 0.2$, we extend them reasonably to encompass lower redshifts ($z \leq 0.2$). In the following, we present our methodology of generating the background galaxy population in detail to enable readers to reproduce our results. We also make our script publicly available at \url{https://github.com/krittisharma/frb_host_sharma2024}.}

To generate a population of $N$ background galaxies at a given redshift $z$, we first sample $N$ stellar masses in the range $\log \mathrm{M}_\ast \in [7, 12]$ from the stellar mass function constructed using the continuity model fits for the redshift evolution~\citep{2020ApJ...893..111L}. \ks{Leja et al. (2020)~\citep{2020ApJ...893..111L} parameterized the redshift evolution of the galaxy stellar mass function, $\Phi(\mathcal{M}, z)$, as a sum of two Schechter functions, where a single Schechter function is written as}

\begin{equation}
    \ks{\Phi(\mathcal{M}) = \ln (10) \phi_\ast 10^{(\mathcal{M}-\mathcal{M}_\ast)(\alpha + 1)} \mathrm{exp}(10^{(\mathcal{M}-\mathcal{M}_\ast)}),}
\end{equation}

\ks{for a given $\phi_\ast$, $\mathcal{M}_\ast = \log M_\ast$ and $\alpha$. For the sum of two Schechter functions, there are a total of five parameters: $\phi_1$, $\phi_2$, $\mathcal{M}_\ast$, $\alpha_1$ and $\alpha_2$. The redshift evolution of $\phi_1$, $\phi_2$ and $\mathcal{M}_\ast$ is modeled with a quadratic equation in redshift as}

\begin{equation}
    \ks{\rho_i(z) = \lambda_{i,0} + \lambda_{i,1}z + \lambda_{i,2}z^2,}
    \label{eqn:redshift_evol}
\end{equation}

\ks{where $\lambda_{i,j}$ are the continuity model parameters and $\alpha_1$, $\alpha_2$ are assumed to be redshift independent to limit degenerate solutions. We refer the reader to Leja et al. (2020)~\citep{2020ApJ...893..111L} for the values of these parameters in the continuity models (listed in their Fig.~3). Although Leja et al. (2020)~\citep{2020ApJ...893..111L} presents model fits in the mass completeness regime for redshifts $z > 0.2$, the parametric representation of the models facilitate extrapolation to lower stellar masses and lower redshifts.}

\ks{After sampling $N$ galaxy stellar masses from the mass function at redshift $z$, we determine the SFR for each stellar mass by referencing the density distribution of galaxies $\rho (\log \mathrm{M}_\ast, \log \mathrm{SFR}, z)$. This density distribution has been modeled using normalizing flows by Leja et al. (2022)~\citep{2022ApJ...936..165L}, which provides an uncertainty-deconvolved distribution of galaxies in the SFR$-\log \mathrm{M}_\ast$ plane at a fixed redshift $z$. We refer the readers to Leja et al. (2022)~\citep{2022ApJ...936..165L} for details of the normalizing flows neural network. We use the trained model publicly available at \url{https://github.com/jrleja/sfs_leja_trained_flow}. Therefore, for each galaxy stellar mass $\log \mathrm{M}_\ast$ at redshift $z$, we sample a SFR from the probability distribution $\rho (\log \mathrm{SFR} | \log \mathrm{M}_\ast, z)$ in the range $\log \mathrm{SFR} \in [-5, 3]$.}

\ks{In the following, we discuss how we sample a SFR in the instances where either the stellar mass falls below mass-completeness thresholds or the redshift is $z \leq 0.2$. Since our methodology for the same involves using the location of the center of star-forming main sequence (SFMS), we first elaborate on it. Leja et al. (2022)~\citep{2022ApJ...936..165L} parameterizes the galaxy SFMS as} 

\begin{equation}
\ks{\log (\mathrm{SFR})= \begin{cases}a\left(\log M-\log M_t\right)+c, & \log M>\log M_t \\ b\left(\log M-\log M_t\right)+c, & \log M \leqslant \log M_t\end{cases}},
\end{equation}

\ks{where $a$ is the slope at high masses, $b$ is the slope at low masses, $c$ is the intercept and $\log M_t$ is the stellar mass where the slope transitions from the low-mass to the high-mass component. The redshift evolution of these parameters is modeled using eqn.~\ref{eqn:redshift_evol}. We refer the reader to Leja et al. (2022)~\citep{2022ApJ...936..165L} for the values of these parameters (listed in their table~1).}

\ks{At a particular redshift ($z$), when the stellar mass ($\log M_\ast$) falls below completeness threshold ($\log M_\mathrm{comp}$), for SFR in the range $\log \mathrm{SFR} \in [-5, 3]$, we first compute the probability distribution of SFRs at the completeness threshold, $p(\log \mathrm{SFR}) = \rho (\log \mathrm{SFR} | \log \mathrm{M}_\mathrm{comp}, z)$ and the offset between the center of SFMS at $M_\mathrm{comp}$ and $M_\ast$, which we denote as $\Delta_\mathrm{SFMS} (z, M_\mathrm{comp} \rightarrow M_\ast)$. We then sample a SFR from the probability distribution adjusted to match the corresponding center of the SFMS at $\log M_\ast$ as $p_{\Delta}(\log \mathrm{SFR}) = p\left[ \log \mathrm{SFR} + \Delta_\mathrm{SFMS} (z, M_\mathrm{comp} \rightarrow M_\ast) \right]$.}

\ks{Since the center of the SFMS is parametrically formulated for redshift evolution, we can confidently extrapolate the galaxy models to low redshifts, as detailed below. In order to compute a SFR corresponding to stellar mass $\log M_\ast$ at a particular redshift $z \leq 0.2$, we first compute the probability distribution of SFRs for stellar mass $\log M_\ast$ at $z^\prime = 0.25$, $p(\log \mathrm{SFR}) = \rho(\log \mathrm{SFR} | \log M_\ast, z^\prime)$ and the offset between the center of SFMS for $\log M_\ast$ at $z$ and $z^\prime$, which we denote as $\Delta_\mathrm{SFMS}(z^\prime \rightarrow z, M_\ast)$. We then sample a SFR from the probability distribution adjusted to match the corresponding center of the SFMS at $z$ as $p_{\Delta}(\log \mathrm{SFR}) = p\left[\log \mathrm{SFR} + \Delta_\mathrm{SFMS} (z^\prime \rightarrow z, M_\ast) \right]$.}

\ks{Next, we quantify the uncertainty from the extrapolation into the low-mass regime and into the low-redshift regime. Note that the population model already includes the effect of cosmic variance (CV) on the high-mass regime, which is small because we assume a smooth redshift evolution, which mitigates CV; and COSMOS is a wide-area survey. We perform a simple calculation of these uncertainties by testing two extreme scenarios - (1) for the low-redshift extrapolation, we test the (close to the worst-case) scenario where there is no redshift evolution, (2) for the low-mass extrapolation, we test the (close to the worst-case) scenario where the power-law index has additional 10\% uncertainty right when the extrapolation begins. We include a few sanity checks of these tests in Supplementary Fig.~\ref{fig:extreme_cases_sanity_check} for reference. Note that the evolution of mass function from $z = 0.2$ to $z = 0$ is $\lesssim 0.1$~dex (see Supplementary Fig.~\ref{fig:extreme_cases_sanity_check}a) and the additional 10\% uncertainties on the low mass end leads to relatively higher number of low-mass galaxies with higher SFRs (see Supplementary Fig.~\ref{fig:extreme_cases_sanity_check}b and \ref{fig:extreme_cases_sanity_check}c). The results from our tests of these two extreme worst-case scenarios are shown in Supplementary Fig.~\ref{fig:additional_uncertainties_no_z_evolution} and we discuss these in detail below.}

\ks{In the first test (extreme scenario \#1), we find that the curve of stellar mass-weighted background distribution changes negligibly, which is primarily because the change in galaxy stellar mass function from redshift $z = 0.2$ to $z = 0.0$ is small (see Supplementary Fig.~\ref{fig:extreme_cases_sanity_check}a). The SFR-weighted background distribution shifts by $\sim 0.1$~dex. These shifts lead to a $p$-value of $< 0.001$ against mass-weighted distribution and 0.049 against SFR-weighted distribution. In the second test (extreme scenario \#2), as expected, we find that the stellar mass-weighted distribution does not change because for additional errors in the power-law index of the galaxy main-sequence, only SFRs are impacted and stellar mass-weighted stellar mass distribution remains the same. We find the the SFR-weighted distribution shifts by $< 0.1$~dex at the low mass end. This is primarily because a higher uncertainty on the power-law index of the galaxy star-forming main sequence results in relatively more low mass galaxies with higher SFRs (see Supplementary Fig.~\ref{fig:extreme_cases_sanity_check}b and \ref{fig:extreme_cases_sanity_check}c). Since the shifts in the SFR-weighted curve is $\lesssim 0.1$~dex in both the cases, the uncertainties from these extreme and worst-case scenarios are small, thus quantifying the effects of extrapolation. This implies that our results are not impacted by the additional uncertainties from the extrapolation.}

Finally, we adjust this simulated background population to account for optical selection effects in our sample of FRB host galaxies. We retain only those galaxies with $r$-band magnitudes $\leq 23.5$. To convert stellar mass to $r$-band magnitudes, we utilize estimates for mass-to-light ratios ($M/L$) from COSMOS-2015 and 3D-HST galaxy catalogs, accounting for variations in the $M/L$ distributions across different redshift ranges. Specifically, we use $\log(M/L) \sim \mathcal{N}(0, \sigma)$, where $\sigma = 0.2$ for $z \in [0, 0.2]$, $\sigma = 0.26$ for $z \in [0.2, 0.4]$, and $\sigma = 0.3$ for $z \in [0.4, 0.7]$.

\textbf{FRB Host Galaxies:} For comparing stellar population properties of the host galaxies, together with our 26 secure host associations, we include the refined host properties of 23 FRBs reported by Gordon et al.~\citep{2023arXiv230205465G} of previously published FRBs~\protect\citemethods{2022AJ....163...69B, 2020ApJ...895L..37B, 2020ApJ...903..152H} discovered by Commensal Real-Time ASKAP Fast-Transients (CRAFT)~\protect\citemethods{2010PASA...27..272M} survey on the Australian Square Kilometer Pathfinder (ASKAP)~\protect\citemethods{2016PASA...33...42M}, the More TRansients and Pulsars (MeerTRAP)~\protect\citemethods{2022MNRAS.514.1961R} project on the MeerKAT radio telescope~\protect\citemethods{2016mks..confE...1J}, the Canadian Hydrogen Intensity Mapping Experiment (CHIME)~\protect\citemethods{2019ApJ...885L..24C}, Arecibo~\protect\citemethods{2014ApJ...790..101S, 2016ApJ...833..177S} and Parkes~\protect\citemethods{2019MNRAS.486.3636P}. The selection criterion of the Gordon et al.~\citep{2023arXiv230205465G} sample particularly excluded FRB 20190614D and FRB 20190523A due to their low PATH association probabilities. Furthermore, they also exclude FRB 20200120E and the Galactic source SGR J1935+2154 because of their low burst spectral energies. We also include the recently reported 4 host galaxies of non-repeating FRBs discovered by CHIME in the local volume with low extragalactic dispersion measures and high galactic latitudes~\citep{2023arXiv231010018B}. All of these host galaxies have been modeled using similar techniques as we use in this work, thus avoiding any modeling biases within the FRB host galaxies sample. Therefore, our FRB host galaxies properties sample consists of 5 repeaters and 47 apparent non-repeaters. For constructing the FRB offset distribution, together with our 26 secure host associations, we use published FRB offset measurements~\protect\citemethods{2021ApJ...917...75M, 2023arXiv231201578W}. Mannings et al.~\protect\citemethods{2021ApJ...917...75M} used high-resolution Hubble Space Telescope ultraviolet and infrared imaging to study the host galaxy morphology and quantify the galactocentric offsets of 10 FRBs. Woodland et al.~\protect\citemethods{2023arXiv231201578W} expanded this sample of FRB galactocentric offsets with adaptive optics-aided diffraction-limited NIR imaging of 4 new FRB host galaxies.

\textbf{FRBs in Low-Mass Galaxies:} Our selection criterion of $r$-band magnitude $\lesssim 23.5$~mag excludes one low-mass ($M_*\lesssim10^9 M_\odot$) host (FRB 20121102A) from the sample. There is a dearth of low-mass FRB hosts at $z<0.2$, despite a lack of expected radio selection bias for FRBs in low-mass, low-redshift galaxies. The mean H$\alpha$ luminosity of low-mass galaxies is $\sim10^2\times$ smaller than higher mass star-forming galaxies~\protect\citemethods{2008ApJS..178..247K}, suggesting that the DM contributions of these galaxies' ISMs should be correspondingly small~\citep{2022ApJ...934...71O}. Based on the DM and scattering of pulsars in the Magellanic Clouds, Ocker et al.\citep{2022ApJ...934...71O} predict typical DMs $\lesssim 100$ pc cm$^{-3}$ and scattering timescales $< 1$ ms at 1 GHz from the ISMs of low-mass galaxies, well within the selection constraints of DSA-110 and all other FRB surveys.
 
An alternative selection bias against finding FRBs in low-mass galaxies could be induced if circumsource environments of FRBs in low-mass galaxies are systematically more extreme. A handful of FRBs have been localized to low-mass galaxies with elevated host DM~\protect\citemethods{2023ApJ...948...67B, 2022Natur.606..873N, 2017ApJ...834L...7T}, which necessitates a dense compact region near the source to account for the entirety of the DM excess. This is unlike typical FRBs in our sample (Connor et al.), and challenges the notion that these FRBs are similar to the broader FRB population, thus underscoring their distinctiveness~\citep{2022ApJ...934...71O}.

\textbf{Host Galaxies of Other Transients:} We compile the core-collapse supernovae (CCSNe) host galaxy sample from Schulze et al.~\citep{2021ApJS..255...29S}, which consists of 831 spectroscopically-classified CCSNe discovered by the Palomar Transient Factory (PTF) in a blind search with a median redshift of $z = 0.04$ ($1\sigma$ inter-quartile range of $0.02-0.08$). In the construction of this sample, Schulze et al. excluded 12 transients with uncertain CCSNe spectroscopic classifications and 5 hostless CCSNe with $< 2\sigma$ detections in archival data and rest-frame $R$-band luminosities $M_R \gtrsim -13.6$~mag, thus comprising the faintest CCSNe host galaxies discovered by the PTF. Despite these minor limitations in completeness, their study harnesses the significant statistical power offered by their large sample size to convincingly affirm that Type II CCSNe effectively trace star-formation in the universe. The host galaxies in this sample were modeled with a $\tau$-linear exponentially declining SFH with an average photometric coverage spanning far-ultraviolet to the mid-infrared, thus yielding fairly robust stellar mass and SFR measurements. We also retrieve the CCSNe projected galactocentric offsets reported by Schulze et al.~\citep{2021ApJS..255...29S} and host-normalized offsets reported by Kelly \& Kirshner~\protect\citemethods{2012ApJ...759..107K}. 

For superluminous supernovae (SLSNe), we use 55 host galaxies reported in the Schulze et al.~\citep{2021ApJS..255...29S} PTF sample with a median redshift of $z = 0.18$ ($1\sigma$ inter-quartile range of $0.11-0.28$). This sample also excluded 5 hostless SLSNe between $z = 0.4 - 1$. We use the SLSNe offset measurements from Hubble Space Telescope imaging for 16 sources reported in Lunnan et al.~\protect\citemethods{2015ApJ...804...90L}. We retrieve host galaxy properties for 14 long-duration gamma-ray bursts (lGRBs) from Vergani et al.~\protect\citemethods{2015A&A...581A.102V} and 17 lGRBs from Taggart \& Perley~\citep{2021MNRAS.503.3931T}. Vergani et al. reported the host galaxy properties of $z < 1$ lGRBs in the BAT6 complete sample, measured using an exponentially-declining SFH. Taggart \& Perley's sample consists of all $z < 0.3$ lGRBs discovered before 2018 with an associated optical counterpart to ensure robust host associations. The host galaxies in this work were modeled using an exponentially declining SFH. The combined sample has a median redshift of $z = 0.28$ ($1\sigma$ inter-quartile range of $0.09-0.62$). For offset distribution analysis, we use projected galactocentric offsets measured by Blanchard et al.~\protect\citemethods{2016ApJ...817..144B} for a complete sample of 79 lGRBs localized by their afterglow detections. The host galaxies sample of 119 Type Ia supernovae discovered by the SNfactory reported in Childress et al.~\protect\citemethods{2013ApJ...770..108C} has a median redshift of $z = 0.07$ ($1\sigma$ inter-quartile range of $0.03-0.12$). They measure stellar mass using the magnitudes and mass-to-light ratios computed using galaxy colors. The SFR was measured using stellar population synthesis models. For constructing offset distributions, we use the measurements reported by Uddin et al.~\protect\citemethods{2020ApJ...901..143U} for the complete sample of 113 nearby Type Ia supernovae discovered by the Carnegie Supernova Project. We use the host galaxy properties reported by Nugent et al.~\protect\citemethods{2022ApJ...940...57N} for 31 short-duration gamma-ray bursts (sGRBs) with robust host associations and spectroscopic redshifts with a median redshift of $z = 0.46$ ($1\sigma$ inter-quartile range of $0.20-0.78$). These measurements were made by assuming a delayed-$\tau$ exponentially declining SFH. We use the corresponding galactocentric offsets presented in Fong et al.~\citep{2022ApJ...940...56F}. We retrieve host galaxy properties and galactocentric offsets of ultra-luminous X-ray (ULX) sources from Kovlakas et al.~\protect\citemethods{2020MNRAS.498.4790K}. We specifically only include reliably-classified, non-nuclear sources to avoid confusion with AGNs. We further impose a minimum X-ray luminosity threshold of $10^{39}$~erg s$^{-1}$ and only include sources within a luminosity distance of 40~Mpc to avoid source confusion.

\textbf{Comparison Methods:} Since the redshift distribution of different classes of transients is significantly different, we correct their measured host galaxy stellar mass and SFR for redshift evolution~\protect\citemethods{2021ApJ...907L..31B}. To ensure that the statistical properties of each transient are representative of galaxies at redshift zero, for a transient host galaxy with stellar mass M$_\ast^{z=z_t}$ at $z = z_t$, we first compute the equivalent stellar mass, M$_\ast^{z=0}$, at redshift $z=0$ such that the fraction of total star-formation in the universe below that stellar mass remains the same:

\begin{equation}
\frac{\int_0^{\mathrm{M}_\ast^{z=0}} \Phi (\mathrm{M}_\ast, z=0)~\mathrm{SFR}(\mathrm{M}_\ast, z=0)~d\mathrm{M}_\ast}{\int_0^{\infty} \Phi (\mathrm{M}_\ast, z=0)~\mathrm{SFR}(\mathrm{M}_\ast, z=0)~d\mathrm{M}_\ast} = \frac{\int_0^{\mathrm{M}_\ast^{z=z_t}} \Phi (\mathrm{M}_\ast, z=z_t)~\mathrm{SFR}(\mathrm{M}_\ast, z=z_t)~d\mathrm{M}_\ast}{\int_0^{\infty} \Phi (\mathrm{M}_\ast, z=z_t)~\mathrm{SFR}(\mathrm{M}_\ast, z=z_t)~d\mathrm{M}_\ast}.
\end{equation}

Since the evolution in the log-normal distribution of galaxies about the SFMS is negligible over the redshift range of interest~\protect\citemethods{2014ApJS..214...15S}, we measure the offset of the host galaxy from the SFMS at the redshift of the transient, and then compute the equivalent SFR at $z=0$ using this offset and ${\mathrm{M}_\ast^{z=0}}$. Having corrected the galaxy properties for redshift evolution, we then perform Monte Carlo simulations to compare distributions of FRB properties with other galaxy samples. We generate $1,000$ samples for each transient's measured property, assuming a normal distribution centered at the median value with a width of reported $1\sigma$ asymmetric errors. We then perform two-sample KS-tests to compute the $p$-value to test the null hypothesis that the two distributions are drawn from the same parent distribution. We repeat this evaluation for all the $1,000$ Monte Carlo simulations and report the median $p$-value.

We note that these corrections for redshift evolution hinge on a number of assumptions. Therefore, we complement the KS-test $p$-values with an independent test statistic to compare different classes of transients without the need for redshift correction. For each event $E_i^j$ of a particular transient class $i$, at redshift $z_i^j$, we simulate $10,000$ background galaxies at the same redshift as the transient. We then compute the fraction of the background population at that redshift with the property $q$ (say, stellar mass or SFR) less than the property of our transient, denoted as $f_{i,q}^j$. Having computed how extreme the locations of these events are with respect to the background galaxy population at their redshifts, we define the test statistic as:

\begin{equation}
    \mathrm{TS}_{q} = -\sum_{j = 0}^{N_i} \log f_{i,q}^j,
\end{equation}

where $N_i$ is the number of events in transient class $i$. The closer the values of this test statistic to the transient under consideration, the higher the similarity between them. We compute these test statistics for $1,000$ Monte Carlo simulations for each transient event, assuming a normal distribution centered at the median value with asymmetric $1\sigma$ errors, and report the median values of the test statistic.

\setlength{\parskip}{12pt}

\noindent{\bfseries \large Dominant Ionization Mechanism}
\setlength{\parskip}{3pt}

The measured nebular emission line fluxes can be used to identify the principal excitation mechanism in the galaxies, such as HII regions (photoionization by O and B stars), planetary nebulae (photoionization by stars much hotter than O-type stars), photoionization by a power-law continuum (Seyfert galaxies) and excitation by shock-wave heating (LINER galaxies). This analysis is an important check of the validity of the SED modeling framework described above, which, for example, assumes that an AGN does not dominate the optical light of a galaxy. The classic Baldwin, Phillips \& Terlevich empirical optical emission-line diagnostic diagrams use the H$\beta$ $\lambda4861$, [OIII] $\lambda5007$, H$\alpha$ $\lambda6563$, [NII] $\lambda 6584$, [SII] $\lambda \lambda6717,31$ and [OI] $\lambda6300$ emission line fluxes to discriminate between different excitation mechanisms~\citep{1981PASP...93....5B}. We use the emission line fluxes measured from stellar continuum subtracted spectra to determine the dominant source of ionization in FRB host galaxies. For host galaxies that lack spectral coverage, we use the emission line measurements as predicted by our best posterior SED fit. We use the theoretical maximum starburst line~\protect\citemethods{2001ApJ...556..121K}, the pure star-forming galaxies vs Seyfert-H II composite objects classification line~\protect\citemethods{2003MNRAS.346.1055K}, and the Seyfert vs LINER classification lines~\citep{2006MNRAS.372..961K} to classify our FRB host galaxies into star-forming, composites, Seyferts and LINERs. The emission line ratios and the classification lines are plotted in Extended Data Fig.~\ref{fig:BPT_analysis}. We also plot the local emission line galaxies with signal-to-noise ratio, SNR $\geq 10$ from SDSS for reference~\citep{2015ApJS..219...12A}. \ks{We note a systematic offset between the star-forming locus and DSA FRBs in Extended Data Fig.~\ref{fig:BPT_analysis}a. This may be due to leakage of H$\alpha$ flux to [NII]. We believe so because with the other two emission line ratio metrics (see Extended Data Fig.~\ref{fig:BPT_analysis}b and \ref{fig:BPT_analysis}c), the DSA FRBs are indeed consistent with the locus of star-forming background galaxies.}

The host galaxies of FRB \frbjackie, FRB \frbnina, and FRB \frbjuan are classified as LINER galaxies, while those of FRB \frbphineas and FRB \frbcharlotte are classified as Seyfert galaxies. \ks{The host galaxy of FRB~\frbcharlotte is a massive star-forming spiral galaxy at its redshift, which are relatively rare. This galaxy exhibits a post-starburst star-formation history, with a rapid decline in its SFR over the past 100 Myr. The BPT emission line diagnostics indicate LINER activity, and the stellar mass of the galaxy suggests the presence of a supermassive black hole (SMBH) with a mass of $\log M_{\mathrm{BH}} = 8.06 \pm 0.31$~\protect\citemethods{2020ARA&A..58..257G}. Given these characteristics, it is plausible that the SMBH is contributing to the observed quenching of star formation via AGN feedback mechanisms. Quenching timescales derived from IllustrisTNG simulations, which incorporate both quasar mode and radio mode AGN feedback, span a broad range from hundreds of Myr to tens of Gyr~\protect\citemethods{2022MNRAS.511.6126W}. Specifically, for galaxies with a similar stellar mass and central SMBH mass, the expected quenching timescale is $\sim 2$~Gyr. This inferred timescale is potentially consistent with the observed reduction in SFR, provided the star formation continues to decline at the rate observed over the past 100 Myr. Thus, while SMBHs are not the sole mechanism for quenching in galaxies, the evidence from both observational diagnostics and simulations supports the hypothesis that AGN feedback may be a significant contributor to the quenching observed in the host galaxy of FRB~\frbcharlotte.} The $\tau$-linear exponentially-declining SFHs of the host galaxies of FRB \frbjackie, FRB \frbjuan, and FRB \frbphineas with low specific SFRs and significant f$_{\mathrm{AGN}}$ estimated by our SED analysis also support the presence of an AGN component in these galaxies. Most FRB hosts align with the star-forming locus of background galaxies, reflecting their high degree of star-formation revealed by SED analysis. \ks{This is consistent with the results of Eftekhari et al. (2023)~\protect\citemethods{2023ApJ...958...66E}, Bhandari et al. (2022)~\protect\citemethods{2022AJ....163...69B} and  Heintz et al. (2020)~\protect\citemethods{2020ApJ...903..152H}.} We statistically quantify these results by performing two-sample two-dimensional KS-tests, where we test the null hypothesis that for a given ionization mechanism, both the FRBs and the background SDSS galaxy population are drawn from the same underlying population. The results from these tests are summarized in Supplementary Table~\ref{table:BPT_analysis}. Employing the same classification scheme for both FRB hosts and background galaxies and comparing FRB hosts with the entire galaxy population, we find that the $p$-value ($p_{\mathrm{KS}}$) exceeds 0.05 for all three classification schemes. This suggests that we cannot statistically rule out the possibility that FRB host galaxies are consistent with the background population of emission line galaxies. Furthermore, FRB host galaxies significantly differ from sub-populations of AGNs (LINERs and Seyferts) with a confidence level exceeding $6\sigma$. However, they are statistically consistent with an underlying population of star-forming+LINER (\ks{consistent with Eftekhari et al. (2023)~\protect\citemethods{2023ApJ...958...66E} results}) and star-forming+Seyfert galaxies ($p_{\mathrm{KS}} \gtrsim 0.05$) across most classification schemes.

\setlength{\parskip}{12pt}

\noindent{\bfseries \large WISE Color-Color Galaxy Classification}
\setlength{\parskip}{3pt}

The median redshift of FRBs is $z = 0.28$. At such high redshifts, it's difficult to get robust morphological classifications with archival ground-based optical imaging data. Indeed, we detect spiral arm features for host galaxies of only 6 events: FRB \frbmark, FRB \frbmizu, FRB \frbzach, FRB \frbsquanto, FRB \frbcharlotte, and FRB \frbmikayla (see Fig.~\ref{fig:host_cutouts}). Hence, we use the WISE color-color classification~\citep{2010AJ....140.1868W} to obtain tentative galaxy classifications (see Extended Data Fig.~\ref{fig:wise_color_color}). We overlay the available WISE data for 22 FRB host galaxies. The stars and early-type galaxies have W1-W2 color close to zero, while the spiral galaxies are red in W2-W3, and ULIRGs tend to be red in both colors. \ks{The WISE color-color galaxy classifications of FRB hosts are broadly consistent with BPT classifications. Majority of FRB host galaxies that were classified as spiral galaxies or luminous infrared galaxies in WISE color-color galaxy classification, are also classified as star-forming or HII regions-AGN composites in BPT diagram.} The majority of FRB host galaxies in our sample are consistent with the locus of spiral galaxies. The only candidate elliptical galaxy is the host of FRB \frbjuan, which has a $\tau$-linear exponentially-declining SFH with a low specific SFR of $0.001_{-0.001}^{+0.001}$ Gyr$^{-1}$, quiescent classification as per its low degree of star-formation \ks{and is a composite/LINER according to BPT classifications}. There are hints for AGN activity in 5 FRBs from our sample: FRB \frbansel (LINER), FRB \frbpingu (starburst/LINER), FRB \frbalex (Seyfert), FRB \frboran (Seyfert), and FRB \frbcharlotte (W1-W2 $\sim 0.5$). \ks{These classifications are consistent with their BPT diagnostics, where FRB \frbansel, FRB \frbpingu, FRB \frbalex and FRB \frboran were classified as AGN-H II regions composites and FRB \frbcharlotte was classified as a borderline case of Composites and Seyferts.}

\setlength{\parskip}{12pt}

\noindent{\bfseries \large Stellar Population Properties}
\setlength{\parskip}{3pt}

We compare the inferred \ks{host} characteristics of repeaters and non-repeaters. The comparison between repeaters, non-repeaters, all FRBs, and the background galaxy population of their derived galaxy properties, such as the stellar mass (log M$_\ast$), SFR averaged over the recent 100~Myr (log SFR$_{\mathrm{100~Myr}}$), specific SFR (sSFR), mass-weighted age ($t_m$), stellar metallicity (log Z/Z$_\odot$), and dust attenuation of old stellar light (A$_{\mathrm{V,~old}}$), is shown in Extended Data Fig.~\ref{fig:all_params_box_plot}. These distributions were constructed using $1,000$ samples of the posterior distributions for each galaxy. The median stellar mass of the host galaxies of repeaters is $\log \mathrm{M}_\ast = 9.64$ ($1\sigma$ inter-quartile range of $9.07$ to $10.18$), while for non-repeaters, the median stellar mass is $\log \mathrm{M}_\ast = 10.20$ ($1\sigma$ inter-quartile range of $9.48$ to $10.93$). We observe that the host galaxies of repeaters are relatively less massive than that of non-repeaters. As a consequence of the mass-metallicity prior in our SED modeling, the metallicity of the host galaxies of repeaters ($\log{\mathrm{Z}/\mathrm{Z}_\odot} = -1.16$, $1\sigma$ inter-quartile range of $-1.79$ to $-0.58$) is also lower than that of non-repeaters ($\log{\mathrm{Z}/\mathrm{Z}_\odot} = -0.45$, $1\sigma$ inter-quartile range of $-1.06$ to $0.07$). Since the SFR ranges probed by repeaters ($ \mathrm{SFR}_\mathrm{100~Myr} = 0.96$ M$_\odot$yr$^{-1}$, $1\sigma$ inter-quartile range of $0.03$ to $2.57$ M$_\odot$yr$^{-1}$) and non-repeaters ($ \mathrm{SFR}_\mathrm{100~Myr} = 1.12$ M$_\odot$yr$^{-1}$, $1\sigma$ inter-quartile range of $0.84$ to $10.90$ M$_\odot$yr$^{-1}$) are comparable, the sSFR of repeaters ($ \mathrm{sSFR} = 0.15$ Gyr$^{-1}$, $1\sigma$ inter-quartile range of $0.01$ to $0.69$ Gyr$^{-1}$) is higher than that of non-repeaters ($ \mathrm{sSFR} = 0.07$ Gyr$^{-1}$, $1\sigma$ inter-quartile range of $0.05$ to $0.33$ Gyr$^{-1}$). To quantify these differences, we perform a two-sample KS-test between stellar properties of the host galaxies of repeaters and non-repeaters. We find $p_{\mathrm{KS}} \geq 0.05$ on comparing $\log \mathrm{M}_\ast$, SFR, sSFR, $\log \mathrm{Z}/\mathrm{Z}_\odot$, A$_\mathrm{V}$, and t$_\mathrm{m}$. Therefore, we find no statistical evidence for differences in host galaxy properties of repeaters and non-repeaters. \ks{These conclusions are similar to the recent results from Gordon et al. (2023)~\citep{2023arXiv230205465G}, where they also find no statistical difference between the hosts of repeaters and non-repeaters, as well as the observation that repeaters may exist in relatively lower mass galaxies than non-repeaters.}

\setlength{\parskip}{12pt}

\noindent{\bfseries \large Galaxy Star-Forming Main Sequence}
\setlength{\parskip}{3pt}

In Extended Data Fig.~\ref{fig:gal_sfms_tm_ssfr}, we show the distribution of FRB host galaxies in the log M$_\ast$ - log SFR$_{\mathrm{100~Myr}}$ space, together with the distribution of background galaxies from 3D-HST~\protect\citemethods{2014ApJS..214...24S} and COSMOS-2015~\protect\citemethods{2016ApJS..224...24L} catalogs in three redshift bins to minimize the effect of the SFMS evolution. The distribution of FRB host offset from the galaxy SFMS at its redshift is centered at $-0.1$~dex with a Gaussian scatter of $0.7$~dex, which is consistent with the scatter of background galaxies about the SFMS. \ks{This observation aligns with the main findings of Gordon et al. (2023)~\citep{2023arXiv230205465G}, who discovered that the host galaxies of FRBs match the characteristics of star-forming galaxies.} Next, we systematically classify our host galaxies as star-forming ($\mathcal{D} > 1/3$), transitioning ($1/20 < \mathcal{D} < 1/3$), and quiescent ($\mathcal{D} < 1/20$) based on their degree of star-formation $\mathcal{D} = \mathrm{sSFR}(z) \times t_{\mathrm{H}}(z)$, where $\mathcal{D}$ is the mass-doubling number~\protect\citemethods{2022ApJ...926..134T}, and $t_{\mathrm{H}}(z)$ is the age of the universe at the redshift of the galaxy. Out of 30 FRB hosts published in this work, 21 are classified as star-forming, 7 as transitioning, and 2 as quiescent galaxies. \ks{The observed majority of star-forming galaxies in our sample is consistent with the significant star-forming galaxies representation in the Gordon et al. (2023)~\citep{2023arXiv230205465G} FRB host galaxies sample.} The quiescent state of the host galaxy of FRB \frbjackie may be a consequence of it being a member of a massive galaxy cluster~\citep{2023ApJ...950..175S}. The quiescent host galaxy of FRB \frbjuan was also classified as a candidate elliptical galaxy in WISE color-color morphological classifications. The quiescent fraction of FRB host galaxies in $z \leq 0.1$ and $0.1 < z \leq 0.3$ bins are 1/9 and 2/23, respectively, which are lower than expected for the background galaxy population in the mass range probed by FRB host galaxies~\citep{2022ApJ...936..165L}.

\setlength{\parskip}{12pt}

\noindent{\bfseries \large Star Formation Histories and Age of the Stellar Population}
\setlength{\parskip}{3pt}

The SFH of the host galaxies of transients can be crucial in inferring the age of their progenitors. The timing and duration of past episodes of star-formation shed light on the processes that shaped their evolution. The galaxies may exhibit different patterns of star-formation over their lifetimes, ranging from intense bursts of star-formation to more gradual and steady processes, which are aptly captured by non-parametric SFHs inferred from the galaxy SED fits. The constrained SFH of FRB host galaxies published in this work are shown in Supplementary Fig.~\ref{fig:host_sfh}. Since most of the FRB host galaxies are 0.1-1~L$_\ast$ galaxies (see Fig.~\ref{fig:frbs_are_biased_tracer_of_sf_in_universe}), it is not surprising that 50\% of the FRB hosts have a delayed-$\tau$ and 16\% of FRB hosts have a $\tau$-linear exponentially-declining SFH. The occurrence of FRBs in galaxies with rejuvenating (10\%), post-starburst (16\%), and rising (8\%) SFHs adds to the diversity of possible delays from the star-formation events. \ks{The majority of exponentially declining SFHs, with a wide variety of other SFHs was also observed by Gordon et al. (2023)~\citep{2023arXiv230205465G}.} This diversity underscores the complex nature of the environments in which FRBs occur and highlights the importance of understanding the detailed SFH of their host galaxies in unraveling the mysteries of FRB sources. \ks{Similar to Gordon et al. (2023)~\citep{2023arXiv230205465G}}, we find that the distribution of FRB hosts, when compared to the background galaxy population in log sSFR - $t_m$ space, which are moments of the SFH, is consistent with the background population (see Extended Data Fig.~\ref{fig:gal_sfms_tm_ssfr}).

\setlength{\parskip}{12pt}

\noindent{\bfseries \large Comparison with Other Transients}
\setlength{\parskip}{3pt}

\textbf{Constructing the Offset Distributions.} 
We compare the offset distribution of FRBs with respect to the host-galaxy light with different classes of transients. We compute the projected physical offsets and the corresponding host normalized offsets using the measurements done for PATH analysis. These offsets for host galaxies in our sample, together with other transients, are plotted in Extended Data Fig.~\ref{fig:offset_distribution_compare_with_other_transients}. Here, we also include the projected offset measurements for 14 FRBs from the literature~\protect\citemethods{2021ApJ...917...75M, 2023arXiv231201578W}. We plot the errors from the 90\% localization region of FRBs where we sample $1,000$ FRB locations within the localization region, while ensuring that we get an equal number of samples away from and closer to the center of the galaxy to avoid any biases from the shape of the localization region. The median offset of FRB location from the center of the host galaxies is 5.54~kpc (1.25 effective galactic radii). The results from two-sample KS-tests between FRBs and different classes of transients are tabulated in Supplementary Table~\ref{table:compare_with_other_transients}.

\textbf{Constructing the Galaxy Properties Distributions.} Comparison of the stellar population properties of the host galaxies can also yield important insights on the similarity of FRBs with different transient classes~\protect\citemethods{2020ApJ...896..142B, 2019ApJ...886..110M, 2017ApJ...843...84N}. Since the redshift distribution of different classes of transients is significantly different, we correct their measured stellar mass and SFR for redshift evolution~\protect\citemethods{2021ApJ...907L..31B}. We directly compare different classes of transients in Extended Data Fig.~\ref{fig:compare_with_other_transients_at_zero_redshift} and report the results from two-sample KS-tests in Supplementary Table~\ref{table:compare_with_other_transients}. Since the conclusions from comparisons of distributions of galaxy properties corrected for redshift evolution hinge on galaxy evolution assumptions, we constructed an independent test statistic to compare transients without the need of redshift correction. We also list the test statistic for all transient classes under consideration in Supplementary Table~\ref{table:compare_with_other_transients}, and we find that conclusions from these test statistics are consistent with our conclusions from KS-tests performed on the redshift-corrected distribution of the host properties.

\ks{\textbf{Inferring the Similarities with Transients:} We find that while the SFR distribution of long-duration GRBs are consistent with FRBs ($p_\mathrm{KS} > 0.05$), other host galaxy properties and offset distribution of FRBs are not consistent with SLSNe and long-duration GRBs ($p_\mathrm{KS} < 0.001$), which are concentrated in the central, star-forming regions of their low-mass host galaxies. Similar conclusions were reached by Bhandari et al. (2020)~\protect\citemethods{2020ApJ...895L..37B}, where they also found that FRBs occur in systematically more massive galaxies than SLSNe and long-duration GRBs. Furthermore, Bhandari et al. (2020)~\protect\citemethods{2020ApJ...895L..37B} and Mannings et al. (2021)~\protect\citemethods{2021ApJ...917...75M} also found that FRBs occur at systematically extended locations within their host galaxies, thus providing evidence that FRBs probably do no originate from sources formed by the deaths of massive stripped envelope stars. Recent work by Woodland et al. (2023)~\protect\citemethods{2023arXiv231201578W} also rejected similarities with SLSNe and long-duration GRBs offsets with high statistical significance.}

\ks{While the projected offset and SFR distributions of ULX sources are consistent with FRBs, they occur in systematically more massive galaxies than FRBs, thus ruling out any similarities with FRBs at high statistical significance ($p_\mathrm{KS} \approx 10^{-6}$). These conclusions are consistent with the previous studies~\protect\citemethods{2020ApJ...895L..37B}.}

\ks{Comparing with CCSNe host galaxy properties, we find that although the SFR and sSFR distributions of CCSNe are consistent with FRBs ($p_\mathrm{KS} > 0.05$), there is a deficit of low-mass galaxies, where we ruled out similarities in stellar mass distributions with more than 95\% confidence ($p_\mathrm{KS}=0.003$). Although the similarities in SFR and sSFR distributions have been observed previously by  Bhandari et al. (2020)~\protect\citemethods{2020ApJ...895L..37B}, in contrast to their results, we find that the CCSNe host stellar mass distribution is not consistent with FRBs, with a deficit of low-mass galaxies. In the Main text, we infer this low-mass host galaxy deficit as a preference of FRBs for metal rich environments, thus supporting the scenario where FRB sources are formed in CCSNe of stellar merger remnants, which are expected to have the requisite magnetic fields to form magnetars. Furthermore, comparing the offset distributions, we find that FRBs occur at systematically larger offsets than typical CCSNe, thus ruling out similarities with more than 95\% confidence ($p_\mathrm{KS} = 0.014$). This is in contrast to previous works, where Bhandari et al. (2020)~\protect\citemethods{2020ApJ...895L..37B}, Mannings et al. (2021)~\protect\citemethods{2021ApJ...917...75M} and Woodland et al. (2023)~\protect\citemethods{2023arXiv231201578W} claim that FRB offsets are consistent with CCSNe. We infer these relatively larger offsets as FRB sources being preferentially formed in stellar mergers, which are a delayed CCSNe formation channel.}

\ks{Since there is no observational evidence of AIC events and binary white dwarf mergers, we assume that the offset distribution of Type Ia supernovae is representative of AIC events and binary white dwarf mergers. Due to small natal kicks of the neutron stars formed through AIC or mergers of white dwarfs, they may not occur with as large offsets as magnetars formed via binary neutron star mergers. We find that the FRB host galaxy properties and offset distribution are consistent with Type Ia supernovae ($p_\mathrm{KS} > 0.05$). These results are consistent with the results from Bhandari et al. (2020)~\protect\citemethods{2020ApJ...895L..37B}, Mannings et al. (2021)~\protect\citemethods{2021ApJ...917...75M} and Woodland et al. (2023)~\protect\citemethods{2023arXiv231201578W}, thus allowing for FRB sources that invoke white dwarf progenitors. Additionally, we find that the host-normalized Type Ia supernovae offsets are systematically at lower physical offsets than our large FRBs sample. This may imply that FRBs select for higher offset subset of Type-Ia supernovae, but any differences in their progenitors at large and small offsets are not characterized in the literature.}

\ks{The larger offsets of FRBs as compared to CCSNe and Type Ia supernovae may as well hint at different formation channels, where larger offsets support delayed formation channels either involving kicks or where the progenitor lifetime is long enough to allow the short-duration GRB sources to travel to outskirts of their host galaxies~\cite{2022ApJ...940...56F}. Similar to Bhandari et al. (2020)~\protect\citemethods{2020ApJ...895L..37B}, Mannings et al. (2021)~\protect\citemethods{2021ApJ...917...75M} and Woodland et al. (2023)~\protect\citemethods{2023arXiv231201578W}, we find that sGRB offset distribution is consistent with FRBs ($p_\mathrm{KS} > 0.05$). However, in contrast to Bhandari et al. (2020)~\protect\citemethods{2020ApJ...895L..37B}, we find that although sGRB SFR and sSFR distributions are consistent with FRBs, the similarity of the stellar mass distribution with FRBs is ruled out at high confidence ($p_\mathrm{KS} = 0.002$).}

\newpage

\vspace{0.5cm} 
\setlength{\parskip}{17pt}%
\bibliographystylemethods{naturemag}
\bibliographymethods{manuscript}
\vspace{-0.5cm}

\newpage

\noindent{\bfseries \LARGE Extended Data}\setlength{\parskip}{12pt}

\newcolumntype{C}{>{\footnotesize\raggedright\arraybackslash}c}

\newcolumntype{L}{>{\footnotesize\raggedright\arraybackslash}l}

\newcolumntype{R}{>{\footnotesize\raggedright\arraybackslash}r}

\begin{table*}[ht!]
    \setlength{\tabcolsep}{3.2pt}
    \centering
    \captionsetup{labelformat=tablelabel}
    \caption{Basic FRB properties.}
    \begin{tabular}{LRRRRRRRLRR}
        \toprule
        
        FRB & RA (FRB) & Decl. (FRB) & RA (Host) & Decl. (Host) & P$_\mathrm{host}$ & $z$ & M$_{\mathrm{AB}}$ & Filter & E(B-V) & Ref. \\
        & [J2000] & [J2000] & [J2000] & [J2000] & & & [mag] & & & \\

        \hline
        
        \frbmark & 02:08:42.70 & +71:02:06.94 & 02:08:40.86 & +71:02:09.15 & 0.99 & 0.0112 & 13.25 & PS1 $r$ & 0.806 & \protect\citemethods{2023arXiv230101000R}$^,$\citep{law2023deep} \\

        (Mark) & $\pm$ 0.58 & $\pm$ 0.55 & & & & $\pm$ 0.0001 & $\pm$ 0.01 & & & \\
        
        \frbmizu & 09:35:56.15 & +73:17:04.80 & 09:35:56.29 & 73:17:05.80 & 0.99 & 0.0368 & 15.41 & BASS $r$ & 0.040 & This Work \\

        (Mizu) & $\pm$ 1.10 & $\pm$ 1.00 & & & & $\pm$ 0.0001 & $\pm$ 0.01 & & & \\

        \frbzach & 20:40:47.89 & +72:52:56.38 & 20:40:47.42 & +72:52:57.90 & 0.97 & 0.0433 & 16.15 & PS1 $r$ & 0.813 & \citep{law2023deep} \\

        (Zach) & $\pm$ 0.63 & $\pm$ 0.54 & & & & $\pm$ 0.0001 & $\pm$ 0.01 & & & \\

        \frbjackie & 18:50:40.80 & +70:14:37.80 & 18:50:41.92 & +70:14:33.95 & 0.97 & 0.0894 & 16.51 &  BASS $r$ & 0.063 & \citep{2023ApJ...950..175S}$^,$\citep{law2023deep} \\

        (Jackie) & $\pm$ 2.20 & $\pm$ 1.50 & & & & $\pm$ 0.0001 & $\pm$ 0.21 & & & \\

        \frbleonidas & 15:27:39.90 & +70:58:05.20 & 15:27:39.80 & +70:58:05.62 & 0.99 & 0.0939 & 19.10 & BASS $r$ & 0.023 & This Work \\

        (Leonidas) & $\pm$ 1.40 & $\pm$ 0.90 & & & & $\pm$ 0.0002 & $\pm$ 0.20 & & & \\
        
        \frbelektra & 18:48:13.63 & +73:20:12.89 & 18:48:13.97 & +73:20:10.68 & 0.97 & 0.1139 & 20.20 & BASS $r$ & 0.077 & \citep{2023ApJ...950..175S}$^,$\citep{law2023deep} \\

        (Elektra) & $\pm$ 0.93 & $\pm$ 0.73 & & & & $\pm$ 0.0001 & $\pm$ 0.19 & & & \\

        \frbtildis & 11:07:08.81 & +72:16:54.64 & 11:07:08.72 & +72:16:53.01 & 0.95 & 0.1270 & 19.94 & BASS $r$ & 0.037 & This Work \\

        (Tildis) & $\pm$ 1.40 & $\pm$ 1.00 & & & & $\pm$ 0.0001 & $\pm$ 0.20 & & & \\

        \frbetienne & 16:01:01.70 & +70:55:07.68 & 16:01:01.63 & +70:55:05.05 & 0.98 & 0.1582 & 19.43 & BASS $r$ & 0.028 & \citep{law2023deep} \\

        (Etienne) & $\pm$ 1.02 & $\pm$0.60 & & & & $\pm$ 0.0002 & $\pm$ 0.20 & & & \\

        \frbnina & 22:48:51.89 & +70:40:52.20 & 22:48:51.71 & +70:40:49.69 & 0.99 & 0.2395 & 18.81 & PS1 $r$ & 0.524 & This Work \\
        (Nina) & $\pm$ 1.20 & $\pm$ 0.90 & & & & $\pm$ 0.0026 & $\pm$ 0.02 & & & \\

        \frbansel & 20:47:55.55 & +72:35:05.89 & 20:47:55.60 & +72:35:06.50 & 0.99 & 0.2414 & 20.74 & PS1 $r$ & 0.543 & \citep{law2023deep} \\

        (Ansel) & $\pm$ 0.78 & $\pm$ 0.69 & & & & $\pm$ 0.0001 & $\pm$ 0.16 & & & \\

        \frbalex & 23:23:29.88 & +72:11:32.59 & 23:23:29.92 & +72:11:30.80 & 0.98 & 0.2481 & 20.45 & PS1 $r$ & 0.438 & \citep{law2023deep} \\

        (Alex) & $\pm$ 1.72 & $\pm$ 1.26 & & & & $\pm$ 0.0001 & $\pm$ 0.13 & & & \\

        \frbisha & 04:45:38.64 & +70:18:26.60 & 04:45:38.82 & +70:18:26.60 & 0.99 & 0.2505 & 21.38 & PS1 $r$ & 0.186 & This Work \\

        (Isha) & $\pm$ 1.40 & $\pm$ 0.90 & & & & $\pm$ 0.0001 & $\pm$ 0.10 & & & \\

        \frbsquanto & 16:10:09.16 & +70:47:06.20 & 16:10:09.586 & +70:47:09.020 & 0.90 & 0.2621 & 18.53 & BASS $r$ & 0.030 & This Work \\

        (Squanto) & $\pm$ 1.20 & $\pm$ 0.60 & & & & $\pm$ 0.0001 & $\pm$ 0.20 & & & \\

        \frbphineas & 11:51:07.52 & +71:41:44.30 & 11:51:07.81 & +71:41:42.99 & 0.97 & 0.2706 & 19.48 & BASS $r$ & 0.015 & This Work \\

        (Phineas) & $\pm$ 0.20 & $\pm$ 0.20 & & & & $\pm$ 0.0001 & $\pm$ 0.19 & & & \\

        \frbcharlotte & 01:24:50.45 & +72:39:14.10 & 01:24:49.67 &  +72:39:11.56 & 0.94 & 0.2764 & 19.01 & WaSP $r$ & 0.423 & This Work \\

        (Charlotte) & $\pm$ 1.00 & $\pm$ 0.60 & & & & $\pm$ 0.0002 & $\pm$ 0.03 & & & \\

        \frbjuan & 18:43:11.69 & +70:31:27.15 & 18:43:12.34 & +70:31:27.06 & 0.99 & 0.2847 & 19.64 & BASS $r$ & 0.057 & \citep{law2023deep} \\

        (Juan) & $\pm$ 1.12 & $\pm$ 0.74 & & & & $\pm$ 0.0001 & $\pm$ 0.19 & & & \\

        \frboran & 21:12:10.76 & +72:49:38.20 & 21:12:10.73 & +72:49:37.76 & 0.99 & 0.300 & 19.91 & WaSP $r$ & 0.806 & This Work \\

        (Oran) & $\pm$ 1.11 & $\pm$ 0.81 & & & & $\pm$ 0.0001 & $\pm$ 0.02 & & & \\

        \frbmikayla & 22:40:06.52 & +70:55:19.82 & 22:40:06.67 & +70:55:20.16 & 0.99 & 0.3015 & 21.86 & WaSP $r$ & 0.585 & This Work \\

        (Mikayla) & $\pm$ 0.90 & $\pm$ 0.80 & & & & $\pm$ 0.0011 & $\pm$ 0.01 & & & \\

        \frbfatima & 15:42:31.10 & +71:08:00.77 & 15:42:31.24 & +71:08:01.18 & 0.99 & 0.3270 & 19.97 & BASS $r$ & 0.806 & This Work \\

        (Fatima) & $\pm$ 1.80 & $\pm$ 1.40 & & & & $\pm$ 0.0001 & $\pm$ 0.19 & & & \\

        \frbishita & 21:30:18.03 & +70:02:27.75 & 21:30:17.28 & 70:02:31.20 & 0.56 & 0.3510 & 20.80 & WIRC $J$ & 0.809 & This Work \\

        (Ishita) & $\pm$ 0.99 & $\pm$ 0.64 & & & & $\pm$ 0.0003 & $\pm$ 0.79 & & & \\
        
        \hline
    \end{tabular}
    \label{table:basic_frb_properties}
\end{table*}

\newpage

\begin{table*}[ht!]
    \ContinuedFloat
    \setlength{\tabcolsep}{3.2pt}
    \centering
    \captionsetup{labelformat=tablelabel}
    \caption{Basic FRB properties {\textit{(cont.)}}.}
    \begin{tabular}{LRRRRRRRLRR}
        \toprule
        
        FRB & RA (FRB) & Decl. (FRB) & RA (Host) & Decl. (Host) & P$_\mathrm{host}$ & $z$ & M$_{\mathrm{AB}}$ & Filter & E(B-V) & Ref. \\
        & [J2000] & [J2000] & [J2000] & [J2000] & & & [mag] & & & \\

        \hline

        \frbgertrude & 04:55:46.96 & +69:55:44.80 & 4:55:46.83 & +69:55:43.32 & 0.99 & 0.3619 & 20.95 & PS1 $r$ & 0.130 & This Work \\

        (Gertrude) & $\pm$ 0.80 & $\pm$ 0.60 & & & & $\pm$ 0.0004 & $\pm$ 0.19 & & & \\

        \frberdos & 10:55:00.30 & +70:21:02.70 & 10:54:59.74 & +70:21:01.16 & 0.63 & 0.3714 & 23.35 & BASS $r$ & 0.020 & This Work \\

        (Erdos) & $\pm$ 1.62 & $\pm$ 0.97 & & & & $\pm$ 0.0002 & $\pm$ 0.20 & & & \\

        \frbfen & 18:16:54.30 & +69:43:21.01 & 18:16:54.30 & +69:43:21.01 & 0.99 & 0.4012 & 23.00 & WASP $r$ & 0.064 & This Work \\

        (Fen) & $\pm$ 1.56 & $\pm$ 0.93 & & & & $\pm$ 0.0003 & $\pm$ 0.02 & & & \\

        \frbpingu & 11:09:26.05 & +72:33:28.02 & 11:09:26.48 & +72:33:28.84 & 0.99 & 0.4525 & 21.36 & BASS $r$ & 0.040 & This Work \\
        (Pingu) & $\pm$ 1.40 & $\pm$ 0.90 & & & & $\pm$ 0.0001 & $\pm$ 0.16 & & & \\

        \frbwhitney & 08:58:52.92 & +73:29:27.00 & 8:58:53.05 & +73:29:27.44 & 0.97 & 0.4780 & 21.20 & PS1 $r$ & 0.029 & \citep{law2023deep} \\

        (Whitney) & $\pm$ 1.90 & $\pm$ 1.40 & & & & $\pm$ 0.0004 & $\pm$ 0.08 & & & \\

        \frbferb & 10:25:53.32 & +03:26:12.57 & 10:25:53.56 & +3:26:12.60 & 0.42 & 0.5310 & 21.38 & BASS $r$ & 0.031 & This Work \\

        (Ferb) & $\pm$ 1.20 & $\pm$ 0.70 & & & & $\pm$ 0.0003 & $\pm$ 0.20 & & & \\

        \frbkoyaanisqatsi & 08:43:29.23 & +72:06:03.50 & 08:43:29.52 & 72:06:03.24 & 0.62 & 0.5422 & 22.84 & BASS $r$ & 0.037 & This Work \\

        (Koy.) & $\pm$ 0.70 & $\pm$ 0.80 & & & & $\pm$ 0.0003 & $\pm$ 0.19 & & & \\

        \frbnihari & 17:10:31.15 & +71:37:36.63 & 17:10:31.15 & +71:37:36.63 & 0.99 & 0.5530 & 22.61 & BASS $r$ & 0.038 & This Work \\

        (Nihari) & $\pm$ 1.54 & $\pm$ 0.93 & & & & $\pm$ 0.0002 & $\pm$ 0.20 & & & \\

        \frbquincy & 14:36:25.34 & +70:05:45.40 & 14:36:25.58 & +70:05:43.60 & 0.97 & 0.6214 & 21.22 & BASS $r$ & 0.019 & \citep{law2023deep} \\

        (Quincy) & $\pm$ 1.36 & $\pm$ 0.76 & & & & $\pm$ 0.0001 & $\pm$ 0.19 & & & \\

        \frbmifanshan & 09:27:51.22 & +72:27:08.34 & 09:27:51.91 & +72:27:07.20 & 0.92 & 0.9750 & 23.17 & BASS $r$ & 0.050 & This Work \\
        
        (Mifanshan) & $\pm$ 0.70 & $\pm$ 0.80 & & & & $\pm$ 0.0006 & $\pm$ 0.22 & & & \\

        \hline
    \end{tabular}
    \footnotesize{{\raggedright N{\scriptsize{OTE}}: The second row for each FRB lists the 90\% localization uncertainties. P$_{\mathrm{host}}$ represents the PATH FRB host association probability computed on the  optical/IR filter listed in the table (see Methods). The listed AB magnitude of the galaxy are computed for the same filter and corrected for the Milky Way galactic dust extinction~\protect\citemethods{2018JOSS....3..695M, 1999PASP..111...63F}. The redshifts of FRB hosts are measured using the \sw{pPXF} software (see Supplementary Table~\ref{table:host_spectroscopy_details} for spectroscopy details). 
    \par}}
\end{table*}

\newpage

\begin{table*}[ht!]
    \centering
    \setlength{\tabcolsep}{3.2pt}
    \captionsetup{labelformat=tablelabel}
    \caption{Derived galaxy properties.}
    \begin{tabular}{LLRRRRRRRRR}
        \toprule
        FRB & Class & log M$_\ast$ & log(Z/Z$_\odot$) & A$_{\rm{V}, \rm{young}}$ & A$_{\rm{V}, \rm{old}}$ & U$_{\mathrm{neb}}$ & SFR$_{\mathrm{20 Myr}}$ & SFR$_{\mathrm{100 Myr}}$ & sSFR & t$_\mathrm{m}$ \\

        & & [M$_\odot$] & & [mag] & [mag] & & [M$_\odot$ yr$^{-1}$] & [M$_\odot$ yr$^{-1}$] & [Gyr$^{-1}$] & [Gyr] \\
        
        & AGN & f$_{\rm{AGN}}$ & $\tau_{\rm{}AGN}$ & q$_{\rm{PAH}}$ & M$_r^0$ & (g-r)$^0$ & (u-r)$^0$ & D$_{\mathrm{L}}$ & d & R$_{\mathrm{e}}$ \\
        
        & & & & & [mag] & [mag] & [mag] & [Mpc] & [kpc] & [kpc] \\

        \hline 

        \frbmark & T & 10.10$_{-0.00}^{+0.00}$ & 0.19$_{-0.00}^{+0.00}$ & 0.36$_{-0.07}^{+0.08}$ & 0.34$_{-0.01}^{+0.01}$ & -3.20$_{-0.02}^{+0.01}$ & 0.38$_{-0.01}^{+0.01}$ & 0.42$_{-0.02}^{+0.02}$ & 0.02$_{-0.00}^{+0.00}$ & 6.12$_{-0.06}^{+0.07}$ \\ 
        (Mark) & Y & 0.01$_{-0.00}^{+0.00}$ & 73.73$_{-4.52}^{+5.20}$ & 5.18$_{-0.44}^{+0.30}$ & -20.24$_{-0.09}^{+0.10}$ & 0.44$_{-0.00}^{+0.00}$ & 1.66$_{-0.01}^{+0.01}$ & 48.86 & 2.14$_{-0.21}^{+0.23}$ & 2.09 \\ 
        
        \frbmizu & T & 10.40$_{-0.00}^{+0.00}$ & -1.12$_{-0.01}^{+0.01}$ & 0.54$_{-0.04}^{+0.02}$ & 0.72$_{-0.01}^{+0.01}$ & -3.01$_{-0.02}^{+0.02}$ & 0.70$_{-0.02}^{+0.01}$ & 0.40$_{-0.03}^{+0.02}$ & 0.01$_{-0.00}^{+0.00}$ & 7.54$_{-0.08}^{+0.14}$ \\ 
        (Mizu) & Y & 0.01$_{-0.00}^{+0.00}$ & 96.59$_{-15.37}^{+24.62}$ & 0.75$_{-0.06}^{+0.08}$ & -20.41$_{-0.05}^{+0.38}$ & 0.89$_{-0.01}^{+0.00}$ & 2.69$_{-0.02}^{+0.02}$ & 163.69 & 0.87$_{-0.60}^{+1.36}$ & 5.48 \\ 
        
        \frbzach & SF & 9.95$_{-0.03}^{+0.03}$ & -0.70$_{-0.05}^{+0.06}$ & 0.11$_{-0.02}^{+0.03}$ & 0.10$_{-0.02}^{+0.02}$ & -3.24$_{-0.02}^{+0.02}$ & 0.49$_{-0.02}^{+0.02}$ & 0.71$_{-0.13}^{+0.12}$ & 0.05$_{-0.01}^{+0.01}$ & 6.12$_{-0.48}^{+0.45}$ \\ 
        (Zach) & Y & 0.00$_{-0.00}^{+0.00}$ & 46.76$_{-29.70}^{+53.66}$ & 3.24$_{-0.95}^{+0.80}$ & -20.33$_{-0.21}^{+0.17}$ & 0.38$_{-0.01}^{+0.01}$ & 2.7$_{-0.02}^{+0.02}$ & 193.53 & 2.21$_{-0.77}^{+0.84}$ & 3.6 \\ 
        
        \frbjackie & Q & 10.70$_{-0.01}^{+0.01}$ & -0.58$_{-0.04}^{+0.03}$ & 0.05$_{-0.02}^{+0.03}$ & 0.05$_{-0.02}^{+0.02}$ & -3.74$_{-0.08}^{+0.07}$ & 0.27$_{-0.01}^{+0.01}$ & 0.25$_{-0.04}^{+0.07}$ & 0.00$_{-0.00}^{+0.00}$ & 7.75$_{-0.18}^{+0.18}$ \\ 
        (Jackie) & Y & 0.04$_{-0.01}^{+0.01}$ & 124.17$_{-22.90}^{+17.46}$ & 3.99$_{-1.69}^{+1.07}$ & -21.40$_{-0.05}^{+0.11}$ & 0.72$_{-0.01}^{+0.01}$ & 2.16$_{-0.01}^{+0.02}$ & 412.95 & 11.54$_{-5.02}^{+5.76}$ & 8.31 \\ 
        
        \frbleonidas & SF & 9.46$_{-0.00}^{+0.00}$ & -0.61$_{-0.01}^{+0.01}$ & 0.00$_{-0.00}^{+0.00}$ & 0.03$_{-0.00}^{+0.00}$ & -3.42$_{-0.00}^{+0.00}$ & 0.09$_{-0.00}^{+0.00}$ & 0.75$_{-0.04}^{+0.04}$ & 0.16$_{-0.01}^{+0.01}$ & 6.85$_{-0.07}^{+0.06}$ \\ 
        (Leonidas) & Y & 0.28$_{-0.01}^{+0.01}$ & 6.95$_{-0.40}^{+0.42}$ & 2.21$_{-1.07}^{+1.39}$ & -18.99$_{-0.12}^{+0.13}$ & 0.43$_{-0.00}^{+0.00}$ & 1.41$_{-0.01}^{+0.02}$ & 435.08 & 1.14$_{-0.81}^{+3.73}$ & 1.53 \\ 
        
        \frbelektra & SF & 9.24$_{-0.04}^{+0.08}$ & -0.44$_{-0.12}^{+0.08}$ & 0.71$_{-0.12}^{+0.15}$ & 0.58$_{-0.11}^{+0.15}$ & -3.31$_{-0.04}^{+0.21}$ & 0.25$_{-0.05}^{+0.06}$ & 0.46$_{-0.06}^{+0.05}$ & 0.17$_{-0.03}^{+0.03}$ & 4.55$_{-0.28}^{+0.29}$ \\ 
        (Elektra) & Y & 0.10$_{-0.02}^{+0.03}$ & 11.95$_{-4.70}^{+5.57}$ & 1.43$_{-0.59}^{+0.81}$ & -18.44$_{-0.31}^{+0.26}$ & 0.56$_{-0.02}^{+0.02}$ & 1.75$_{-0.08}^{+0.07}$ & 534.94 & 5.54$_{-2.51}^{+2.94}$ & 2.24 \\ 
        
        \frbtildis & SF & 9.29$_{-0.03}^{+0.03}$ & -0.81$_{-0.14}^{+0.18}$ & 0.22$_{-0.08}^{+0.09}$ & 0.20$_{-0.07}^{+0.07}$ & -3.48$_{-0.04}^{+0.04}$ & 0.14$_{-0.01}^{+0.02}$ & 0.14$_{-0.03}^{+0.04}$ & 0.04$_{-0.01}^{+0.01}$ & 4.82$_{-0.50}^{+0.56}$ \\ 
        (Tildis) & Y & 0.05$_{-0.02}^{+0.03}$ & 25.26$_{-10.58}^{+19.26}$ & 2.14$_{-1.03}^{+1.19}$ & -18.70$_{-0.24}^{+0.23}$ & 0.49$_{-0.02}^{+0.02}$ & 1.61$_{-0.05}^{+0.03}$ & 601.66 & 3.85$_{-2.55}^{+2.94}$ & 2.43 \\ 
        
        \frbetienne & SF & 9.87$_{-0.01}^{+0.01}$ & -0.37$_{-0.08}^{+0.07}$ & 0.47$_{-0.09}^{+0.09}$ & 0.56$_{-0.03}^{+0.03}$ & -3.24$_{-0.07}^{+0.06}$ & 0.42$_{-0.04}^{+0.04}$ & 1.62$_{-0.21}^{+0.20}$ & 0.14$_{-0.02}^{+0.02}$ & 4.27$_{-0.29}^{+0.27}$ \\ 
        (Etienne) & Y & 0.10$_{-0.01}^{+0.01}$ & 5.34$_{-0.24}^{+0.36}$ & 5.24$_{-0.63}^{+0.76}$ & -20.02$_{-0.42}^{+0.11}$ & 0.53$_{-0.01}^{+0.01}$ & 1.52$_{-0.02}^{+0.02}$ & 764.62 & 7.32$_{-2.40}^{+2.97}$ & 4.56 \\ 
        
        \frbnina & SF & 11.21$_{-0.02}^{+0.03}$ & 0.13$_{-0.07}^{+0.04}$ & 0.88$_{-0.34}^{+0.24}$ & 1.50$_{-0.05}^{+0.06}$ & -3.02$_{-0.26}^{+0.30}$ & 2.85$_{-0.68}^{+0.68}$ & 12.27$_{-4.31}^{+4.71}$ & 0.05$_{-0.02}^{+0.02}$ & 5.49$_{-0.24}^{+0.33}$ \\ 
        (Nina) & Y & 0.00$_{-0.00}^{+0.00}$ & 27.48$_{-15.58}^{+48.06}$ & 0.73$_{-0.15}^{+0.17}$ & -21.15$_{-0.10}^{+0.30}$ & 1.09$_{-0.04}^{+0.02}$ & 2.42$_{-0.11}^{+0.09}$ & 1,214.93 & 10.16$_{-5.10}^{+6.23}$ & 4.75 \\ 
        
        \frbansel & SF & 10.01$_{-0.06}^{+0.06}$ & 0.09$_{-0.08}^{+0.05}$ & 1.54$_{-0.21}^{+0.26}$ & 1.50$_{-0.07}^{+0.06}$ & -3.29$_{-0.05}^{+0.11}$ & 9.62$_{-2.00}^{+2.68}$ & 7.91$_{-0.94}^{+1.26}$ & 0.31$_{-0.07}^{+0.09}$ & 5.15$_{-0.31}^{+0.32}$ \\ 
        (Ansel) & Y & 0.00$_{-0.00}^{+0.00}$ & 39.02$_{-16.22}^{+22.59}$ & 0.78$_{-0.20}^{+0.35}$ & -19.35$_{-0.39}^{+0.76}$ & 0.76$_{-0.02}^{+0.04}$ & 1.58$_{-0.06}^{+0.09}$ & 1,225.88 & 2.52$_{-1.76}^{+4.86}$ & 4.12 \\ 

        \frbalex & SF & 10.14$_{-0.04}^{+0.03}$ & -1.51$_{-0.12}^{+0.07}$ & 1.57$_{-0.32}^{+0.33}$ & 1.87$_{-0.08}^{+0.07}$ & -2.85$_{-0.11}^{+0.12}$ & 2.66$_{-0.65}^{+0.82}$ & 3.17$_{-1.03}^{+1.35}$ & 0.09$_{-0.03}^{+0.04}$ & 3.64$_{-0.43}^{+0.45}$ \\     
        (Alex) & N & --- & --- & --- & -19.84$_{-0.13}^{+0.43}$ & 0.77$_{-0.02}^{+0.02}$ & 1.69$_{-0.06}^{+0.07}$ & 1,264.64 & 7.09$_{-4.65}^{+9.80}$ & 3.70 \\ 
        
        \frbisha & SF & 9.48$_{-0.04}^{+0.04}$ & -0.00$_{-0.22}^{+0.13}$ & 0.28$_{-0.12}^{+0.14}$ & 0.29$_{-0.11}^{+0.10}$ & -3.19$_{-0.06}^{+0.05}$ & 0.13$_{-0.03}^{+0.03}$ & 0.24$_{-0.07}^{+0.08}$ & 0.05$_{-0.01}^{+0.01}$ & 4.16$_{-0.96}^{+0.45}$ \\ 
        (Isha) & N & --- & --- & --- & -18.81$_{-0.64}^{+0.23}$ & 0.83$_{-0.03}^{+0.03}$ & 1.93$_{-0.07}^{+0.08}$ & 1,278.58 & 3.62$_{-2.66}^{+8.74}$ & 2.93 \\ 

        \frbsquanto & T & 11.04$_{-0.01}^{+0.01}$ & 0.19$_{-0.00}^{+0.00}$ & 0.37$_{-0.08}^{+0.03}$ & 0.48$_{-0.01}^{+0.02}$ & -3.08$_{-0.01}^{+0.01}$ & 2.78$_{-0.16}^{+0.08}$ & 4.85$_{-0.21}^{+0.22}$ & 0.03$_{-0.00}^{+0.00}$ & 4.86$_{-0.09}^{+0.10}$ \\ 
        (Squanto) & Y & 0.14$_{-0.03}^{+0.03}$ & 62.70$_{-17.32}^{+27.05}$ & 3.68$_{-0.50}^{+0.51}$ & -21.85$_{-0.19}^{+0.09}$ & 1.3$_{-0.01}^{+0.01}$ & 3.11$_{-0.03}^{+0.03}$ & 1,346.36 & 14.40$_{-4.77}^{+6.78}$ & 14.57 \\ 
        
        \frbphineas & T & 10.76$_{-0.02}^{+0.03}$ & -0.00$_{-0.03}^{+0.02}$ & 0.10$_{-0.05}^{+0.05}$ & 0.10$_{-0.04}^{+0.04}$ & -2.49$_{-0.08}^{+0.09}$ & 0.30$_{-0.03}^{+0.04}$ & 0.46$_{-0.06}^{+0.06}$ & 0.01$_{-0.00}^{+0.00}$ & 7.07$_{-0.40}^{+0.27}$ \\ 
        (Phineas) & Y & 0.24$_{-0.05}^{+0.05}$ & 106.96$_{-26.52}^{+25.17}$ & 4.75$_{-2.13}^{+1.12}$ & -20.88$_{-0.11}^{+0.09}$ & 1.26$_{-0.01}^{+0.01}$ & 2.77$_{-0.03}^{+0.03}$ & 1,396.47 & 7.93$_{-1.27}^{+1.39}$ & 3.98 \\ 
        
        \frbcharlotte & SF & 11.01$_{-0.02}^{+0.02}$ & 0.04$_{-0.05}^{+0.07}$ & 1.87$_{-0.50}^{+0.39}$ & 2.26$_{-0.09}^{+0.07}$ & -3.31$_{-0.24}^{+0.22}$ & 3.27$_{-1.05}^{+1.42}$ & 22.58$_{-6.06}^{+6.23}$ & 0.14$_{-0.04}^{+0.04}$ & 5.64$_{-0.17}^{+0.52}$ \\ 
        (Charlotte) & Y & 0.00$_{-0.00}^{+0.00}$ & 70.81$_{-28.79}^{+43.75}$ & 0.73$_{-0.16}^{+0.19}$ & -19.95$_{-0.15}^{+0.63}$ & 1.17$_{-0.02}^{+0.04}$ & 2.28$_{-0.06}^{+0.11}$ & 1,430.87 & 18.41$_{-5.49}^{+6.58}$ & 10.51 \\ 
        
        \frbjuan & Q & 10.96$_{-0.02}^{+0.02}$ & 0.18$_{-0.02}^{+0.01}$ & 0.21$_{-0.04}^{+0.05}$ & 0.21$_{-0.04}^{+0.05}$ & -3.16$_{-0.34}^{+0.29}$ & 0.19$_{-0.06}^{+0.08}$ & 0.18$_{-0.07}^{+0.10}$ & 0.00$_{-0.00}^{+0.01}$ & 6.74$_{-0.21}^{+0.26}$ \\ 
        (Juan) & Y & 0.79$_{-0.16}^{+0.17}$ & 124.86$_{-23.25}^{+17.34}$ & 5.03$_{-1.14}^{+1.10}$ & -20.93$_{-0.13}^{+0.14}$ & 1.52$_{-0.01}^{+0.01}$ & 3.4$_{-0.07}^{+0.07}$ & 1,480.38 & 14.14$_{-7.16}^{+8.15}$ & 5.48 \\ 
        
        \frboran & T & 10.45$_{-0.03}^{+0.01}$ & -0.12$_{-0.06}^{+0.06}$ & 0.11$_{-0.04}^{+0.04}$ & 0.14$_{-0.04}^{+0.05}$ & -2.88$_{-0.05}^{+0.05}$ & 1.02$_{-0.06}^{+0.07}$ & 1.20$_{-0.17}^{+0.17}$ & 0.03$_{-0.01}^{+0.01}$ & 4.77$_{-0.16}^{+0.20}$ \\ 
        (Oran) & Y & 2.68$_{-0.20}^{+0.18}$ & 123.20$_{-19.96}^{+17.76}$ & 3.31$_{-1.59}^{+1.82}$ & -20.9$_{-0.16}^{+0.09}$ & 0.97$_{-0.01}^{+0.01}$ & 1.87$_{-0.03}^{+0.03}$ & 1,574.97 & 2.08$_{-1.48}^{+7.75}$ & 6.72 \\ 
        
        \frbmikayla & SF & 10.29$_{-0.02}^{+0.0}$ & -2.00$_{-0.00}^{+0.00}$ & 0.32$_{-0.08}^{+0.08}$ & 3.43$_{-0.02}^{+0.02}$ & -3.3$_{-0.01}^{+0.01}$ & 1.52$_{-0.11}^{+0.10}$ & 4.10$_{-0.45}^{+0.60}$ & 0.15$_{-0.02}^{+0.02}$ & 0.60$_{-0.01}^{+0.01}$ \\ 
        (Mikayla) & N & --- & --- & --- & -18.09$_{-0.52}^{+0.76}$ & 1.04$_{-0.00}^{+0.00}$ & 8.89$_{-0.00}^{+0.01}$ & 1,581.64 & 3.74$_{-2.56}^{+6.68}$ & 8.05 \\ 
        
        \frbfatima & T & 10.44$_{-0.04}^{+0.04}$ & -0.72$_{-0.08}^{+0.10}$ & 0.29$_{-0.08}^{+0.09}$ & 0.29$_{-0.06}^{+0.08}$ & -3.10$_{-0.08}^{+0.08}$ & 1.04$_{-0.12}^{+0.15}$ & 0.98$_{-0.24}^{+0.33}$ & 0.02$_{-0.01}^{+0.01}$ & 5.41$_{-0.36}^{+0.29}$ \\ 
        (Fatima) & Y & 0.68$_{-0.38}^{+0.21}$ & 103.25$_{-43.07}^{+31.04}$ & 2.55$_{-1.40}^{+1.82}$ & -20.79$_{-0.31}^{+0.28}$ & 1.19$_{-0.03}^{+0.03}$ & 2.39$_{-0.13}^{+0.13}$ & 1,737.93 & 3.71$_{-2.61}^{+14.44}$ & 1.96 \\ 
        
        \frbishita & SF & 10.08$_{-0.02}^{+0.02}$ & -1.86$_{-0.09}^{+0.13}$ & 0.63$_{-0.06}^{+0.05}$ & 0.72$_{-0.04}^{+0.04}$ & -3.1$_{-0.05}^{+0.02}$ & 0.73$_{-0.05}^{+0.09}$ & 0.68$_{-0.08}^{+0.07}$ & 0.04$_{-0.01}^{+0.00}$ & 5.21$_{-0.15}^{+0.23}$ \\ 
        (Ishita) & N & --- & --- & --- & -19.68$_{-0.12}^{+0.55}$ & 1.27$_{-0.02}^{+0.01}$ & 2.57$_{-0.07}^{+0.03}$ & 1,887.81 & 25.88$_{-6.34}^{+7.44}$ & 3.86 \\ 
        
        \hline
    \end{tabular}
    \label{table:derived_galaxy_properties}
\end{table*}

\newpage

\begin{table*}[ht!]
    \ContinuedFloat
    \centering
    \setlength{\tabcolsep}{3.2pt}
    \captionsetup{labelformat=tablelabel}
    \caption{Derived galaxy properties {\textit{(cont.)}}.}
    \begin{tabular}{LLRRRRRRRRR}
        \toprule
        FRB & Class & log M$_\ast$ & log(Z/Z$_\odot$) & A$_{\rm{V}, \rm{young}}$ & A$_{\rm{V}, \rm{old}}$ & U$_{\mathrm{neb}}$ & SFR$_{\mathrm{20 Myr}}$ & SFR$_{\mathrm{100 Myr}}$ & sSFR & t$_\mathrm{m}$ \\

        & & [M$_\odot$] & & [mag] & [mag] & & [M$_\odot$ yr$^{-1}$] & [M$_\odot$ yr$^{-1}$] & [Gyr$^{-1}$] & [Gyr] \\
        
        & AGN & f$_{\rm{AGN}}$ & $\tau_{\rm{}AGN}$ & q$_{\rm{PAH}}$ & M$_r^0$ & $(g-r)^0$ & $(u-r)^0$ & D$_{\mathrm{L}}$ & d & R$_{\mathrm{e}}$ \\
        
        & & & & & [mag] & [mag] & [mag] & [Mpc] & [kpc] & [kpc] \\ 

        \hline 

        \frbgertrude & T & 10.18$_{-0.03}^{+0.04}$ & 0.10$_{-0.11}^{+0.07}$ & 0.29$_{-0.12}^{+0.13}$ & 0.27$_{-0.11}^{+0.11}$ & -3.25$_{-0.11}^{+0.12}$ & 0.45$_{-0.08}^{+0.10}$ & 0.71$_{-0.16}^{+0.19}$ & 0.03$_{-0.01}^{+0.01}$ & 5.35$_{-0.36}^{+0.52}$ \\ 
        (Gertrude) & Y & 0.65$_{-0.21}^{+0.39}$ & 10.15$_{-3.89}^{+9.04}$ & 5.61$_{-1.19}^{+0.94}$ & -19.8$_{-0.18}^{+0.29}$ & 1.28$_{-0.04}^{+0.07}$ & 2.33$_{-0.13}^{+0.18}$ & 1,956.75 & 8.33$_{-4.48}^{+5.80}$ & 2.76 \\ 

        \frberdos & SF & 10.50$_{-0.01}^{+0.02}$ & -0.69$_{-0.01}^{+0.01}$ & 0.03$_{-0.02}^{+0.02}$ & 3.85$_{-0.02}^{+0.02}$ & -2.9$_{-0.01}^{+0.01}$ & 1.89$_{-0.05}^{+0.05}$ & 2.01$_{-0.07}^{+0.12}$ & 0.04$_{-0.00}^{+0.00}$ & 4.08$_{-0.21}^{+0.40}$ \\ 
        (Erdos) & Y & 0.00$_{-0.00}^{+0.01}$ & 28.81$_{-18.78}^{+58.44}$ & 5.13$_{-1.88}^{+1.22}$ & -17.04$_{-0.66}^{+0.15}$ & 1.5$_{-0.00}^{+0.00}$ & 2.58$_{-0.00}^{+0.00}$ & 2,017.27 & 16.72$_{-10.10}^{+12.85}$ & 1.51 \\ 
        
        \frbfen & SF & 9.7$_{-0.09}^{+0.04}$ & -0.28$_{-0.09}^{+0.06}$ & 1.8$_{-0.19}^{+0.18}$ & 1.54$_{-0.14}^{+0.16}$ & -3.07$_{-0.04}^{+0.04}$ & 3.37$_{-0.82}^{+1.31}$ & 4.31$_{-0.88}^{+1.40}$ & 0.60$_{-0.13}^{+0.18}$ & 4.16$_{-0.36}^{+0.32}$ \\ 
        (Fen) & N & --- & --- & --- & -18.15$_{-0.49}^{+0.34}$ & 1.01$_{-0.03}^{+0.01}$ & 1.46$_{-0.05}^{+0.04}$ & 2,209.70 & 12.07$_{-8.35}^{+13.27}$ & 4.85 \\ 
        
        \frbpingu & SF & 11.13$_{-0.01}^{+0.01}$ & -0.24$_{-0.00}^{+0.01}$ & 0.36$_{-0.09}^{+0.08}$ & 2.18$_{-0.03}^{+0.05}$ & -3.02$_{-0.01}^{+0.02}$ & 3.72$_{-0.42}^{+0.51}$ & 29.91$_{-0.79}^{+0.73}$ & 0.14$_{-0.01}^{+0.01}$ & 3.85$_{-0.06}^{+0.06}$ \\ 
        (Pingu) & Y & 0.00$_{-0.00}^{+0.00}$ & 38.67$_{-22.96}^{+57.25}$ & 1.12$_{-0.36}^{+0.31}$ & -20.08$_{-0.16}^{+0.05}$ & 1.62$_{-0.00}^{+0.00}$ & 2.73$_{-0.01}^{+0.01}$ & 2,549.77 & 12.24$_{-8.65}^{+12.45}$ & 5.68 \\ 
        
        \frbwhitney & SF & 9.98$_{-0.06}^{+0.08}$ & -1.20$_{-0.15}^{+0.14}$ & 0.04$_{-0.03}^{+0.08}$ & 0.04$_{-0.03}^{+0.06}$ & -3.03$_{-0.20}^{+0.14}$ & 0.41$_{-0.08}^{+0.15}$ & 0.97$_{-0.29}^{+0.32}$ & 0.07$_{-0.02}^{+0.02}$ & 2.88$_{-0.67}^{+0.75}$ \\ 
        (Whitney) & Y & 1.49$_{-1.09}^{+0.84}$ & 60.63$_{-26.61}^{+31.91}$ & 5.03$_{-1.00}^{+0.98}$ & -20.94$_{-0.40}^{+0.37}$ & 0.98$_{-0.02}^{+0.02}$ & 1.35$_{-0.03}^{+0.04}$ & 2,722.79 & 4.39$_{-3.06}^{+19.94}$ & 4.59 \\ 
        
        \frbferb & SF & 9.82$_{-0.01}^{+0.0}$ & -0.63$_{-0.00}^{+0.00}$ & 0.8$_{-0.01}^{+0.01}$ & 0.81$_{-0.01}^{+0.01}$ & -3.16$_{-0.00}^{+0.00}$ & 7.27$_{-0.12}^{+0.12}$ & 20.81$_{-0.17}^{+0.20}$ & 2.24$_{-0.04}^{+0.04}$ & 1.27$_{-0.03}^{+0.04}$ \\ 
        (Ferb) & Y & 2.57$_{-0.04}^{+0.04}$ & 33.29$_{-1.15}^{+1.15}$ & 0.93$_{-0.29}^{+0.42}$ & -20.89$_{-0.64}^{+0.26}$ & 0.75$_{-0.00}^{+0.00}$ & 0.97$_{-0.01}^{+0.00}$ & 3,090.42 & 23.10$_{-11.65}^{+13.03}$ & 4.28 \\ 
        
        \frbkoyaanisqatsi & SF & 9.47$_{-0.04}^{+0.07}$ & -1.38$_{-0.10}^{+0.88}$ & 0.17$_{-0.06}^{+0.05}$ & 0.20$_{-0.07}^{+0.06}$ & -2.20$_{-0.24}^{+0.04}$ & 0.56$_{-0.09}^{+0.33}$ & 0.56$_{-0.09}^{+0.16}$ & 0.12$_{-0.01}^{+0.02}$ & 3.37$_{-0.16}^{+0.13}$ \\ 
        (Koy.) & N & --- & --- & --- & -19.15$_{-0.51}^{+0.33}$ & 0.80$_{-0.04}^{+0.05}$ & 1.08$_{-0.07}^{+0.14}$ & 3,169.44 & 8.80$_{-5.32}^{+7.68}$ & 2.97 \\ 
        
        \frbnihari & SF & 10.21$_{-0.04}^{+0.03}$ & -0.44$_{-0.03}^{+0.03}$ & 0.99$_{-0.07}^{+0.05}$ & 1.04$_{-0.08}^{+0.08}$ & -3.32$_{-0.01}^{+0.01}$ & 0.97$_{-0.09}^{+0.09}$ & 1.78$_{-0.23}^{+0.24}$ & 0.07$_{-0.01}^{+0.01}$ & 3.59$_{-0.15}^{+0.15}$ \\ 
        (Nihari) & Y & 2.74$_{-0.25}^{+0.18}$ & 121.44$_{-25.41}^{+19.61}$ & 1.82$_{-0.89}^{+1.25}$ & -19.30$_{-0.30}^{+0.09}$ & 1.18$_{-0.03}^{+0.04}$ & 1.86$_{-0.10}^{+0.11}$ & 3,246.07 & 5.92$_{-4.03}^{+14.96}$ & 5.61 \\ 
        
        \frbquincy & SF & 10.26$_{-0.02}^{+0.02}$ & -0.46$_{-0.05}^{+0.09}$ & 0.02$_{-0.01}^{+0.02}$ & 0.02$_{-0.01}^{+0.02}$ & -3.19$_{-0.03}^{+0.02}$ & 1.27$_{-0.10}^{+0.19}$ & 1.31$_{-0.13}^{+0.15}$ & 0.05$_{-0.01}^{+0.01}$ & 2.59$_{-0.29}^{+0.27}$ \\ 
        (Quincy) & Y & 0.06$_{-0.03}^{+0.06}$ & 46.30$_{-13.01}^{+15.21}$ & 0.86$_{-0.24}^{+0.35}$ & -20.81$_{-0.35}^{+0.09}$ & 0.87$_{-0.02}^{+0.02}$ & 1.16$_{-0.03}^{+0.02}$ & 3,740.05 & 15.15$_{-7.80}^{+12.12}$ & 9.51 \\ 
        
        \frbmifanshan & SF & 10.59$_{-0.10}^{+0.14}$ & -0.39$_{-0.14}^{+0.15}$ & 0.63$_{-0.27}^{+0.31}$ & 0.80$_{-0.26}^{+0.38}$ & -1.88$_{-0.29}^{+0.40}$ & 3.82$_{-1.85}^{+4.31}$ & 5.21$_{-2.52}^{+5.54}$ & 0.08$_{-0.03}^{+0.07}$ & 2.93$_{-0.17}^{+0.21}$ \\ 
        (Mifanshan) & N & --- & --- & --- & -19.46$_{-0.36}^{+0.68}$ & 0.84$_{-0.14}^{+0.09}$ & 1.23$_{-0.20}^{+0.16}$ & 6,518.03 & 26.97$_{-9.14}^{+9.72}$ & 3.81 \\   
        
        \hline
    \end{tabular}

    \footnotesize{{\raggedright N{\scriptsize{OTE}}: The tabulated values are the median and 68\% confidence intervals. $\log \mathrm{M}_{\ast}$ denotes the stellar mass and log(Z/Z$_\odot$) represents stellar metallicity. A$_{\rm{V}, \rm{young}}$ and  A$_{\rm{V}, \rm{old}}$ denotes the dust extinction of young and old stellar light in magnitudes. U$_{\mathrm{neb}}$ is the nebular ionization parameter. SFR$_{\mathrm{20 Myr}}$ and SFR$_{\mathrm{100 Myr}}$ represents the SFR of the galaxy averaged over 20 and 100~Myr respectively. sSFR is the specific SFR and t$_\mathrm{m}$ is the mass-weighted age of the galaxy. The AGN column denotes whether or not an AGN was included in the SED model. f$_{\rm{AGN}}$ and $\tau_{\rm{}AGN}$ are the components of the AGN dust emission model~\protect\citemethods{2008ApJ...685..160N}. q$_{\rm{PAH}}$ describes the grain size distribution of PAHs. M$_r^0$, $(g-r)^0$ and $(u-r)^0$ denotes the rest-frame $r$-band magnitude and colors. D$_{\mathrm{L}}$, d and R$_{\mathrm{e}}$ denotes the luminosity distance, the projected galactocentric offset of the FRB and effective radius of the galaxy.
    \par}}
    
\end{table*}

\newpage
\clearpage
\newpage
\newpage

\begin{figure*}[ht!]
\centering
\includegraphics[width=\textwidth]{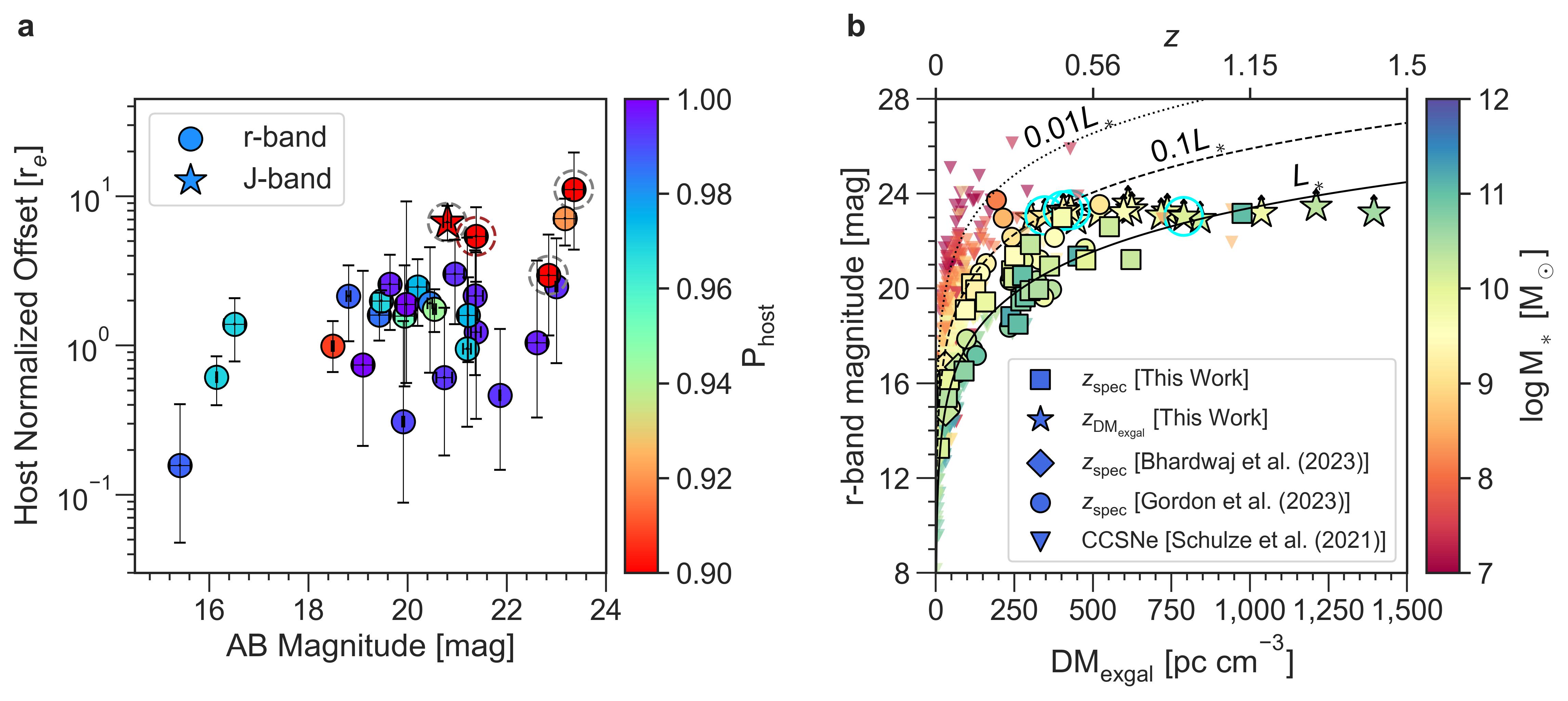}
\caption{{\edfigurelabel{fig:path_analysis}} \textbf{Probabilistic host association for DSA-110 discovered FRBs and sample selection.} The distribution of PATH association probability (P$_{\mathrm{host}}$) in the space of host-normalized offsets and AB magnitudes of 30 FRBs included in our sample is shown in panel \textbf{a}. We use the deepest available $r$-band or $J$-band imaging data for PATH analysis and compute the magnitudes and offsets, along with their measurement errors, using published catalogs or custom software (see Methods). We identified 1 case where P$_{\mathrm{host}}$ is low due to high host-normalized offsets (brown dashed circles) and 3 cases where P$_{\mathrm{host}}$ is low because of multiple plausible host galaxies (gray dashed circles). The panel \textbf{b} shows the distribution of the $r$-band magnitude and redshift of 26 secure FRB hosts published in this work (squares), alongside Gordon et al.~\citep{2023arXiv230205465G} (circles) and Bhardwaj et al.~\citep{2023arXiv231010018B} (diamonds) FRB host galaxies samples with $r$-band magnitude $\lesssim 23.5$ (see Fig.~\ref{fig:frbs_are_biased_tracer_of_sf_in_universe} for a zoom on $z \leq 0.2$ regime). For completeness, we also show the $5\sigma$ $r$-band limiting magnitude and the extragalactic DM of 12 DSA-110 FRBs excluded from this sample (stars) and 4 insecure host associations (stars with cyan circles).}
\end{figure*}

\newpage
\clearpage
\newpage
\newpage

\begin{figure*}[ht!]
    \includegraphics[width=\textwidth]{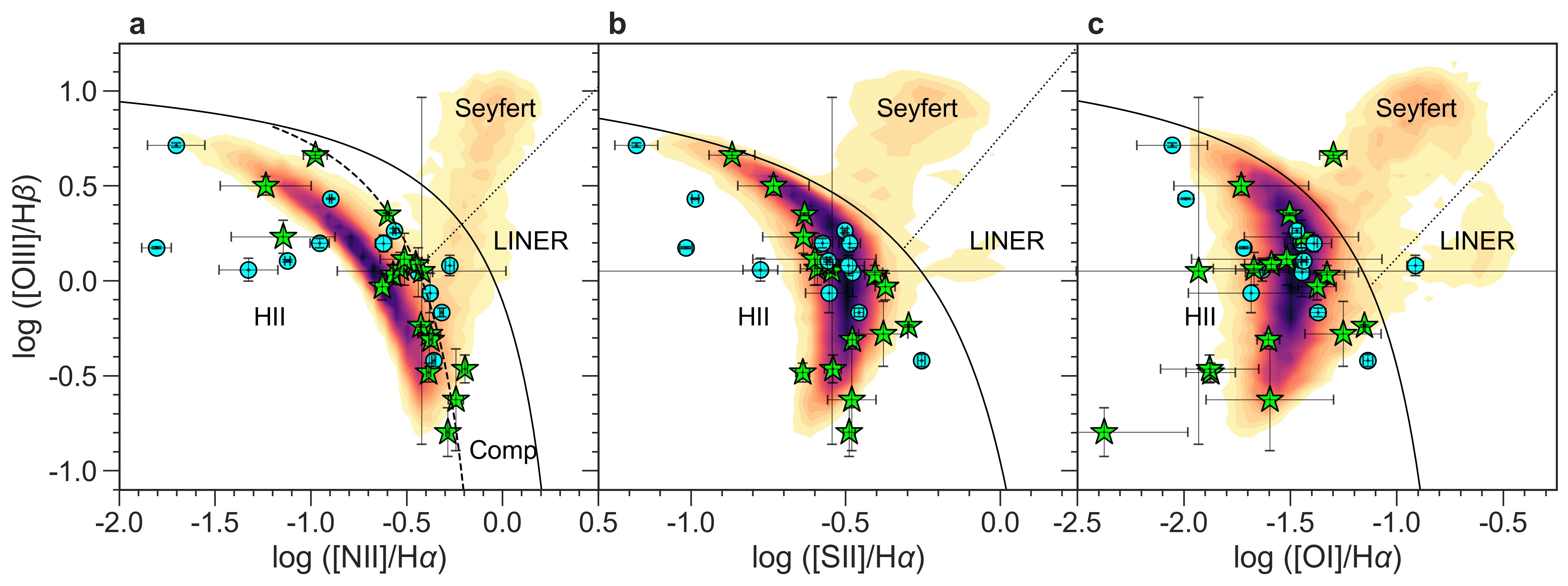}
    \caption{{\edfigurelabel{fig:BPT_analysis}} \textbf{Dominant ionization mechanism in FRB host galaxies.} The Baldwin, Phillips \& Terlevich empirical optical emission-line diagnostic diagrams using the optical line ratios [OI]/H$\alpha$, [SII]/H$\alpha$, [NII]/H$\alpha$, and [OIII]/H$\beta$~\citep{1981PASP...93....5B}. The maximum starburst line~\protect\citemethods{2001ApJ...556..121K}, the pure star-forming galaxies vs Seyfert-H II composite objects classification line~\protect\citemethods{2003MNRAS.346.1055K}, and the Seyfert vs LINER classification lines~\citep{2006MNRAS.372..961K} are shown as solid, dashed and dotted lines respectively. The emission line diagnostics for SDSS galaxies are plotted in the background~\citep{2015ApJS..219...12A} (see Methods). While a majority of the FRB hosts spectra had the data to measure all the emission line ratios (green stars), some of them had limited spectral coverage, either due to a high redshift or due to instrument malfunctions. For these, we used the \sw{Prospector}-derived SEDs to measure the line ratios (cyan circles). The error bars on green stars represent the measurement errors caused by noise in the spectrum, while on cyan circles represent the errors computed using $1,000$ \sw{Prospector}-derived posteriors of the galaxy SED.
}   
\end{figure*}

\newpage
\clearpage
\newpage
\newpage

\begin{figure*}[ht!]
\centering
    \includegraphics[width=0.5\textwidth]{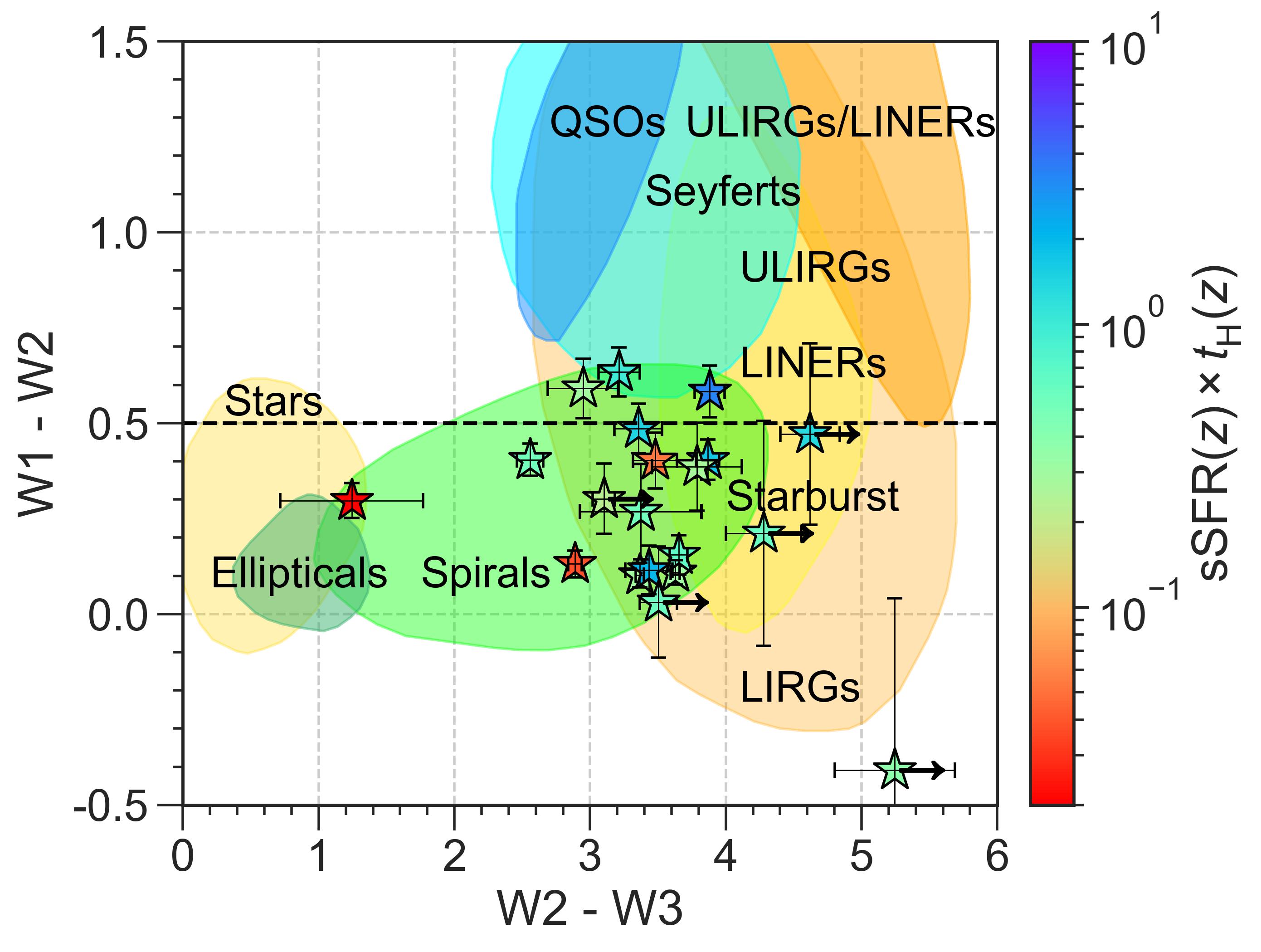}
    \caption{{\edfigurelabel{fig:wise_color_color}} \textbf{WISE color-color galaxy classifications~\citep{2010AJ....140.1868W}.} The mid-infrared colors are retrieved using the publicly-available WISE catalog~\citep{2010AJ....140.1868W}. The shaded regions represent the standard WISE color-color classification regimes~\citep{2010AJ....140.1868W}. The dashed line marks the typical $W1-W2 = 0.5$ boundary used to idenitify AGN activity in galaxies. The markers are colored by the degree of star-formation in FRB host galaxies~\protect\citemethods{2022ApJ...926..134T}.}
\end{figure*}

\newpage
\clearpage
\newpage
\newpage

\begin{figure*}[ht!]
    \includegraphics[width=\textwidth]{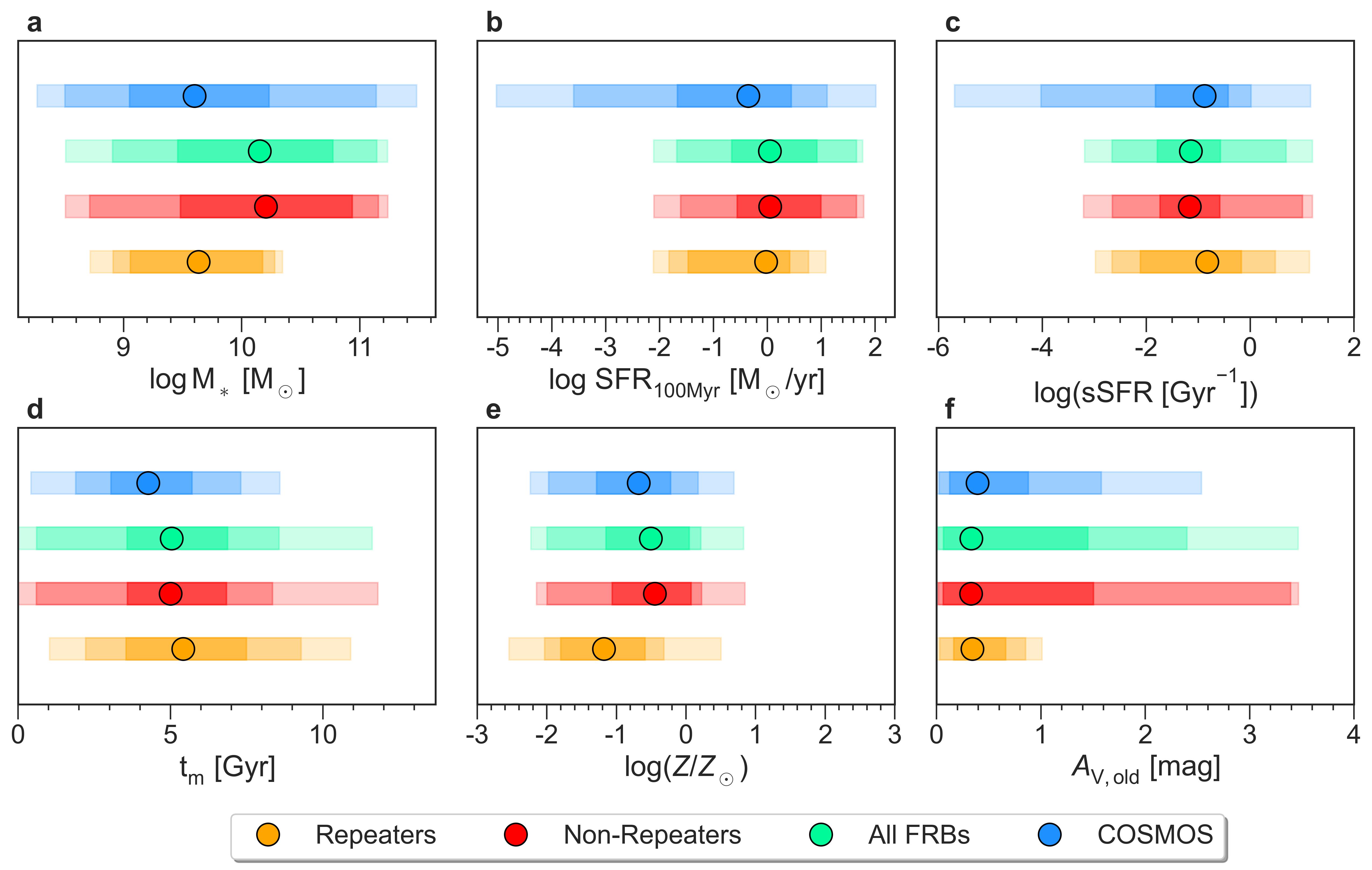}
    \caption{{\edfigurelabel{fig:all_params_box_plot}} \textbf{Distribution of derived galaxy properties}. We compare repeaters (orange), non-repeaters (red), all FRBs (green) and the background galaxy population (blue) from COSMOS-2015~\protect\citemethods{2016ApJS..224...24L}$^,$\citep{2020ApJ...893..111L} and 3D-HST~\protect\citemethods{2014ApJS..214...24S} galaxy catalogs. The FRB population includes our secure host associations and previously published FRBs~\citep{2023arXiv230205465G, 2023arXiv231010018B}. The marker denotes the median together with 1,2,3$\sigma$ bands. We observe that the host galaxies of non-repeaters are relatively more massive, have lower sSFR and higher metallicity (potentially a consequence of enforcing the mass-metallicity relation in the SED modeling) than repeaters. However, KS-tests reveals that these differences are not statistically significant (see Methods).}
\end{figure*}

\newpage
\clearpage
\newpage
\newpage

\begin{figure*}[ht!]
    \centering
    \includegraphics[width=\textwidth]{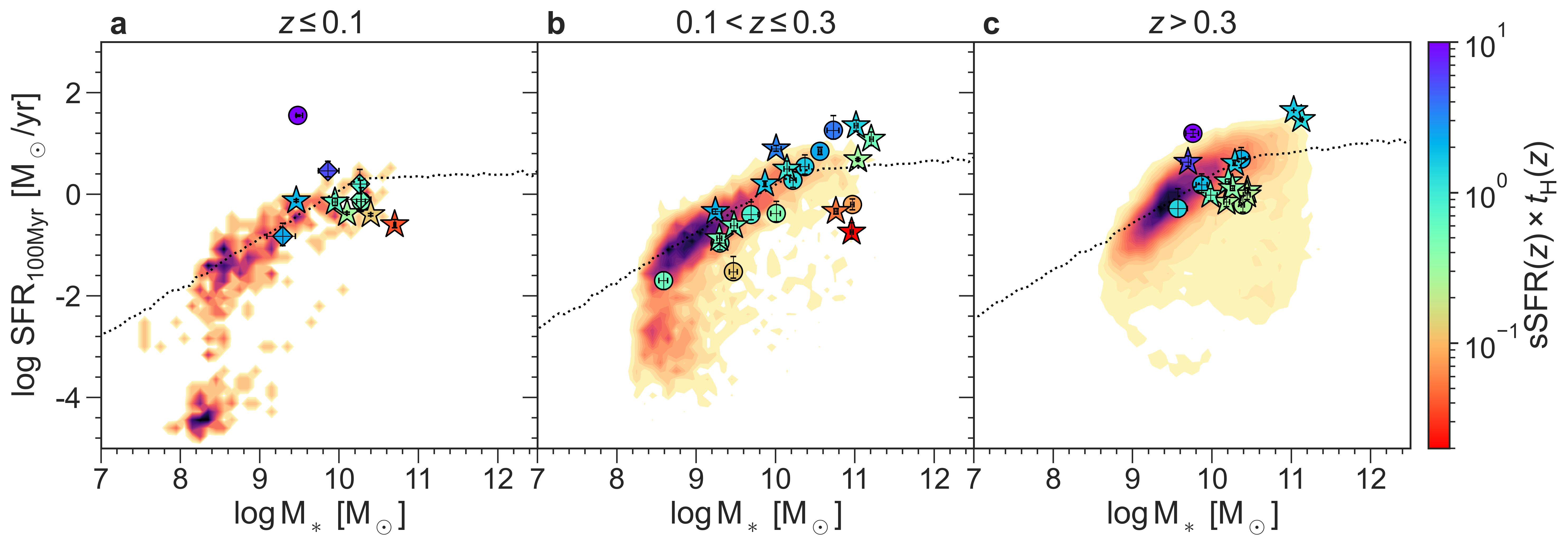}
    \includegraphics[width=\textwidth]{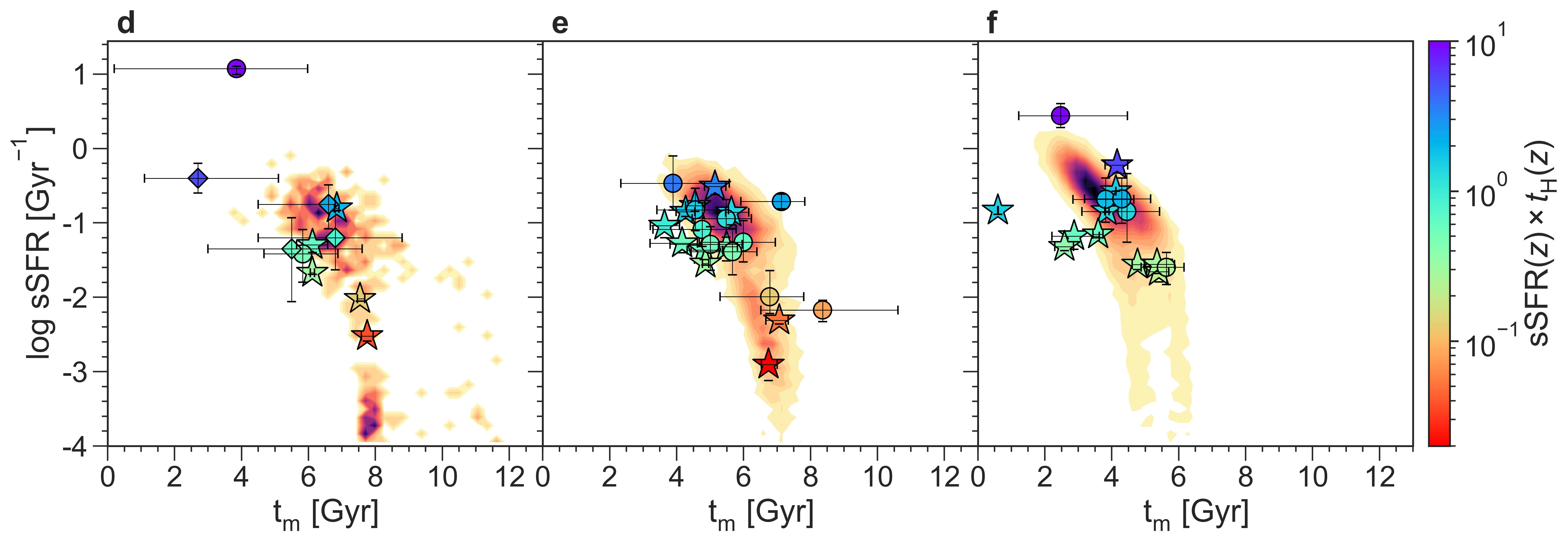}
    \caption{{\edfigurelabel{fig:gal_sfms_tm_ssfr}} \textbf{Comparison of FRB host galaxies with background galaxy population.} We use the COSMOS-2015~\protect\citemethods{2016ApJS..224...24L}$^,$\citep{2020ApJ...893..111L} and 3D-HST~\protect\citemethods{2014ApJS..214...24S} galaxy catalogs to represent the field galaxy population. Along with the FRB host galaxies published in this work (stars), we also include Gordon et al.~\citep{2023arXiv230205465G} (circles) and Bhardwaj et al.~\citep{2023arXiv231010018B} (diamonds) samples. The insecure hosts are marked with cyan circles. The markers are colored by the degree of star-formation~\protect\citemethods{2022ApJ...926..134T}. The error bars are computed using $1,000$ samples from the posterior distributions of measured properties from SED analysis. Panels \textbf{a}, \textbf{b}, \textbf{c} show the comparison with the star-forming main sequence (SFMS) of galaxies in three redshift bins to avoid biases from redshift evolution. For reference, the center of the SFMS at $z = 0.05,~0.2$ and $0.45$ are marked as dotted lines~\citep{2022ApJ...936..165L}. Although we observe a handful of quiescent galaxies, the distribution of offset of each FRB host from the SFMS at their redshift~\citep{2022ApJ...936..165L} is centered at $\sim -0.1$~dex with a scatter of 0.7~dex, which is comparable to the intrinsic scatter of star-forming galaxies about the SFMS, implying that the FRB hosts are consistent with the locus of star-forming galaxies. This similarity with the background population is also seen in comparison of ages and specific SFR in panels \textbf{d}, \textbf{e}, \textbf{f}.}
\end{figure*}

\newpage
\clearpage
\newpage
\newpage

\begin{figure*}[ht!]
\centering
\includegraphics[width=0.5\textwidth]{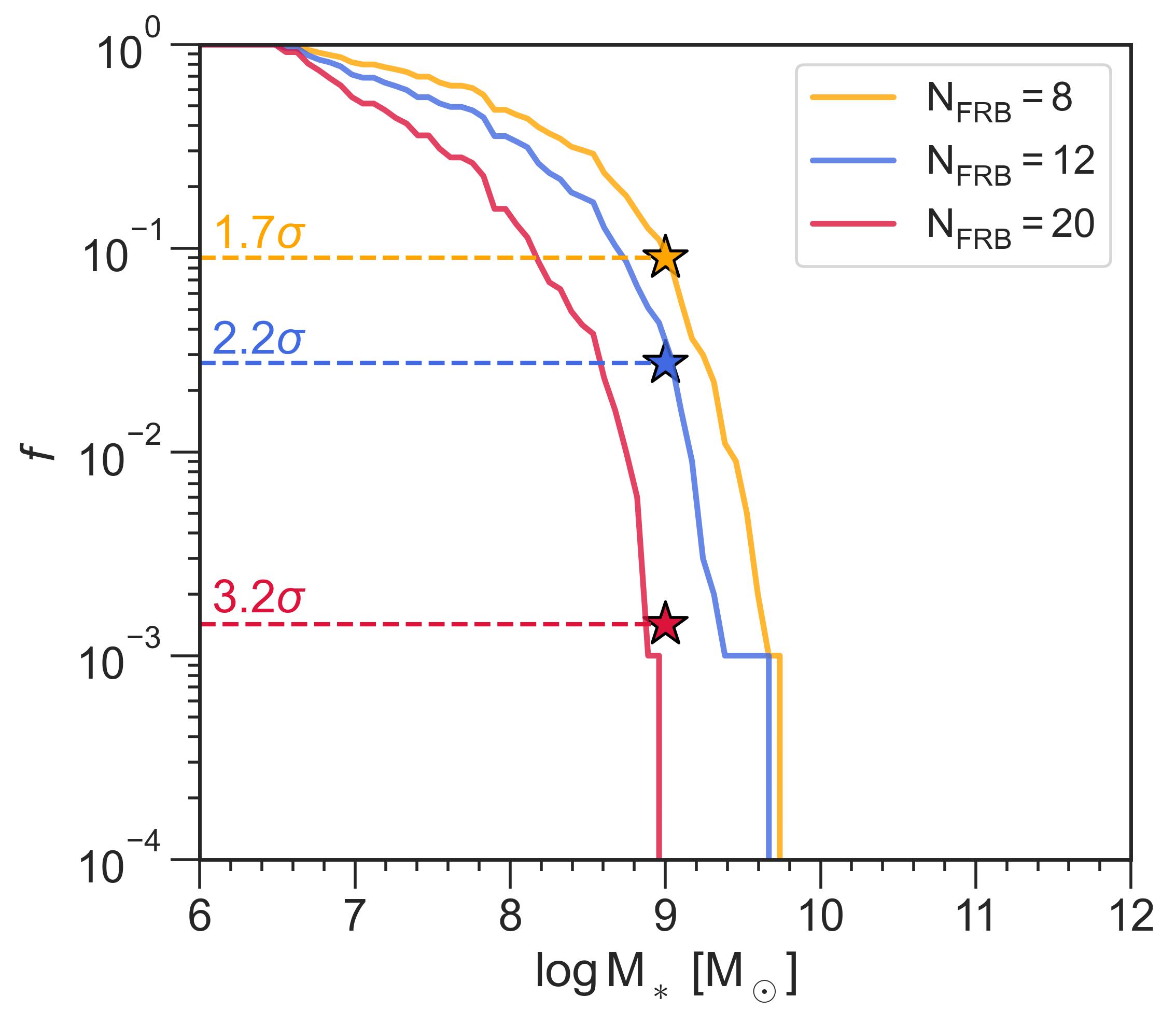}
\caption{{\edfigurelabel{fig:significance_of_dwarfs}} \textbf{The significance of the lack of low-mass galaxies in FRB hosts sample.} We draw $1,000$ samples of size N$_\mathrm{FRB}$ of CCSN host galaxies, where N$_\mathrm{FRB}$ can be the number of FRBs localized by DSA-110 (8, orange), localized by other instruments (12, blue) or all FRBs (20, crimson) at redshifts $z \leq 0.2$. For each stellar mass M$_\ast$, we plot the fraction of these $1,000$ samples with no host galaxy more massive than M$_\ast$. Given the sample size of FRB host galaxies, we find that the CCSN host galaxies above $10^9$~M$_\odot$ are rare with $3.2\sigma$ confidence.}
\end{figure*}

\newpage
\clearpage
\newpage
\newpage

\begin{figure*}[ht!]
\centering
\includegraphics[width=\textwidth]{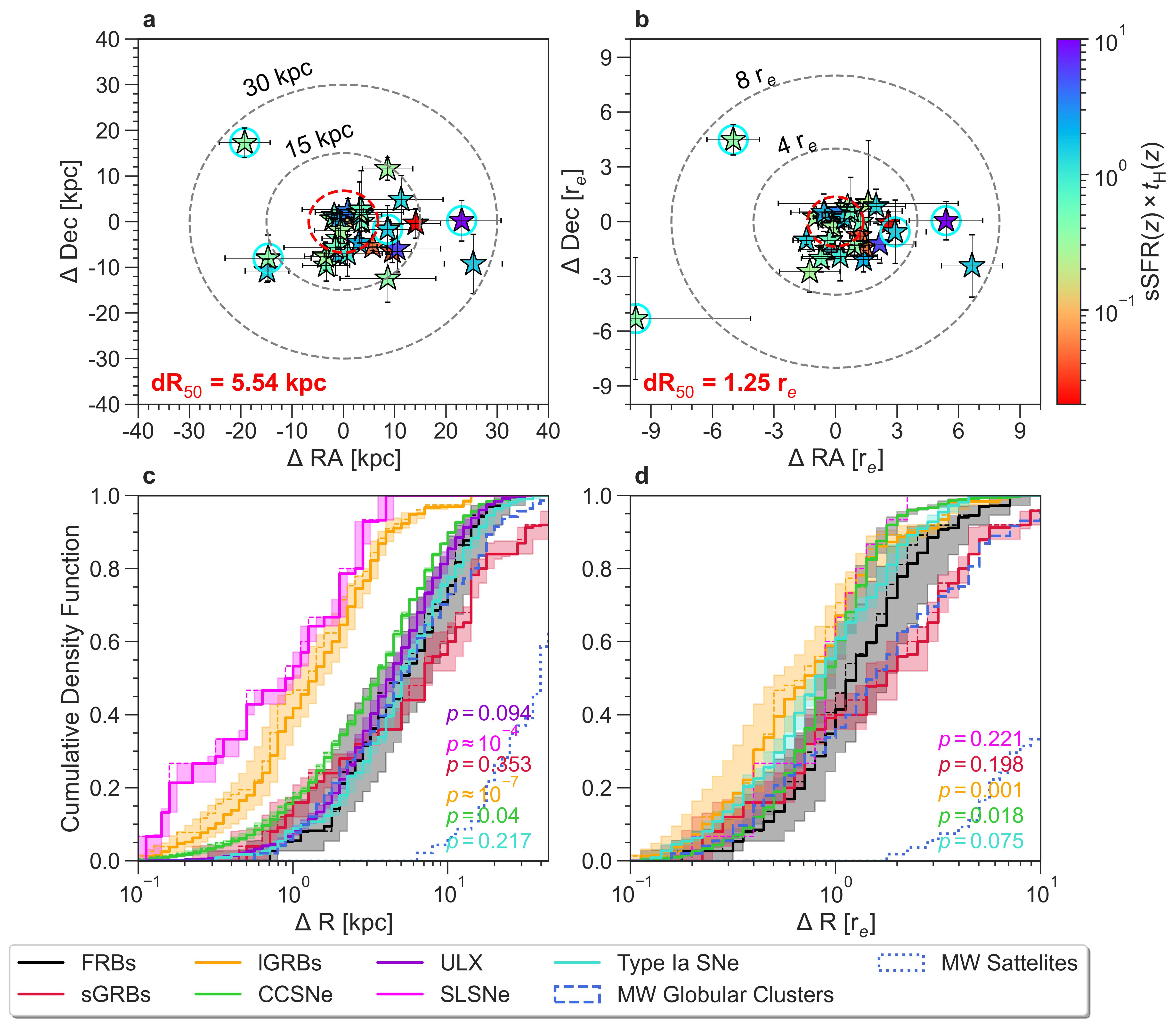}
\caption{{\edfigurelabel{fig:offset_distribution_compare_with_other_transients}} \textbf{Comparison of offset distributions of transients.} Panels \textbf{a} and \textbf{b} display the location of the DSA-110 FRBs (panel \textbf{a}) and host-normalized offsets (panel \textbf{b}) with respect to the center of their host galaxies. The uncertainties correspond to the 90\% localization regions. The markers are colored by their degree of star-formation~\protect\citemethods{2022ApJ...926..134T}. The red dashed circle marks the median of the offset distribution. Majority of the FRBs are within a radius of 15~kpc (3 effective radii). FRBs outside of this radius have insecure host association and we find no correlation between their degree of star-formation and offsets. For comparison with other transients in panels \textbf{c} and \textbf{d}, we include our secure host associations FRBs, together with previously published FRB offset measurements~\protect\citemethods{2021ApJ...917...75M, 2023arXiv231201578W}. We compare FRBs with the galactocentric offsets of superluminous supernovae (SLSNe)~\protect\citemethods{2015ApJ...804...90L}, long-duration GRBs (lGRBs)~\protect\citemethods{2016ApJ...817..144B}, core-collapse supernovae (CCSNe)~\citep{2021ApJS..255...29S}$^,$\protect\citemethods{2012ApJ...759..107K}, short-duration GRBs (sGRBs)~\citep{2022ApJ...940...56F}, Type Ia supernovae~\protect\citemethods{2020ApJ...901..143U} and ultra-luminous X-ray sources (ULX)~\protect\citemethods{2020MNRAS.498.4790K}. The measured values (dashed lines), median (thick lines) and $1\sigma$ errors (shaded regions) computed using 1,000 Monte Carlo samples of offset measurements of the transients assuming a normal distribution with errors as quoted in the literature are plotted. \ks{We also show the galactocentric offsets of satellites~\protect\citemethods{2020ApJ...893...47D} and globular clusters~\protect\citemethods{1996AJ....112.1487H} of the Milky Way.} Radio selection effects may prohibit the discovery of FRBs near the center of the galaxies due to high DMs and scattering timescales. If FRBs were to trace star-formation, then this bias would over-estimate the median of the distribution by $\sim$kpc~\citep{2021arXiv211207639S}.}
\end{figure*}

\newpage
\clearpage
\newpage
\newpage

\begin{figure*}[ht!]
    \includegraphics[width=\textwidth]{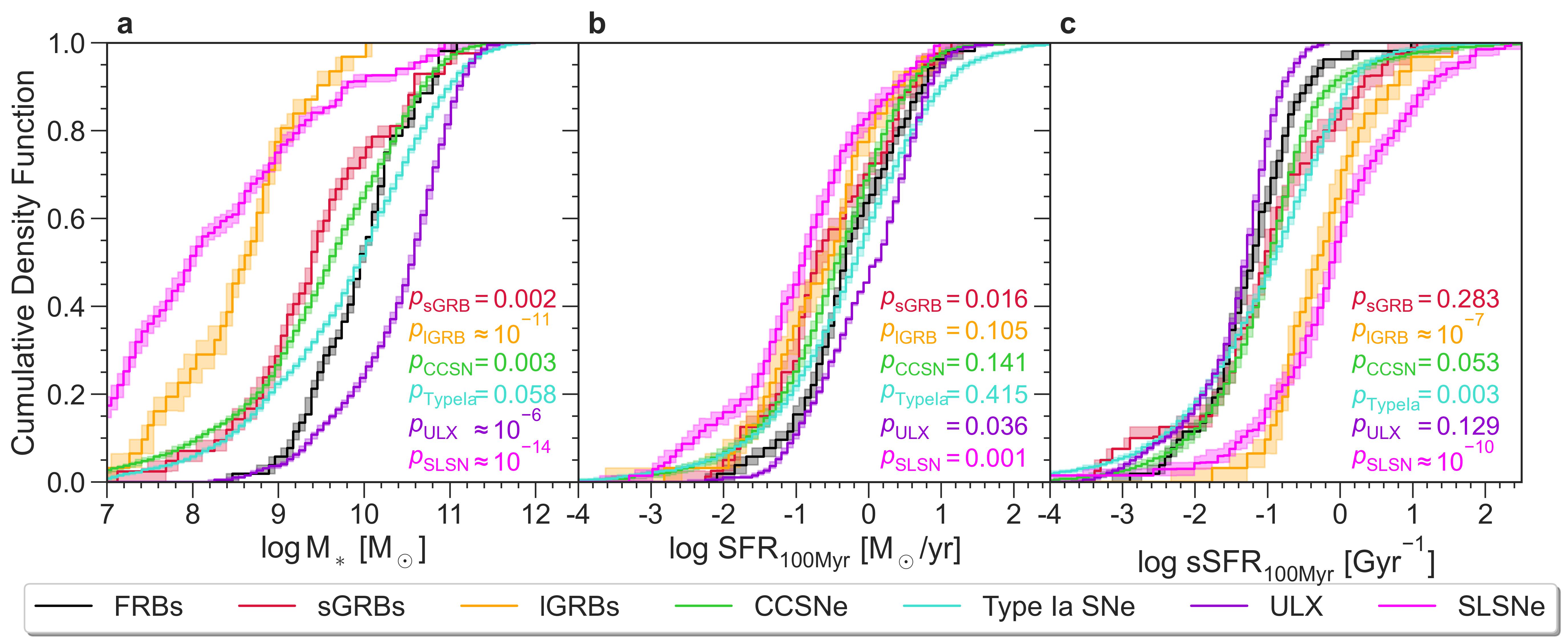}
    \caption{{\edfigurelabel{fig:compare_with_other_transients_at_zero_redshift}} \textbf{Comparison of host galaxy properties of transients.} In order to compare our FRBs with secure host association, together with previously published FRBs~\citep{2023arXiv230205465G, 2023arXiv231010018B}, against Type Ia supernovae~\protect\citemethods{2010ApJ...722..566L}, ultra-luminous X-ray sources (ULX)~\protect\citemethods{2020MNRAS.498.4790K}, superluminous supernovae (SLSNe)~\citep{2021ApJS..255...29S}, core-collapse supernovae (CCSNe)~\citep{2021ApJS..255...29S}, short-duration GRBs (sGRBs)~\protect\citemethods{2022ApJ...940...57N} and long-duration GRBs (lGRBs)~\protect\citemethods{2015A&A...581A.102V}$^,$\citep{2021MNRAS.503.3931T}, we correct for redshift evolution (see Methods). The errors on cumulative distributions are computed using $1,000$ Monte Carlo samples of each measured property of the transients assuming a normal distribution with asymmetric errors as quoted in the literature.
    }
\end{figure*}

\newpage
\clearpage
\newpage
\newpage

\noindent{\bfseries \LARGE Supplementary Information}

\begin{figure*}[ht!]
\centering
\includegraphics[width=0.7\textwidth]{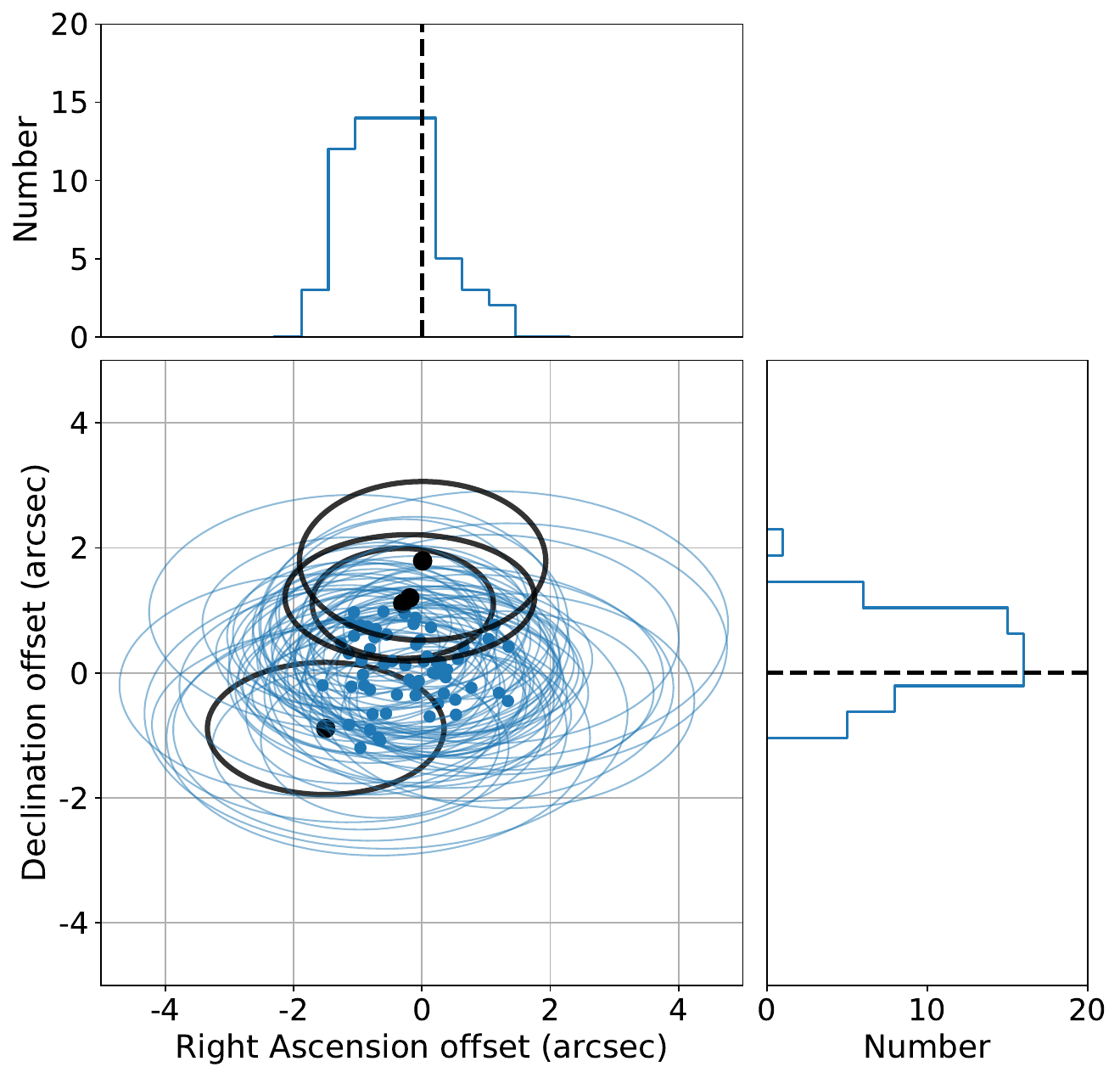}
\caption{{\supfigurelabel{fig:offsets_test}} 
\ks{\textbf{DSA-110 localization offsets of 69 continuum sources with known positions.} The sources were selected from the NVSS catalog as bright ($>400$\,mJy), compact (major axes $<35$\,arcsec), and within 2.1\,deg of the primary-beam center in 35 archived voltage dumps. Each source was localized using standard DSA-110 procedures (see Methods). The image signal-to-noise ratios of sources were between 8 and 20. The central panel shows the derived source-position offsets from the true positions (dots), and 90\% confidence localization ellipses. Only 4/69 sources (highlighted as thick black dots/lines) have localization ellipses that do not contain the true positions. The histograms show the distributions of the right ascension and declination offsets.}}
\end{figure*}

\begin{figure*}[ht!]
\centering
\includegraphics[width=0.8\textwidth]{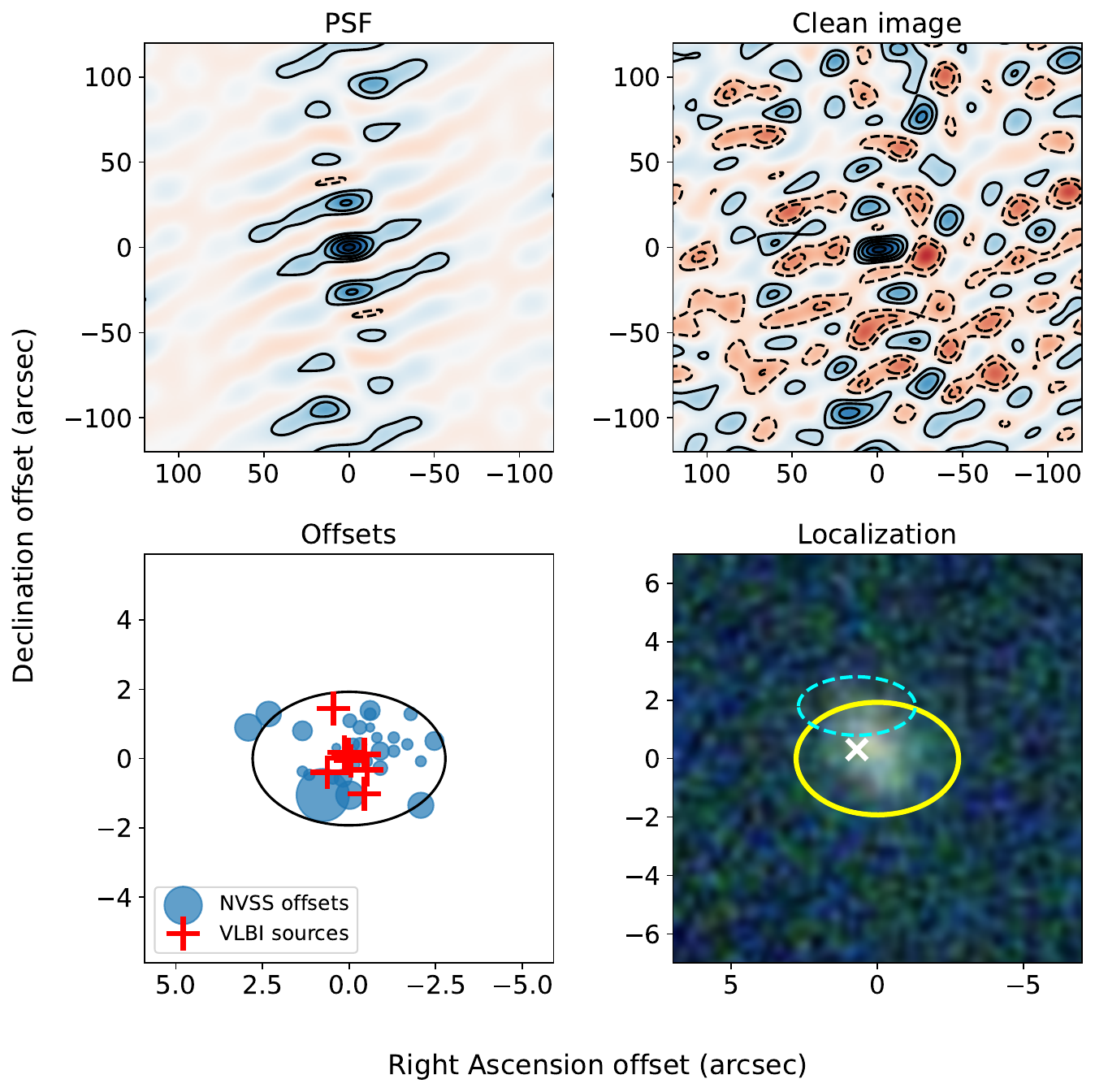}
\caption{{\supfigurelabel{fig:repeater_localization}} 
\ks{\textbf{Revised DSA-110 localization of FRB\,20220912A.} All images are centered on the revised DSA-110 burst position of (J2000) 23h09m04.83s $+$48d42m23.6s. The top panels show the interferometric point-spread function (PSF) and cleaned image of a burst detected on 2022 October 18. Contours are at values of $-0.4$, $-0.2$, 0.2, 0.4, 0.6, and 0.8 relative to the peak. The image signal-to-noise ratio is 8. The bottom-left panel shows the 90\% confidence localization ellipse, and the corrected offsets of adjacent VLBI calibrators (red crosses) and in-field NVSS sources (blue circles with sizes scaled according to the flux densities). The bottom-right panel shows a three-color ($gri$) PS1 image of the host galaxy of FRB\,20220912A, together with the originally published DSA-110 90\% confidence localization ellipse (blue dashed)~\protect\citemethods{2023ApJ...949L...3R}, the VLBI position (white cross)~\protect\citemethods{2024MNRAS.529.1814H}, and the new DSA-110 90\% confidence localization ellipse (yellow solid).}}
\end{figure*}

\newpage
\clearpage
\newpage
\newpage

\begin{figure*}[ht!]
\centering
\includegraphics[width=\textwidth]{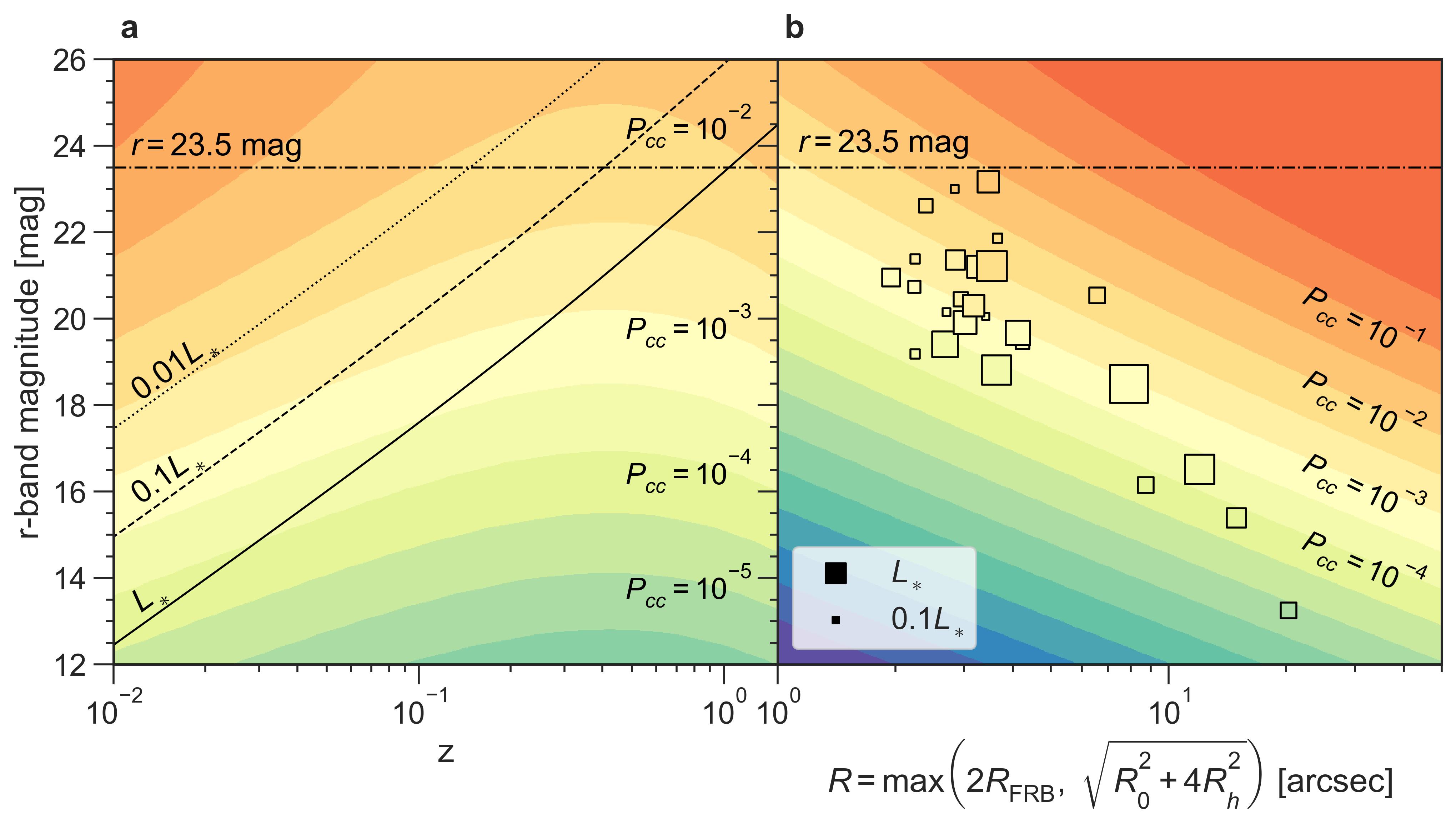}
\caption{{\supfigurelabel{fig:association_capabilities_demonstration}} 
\ks{\textbf{Assessing the host associations of DSA-110 FRBs.} The robustness of FRB host galaxy associations is a function of the localization uncertainties and the depth of optical imaging surveys being utilized~\protect\citemethods{2017ApJ...849..162E}. The contours in both panels show the probability of chance coincidence, $P_{cc}$, as defined by eqn.~\ref{eqn:Pcc}~\protect\citemethods{2017ApJ...849..162E}. In panel (a), for a given r-band magnitude and redshift in the background, we compute the galaxy stellar mass ($M_\ast$) assuming a mass-to-light ratio, $M_\ast/L_\ast = 1$, and then use the galaxy mass-radius relation for late-type galaxies~\protect\citemethods{2014ApJ...788...28V} to compute the representative half-light radius ($R_h$) for a galaxy of mass $M_\ast$. We use $R_0 \approx 2R_h$ and $R_{\mathrm{FRB}} = 3$\arcsecs for generating the upper limits on $P_{cc}$. Without any optical observation biases, our $\leq 3$\arcsecs localizations are sufficient to associate FRBs to $\lesssim 0.01~L_\ast$ galaxies with $P_{cc} \lesssim 0.1$. However, due to limited optical imaging depth of r-band magnitude $\lesssim 23.5$~mag, our association confidence gets limited to $\lesssim 0.1-1~L_\ast$ galaxies with $P_{cc} \lesssim 0.001-0.01$ at high redshifts. However, we assert that these optical biases barely impact low redshifts (e.g., $z \lesssim 0.2$), allowing us to confidently associate FRBs to $\lesssim 0.01~L_\ast$ galaxies with $P_{cc} \lesssim 0.1$. This implies that our host associations at $z \lesssim 0.2$ are robust and the deficit of low-mass galaxies is indeed real. Panel (b) shows the distribution of $P_{cc}$ in the space of r-band magnitude and radius $R = \max \left(2R_{\mathrm{FRB}},~\sqrt{R_0^2 + 4R_h^2}\right)$. We use the actual DSA-110 FRB localizations, galactocentric offsets and galaxies half-light radii for plotting the $P_{cc}$ of 30 FRBs published in this work (squares). The $P_{cc}$ of all our FRB host galaxies is $< 0.1$, thus implying that the association of high redshift FRBs to their hosts is robust. The low $P_{cc}$ is a consequence of the fact that the cosmic volume mapped by the DSA-110 localizations is small and the scenario of existence of rare massive galaxies within or near the localization areas is unlikely, thus strengthening the robustness of our associations.}}
\end{figure*}

\newpage
\clearpage
\newpage
\newpage

\begin{figure*}[ht!]
\centering
\includegraphics[width=\textwidth]{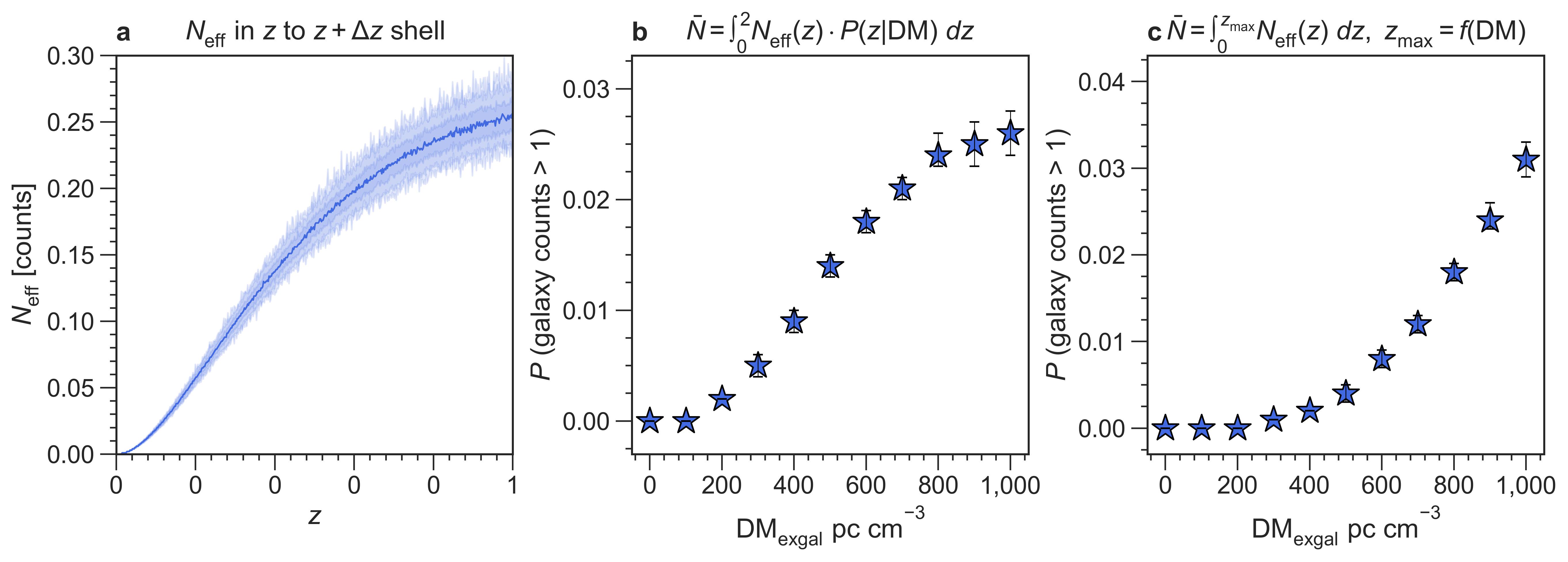}
\caption{{\supfigurelabel{fig:association_capabilities_demonstration_highz}} \ks{\textbf{Quantifying the rarity of existence of multiple candidate galaxies within the cosmic volume permitted by DSA-110 localization uncertainties and FRB dispersion measure.} Panel (a) shows the effective number of galaxy counts~\citep{2020ApJ...893..111L} ($N_\mathrm{eff} (z)$), while accounting for the effects of galaxy clustering~\protect\citemethods{1980lssu.book.....P}, in the comoving volume defined by the redshift range $z$ to $z + dz$ and solid angle defined by double of the maximum localization uncertainties ($\theta = 6$\arcsecs) to present a conservative estimate of galaxy counts (see Methods). We observe that $N_\mathrm{eff}$ monotonically increases with redshift, primarily due to the increasing volume of the comoving conical shell volume element. We use $N_\mathrm{eff}(z)$, together with $P(z|\mathrm{DM}_{\mathrm{exgal}})$ from the DM$_\mathrm{IGM} - z$ relation~\protect\citemethods{2021ApJ...906...49Z} to compute the mean expected galaxy counts, given the FRB extragalactic DM (see Methods). We then sample 10,000 galaxy counts using these means and do a Monte Carlo simulation to compute the fraction of these samples with galaxy counts greater than one. We plot our results from this simulation in panel (b). We observe that for $\mathrm{DM}_\mathrm{exgal} \lesssim 200$~pc~cm$^{-3}$, the probability of existence of more than one galaxy within 6\arcsecs of the localization is $\lesssim 0.20_{-0.04}^{+0.04}$\%. For increasing $\mathrm{DM}_\mathrm{exgal}$, this probability increases, primarily due to the larger cosmic volume probed by the localization region. The maximum probability of existence of more than one galaxy within 6\arcsecs of FRB localization is $\approx 3$\%, implying that out of 30 DSA-110 FRBs, we expect more than one candidate host galaxy for at most $30 \times 0.03 \approx 1$ case, thus quantifying the rarity of this scenario. For completeness, we use another methodology where we compute the maximum possible redshift ($z_\mathrm{max}$, 95th percentile) permitted by the $\mathrm{DM}_\mathrm{exgal}$ and then compute the mean galaxy counts as integrated effective counts out to redshift $z_\mathrm{max}$ (see Methods). We repeat the simulation with this methodology and plot the results in panel (c). We find that the probability of existence of more than one galaxies within 6\arcsecs of the localization is $\lesssim 3.48_{-0.19}^{+0.18}$\%, again quantifying the rarity of this scenario.}}
\end{figure*}

\newpage
\clearpage
\newpage
\newpage

\begin{figure}[ht!]
\centering
\includegraphics[width=0.9\textwidth]{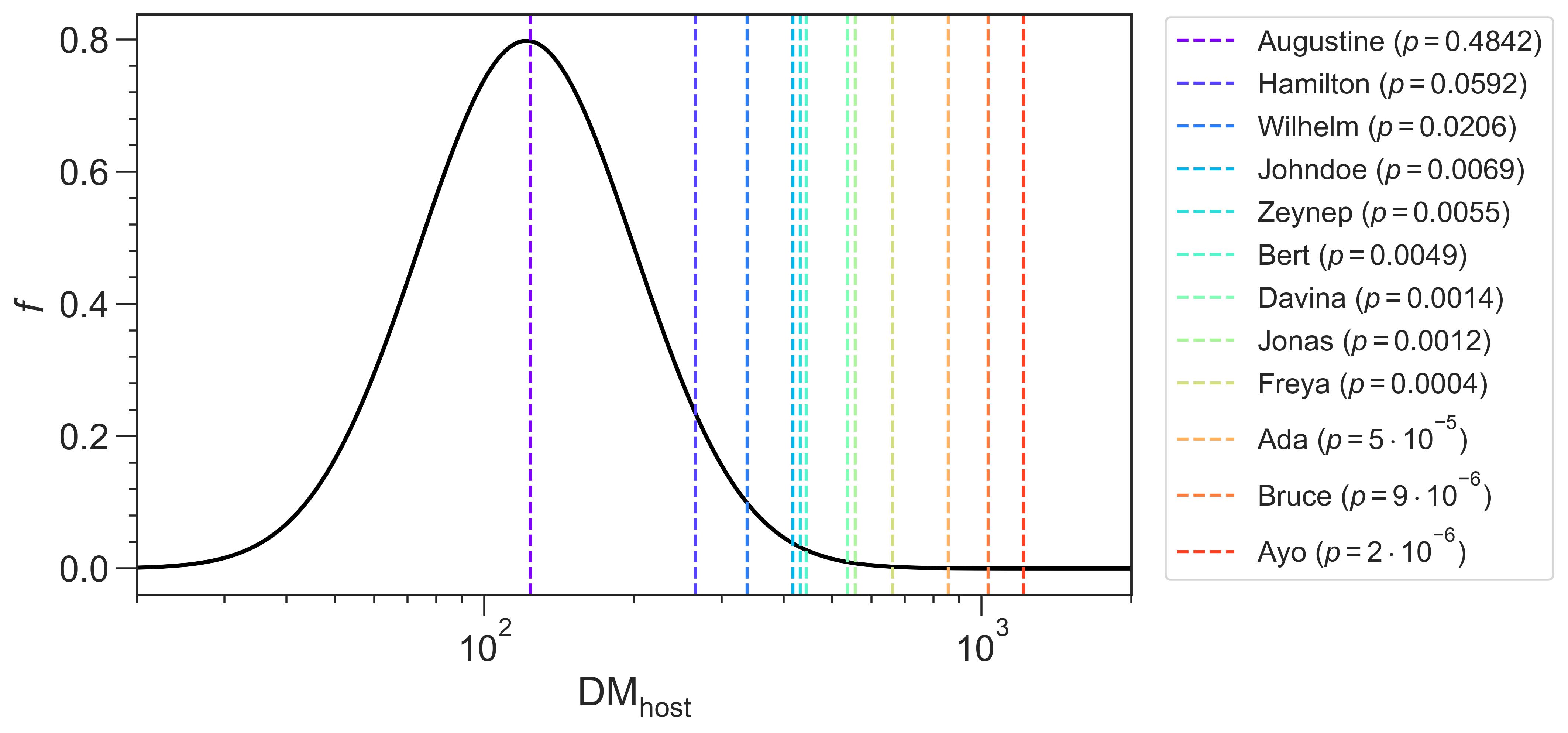}
\caption{{\supfigurelabel{fig:hostDMdistribution}} \ks{\textbf{Quantifying the likelihood of enough low-mass FRB hosts in our hostless FRBs sample to break our low-mass FRB hosts deficit statistics.} We consider the possibility that some of the 12 hostless DSA FRBs, which were not associated with hosts down to an $r$-band magnitude of $\lesssim 23.5$ mag, could actually have hosts within the local universe, specifically in the redshift bin $z \leq 0.2$. To investigate this, we utilized the extragalactic DMs of these FRBs to quantify how rare the excess local $\mathrm{DM}_\mathrm{host}$ would need to be for a sufficient number of $z \leq 0.2$ hosts to exist in this sample of hostless FRBs. Since $\sim$20\% of the star formation in the universe at $z \leq 0.2$ occurs in low-mass galaxies ($\log M_\ast \leq 9$), and the FRB sample has a total of 20 FRBs with $\log M_\ast \geq 9$ at $z \leq 0.2$, we expect 5 low-mass, low-redshift hosts in the hostless sample for FRBs to perfectly trace star formation. Following our companion paper by Connor et al., we assumed a log-normal distribution for the host contribution to the DM with parameters $\mu_\mathrm{host} = 4.8$ and $\sigma_\mathrm{host} = 0.5$. We computed the $p$-values for each FRB as the probability $P(\mathrm{DM}_\mathrm{host} > \mathrm{DM}_\mathrm{exgal} - \langle \mathrm{DM}_\mathrm{IGM}(z=0.2) \rangle)$, where we subtracted the median DM contribution from the IGM at $z = 0.2$ of $\mathrm{DM}_\mathrm{IGM} = 180$~pc~cm$^{-3}$. Under the hypothesis that some of these 12 FRBs may exist at $z \leq 0.2$, these $\mathrm{DM}_\mathrm{host}$ values would represent an extreme scenario. We computed the probability that 5 of 12 hostless FRBs are low-mass, low-redshift hosts as the sum of the product of $p$-values of all possible combinations of 5 FRBs out of the 12 hostless candidates. We found that the probability of this scenario is $\approx 10^{-7}$, thus implying that the probability of 5 of these 12 hostless FRBs being part of our local-universe sample is exceedingly low. This analysis demonstrates that the lack of low-mass hosts is not merely a selection effect and supports the conclusion that the deficit of low-mass hosts among the DSA-discovered FRBs is not due to observational limitations but reflects a genuine absence of such hosts in our sample.}}
\end{figure}

\begin{figure*}[ht!]
\centering
\includegraphics[width=0.3225\textwidth, height=4.2cm]{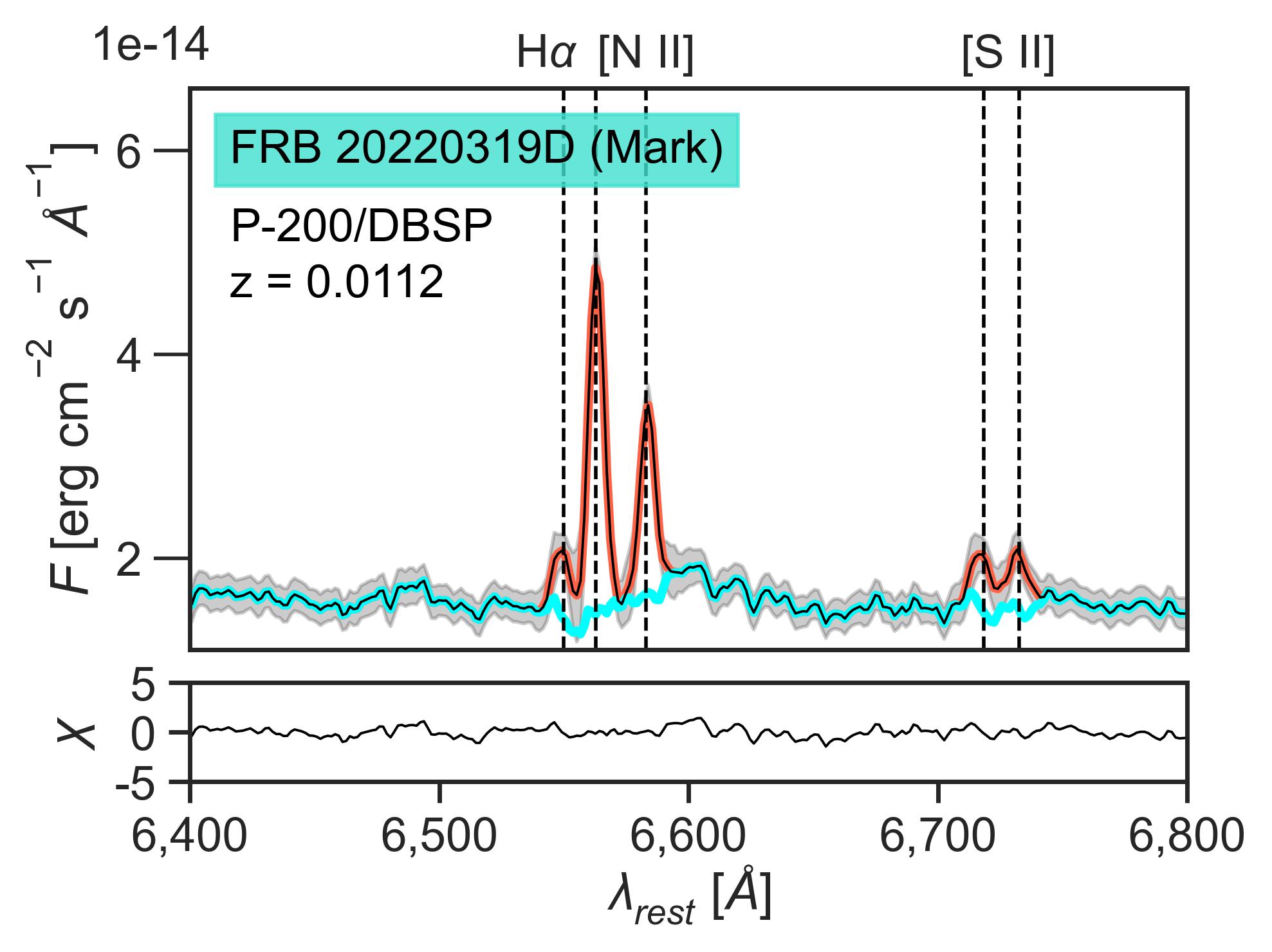}
\includegraphics[width=0.3225\textwidth, height=4.2cm]{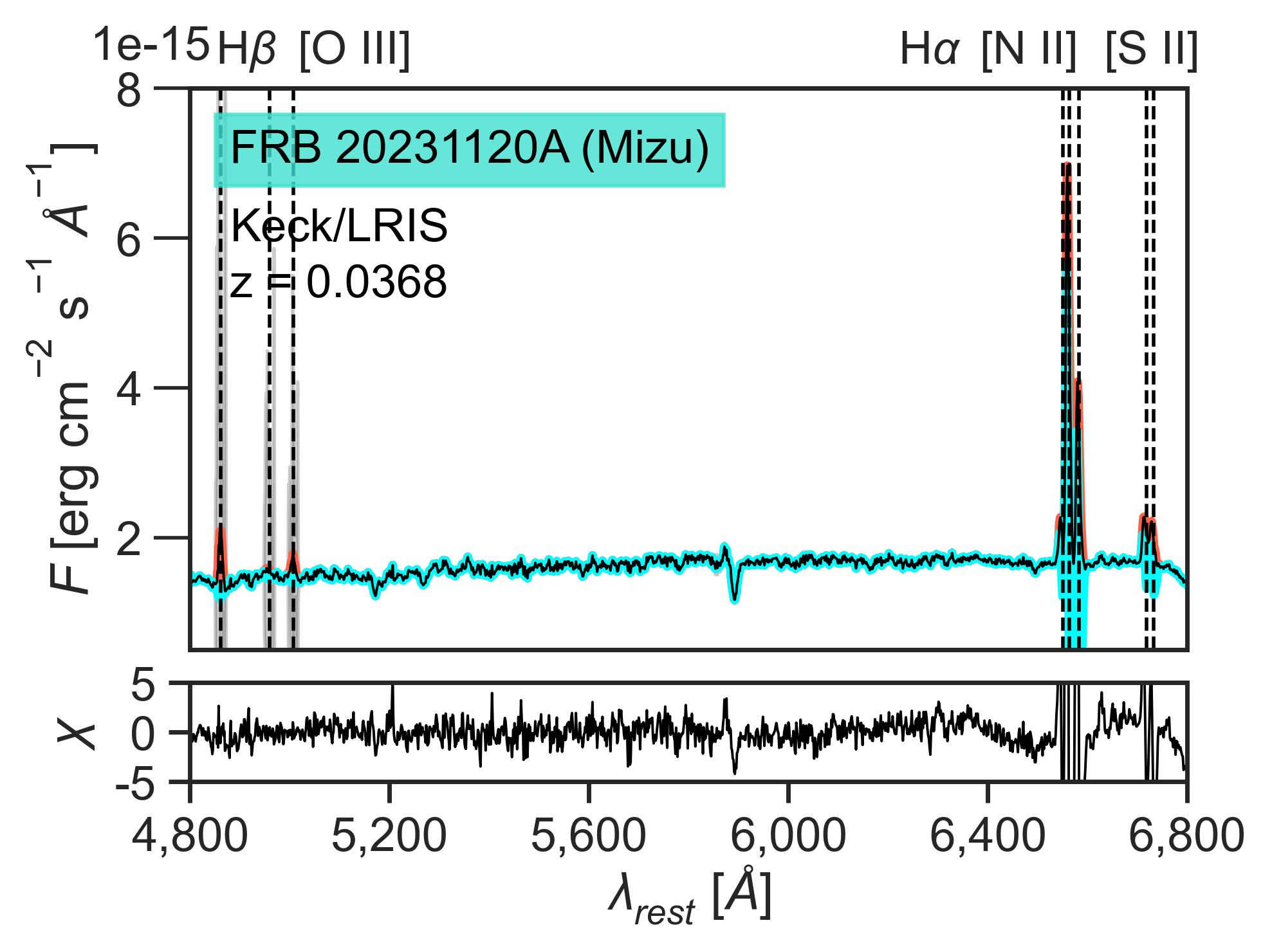}
\includegraphics[width=0.3225\textwidth, height=4.2cm]{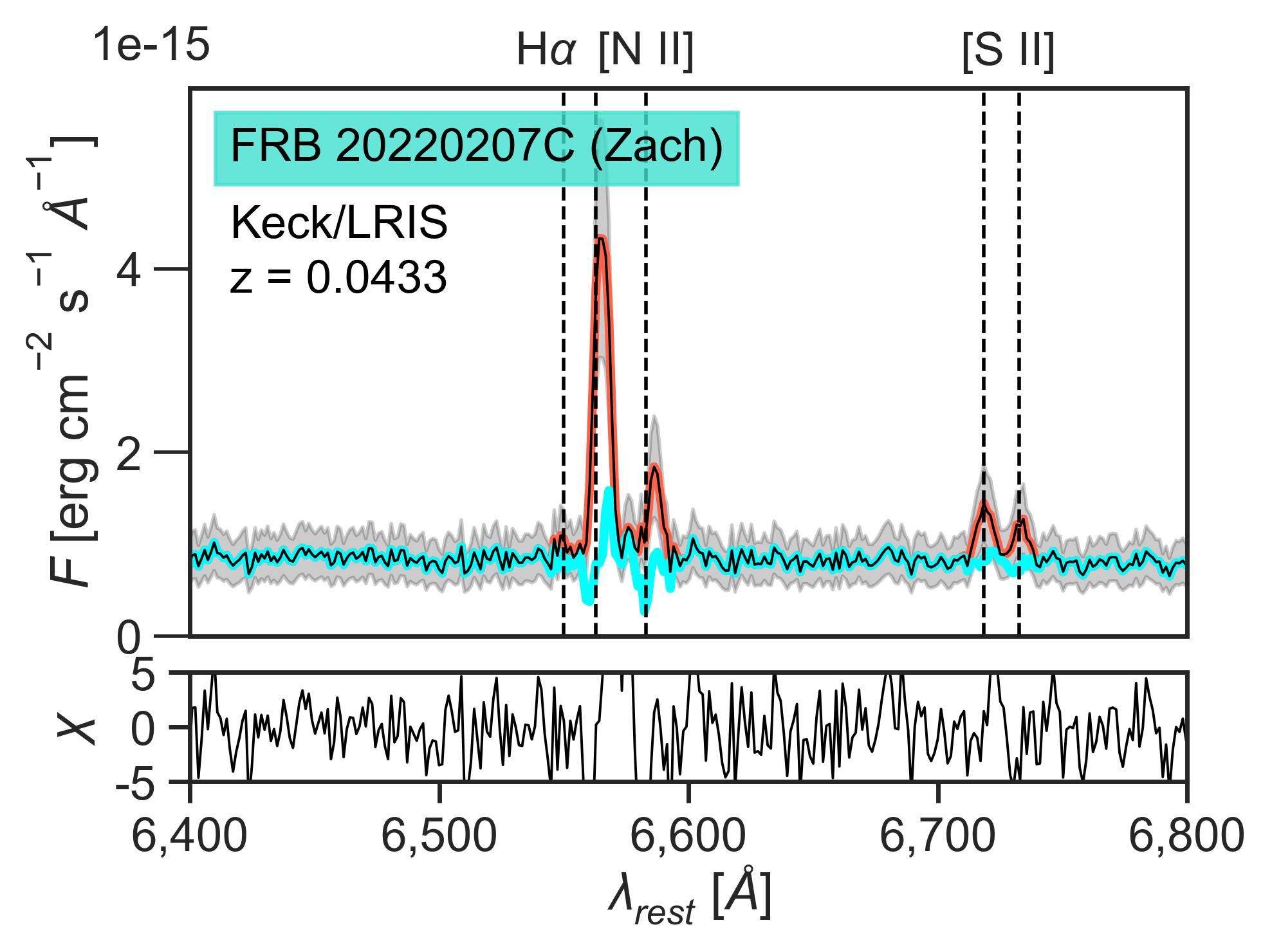}
\includegraphics[width=0.3225\textwidth, height=4.2cm]{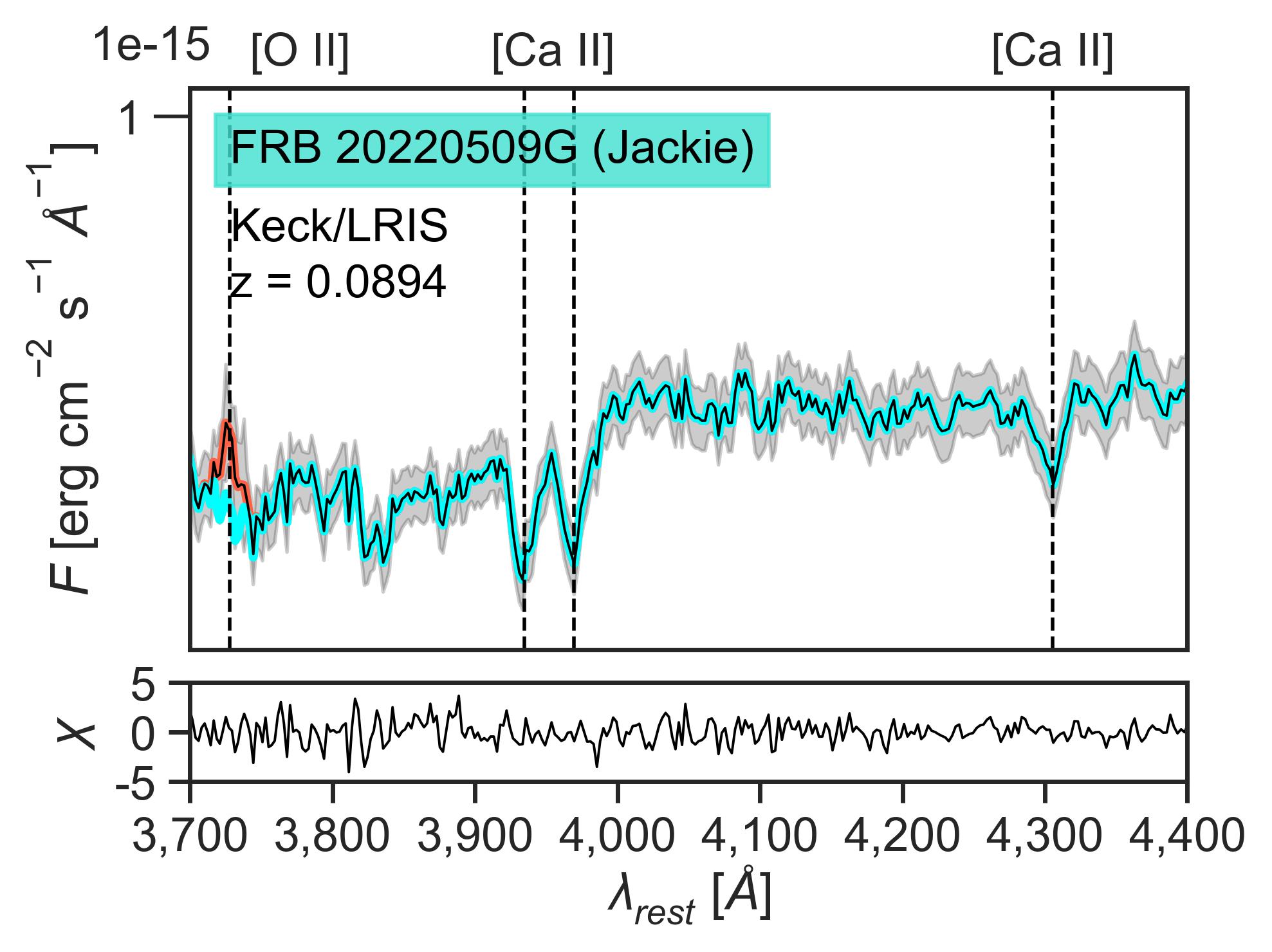}
\includegraphics[width=0.3225\textwidth, height=4.2cm]{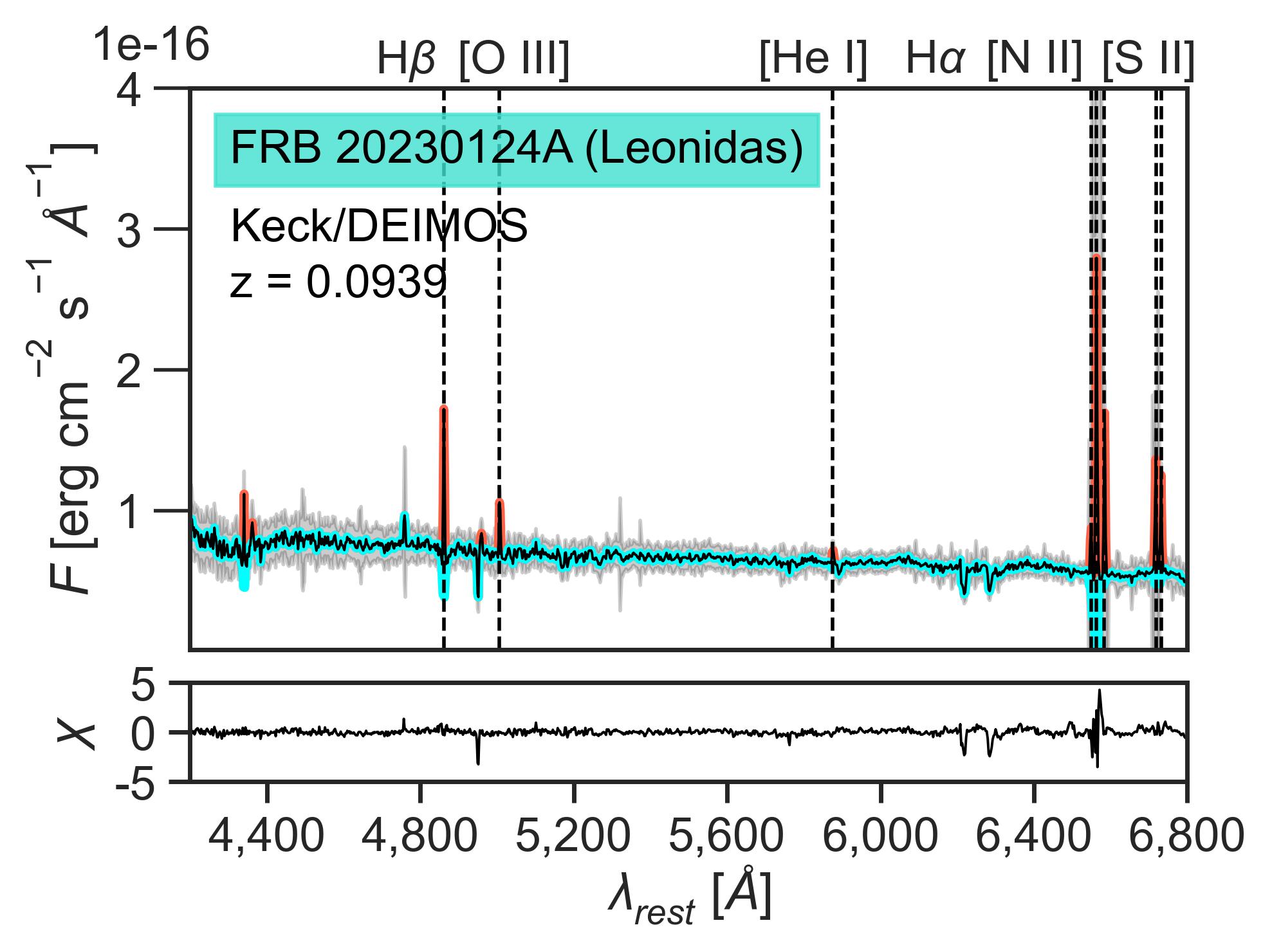}
\includegraphics[width=0.3225\textwidth, height=4.2cm]{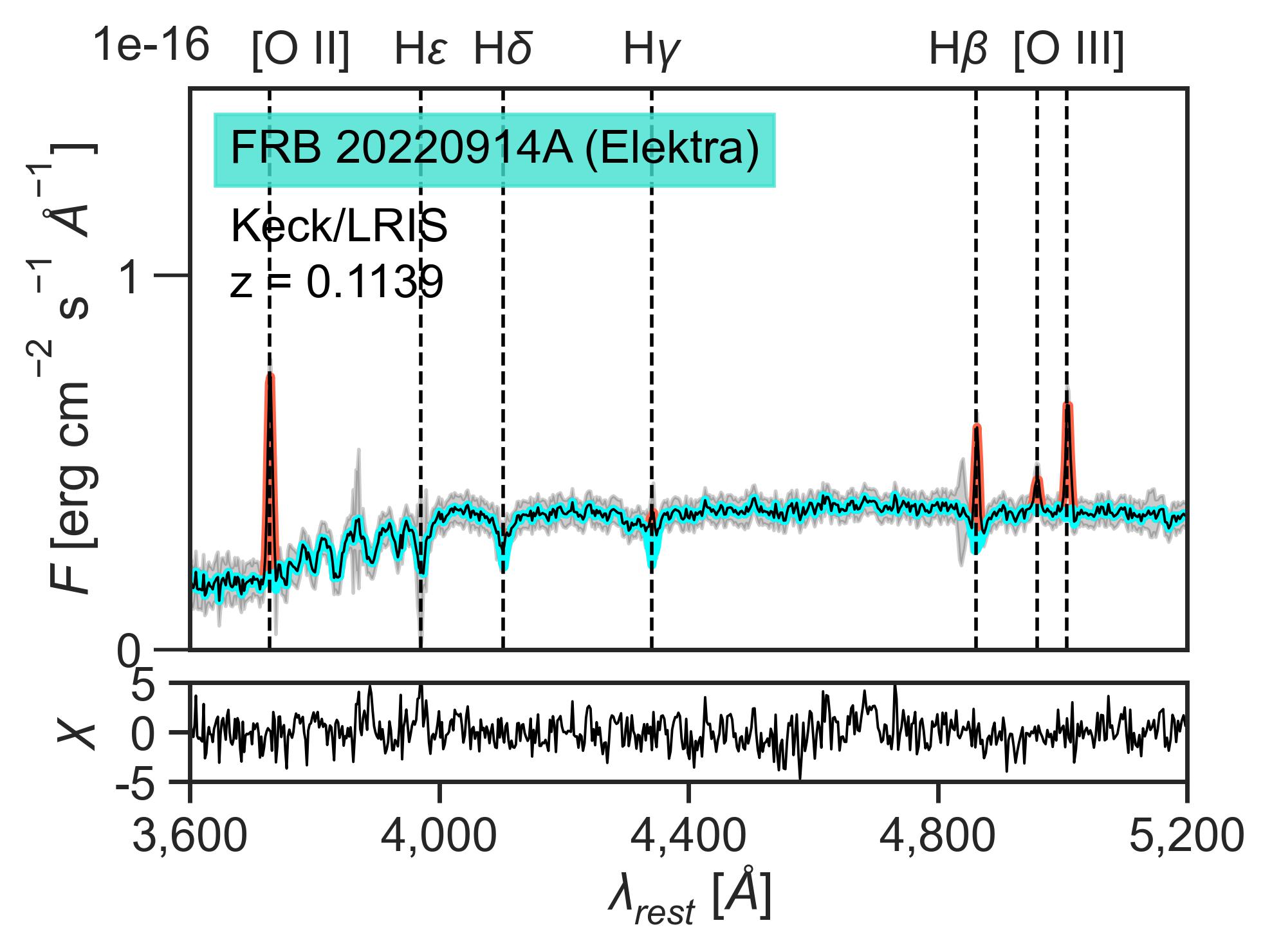}
\includegraphics[width=0.3225\textwidth, height=4.2cm]{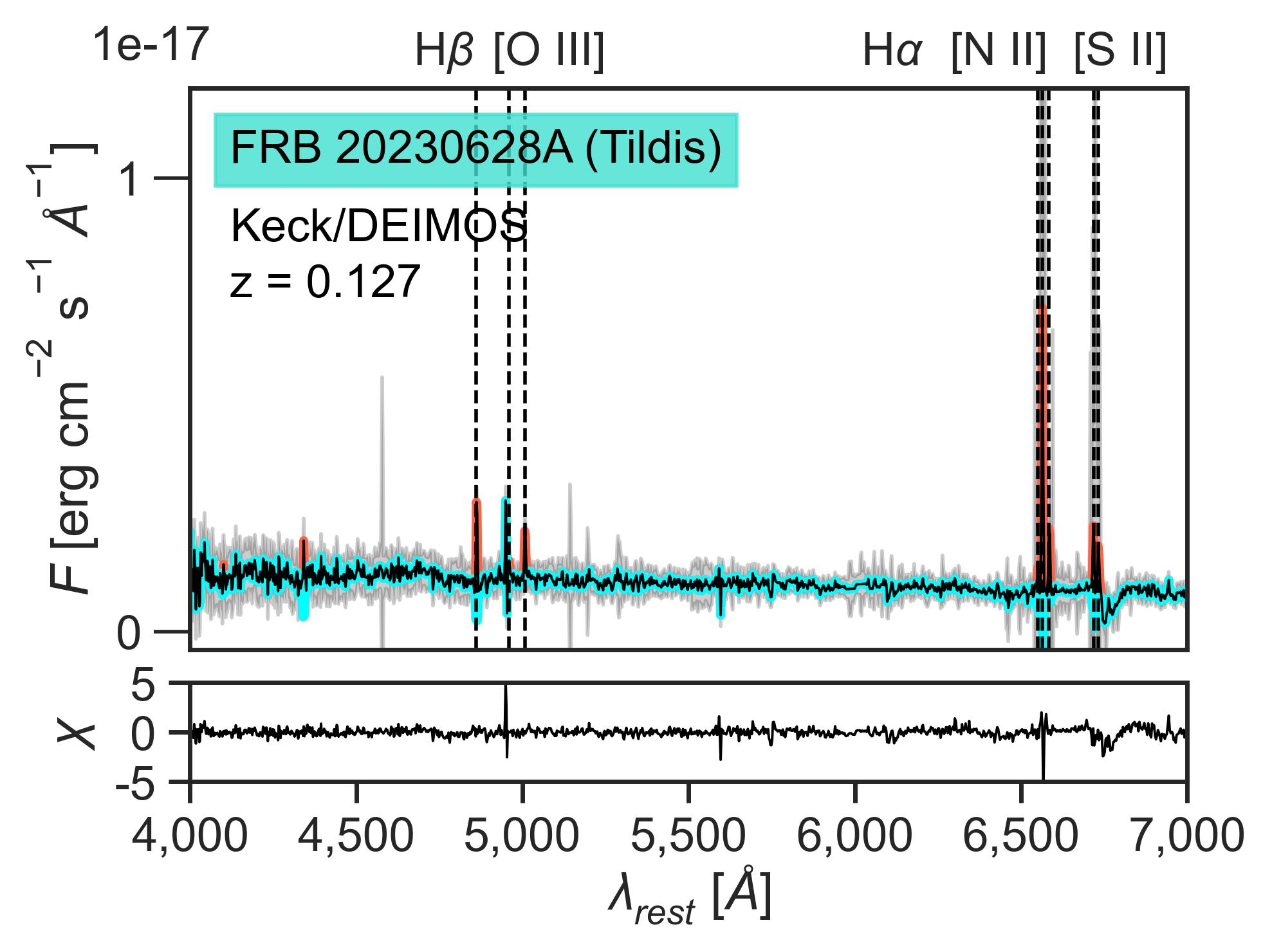}
\includegraphics[width=0.3225\textwidth, height=4.2cm]{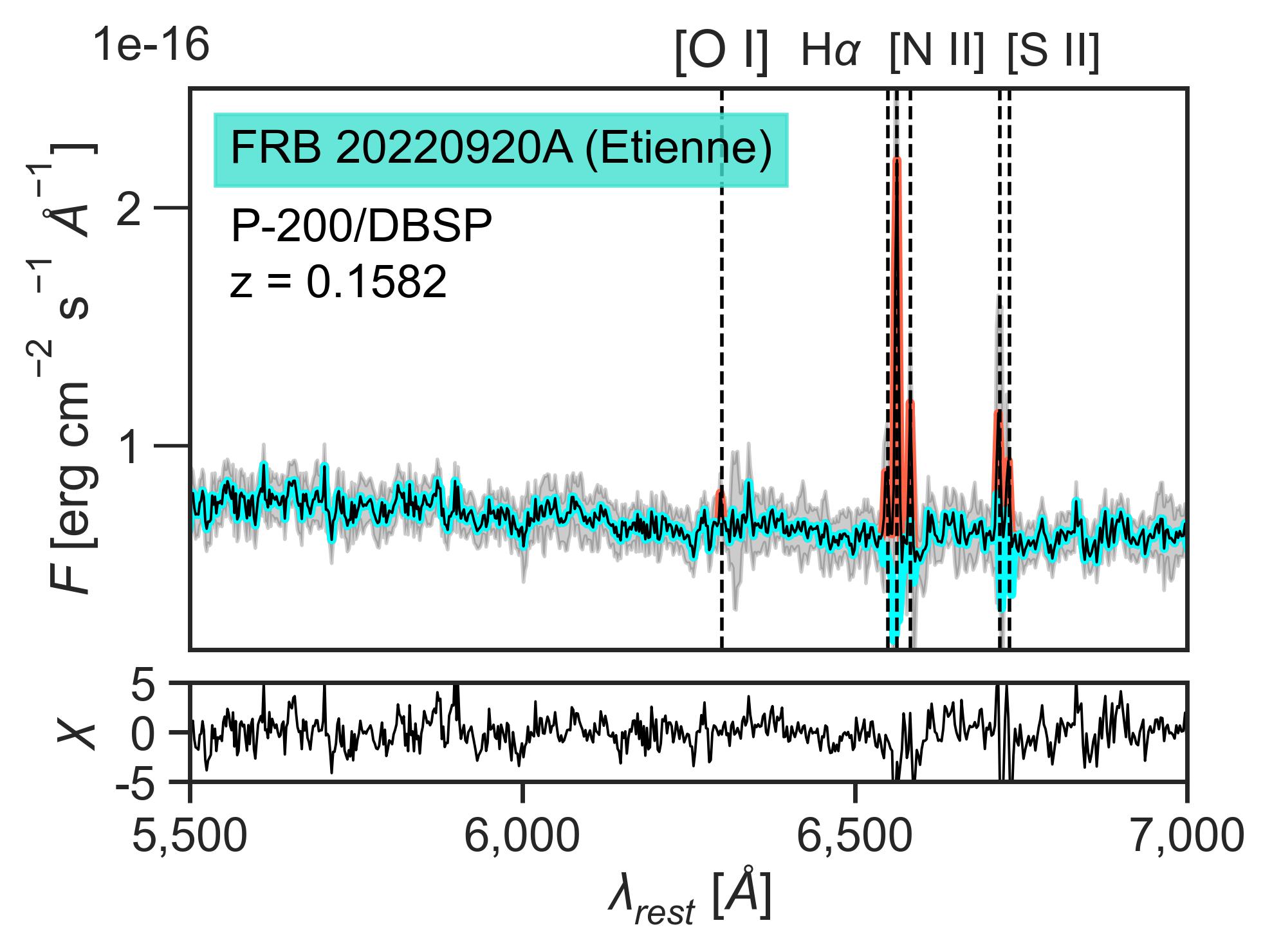}
\includegraphics[width=0.3225\textwidth, height=4.2cm]{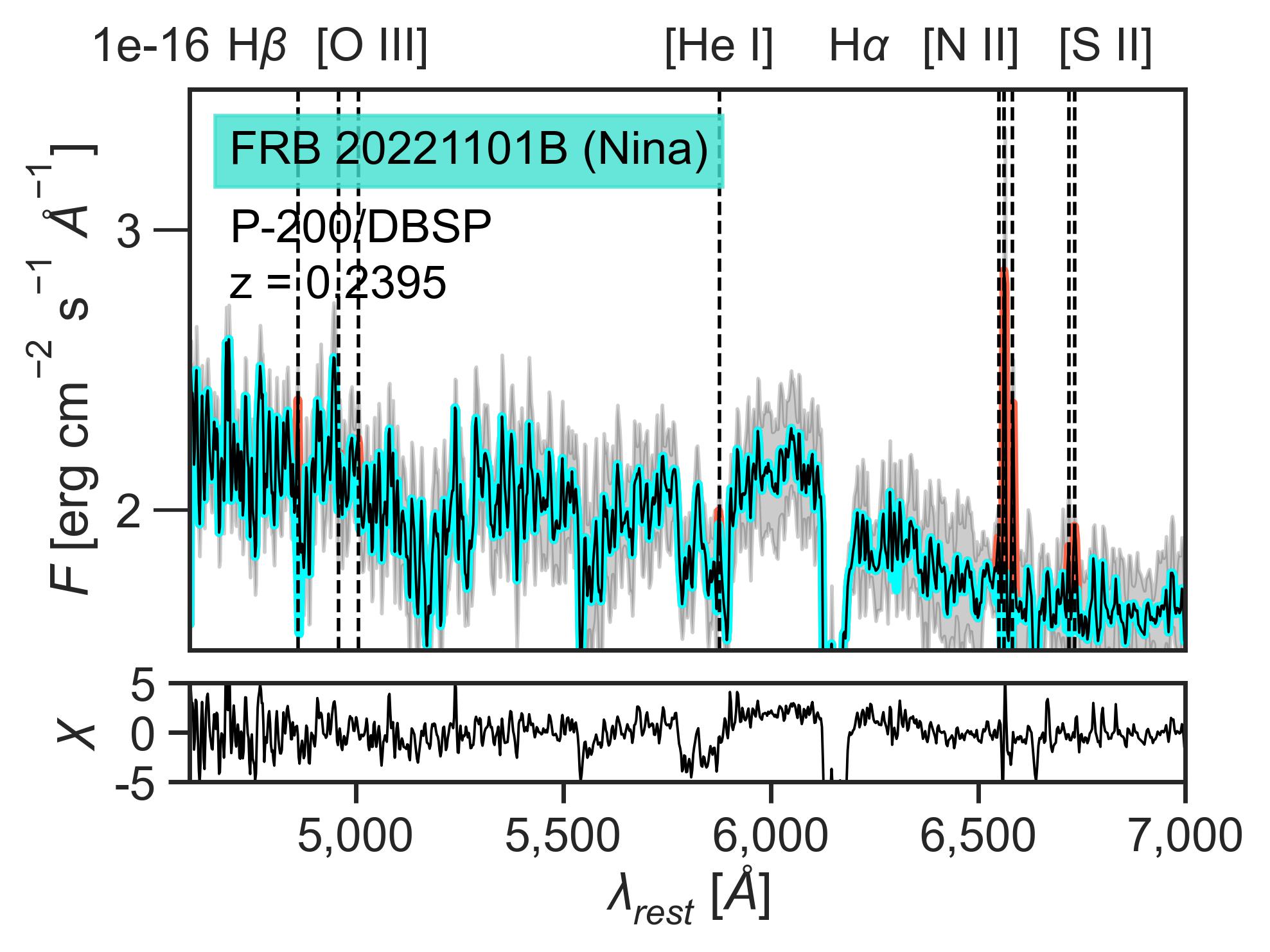}
\includegraphics[width=0.3225\textwidth, height=4.2cm]{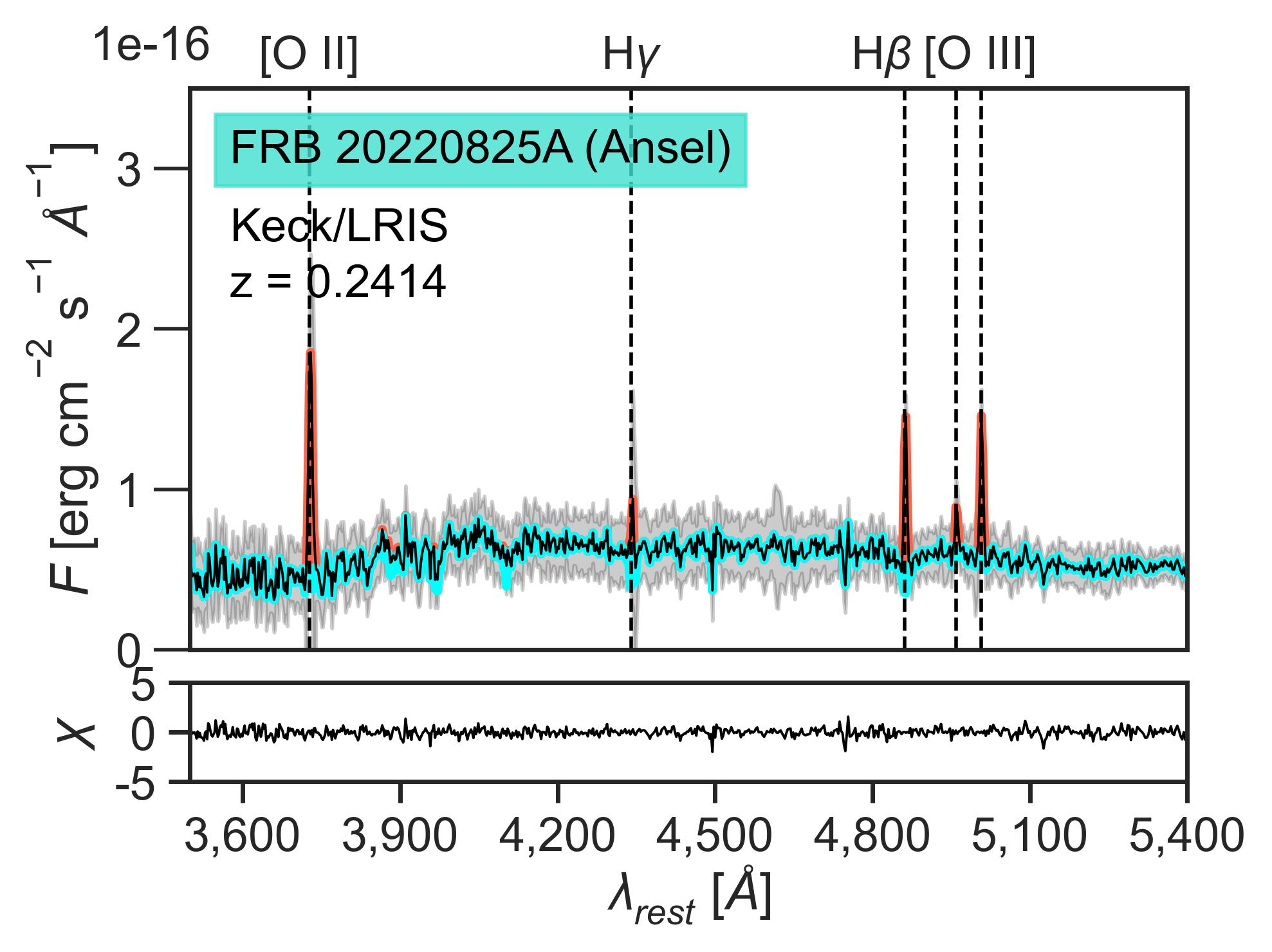}
\includegraphics[width=0.3225\textwidth, height=4.2cm]{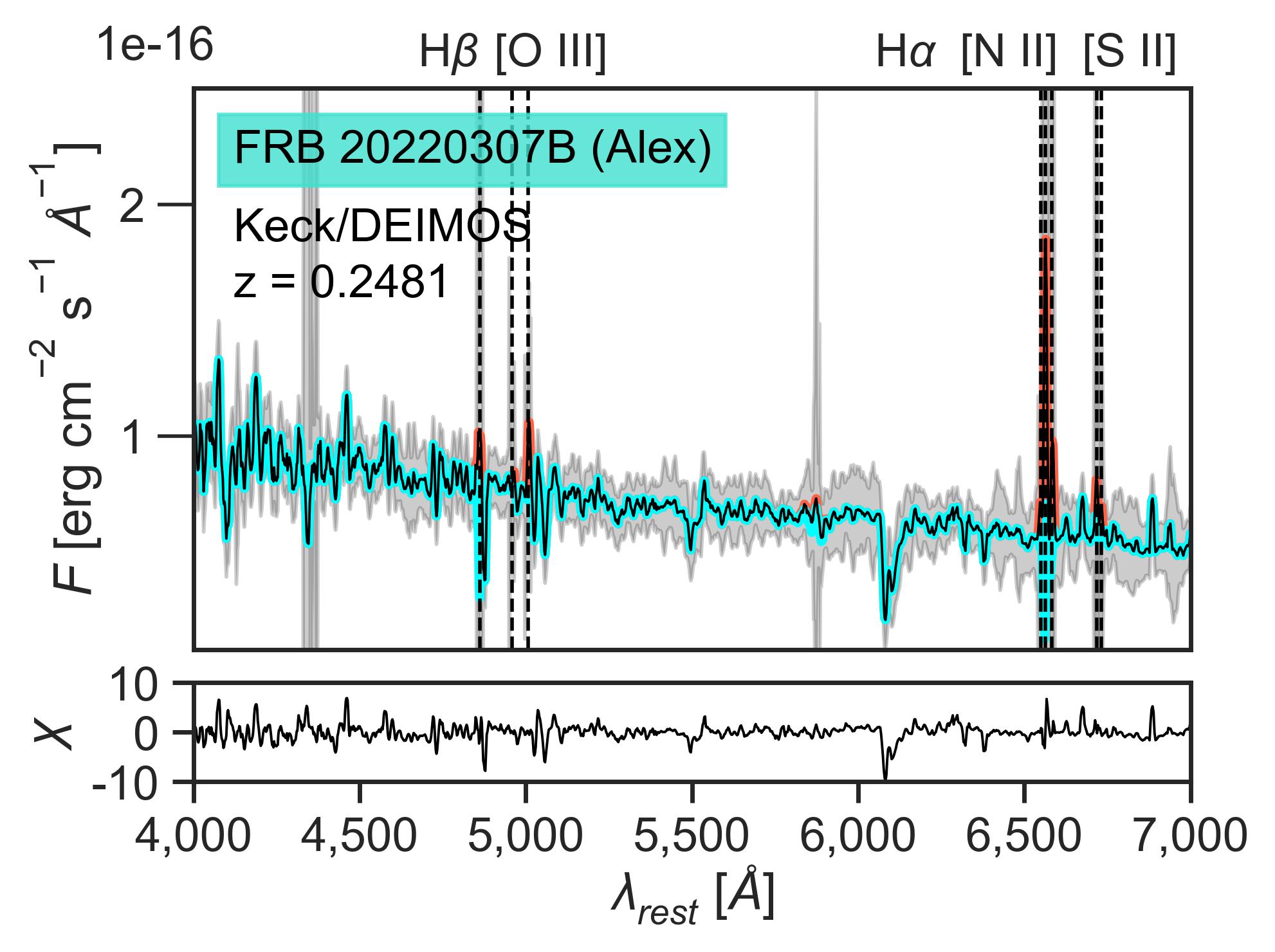}
\includegraphics[width=0.3225\textwidth, height=4.2cm]{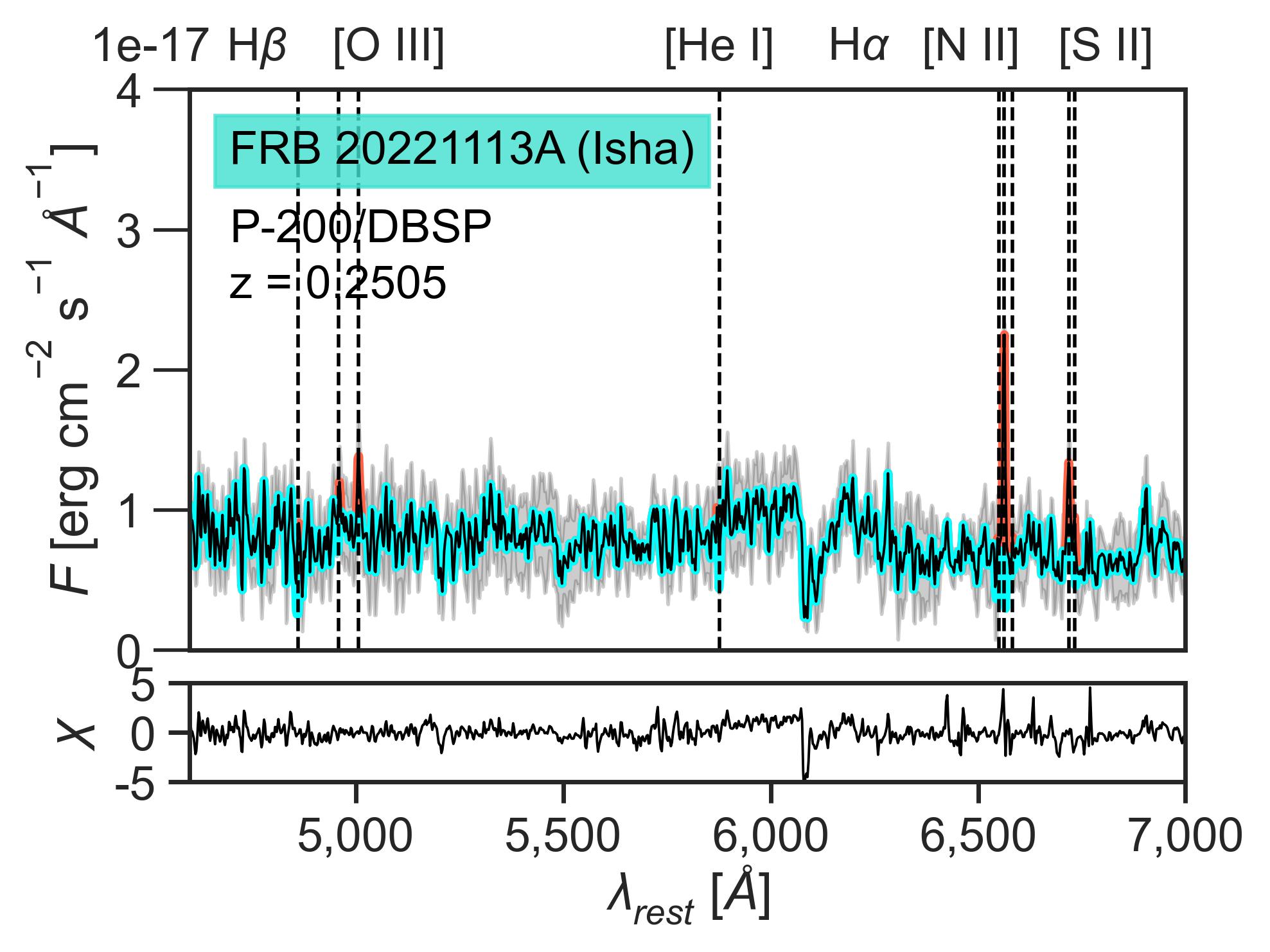}
\includegraphics[width=0.3225\textwidth, height=4.2cm]{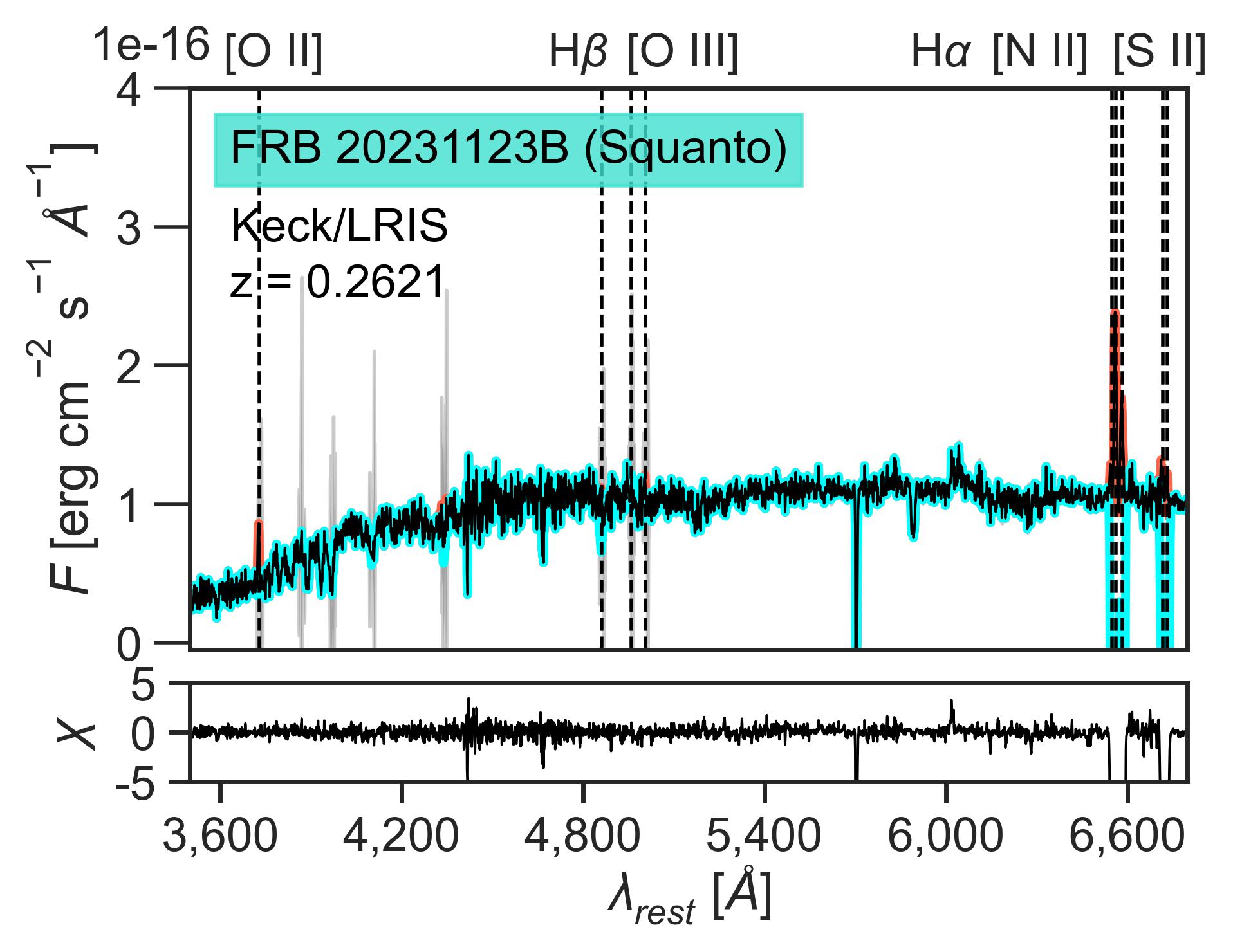}
\includegraphics[width=0.3225\textwidth, height=4.2cm]{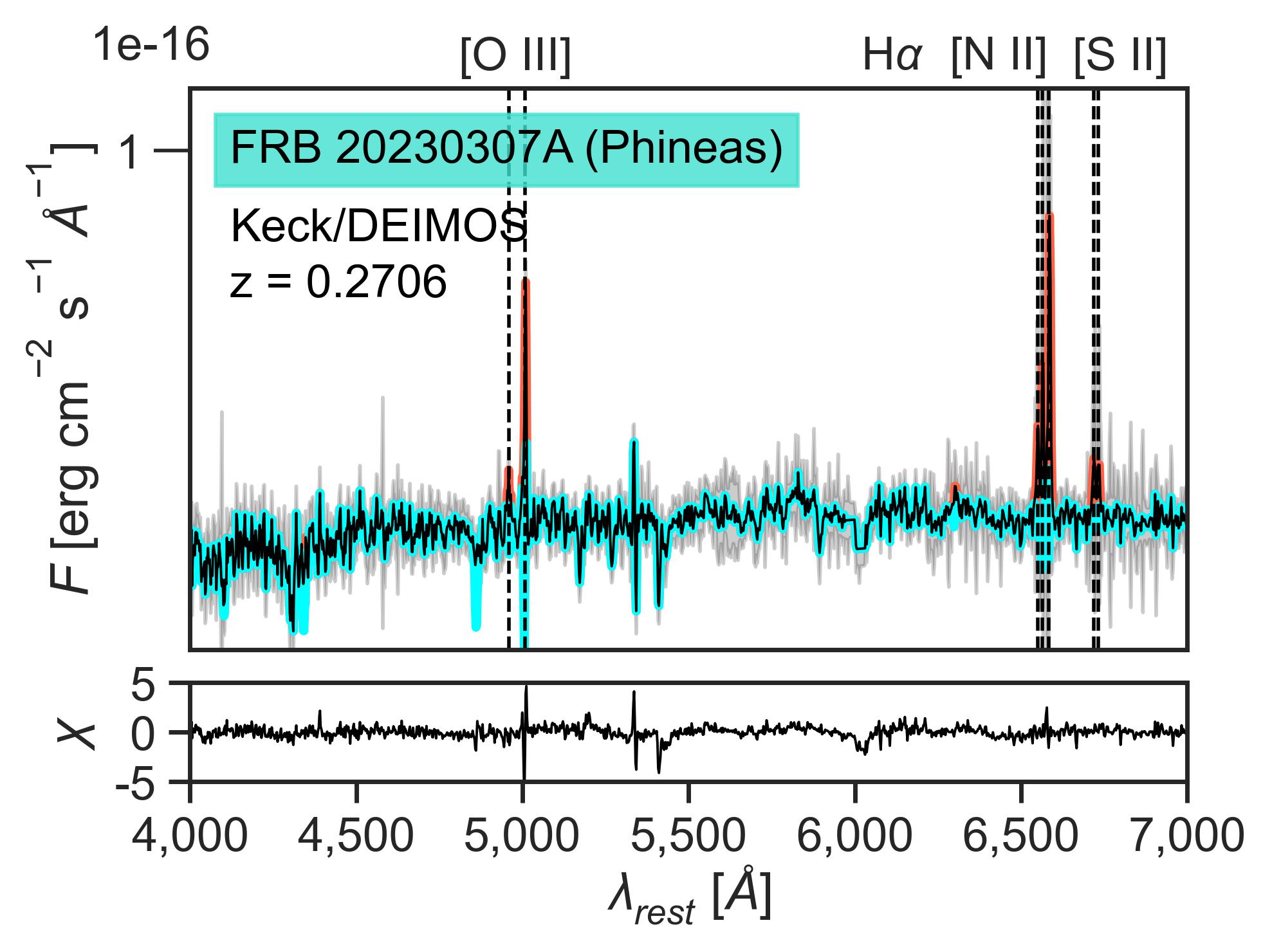}
\includegraphics[width=0.3225\textwidth, height=4.2cm]{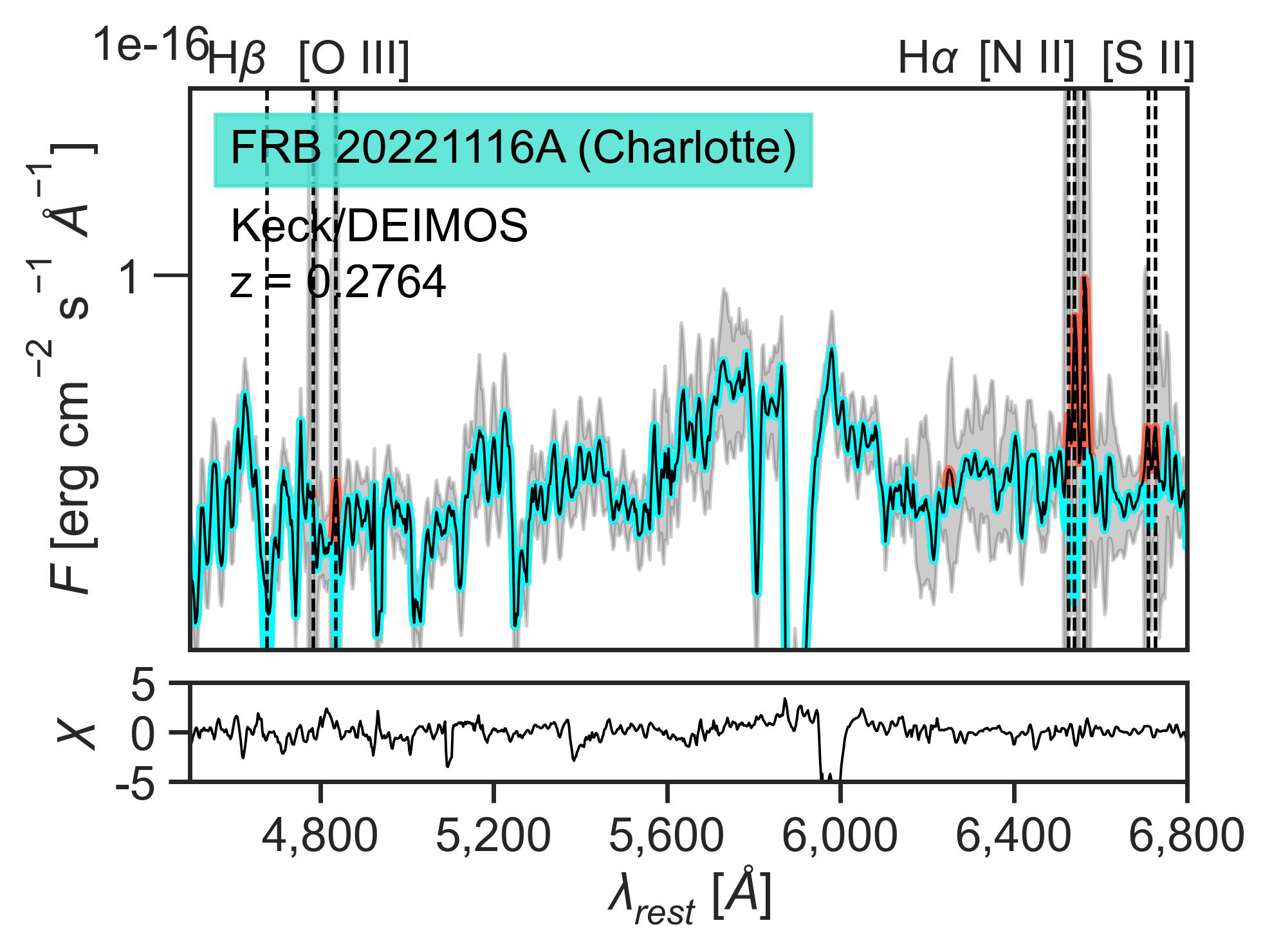}
\caption{{\supfigurelabel{fig:host_spectra}} \textbf{Spectroscopic observations obtained with Keck:I/LRIS, Keck:II/DEIMOS and P-200/DBSP.} We scale the spectra to match galactic-extinction corrected photometry using \sw{Prospector}. The \sw{pPXF} fits to stellar continuum (cyan) and nebular emission (red) are included in all panels.}
\end{figure*}

\newpage
\clearpage
\newpage
\newpage

\begin{figure*}[ht!]
\centering
\ContinuedFloat
\includegraphics[width=0.3225\textwidth, height=4.2cm]{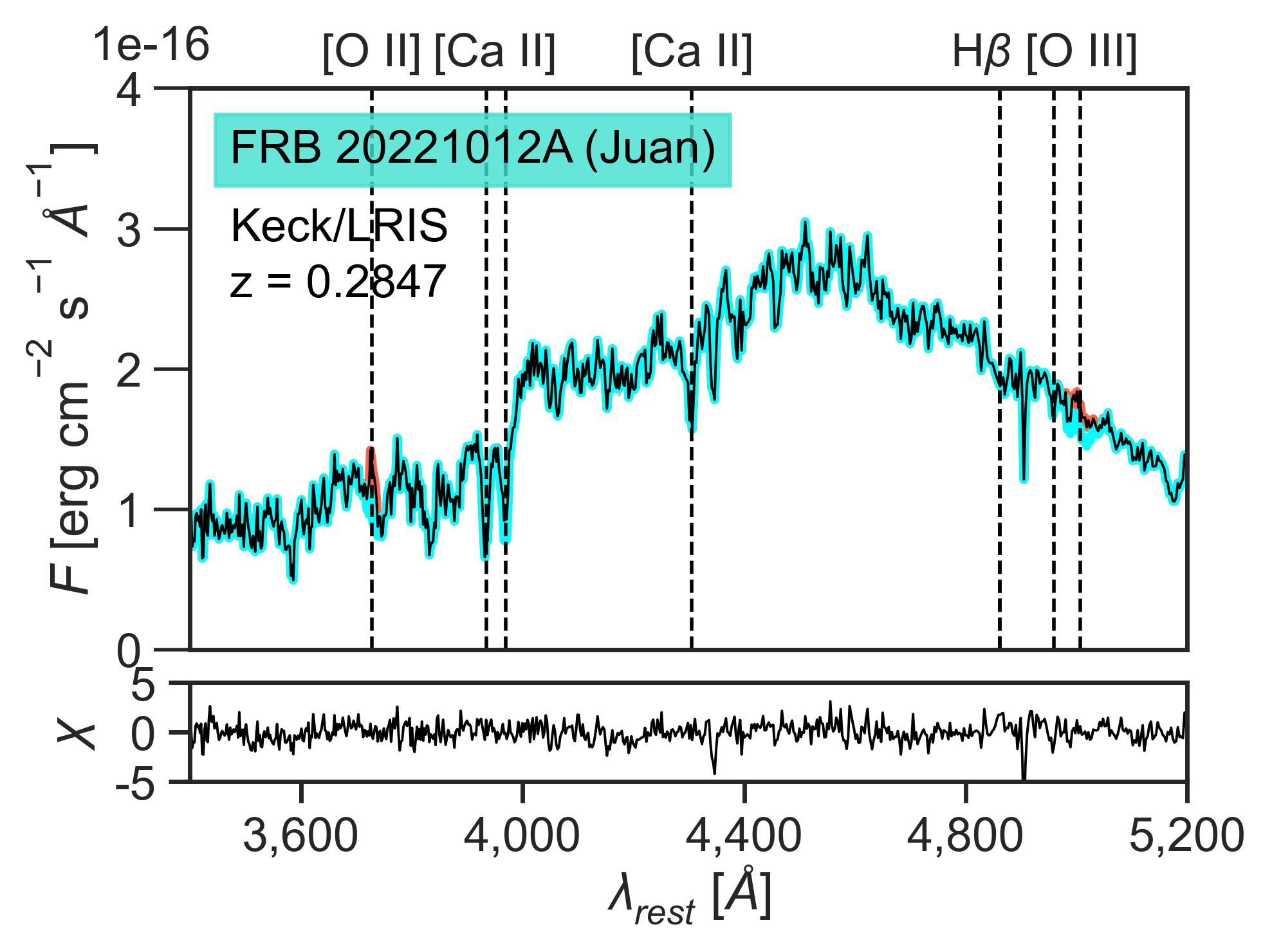}
\includegraphics[width=0.3225\textwidth, height=4.2cm]{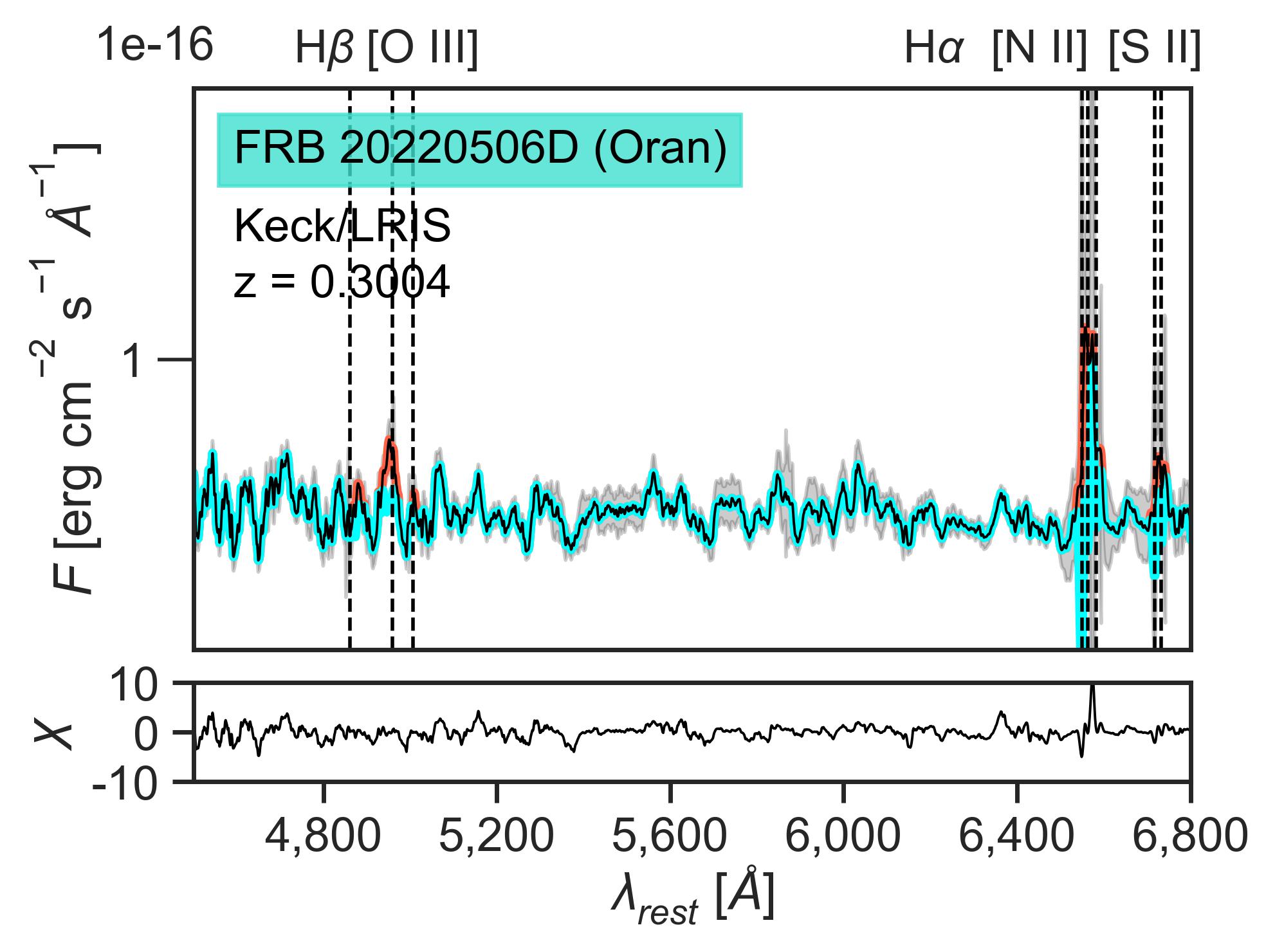}
\includegraphics[width=0.3225\textwidth, height=4.2cm]{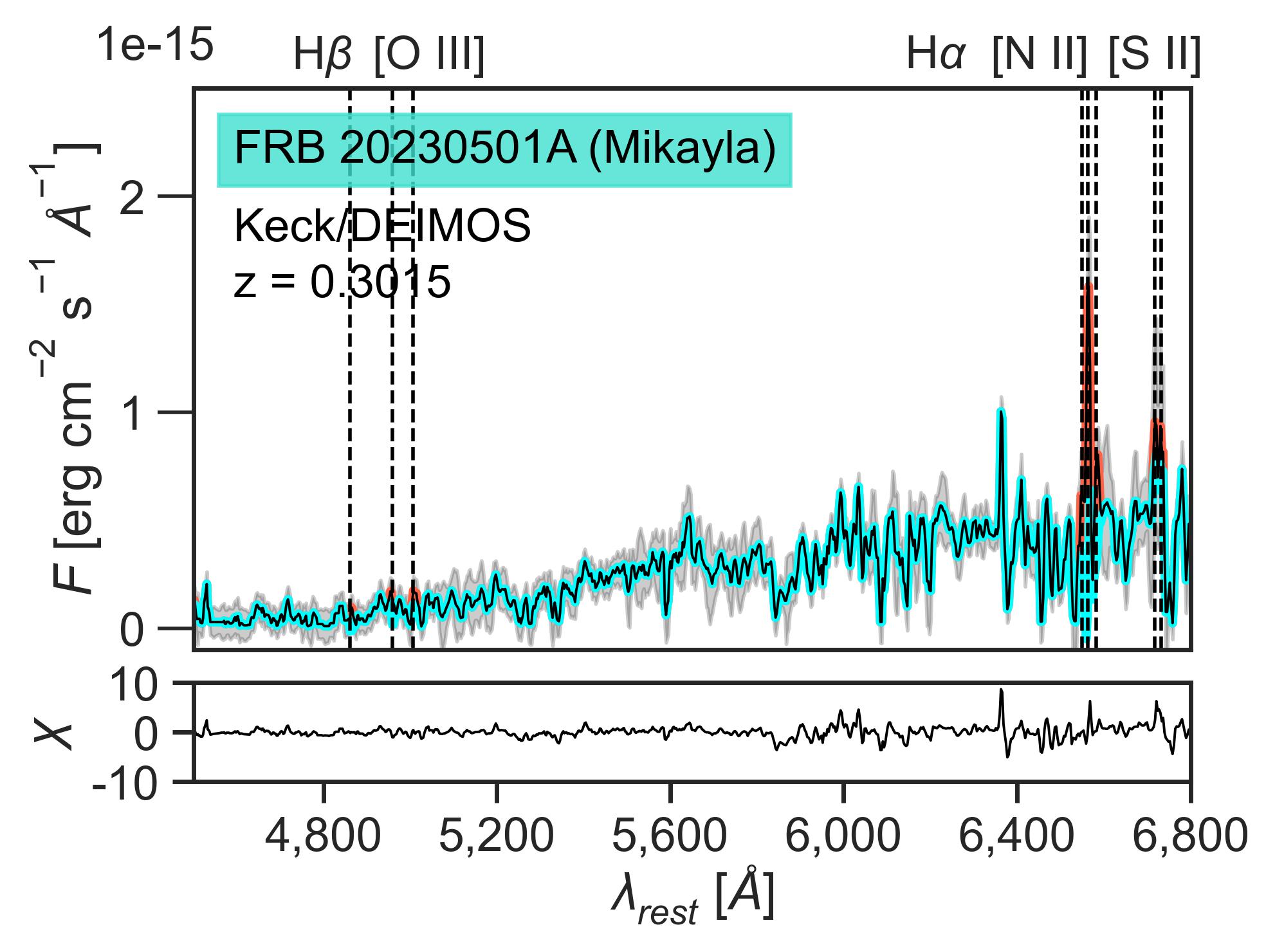}
\includegraphics[width=0.3225\textwidth, height=4.2cm]{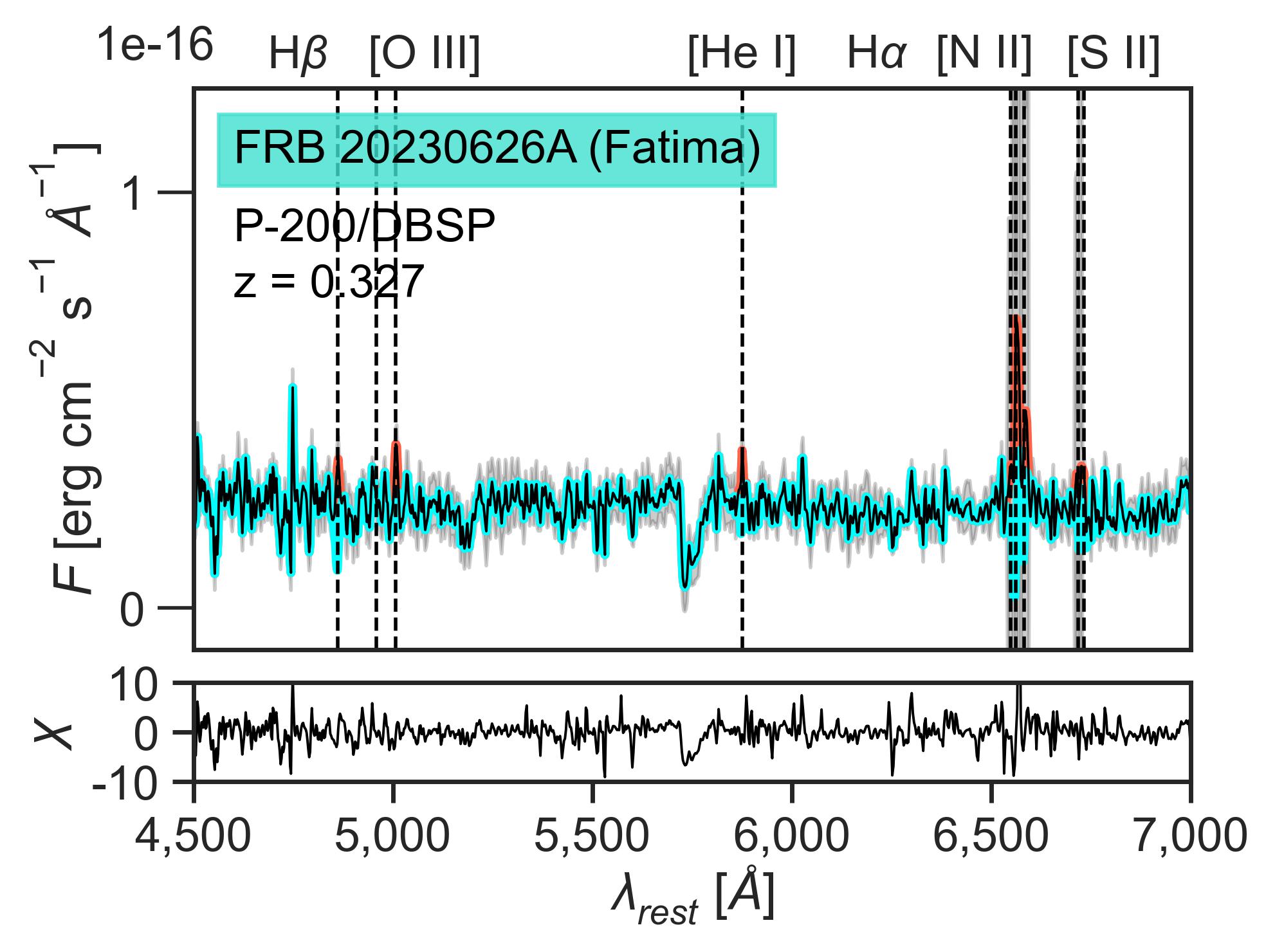}
\includegraphics[width=0.3225\textwidth, height=4.2cm]{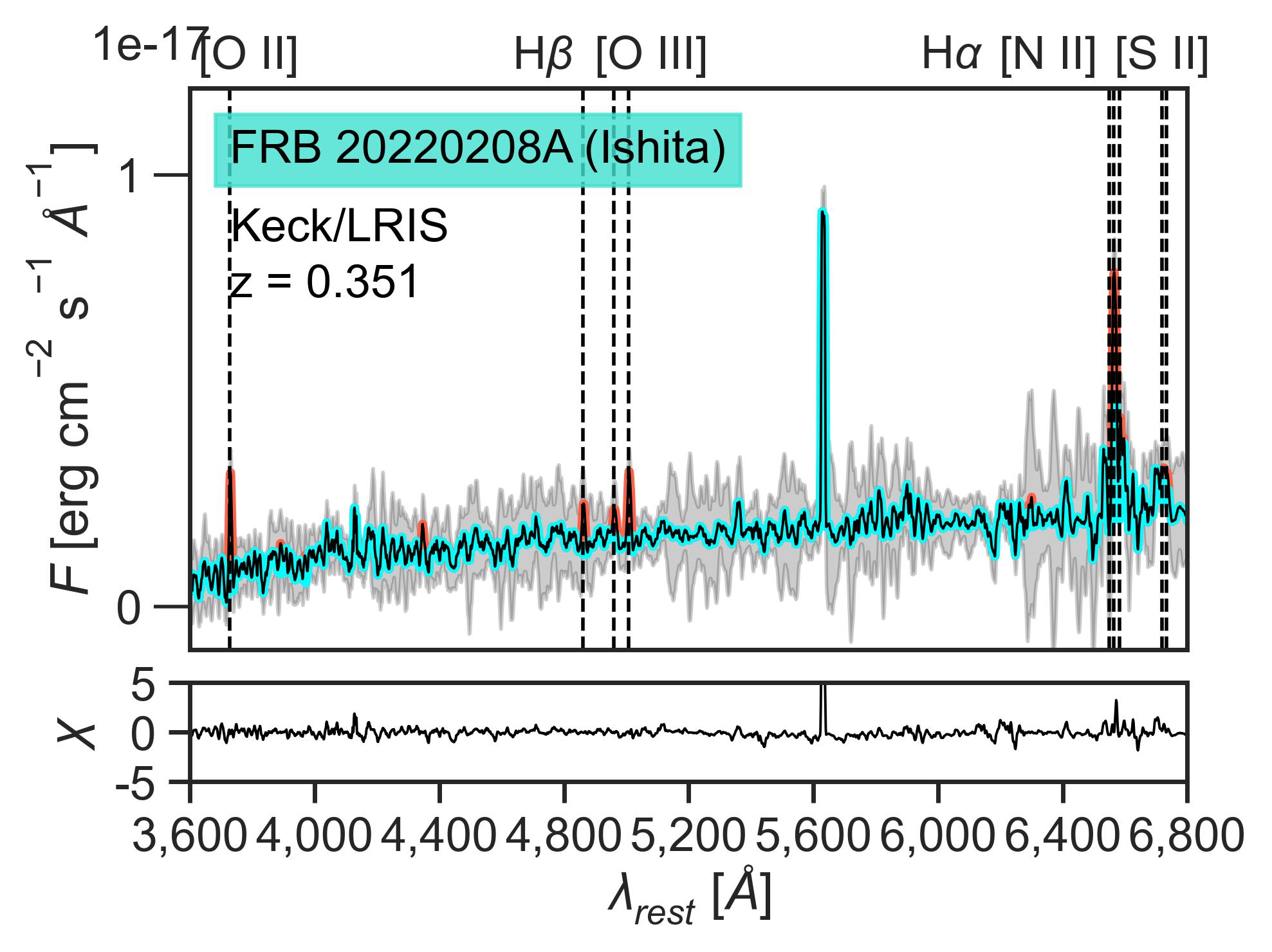}
\includegraphics[width=0.3225\textwidth, height=4.2cm]{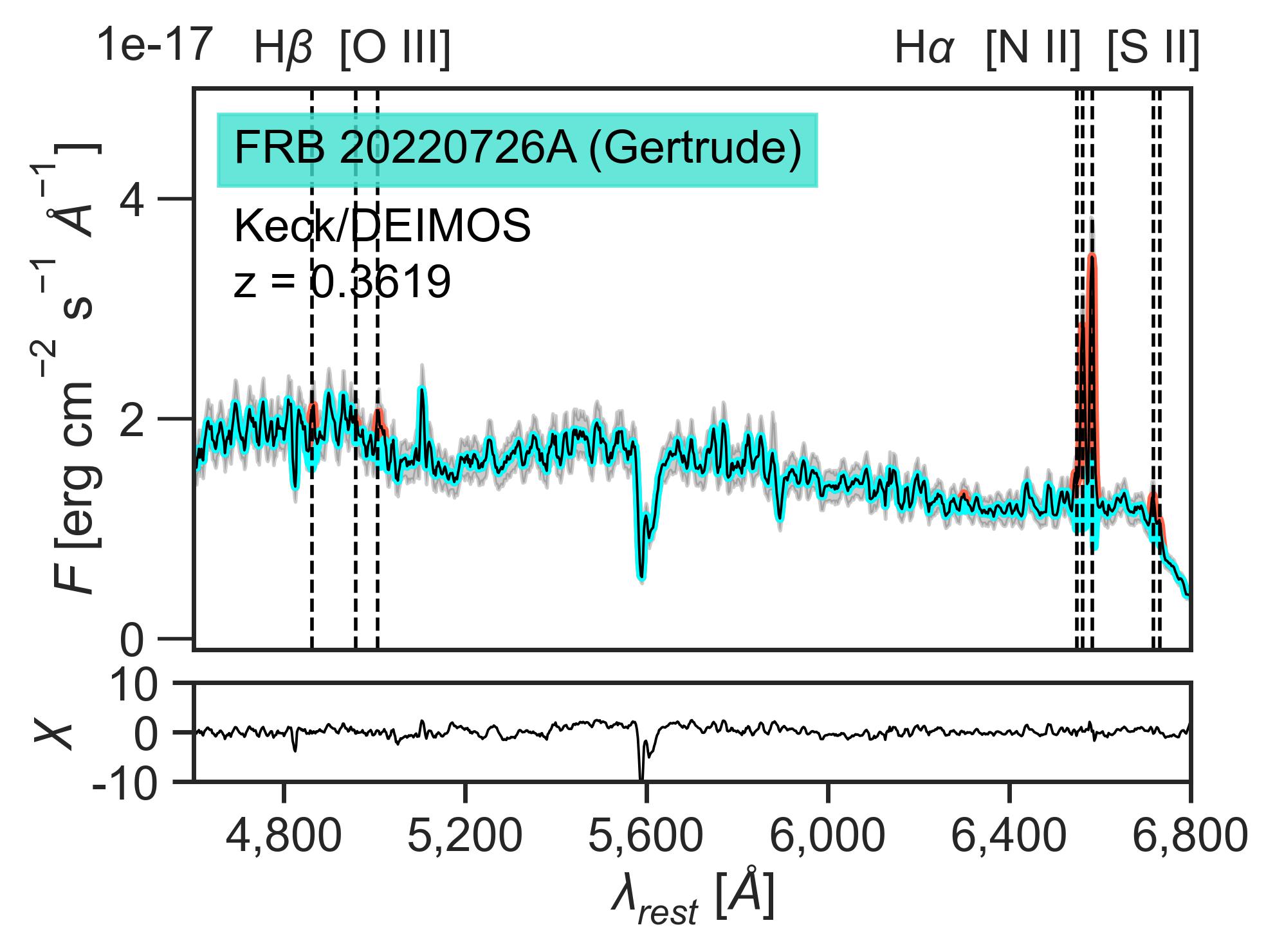}
\includegraphics[width=0.3225\textwidth, height=4.2cm]{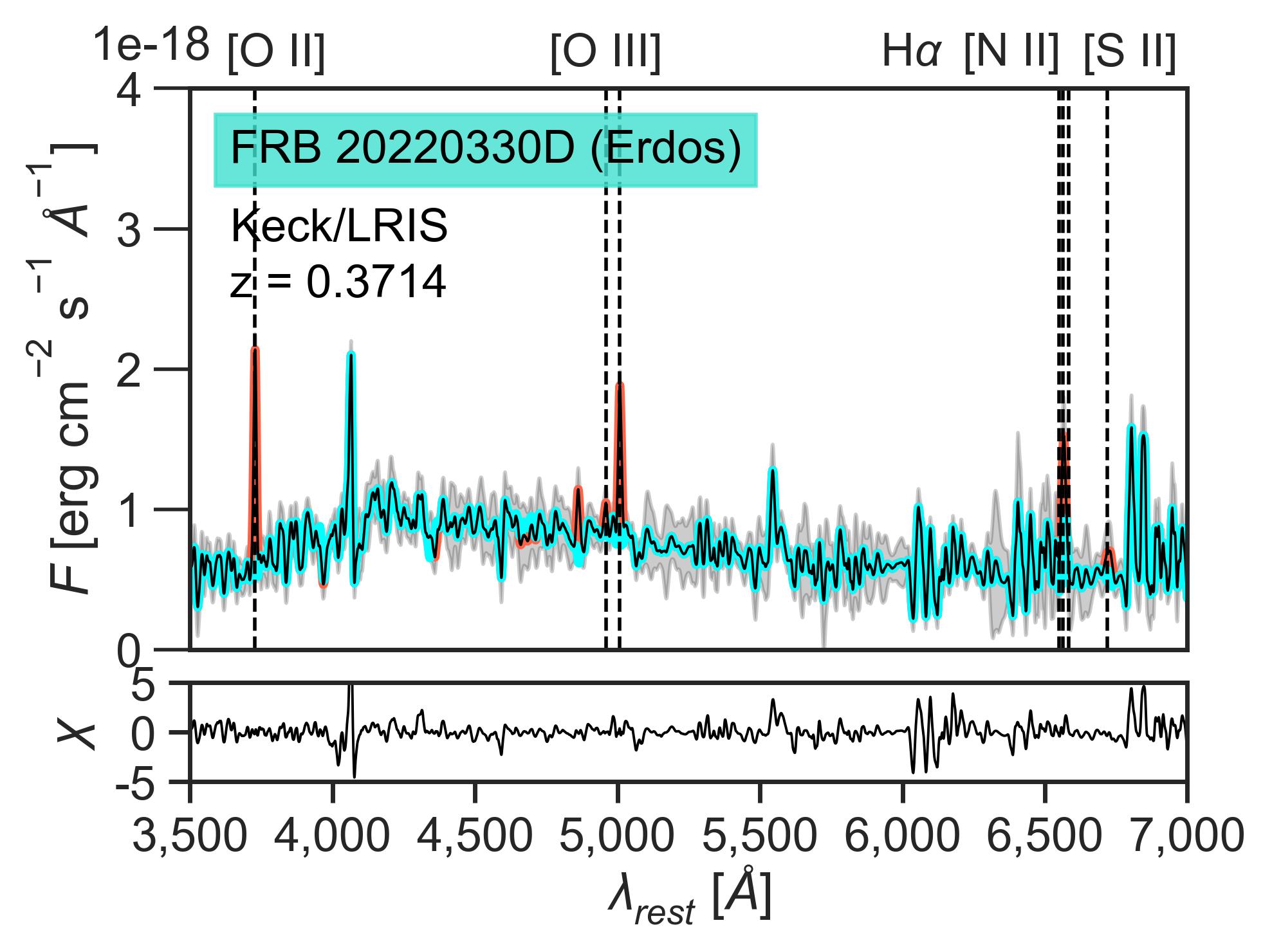}
\includegraphics[width=0.3225\textwidth, height=4.2cm]{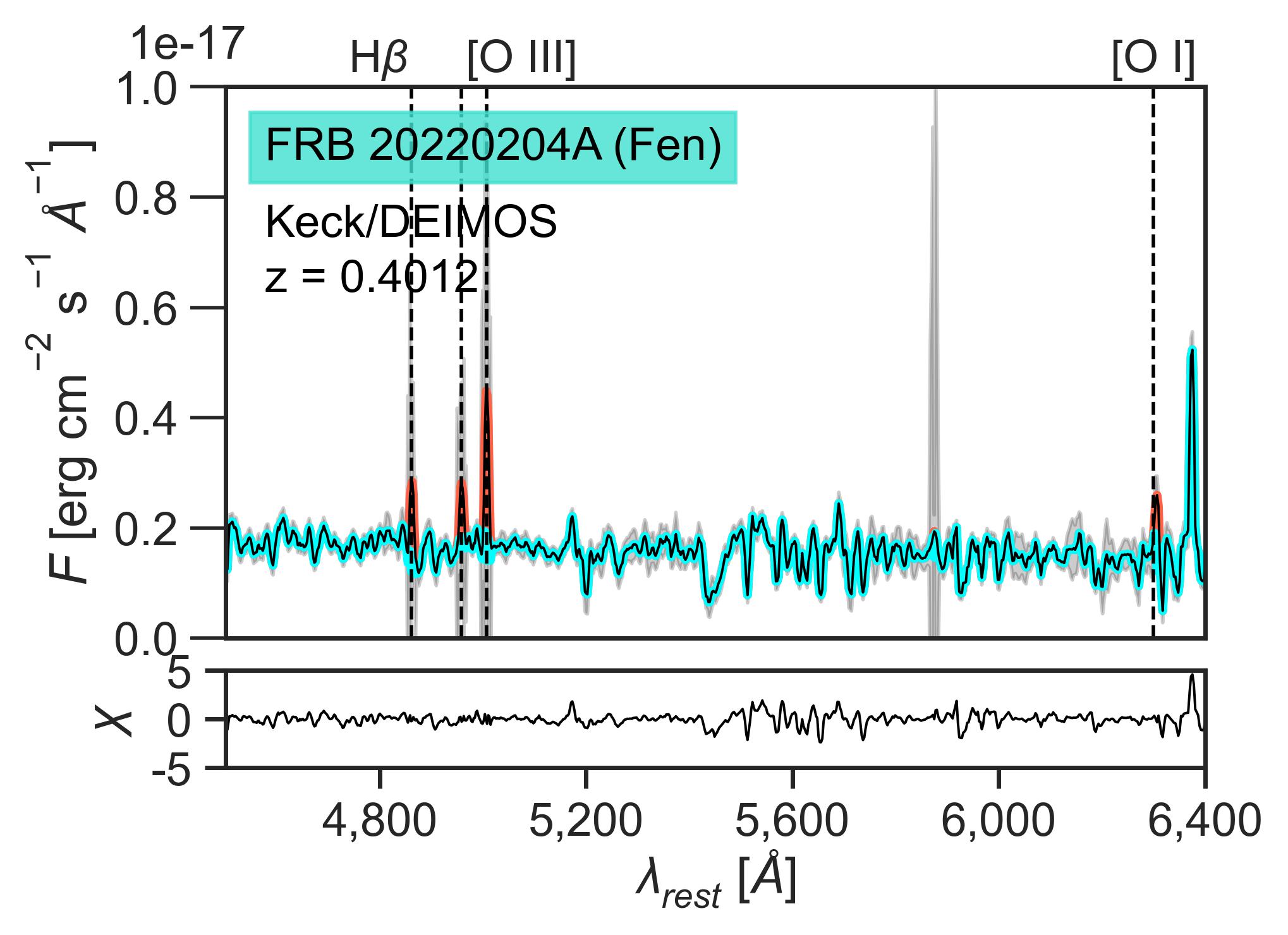}
\includegraphics[width=0.3225\textwidth, height=4.2cm]{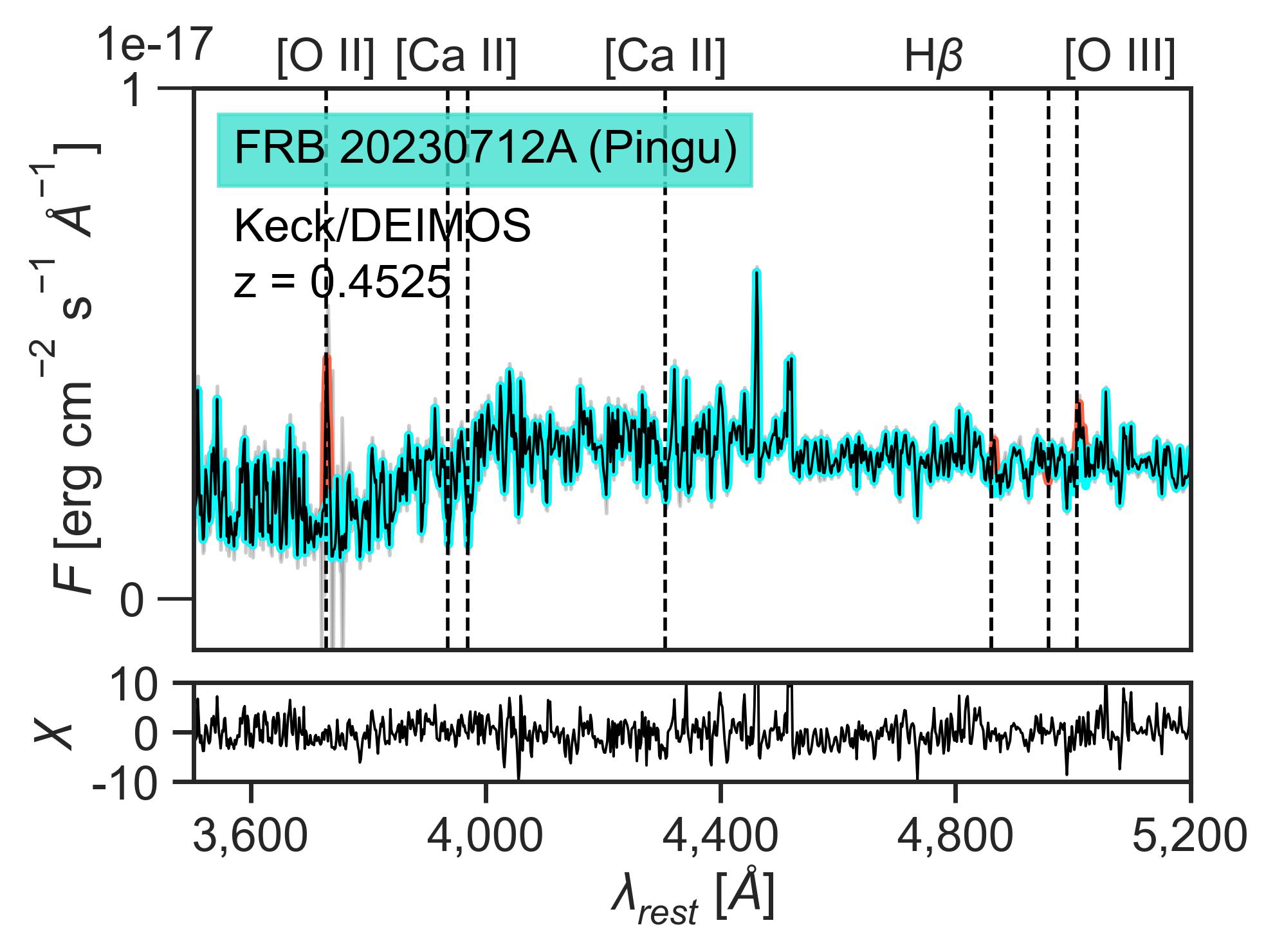}
\includegraphics[width=0.3225\textwidth, height=4.2cm]{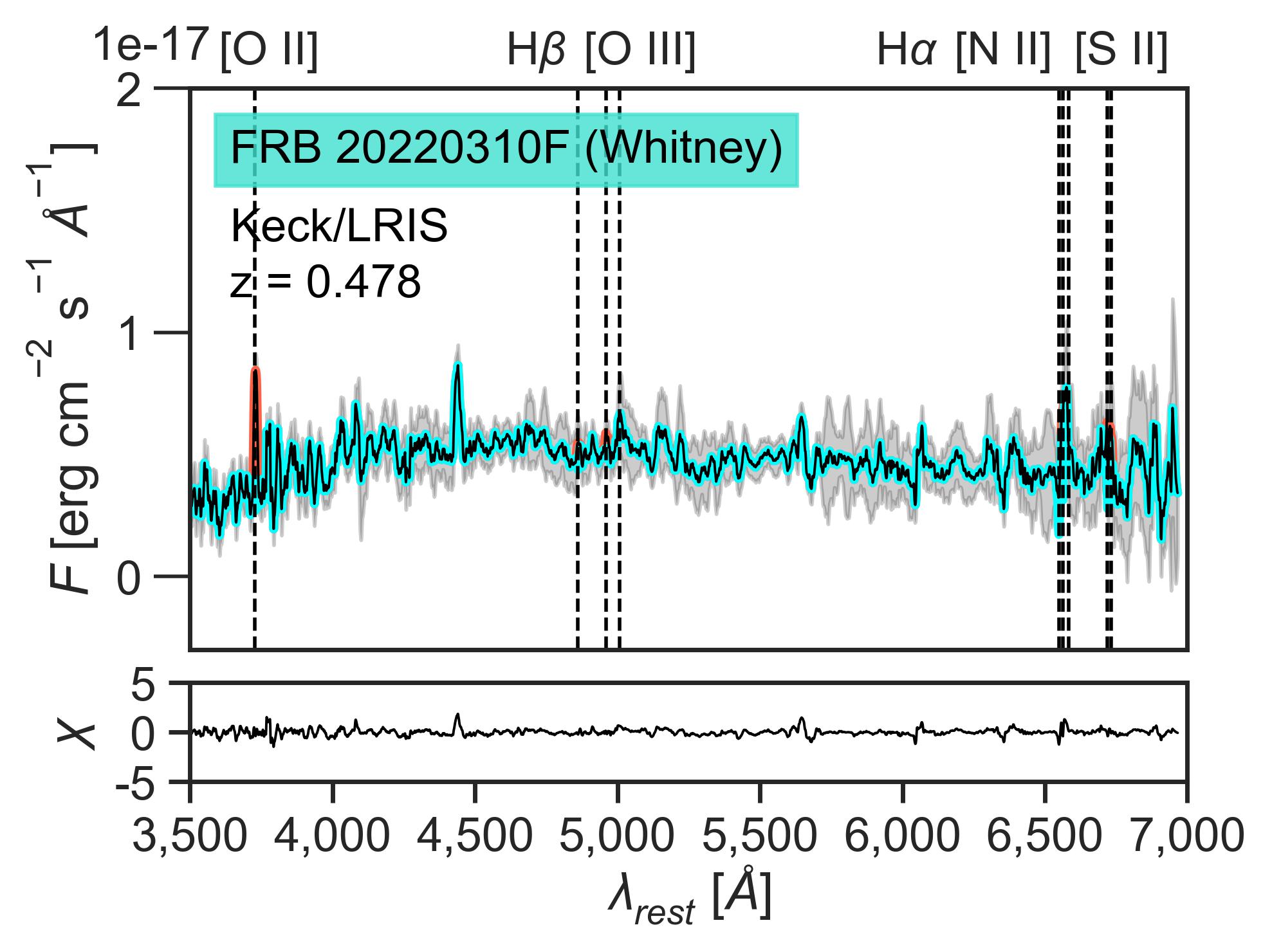}
\includegraphics[width=0.3225\textwidth, height=4.2cm]{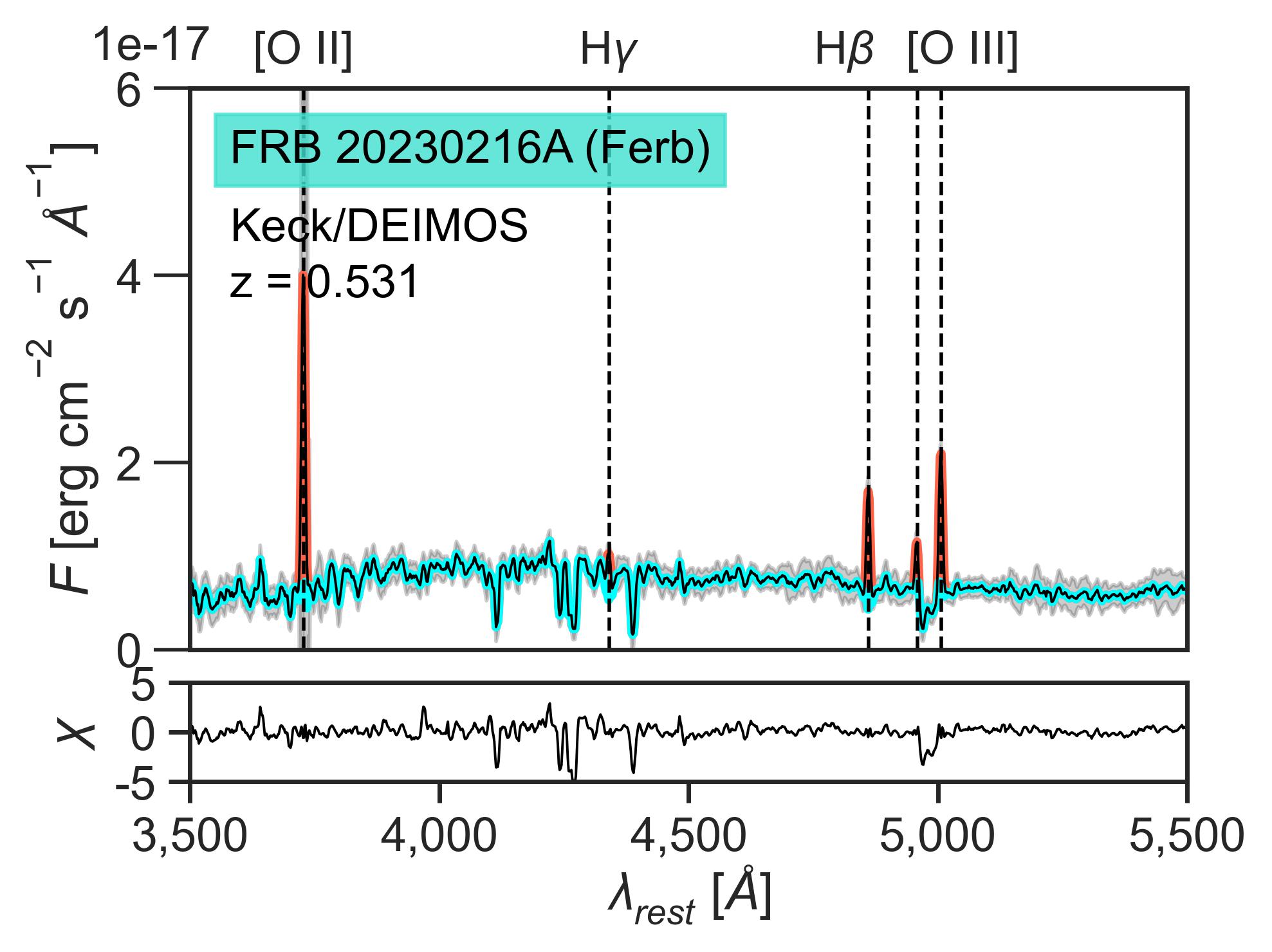}
\includegraphics[width=0.3225\textwidth, height=4.2cm]{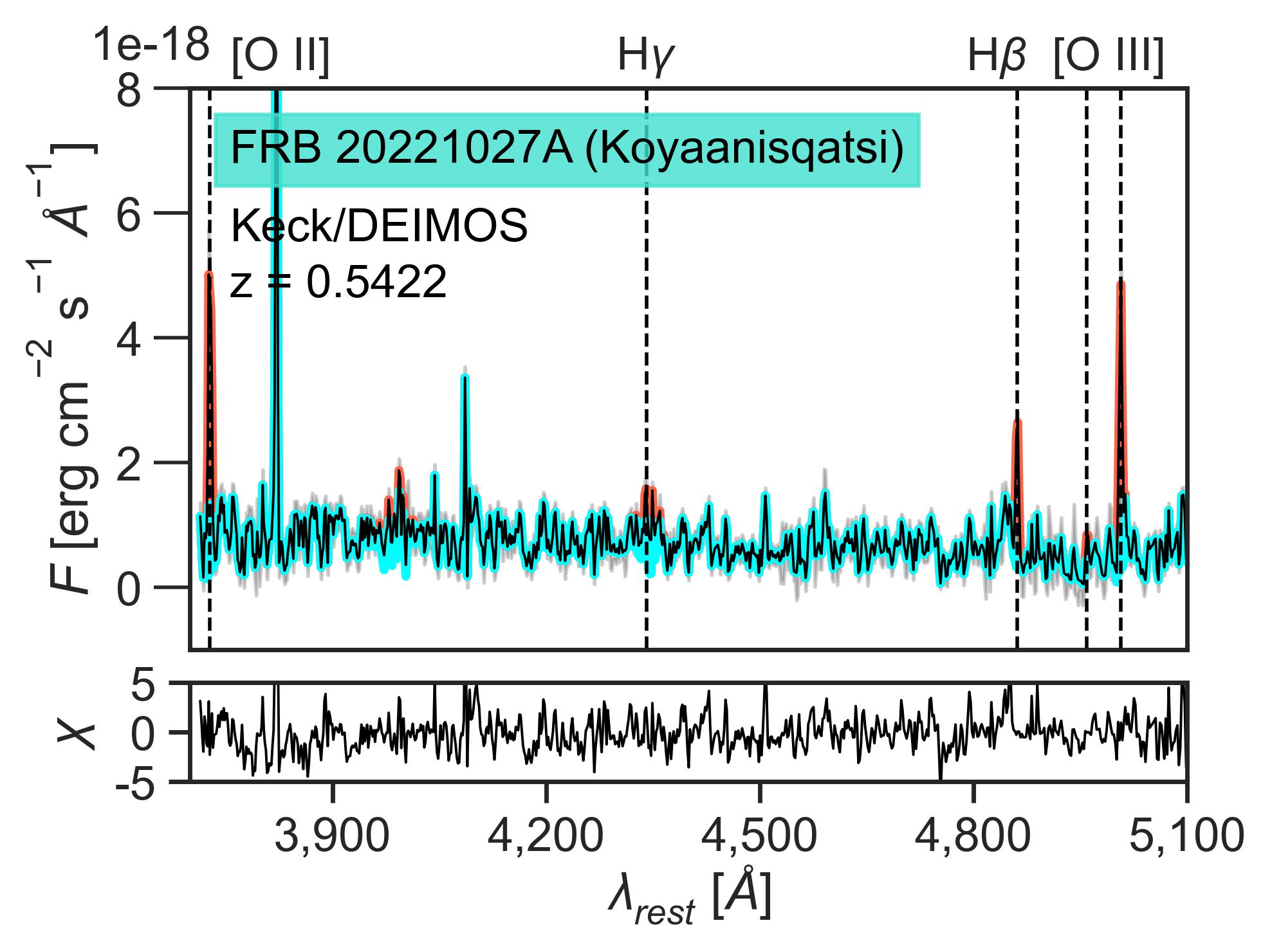}
\includegraphics[width=0.3225\textwidth, height=4.2cm]{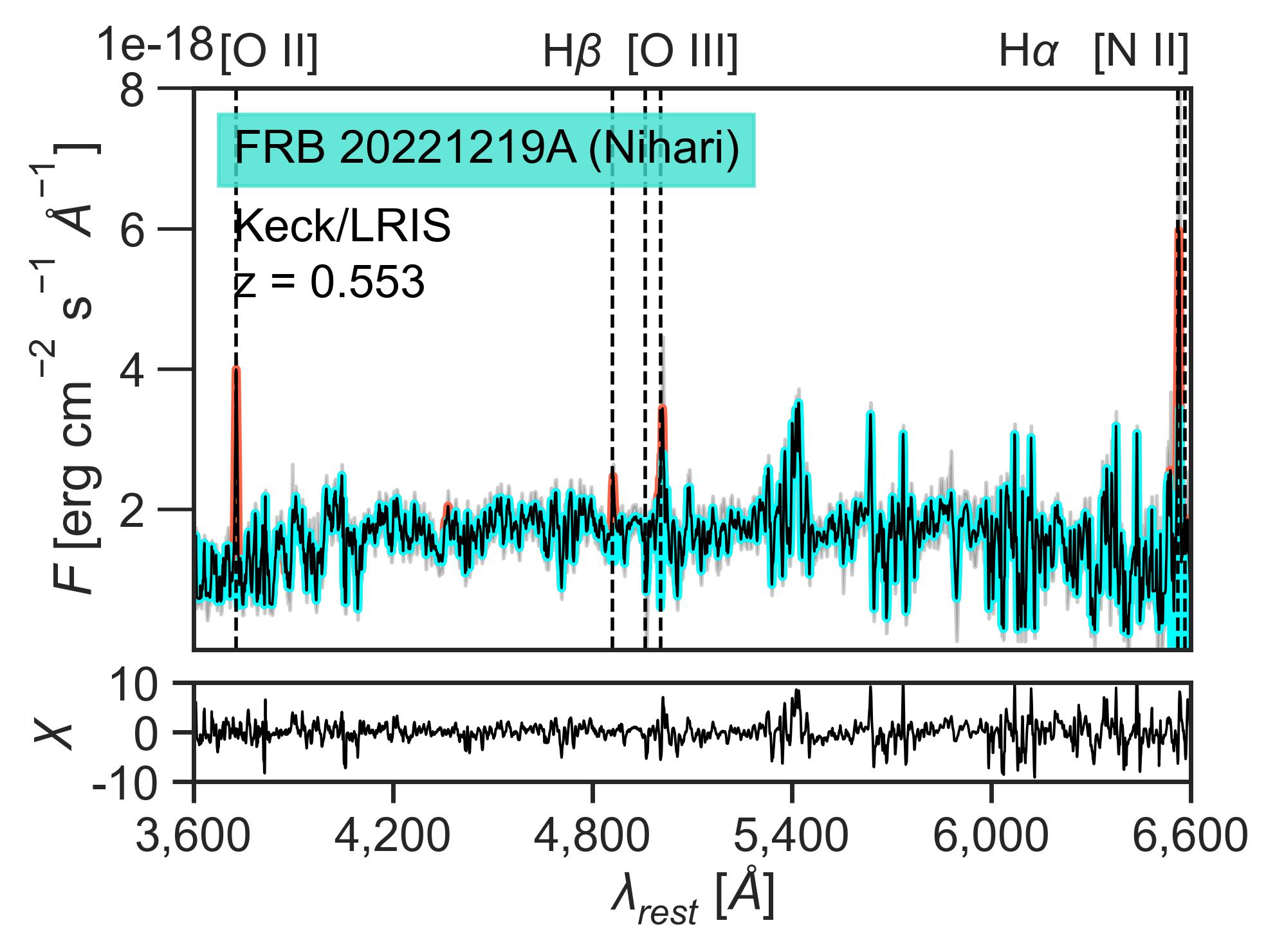}
\includegraphics[width=0.3225\textwidth, height=4.2cm]{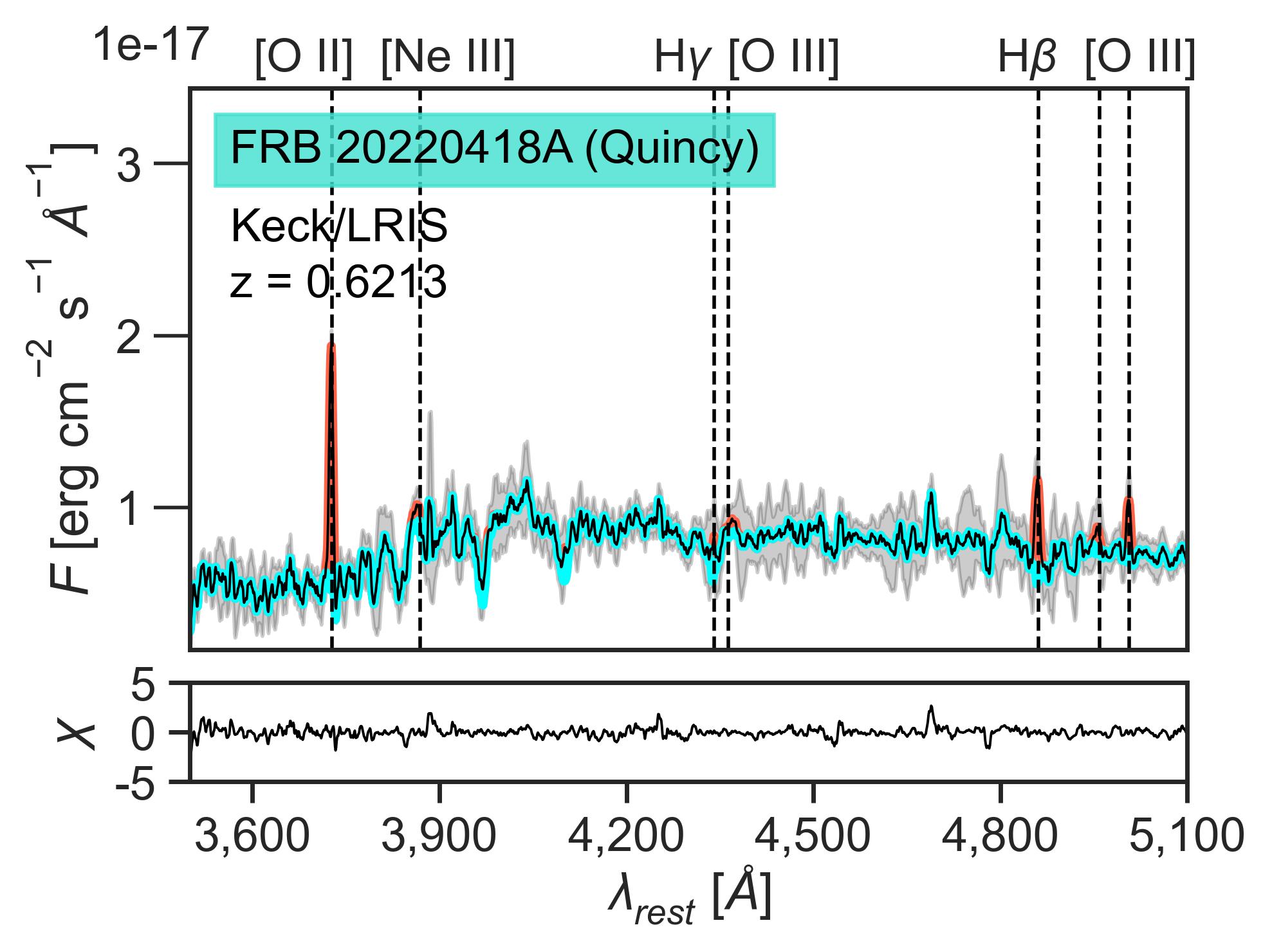}
\includegraphics[width=0.3225\textwidth, height=4.2cm]{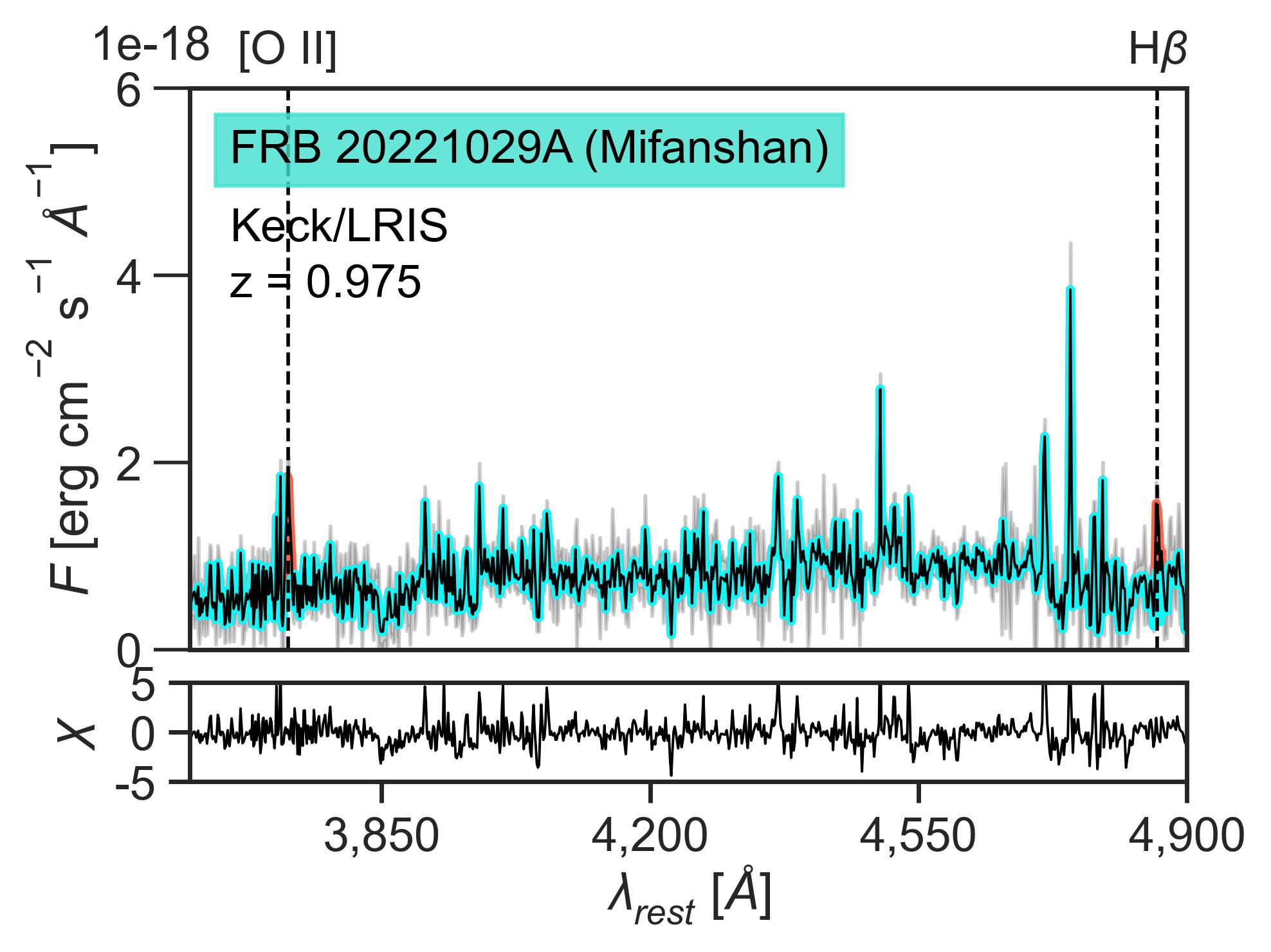}
\caption{\textbf{Supplementary Fig.\ref{fig:host_spectra}:} \textit{(Cont.)}.}
\end{figure*}

\newpage
\clearpage
\newpage
\newpage

\begin{figure*}[!ht]
\centering
\includegraphics[width=\textwidth]{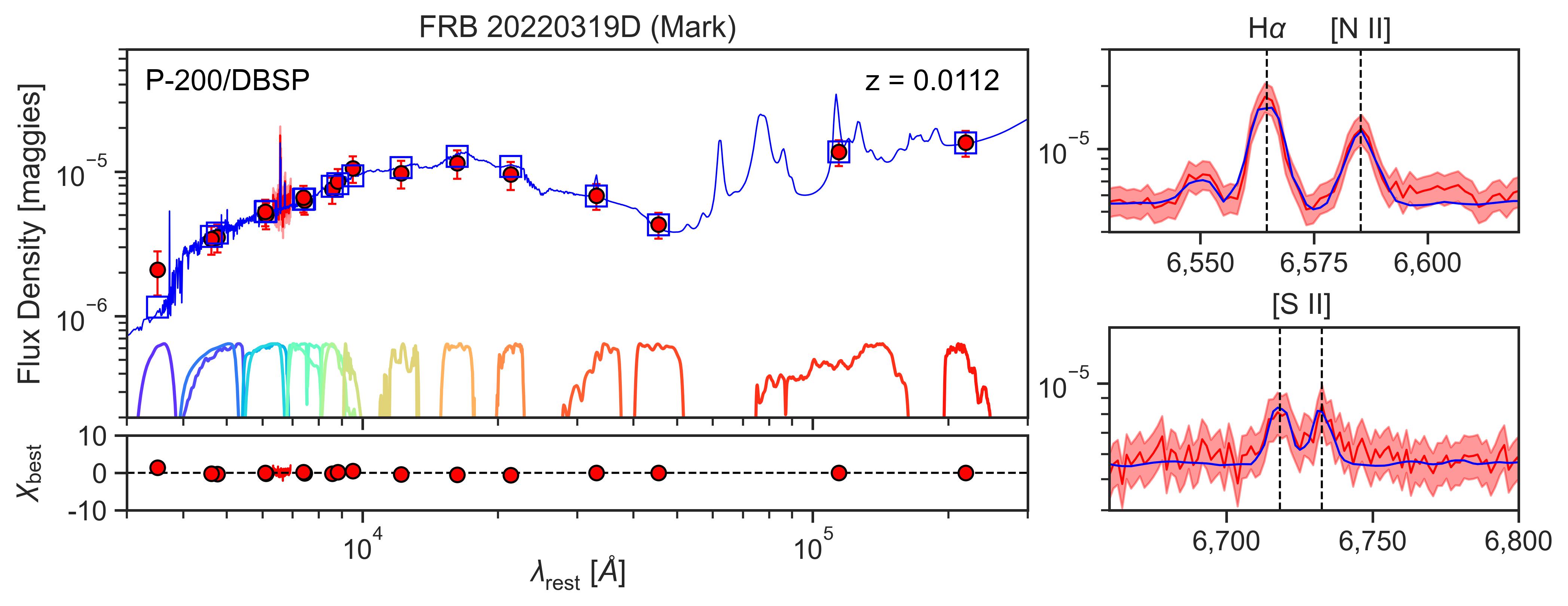}
\includegraphics[width=\textwidth]{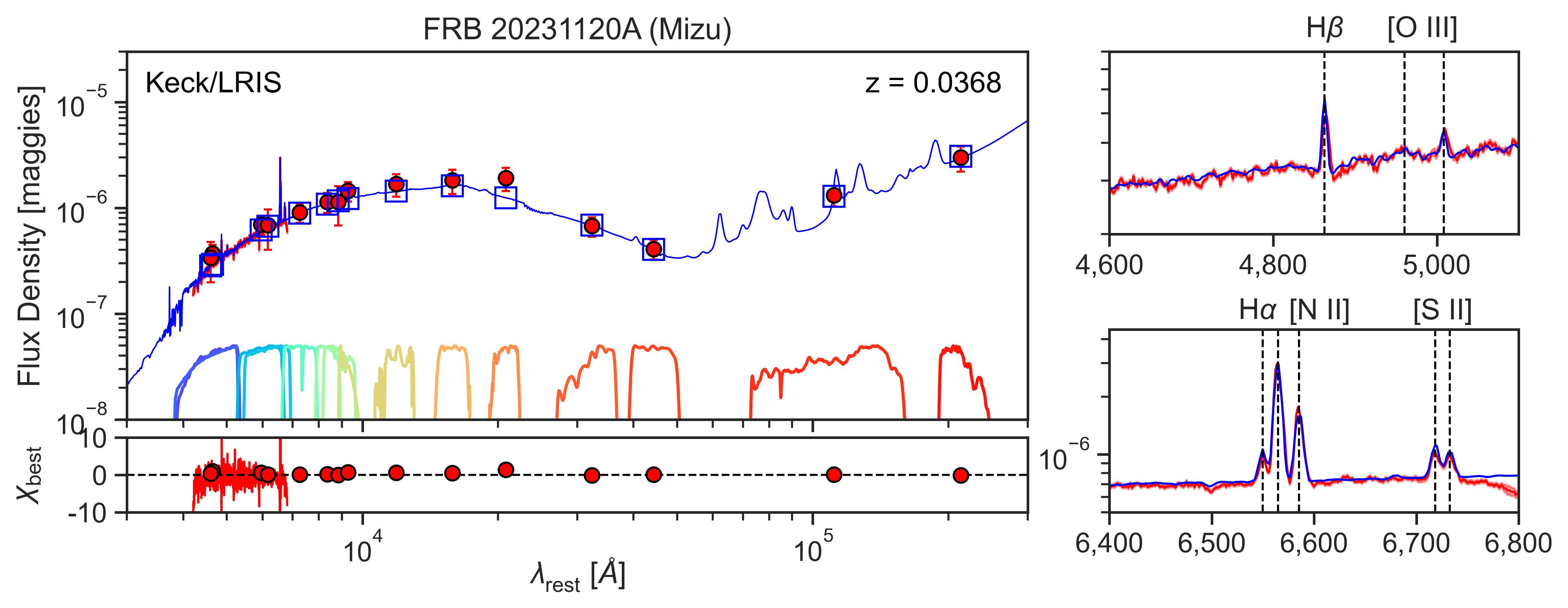}
\includegraphics[width=\textwidth]{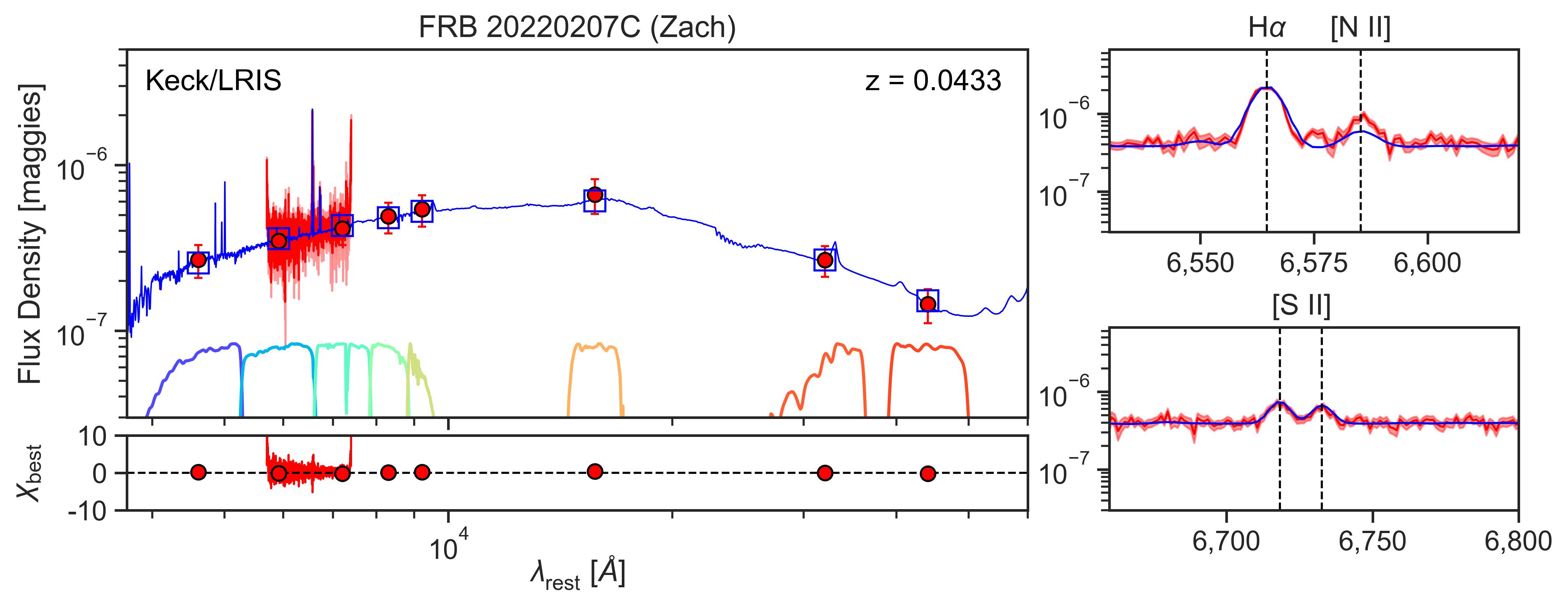}
\caption{{\supfigurelabel{fig:host_seds}} \textbf{Modeling the spectral energy distribution of galaxies.} The non-parametric SFH SED fits to multi-band photometry and spectrum (red) of all FRB hosts derived using \sw{Prospector} (see Methods). Panel (a) shows the SED fit with residuals for the best posterior sample (blue). The plots zoomed at the nebular emission features in the spectrum indicate the accuracy of our SED fits. 
}
\end{figure*}

\begin{figure*}[ht!]
\ContinuedFloat
\centering
\includegraphics[width=\textwidth]{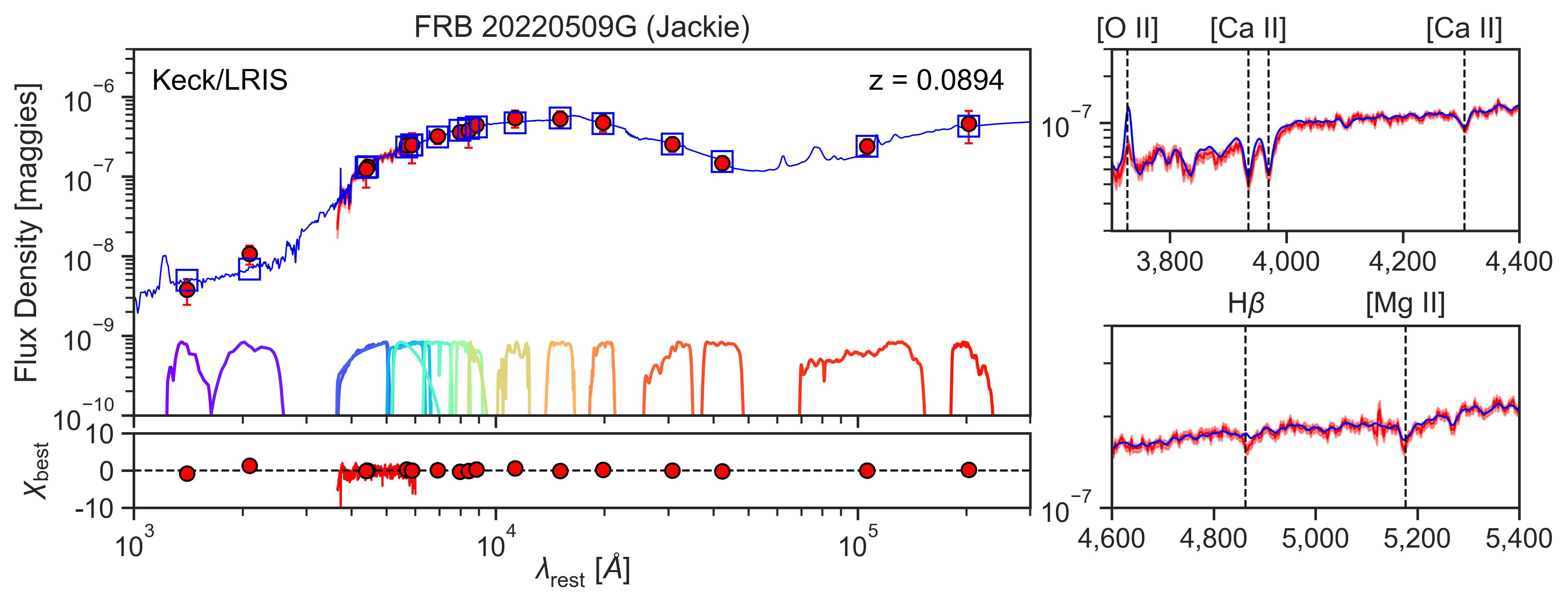}
\includegraphics[width=\textwidth]{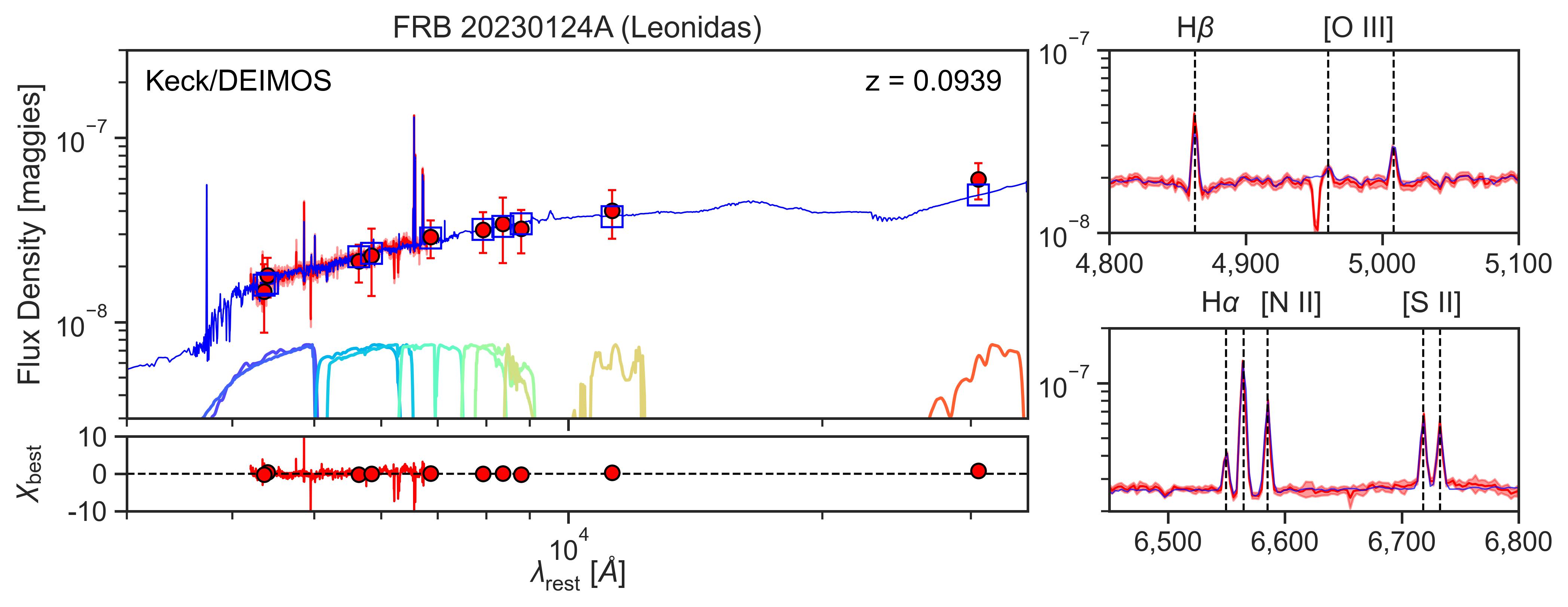}
\includegraphics[width=\textwidth]{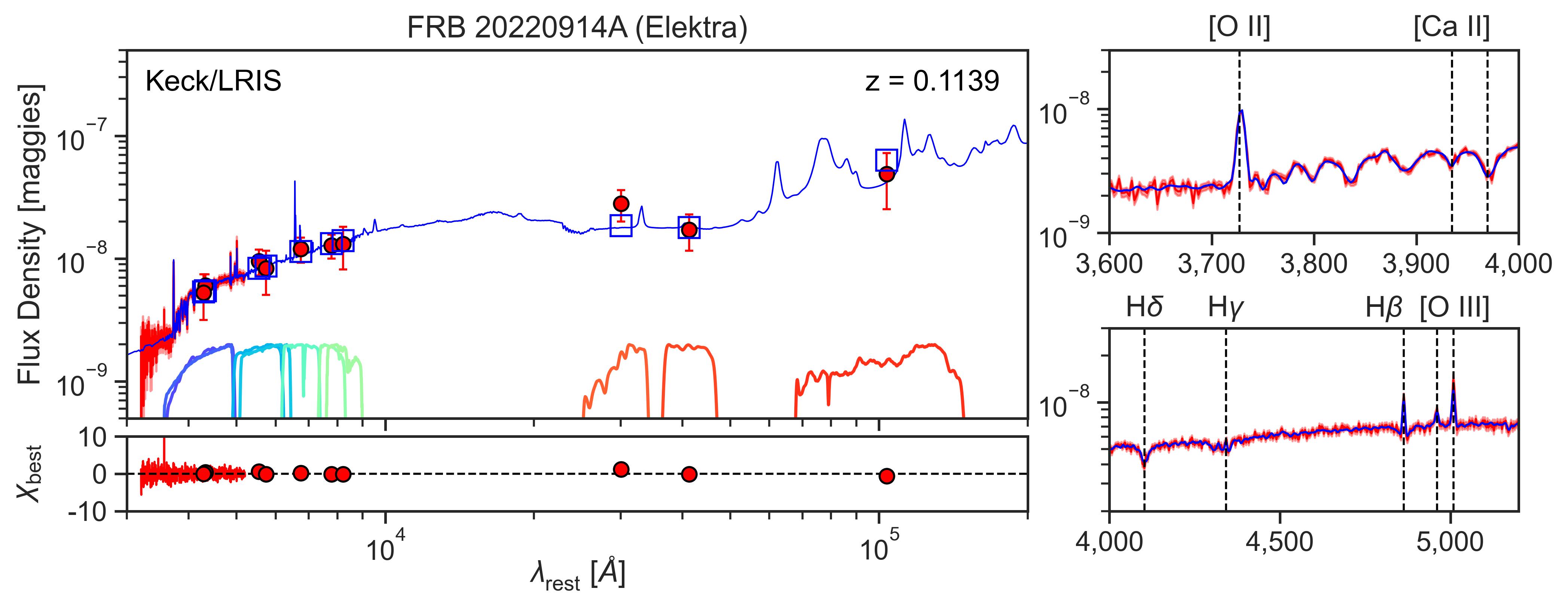}
\caption{\textbf{Supplementary Fig.\ref{fig:host_seds}:} \textit{(Cont.)}.}
\end{figure*}

\begin{figure*}[ht!]
\ContinuedFloat
\centering
\includegraphics[width=\textwidth]{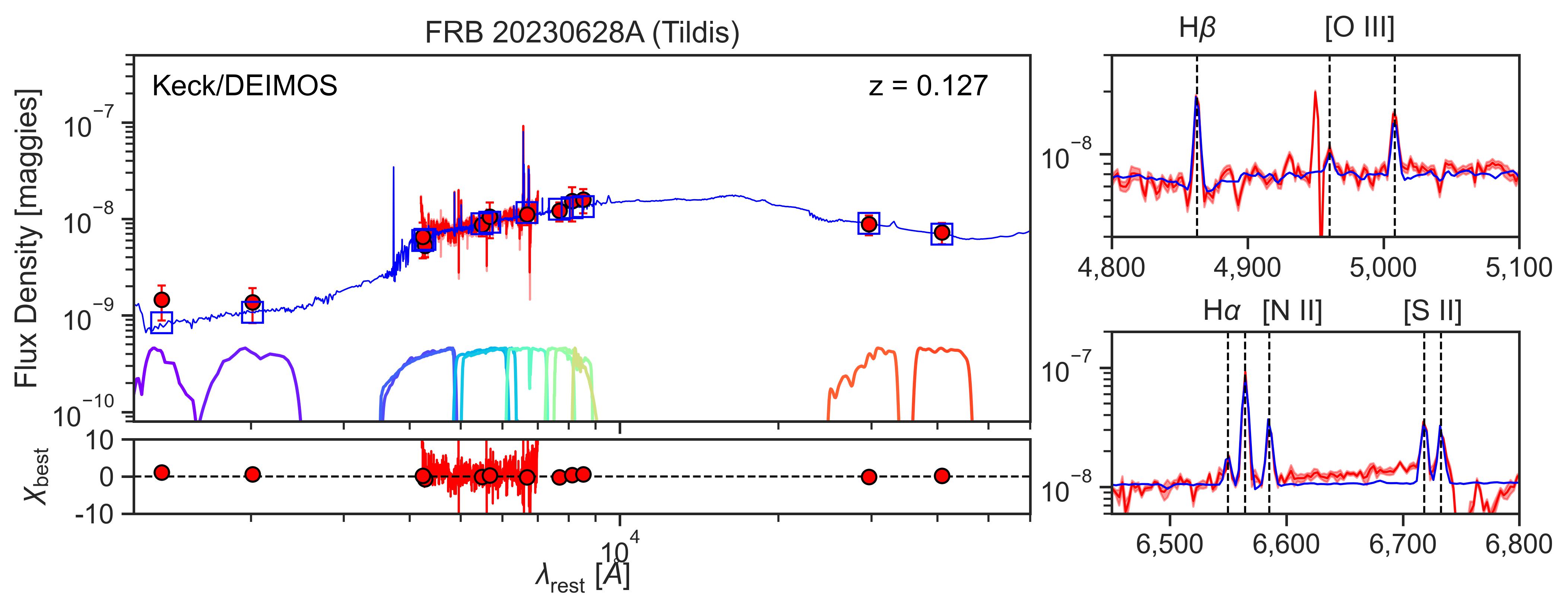}
\includegraphics[width=\textwidth]{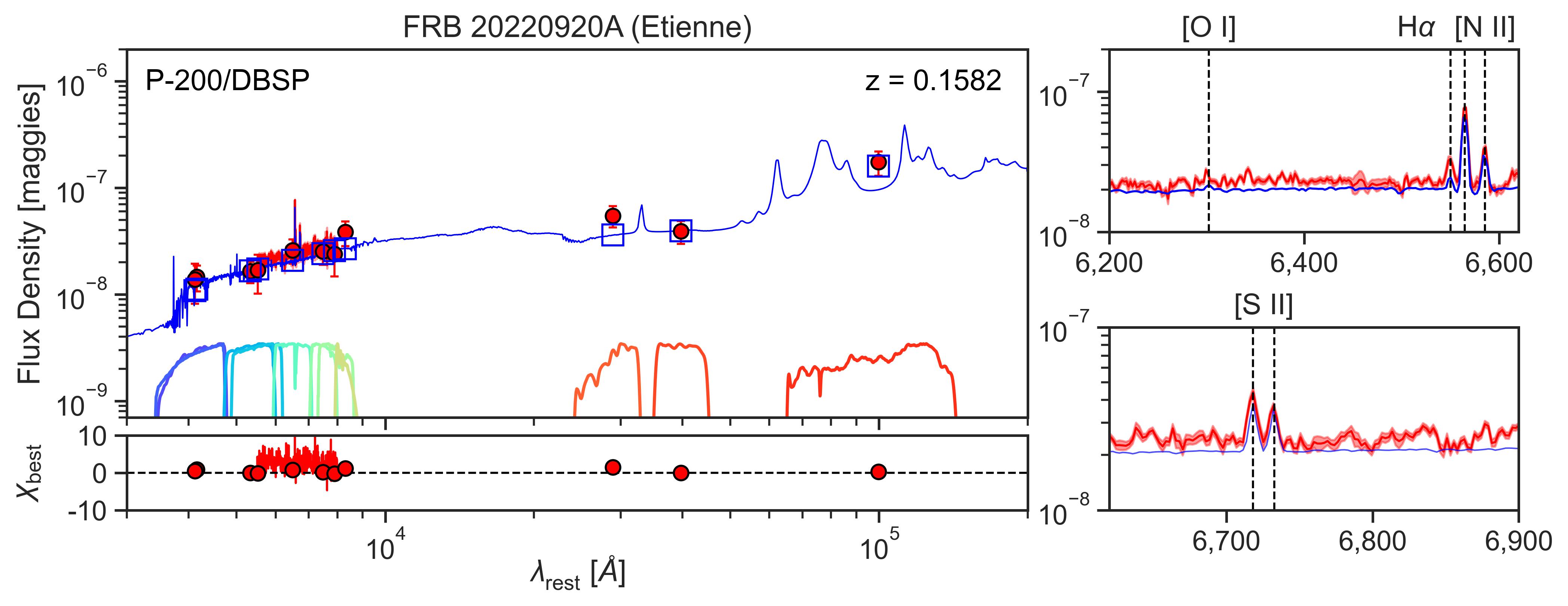}
\includegraphics[width=\textwidth]{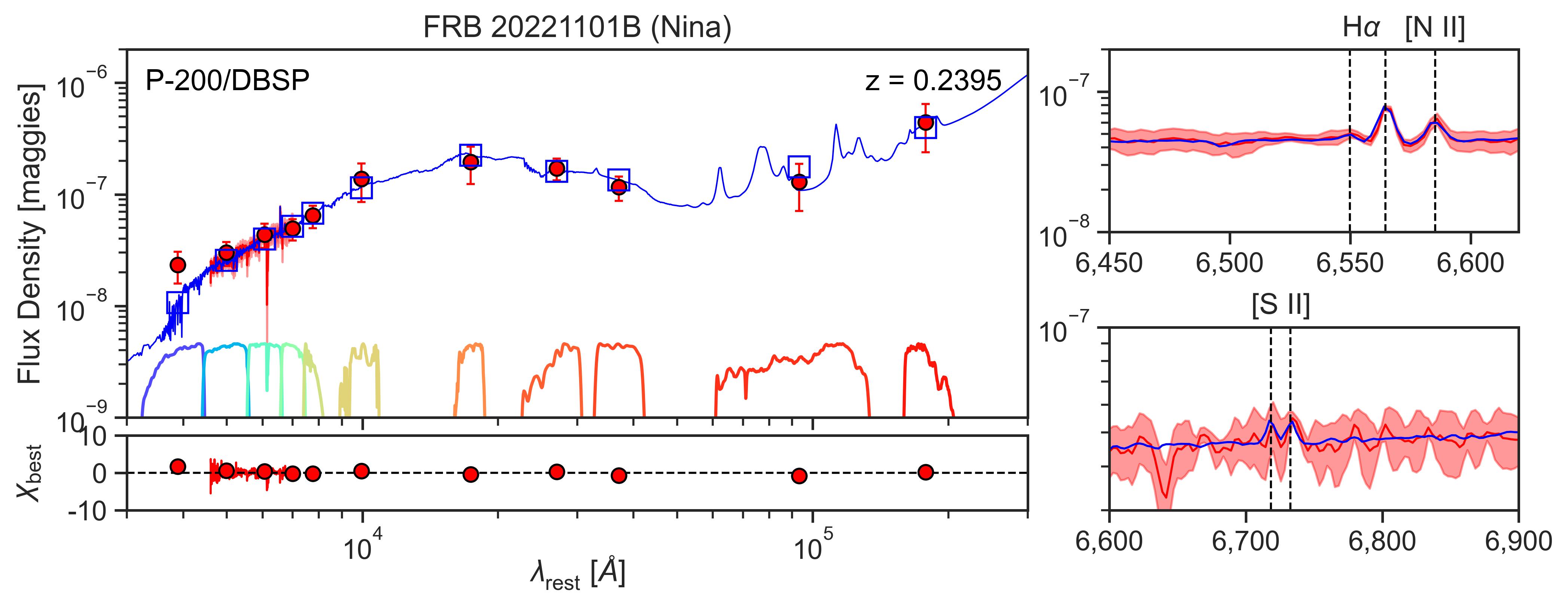}
\caption{\textbf{Supplementary Fig.\ref{fig:host_seds}:}  \textit{(Cont.)}.}
\end{figure*}

\begin{figure*}[ht!]
\ContinuedFloat
\centering
\includegraphics[width=\textwidth]{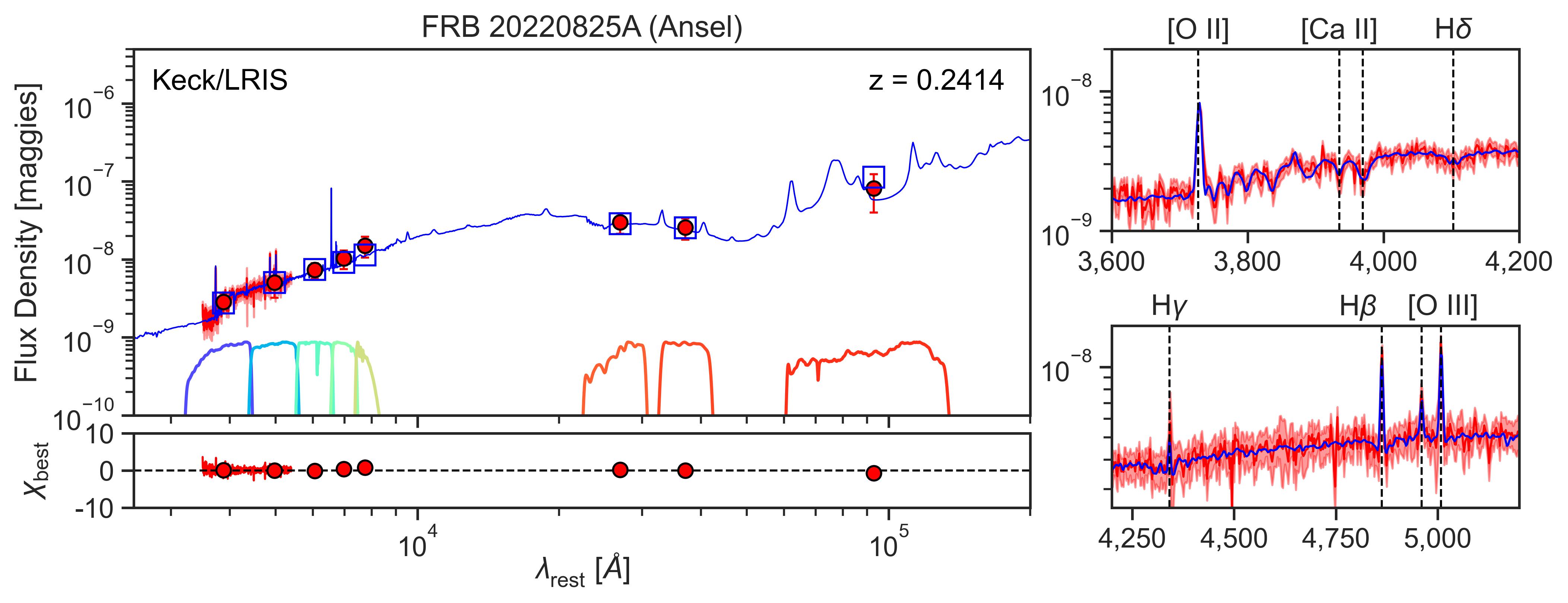}
\includegraphics[width=\textwidth]{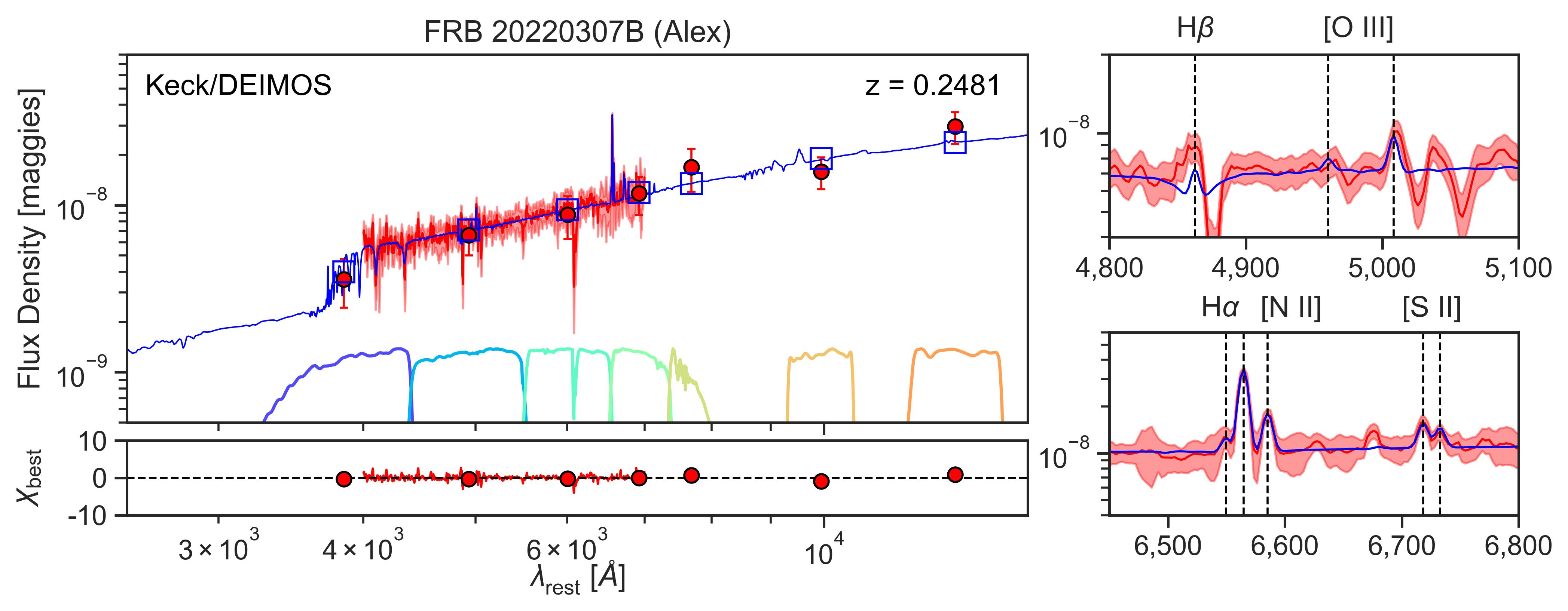}
\includegraphics[width=\textwidth]{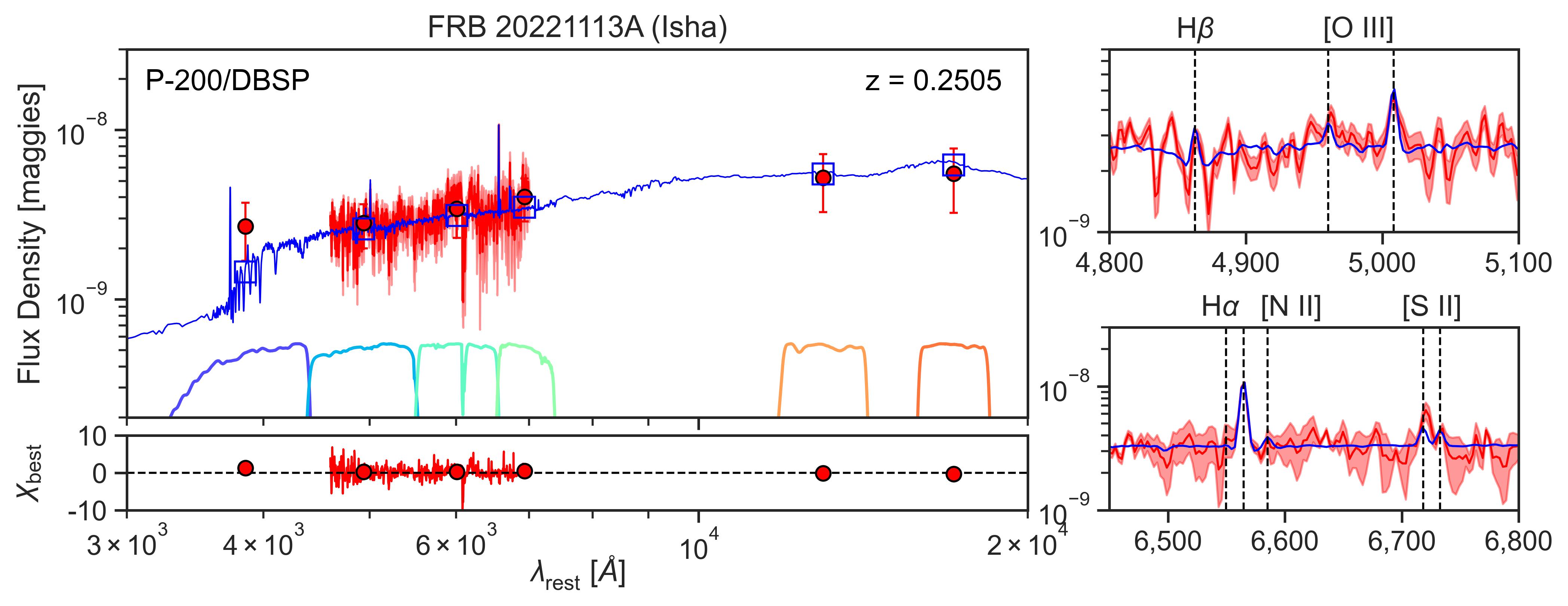}
\caption{\textbf{Supplementary Fig.\ref{fig:host_seds}:}  \textit{(Cont.)}.}
\end{figure*}

\begin{figure*}[ht!]
\ContinuedFloat
\centering
\includegraphics[width=\textwidth]{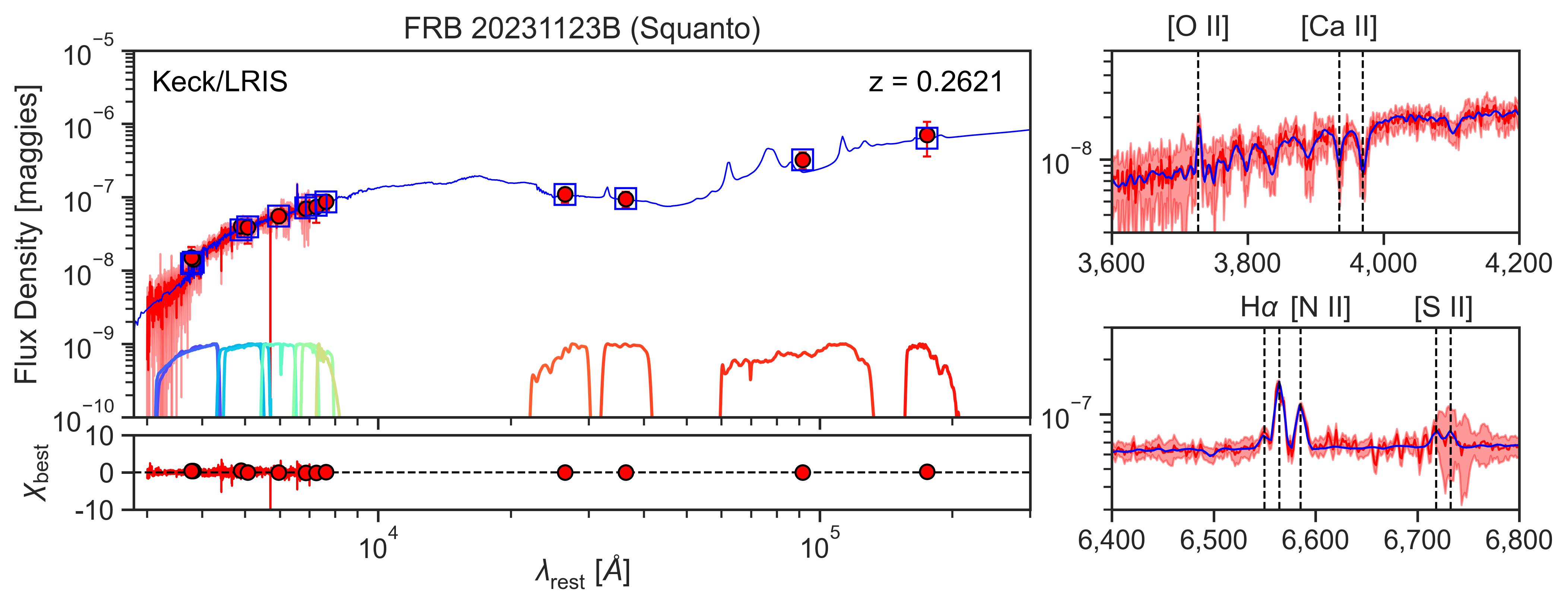}
\includegraphics[width=\textwidth]{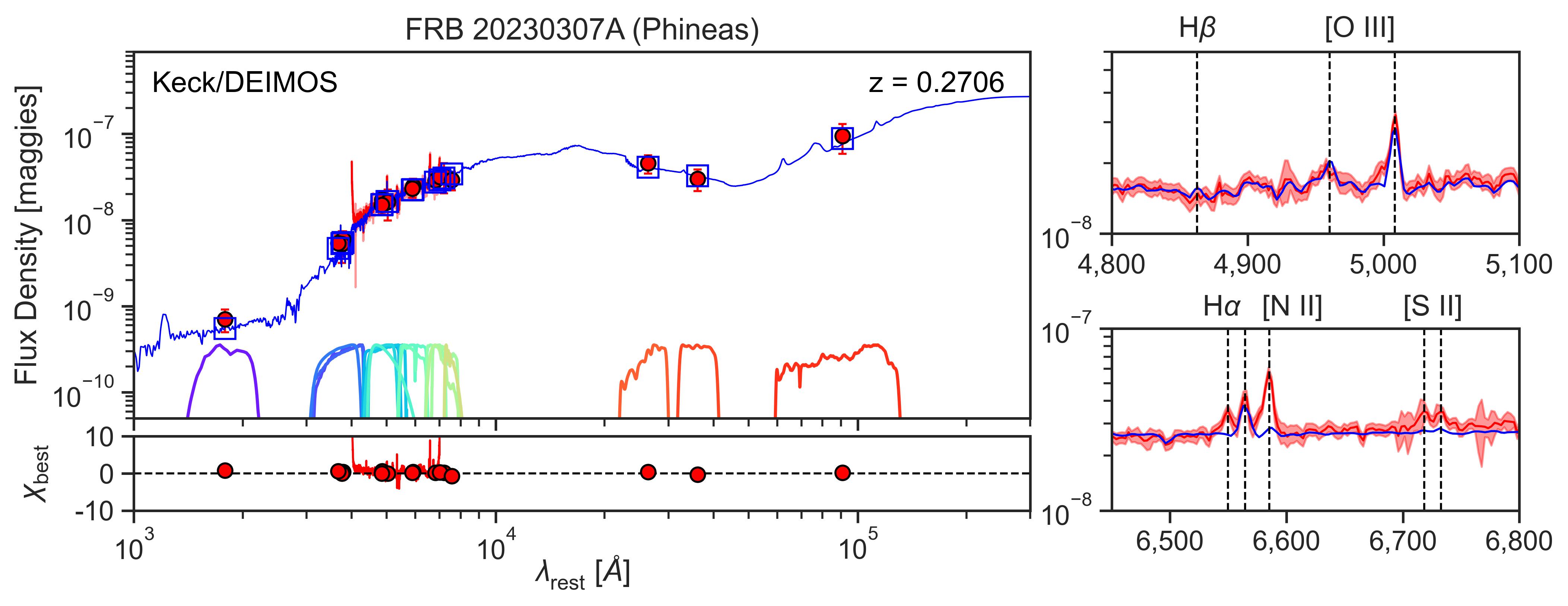}
\includegraphics[width=\textwidth]{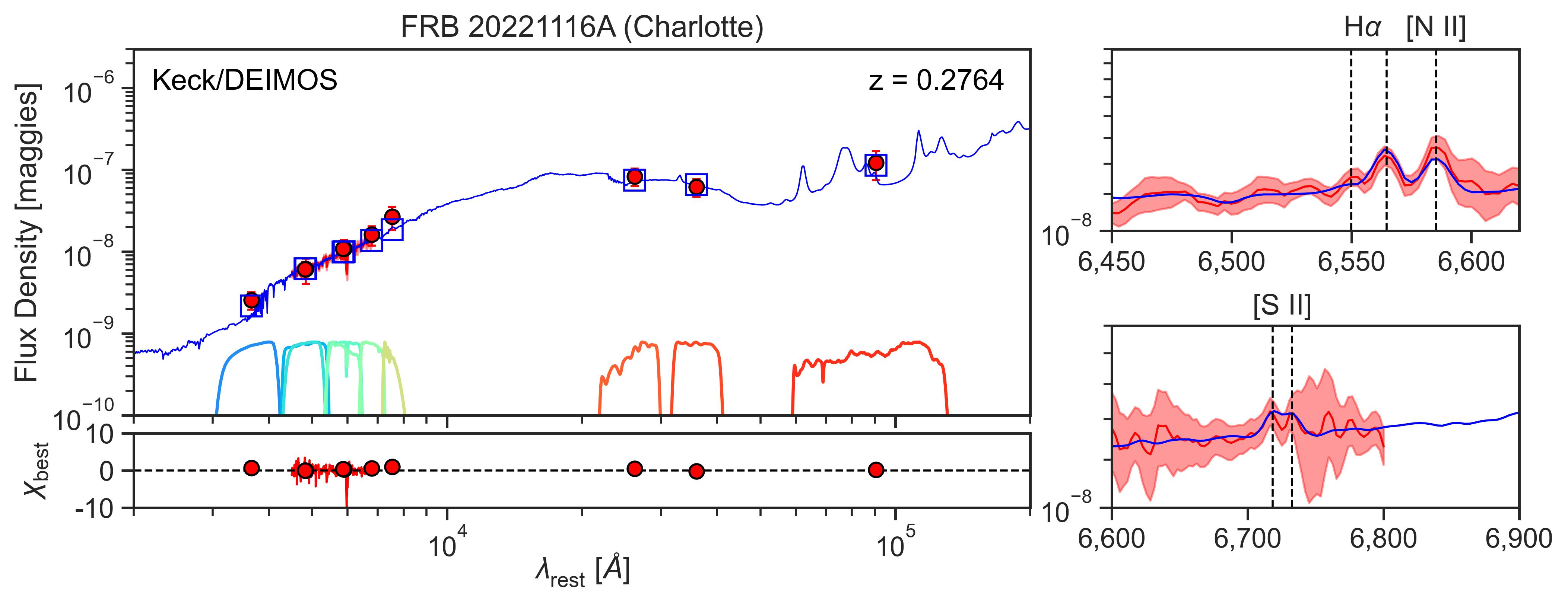}
\caption{\textbf{Supplementary Fig.\ref{fig:host_seds}:}  \textit{(Cont.)}.}
\end{figure*}

\begin{figure*}[ht!]
\ContinuedFloat
\centering
\includegraphics[width=\textwidth]{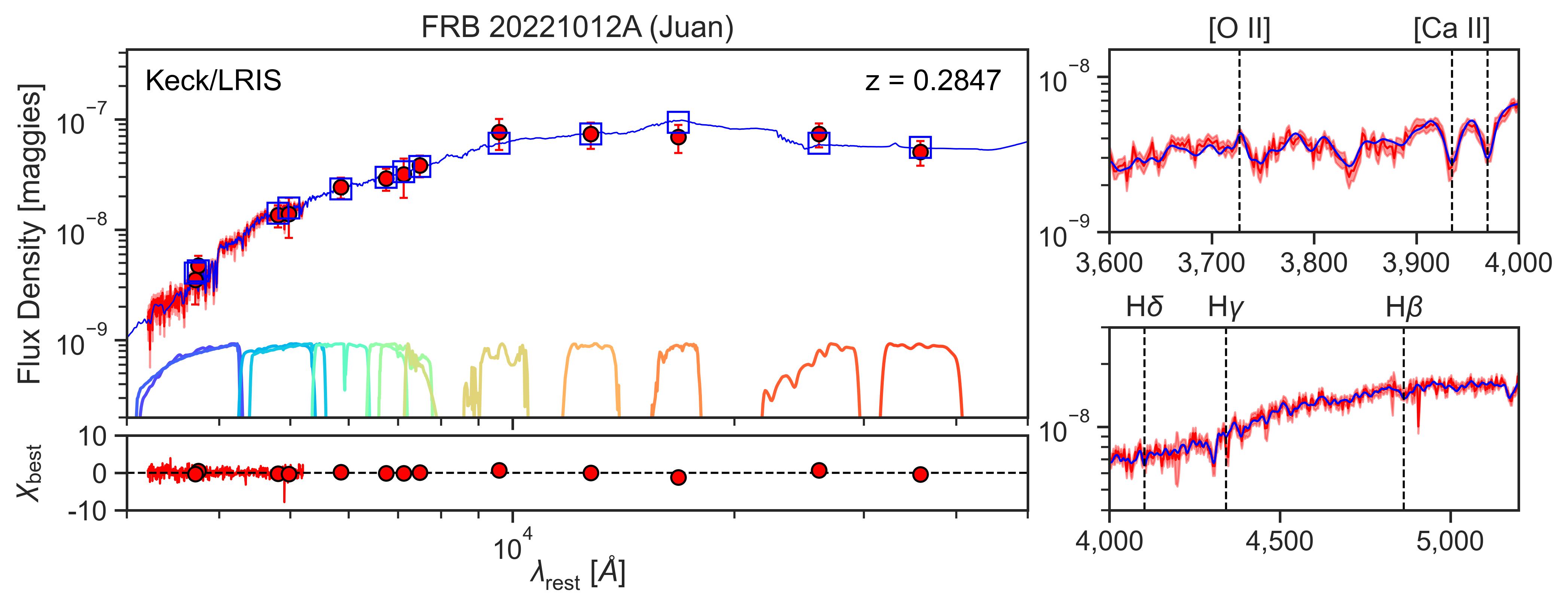}
\includegraphics[width=\textwidth]{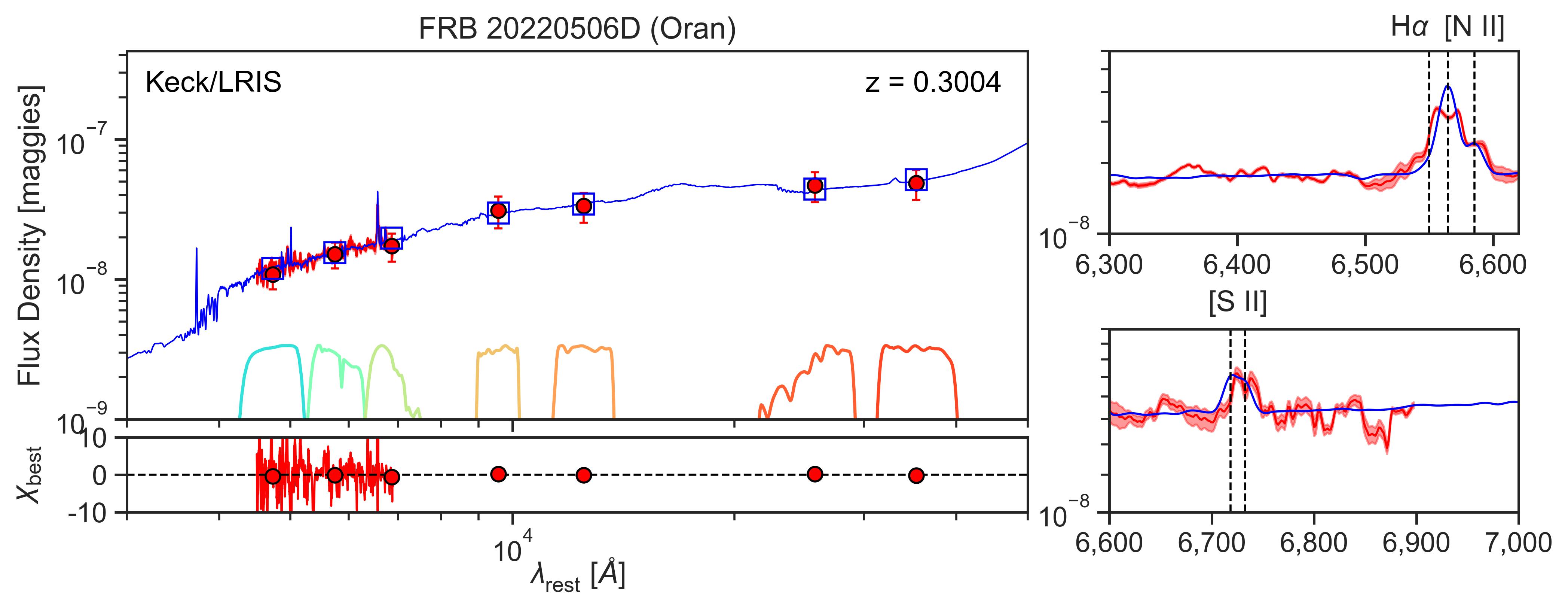}
\includegraphics[width=\textwidth]{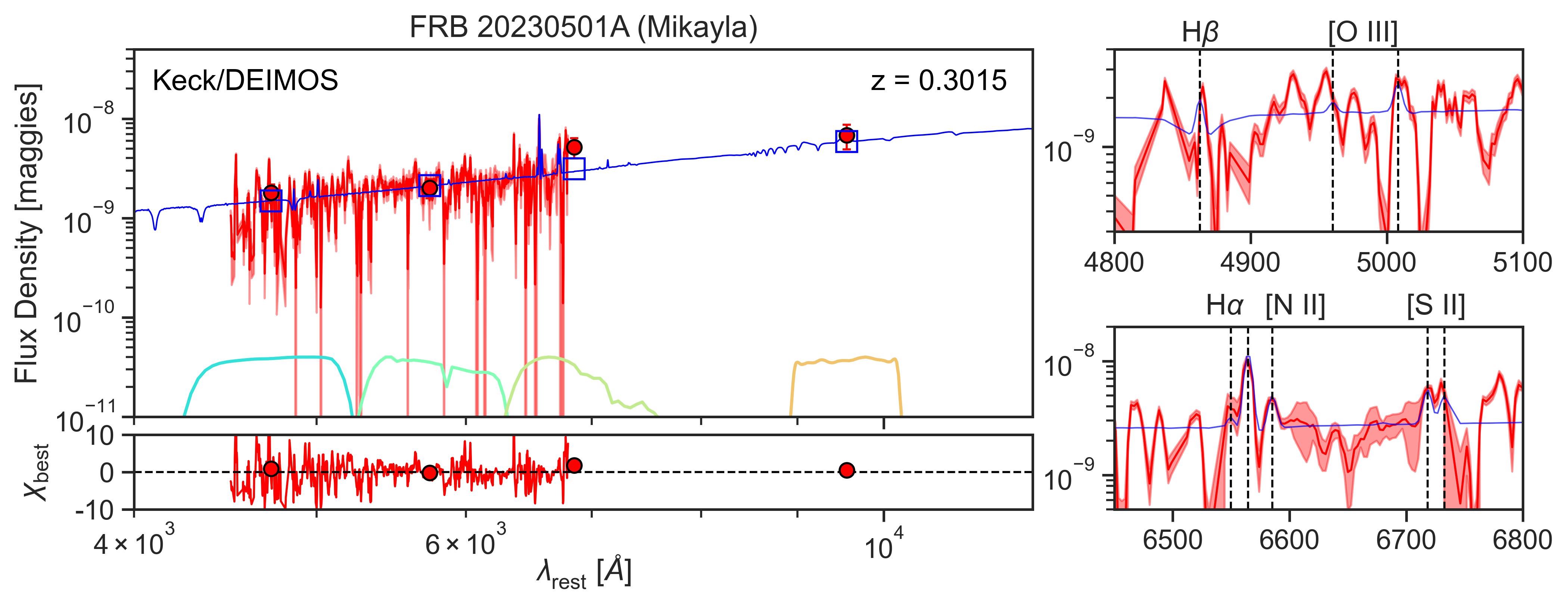}
\caption{\textbf{Supplementary Fig.\ref{fig:host_seds}:}  \textit{(Cont.)}.}
\end{figure*}

\begin{figure*}[ht!]
\ContinuedFloat
\centering
\includegraphics[width=\textwidth]{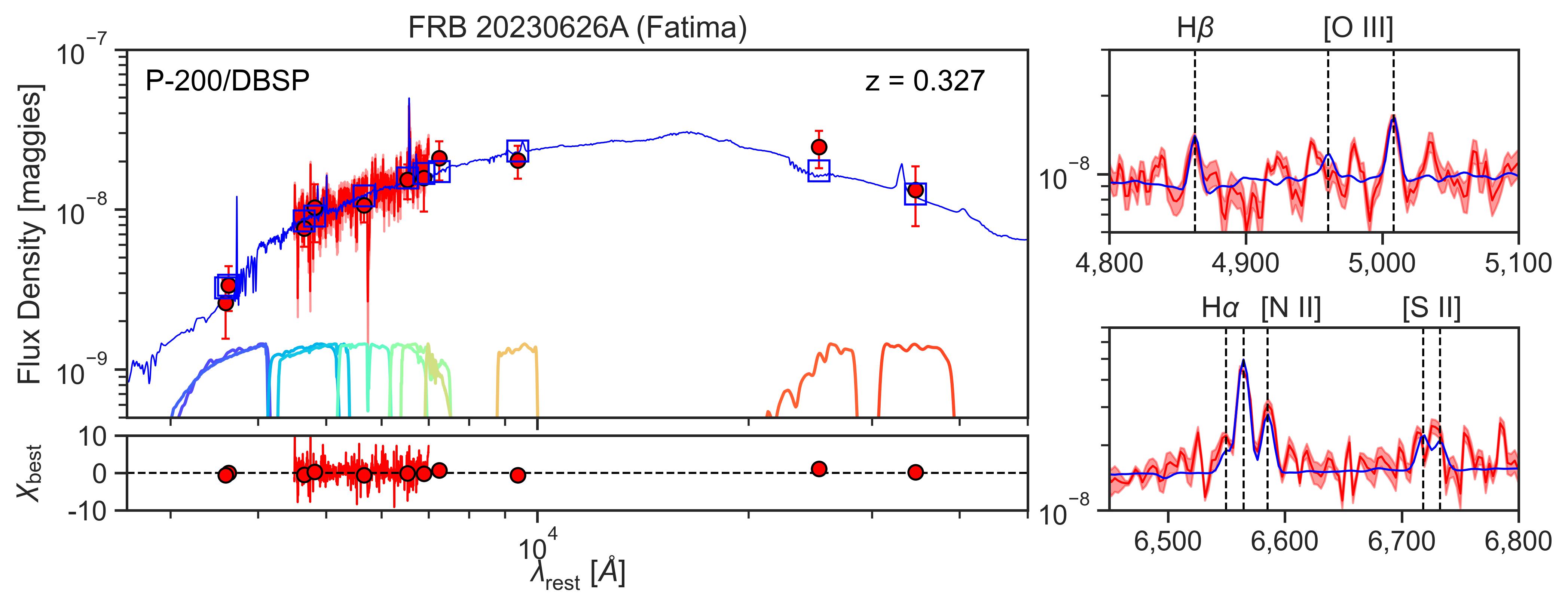}
\includegraphics[width=\textwidth]{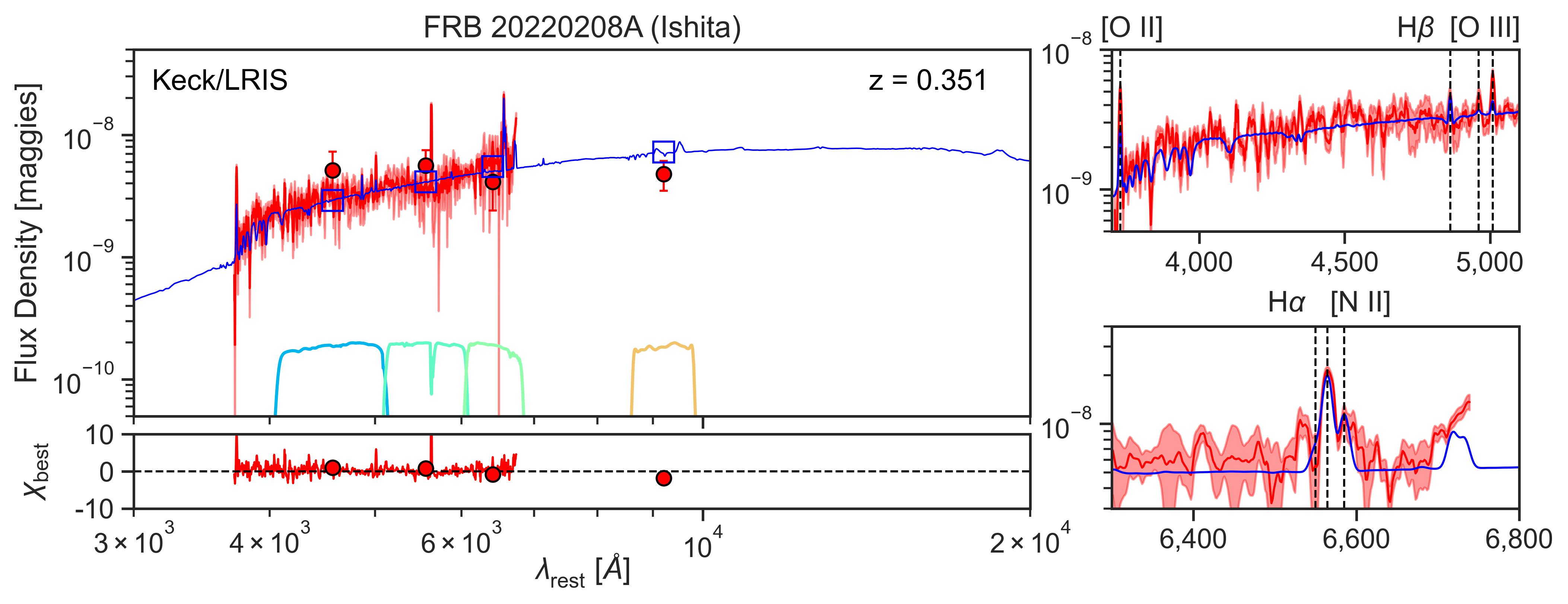}
\includegraphics[width=\textwidth]{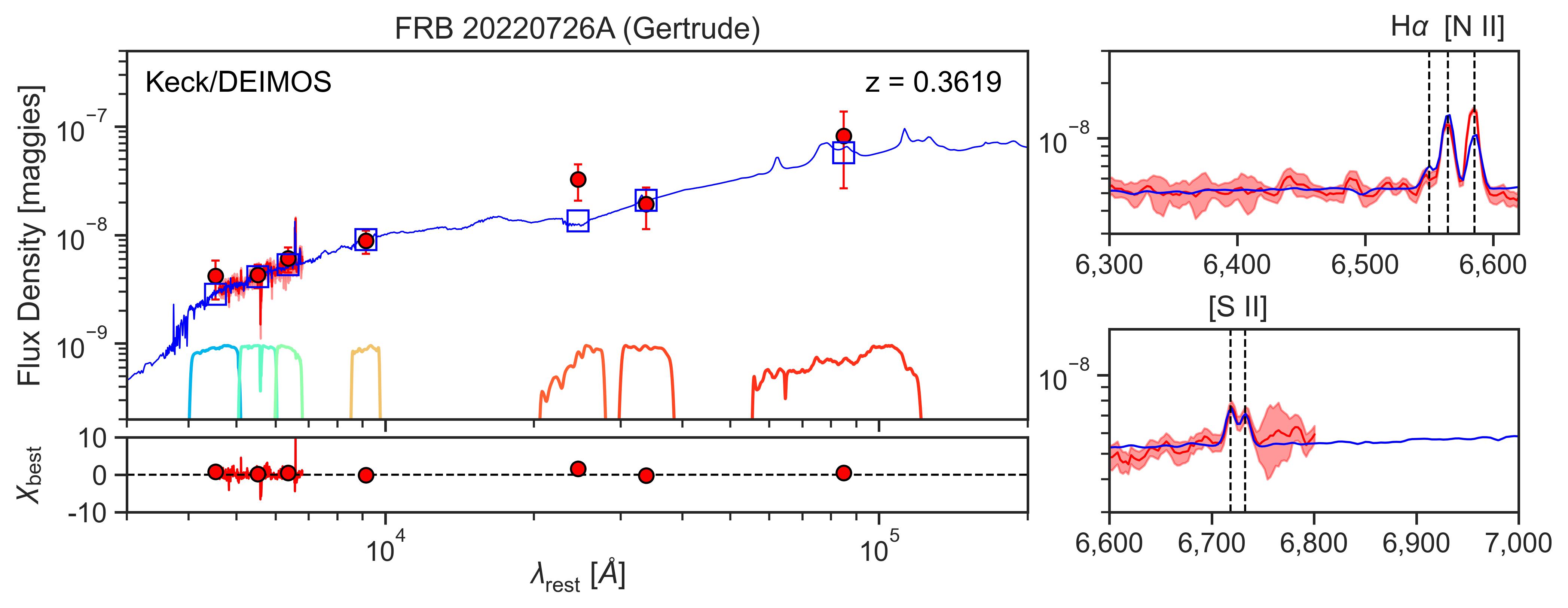}
\caption{\textbf{Supplementary Fig.\ref{fig:host_seds}:}  \textit{(Cont.)}.}
\end{figure*}

\begin{figure*}[ht!]
\ContinuedFloat
\centering
\includegraphics[width=\textwidth]{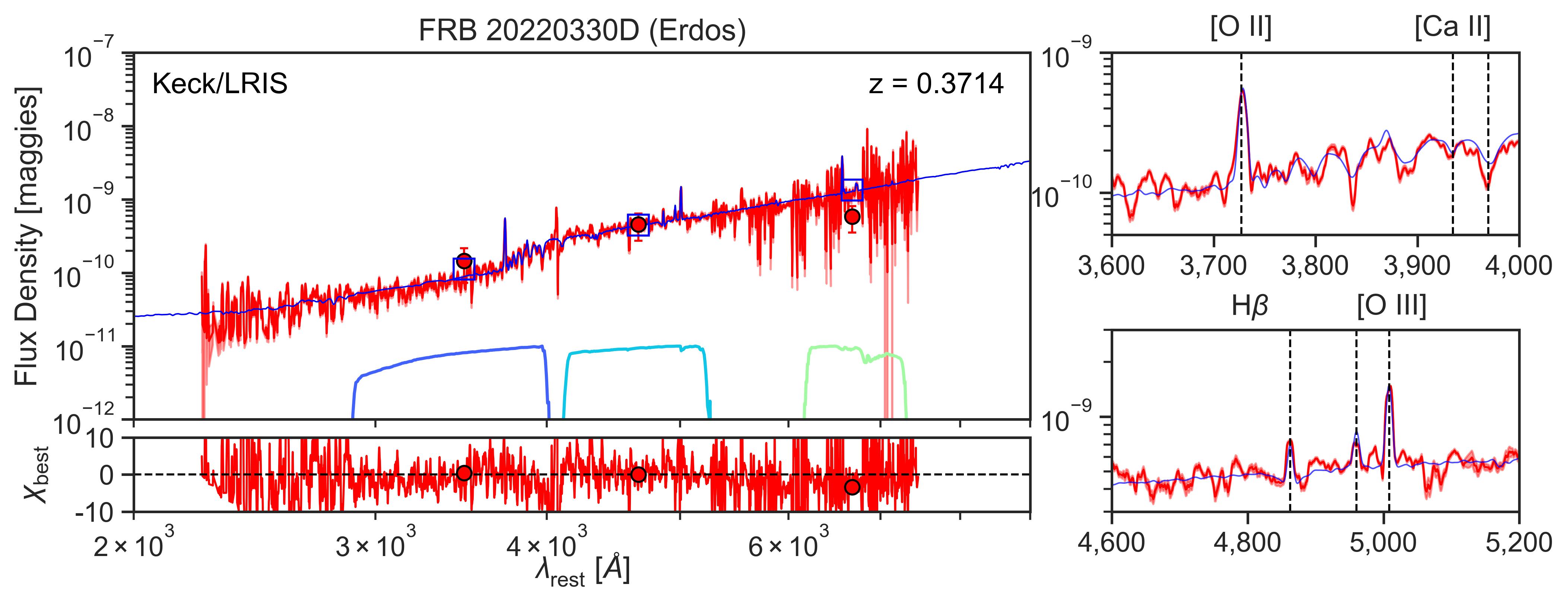}
\includegraphics[width=\textwidth]{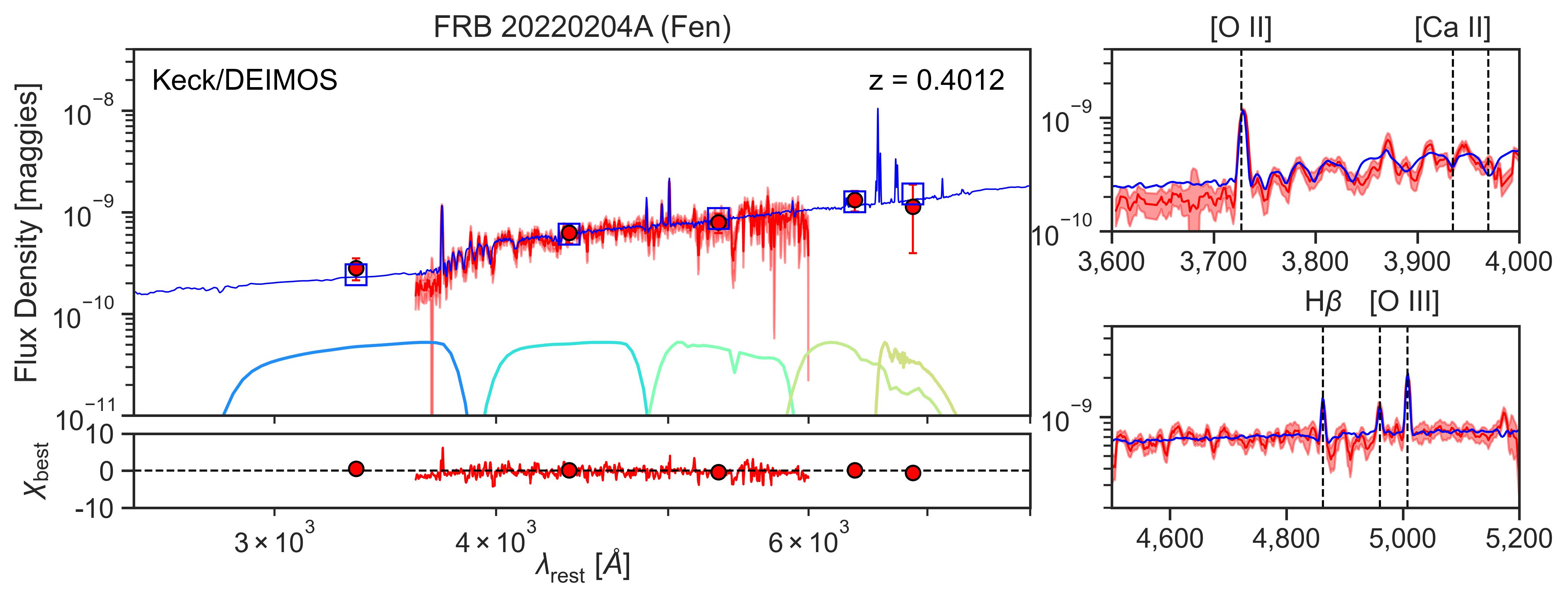}
\includegraphics[width=\textwidth]{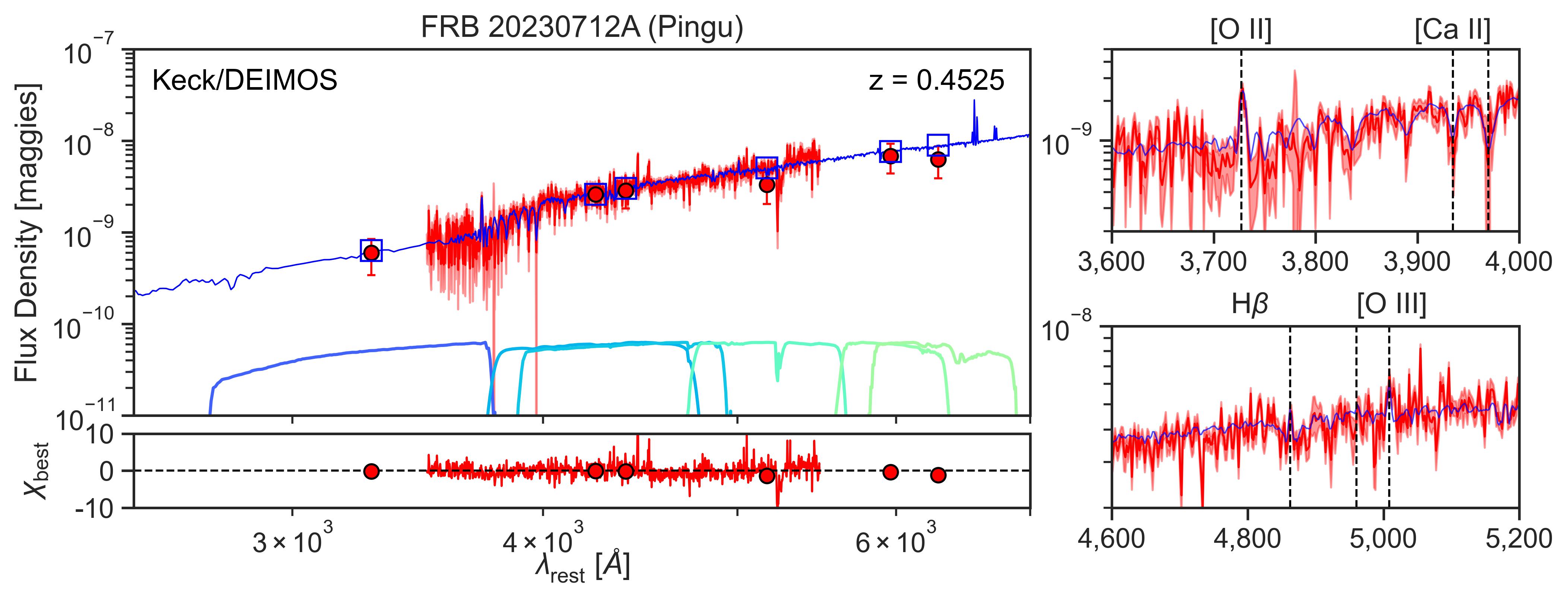}
\caption{\textbf{Supplementary Fig.\ref{fig:host_seds}:}  \textit{(Cont.)}.}
\end{figure*}

\begin{figure*}[ht!]
\ContinuedFloat
\centering
\includegraphics[width=\textwidth]{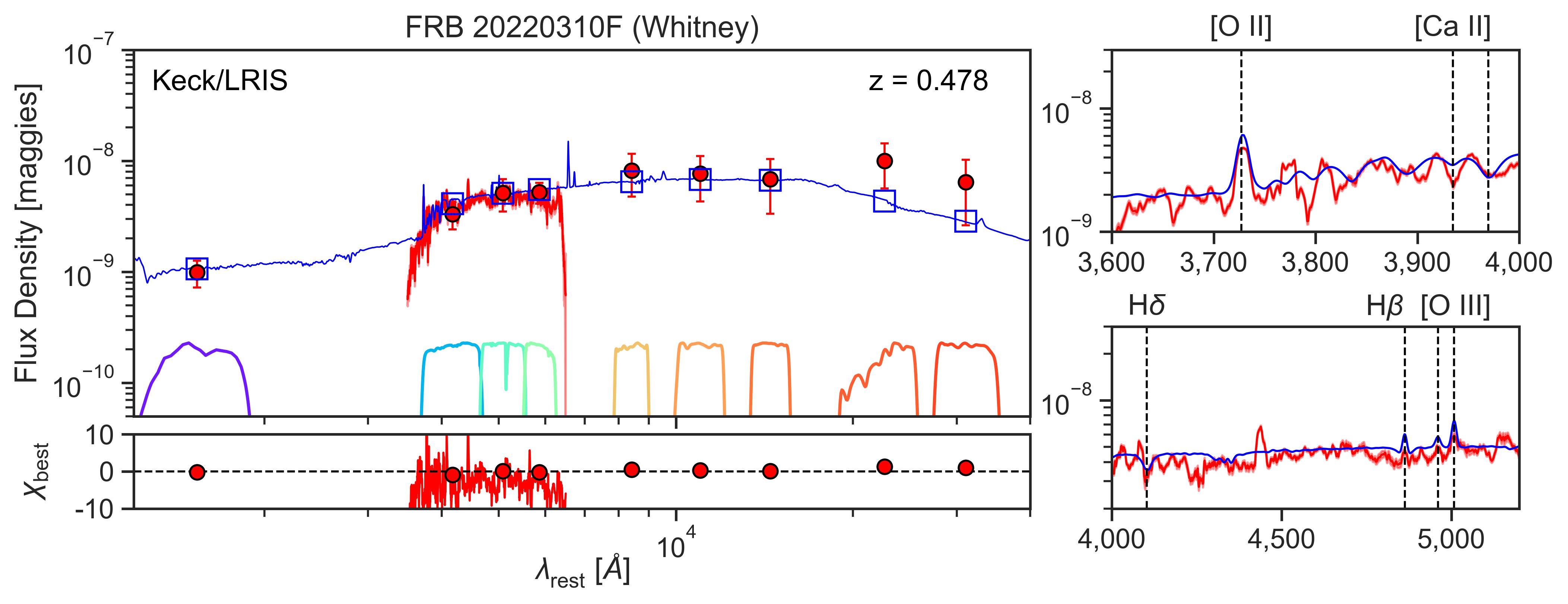}
\includegraphics[width=\textwidth]{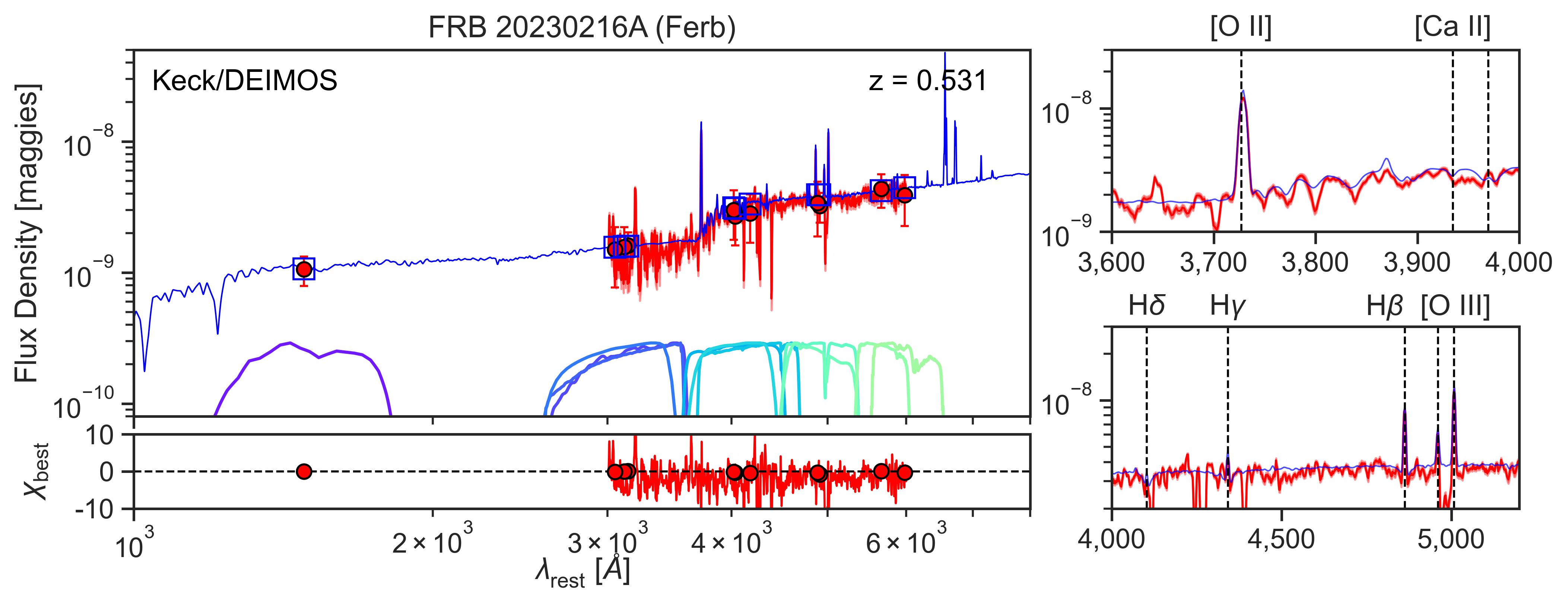}
\includegraphics[width=\textwidth]{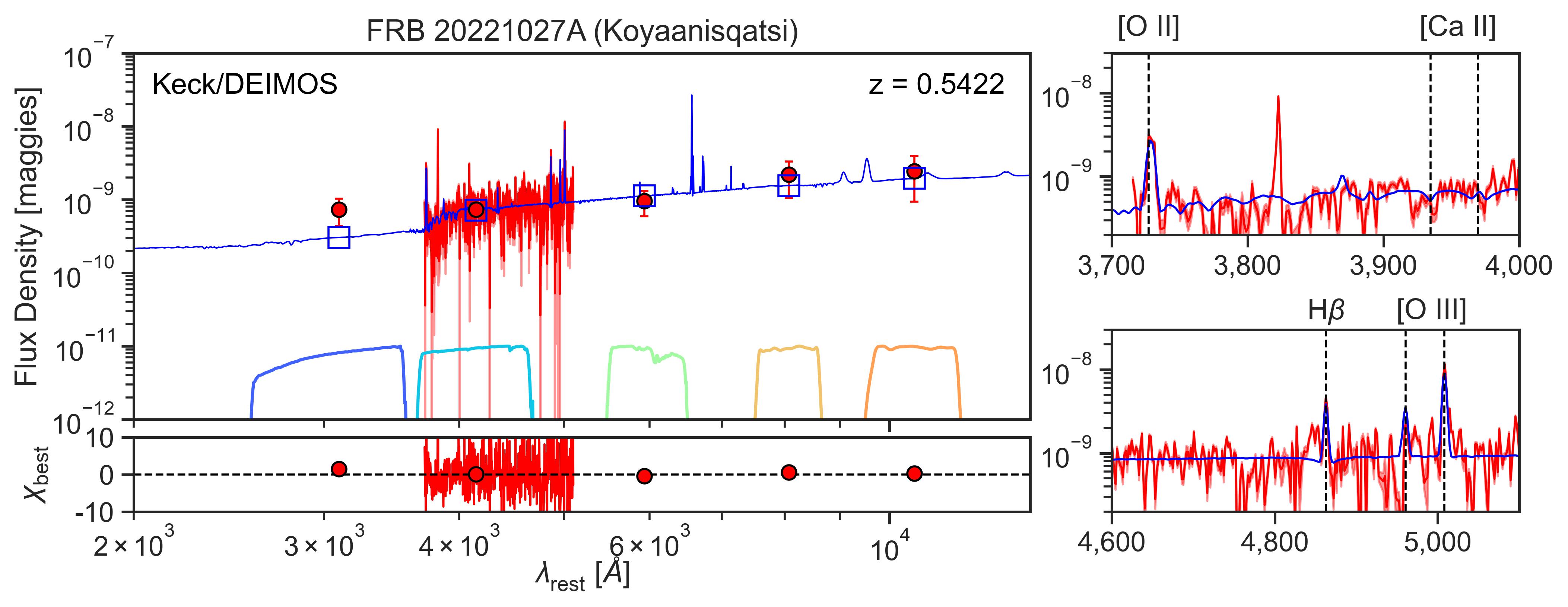}
\caption{\textbf{Supplementary Fig.\ref{fig:host_seds}:}  \textit{(Cont.)}.}
\end{figure*}

\begin{figure*}[ht!]
\ContinuedFloat
\centering
\includegraphics[width=\textwidth]{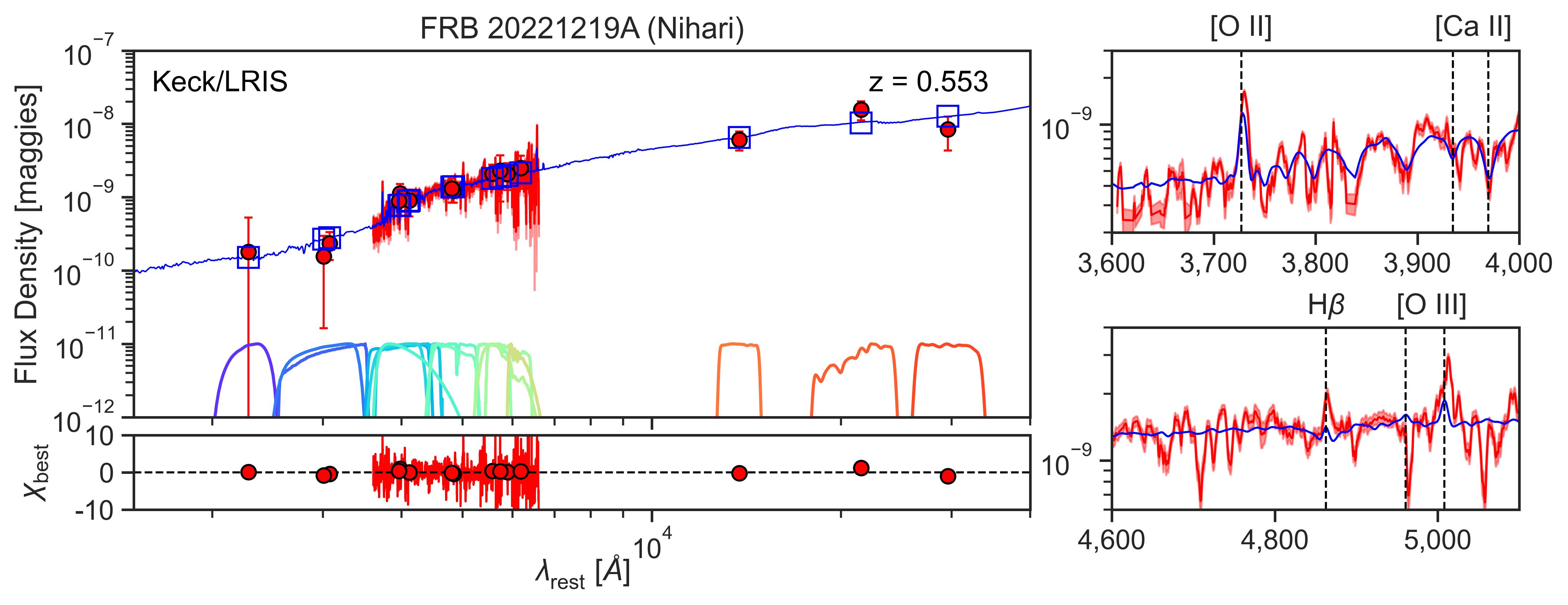}
\includegraphics[width=\textwidth]{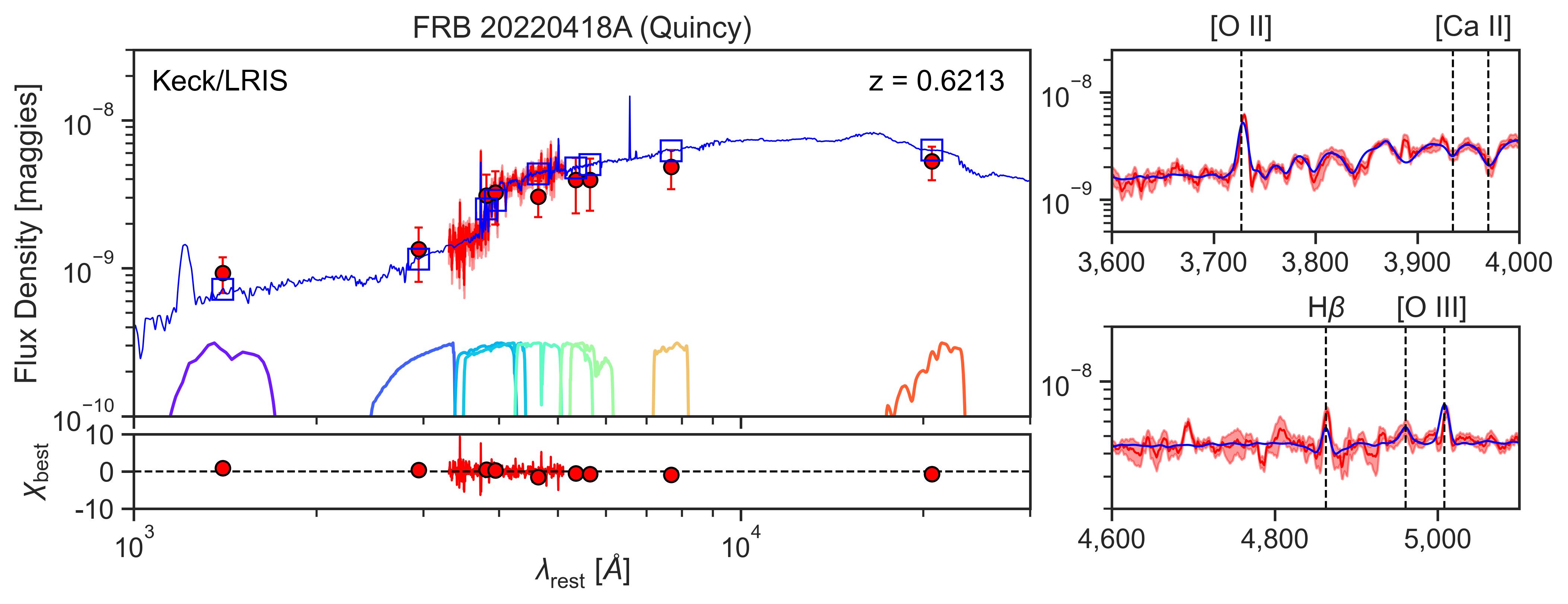}
\includegraphics[width=\textwidth]{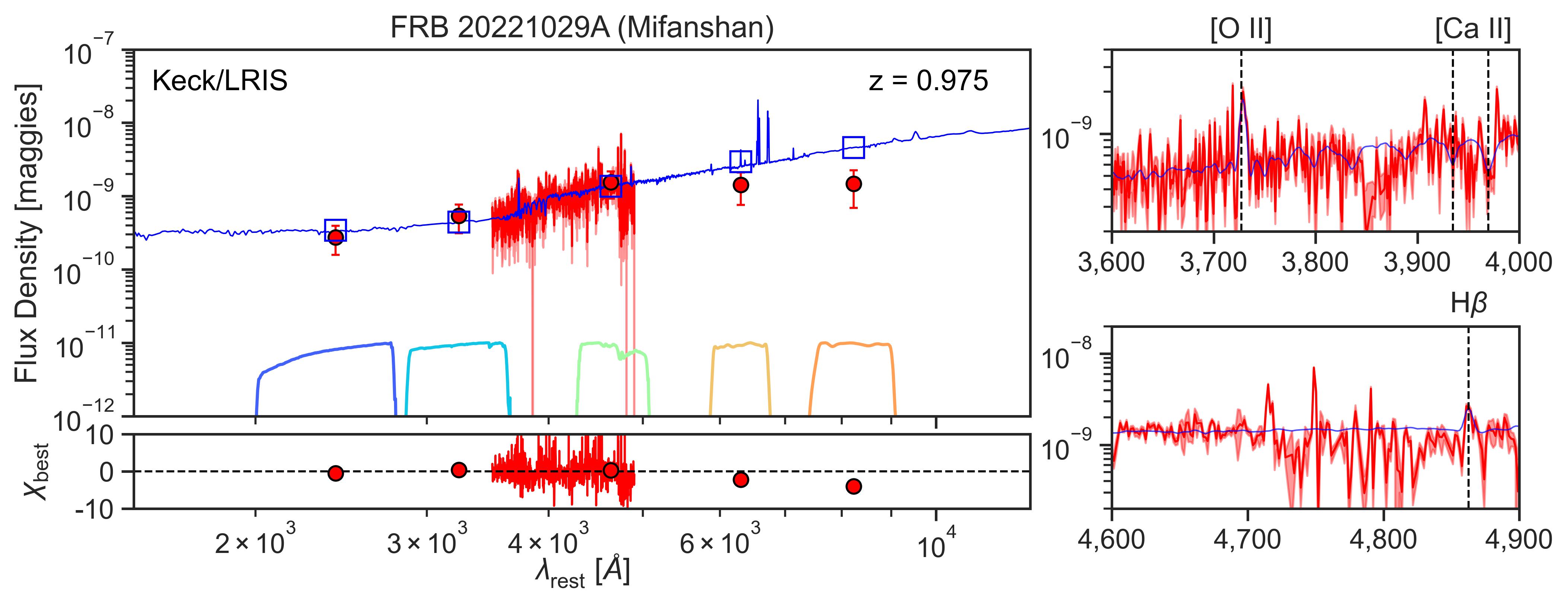}
\caption{\textbf{Supplementary Fig.\ref{fig:host_seds}:}  \textit{(Cont.)}.}
\end{figure*}

\begin{figure*}[!ht]
\centering
\includegraphics[width=0.9\textwidth]{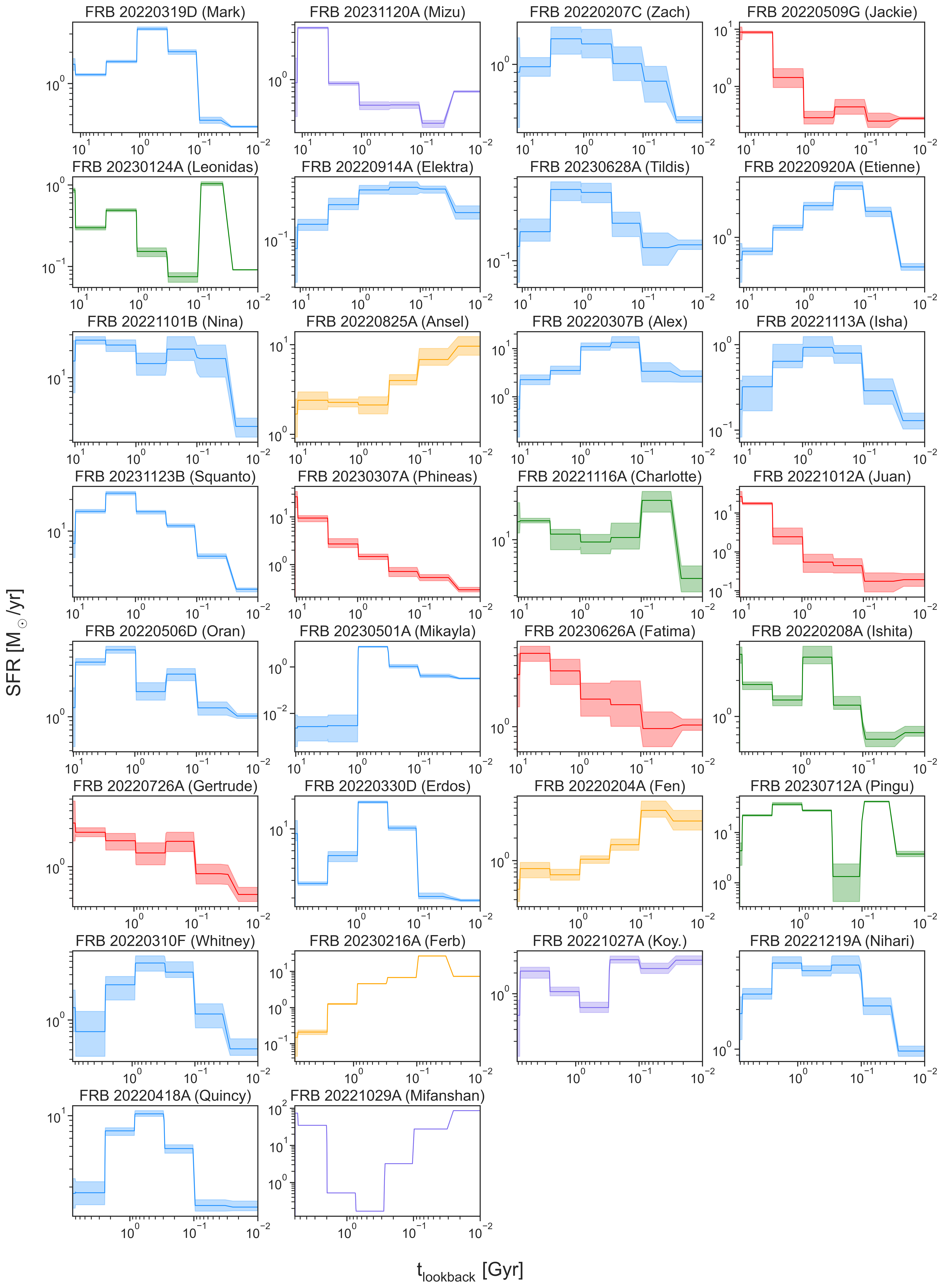}
\caption{{\supfigurelabel{fig:host_sfh}} \textbf{Star formation histories of FRB host galaxies.} The SFHs are plotted backwards from the age of the universe at the redshift of the galaxy to present day (t$_{\mathrm{lookback}} = 0$). We color code all the SFHs by their type: delayed-$\tau$ exponentially declining (blue), $\tau$-linear exponentially-declining (red), rejuvenating (purple), post-starburst (green) and rising (orange).}
\end{figure*}

\newpage
\clearpage
\newpage
\newpage

\begin{figure*}[ht!]
\centering
\includegraphics[width=\textwidth]{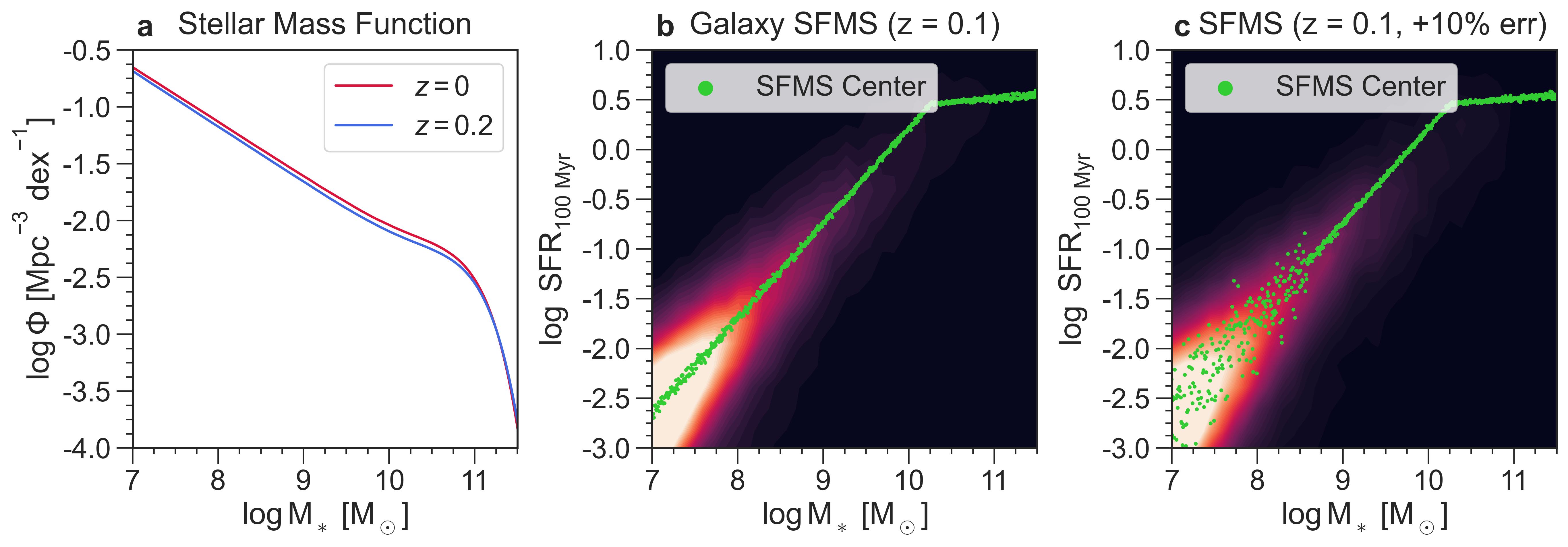}
\caption{{\supfigurelabel{fig:extreme_cases_sanity_check}} \ks{\textbf{Verifying the two extreme worst-case scenarios to quantify uncertainties on the background galaxy population at low redshifts and low stellar mass.} Panel (a) shows that the evolution of the galaxy mass function between redshift $z=0$ and $z=0.2$ is $\lesssim 0.1$~dex. Panel (b) shows the galaxy star-forming main sequence (green) with scatter from the fits in Leja et al. (2022) and the distribution of galaxies in the stellar mass - SFR space in the background, as sampled using our methodology. Panel (c) shows the distribution sampled with additional 10\% uncertainties on the galaxy star-forming main sequence. With these additional 10\% uncertainties, there are marginally more low-mass galaxies with higher star-formation rates than the case with lower uncertainties. Overall, the sampled background galaxy distribution in panels (b) and (c) are quite similar.}}
\end{figure*}

\newpage
\clearpage
\newpage
\newpage

\begin{figure*}[ht!]
\centering
\includegraphics[width=0.465\textwidth]{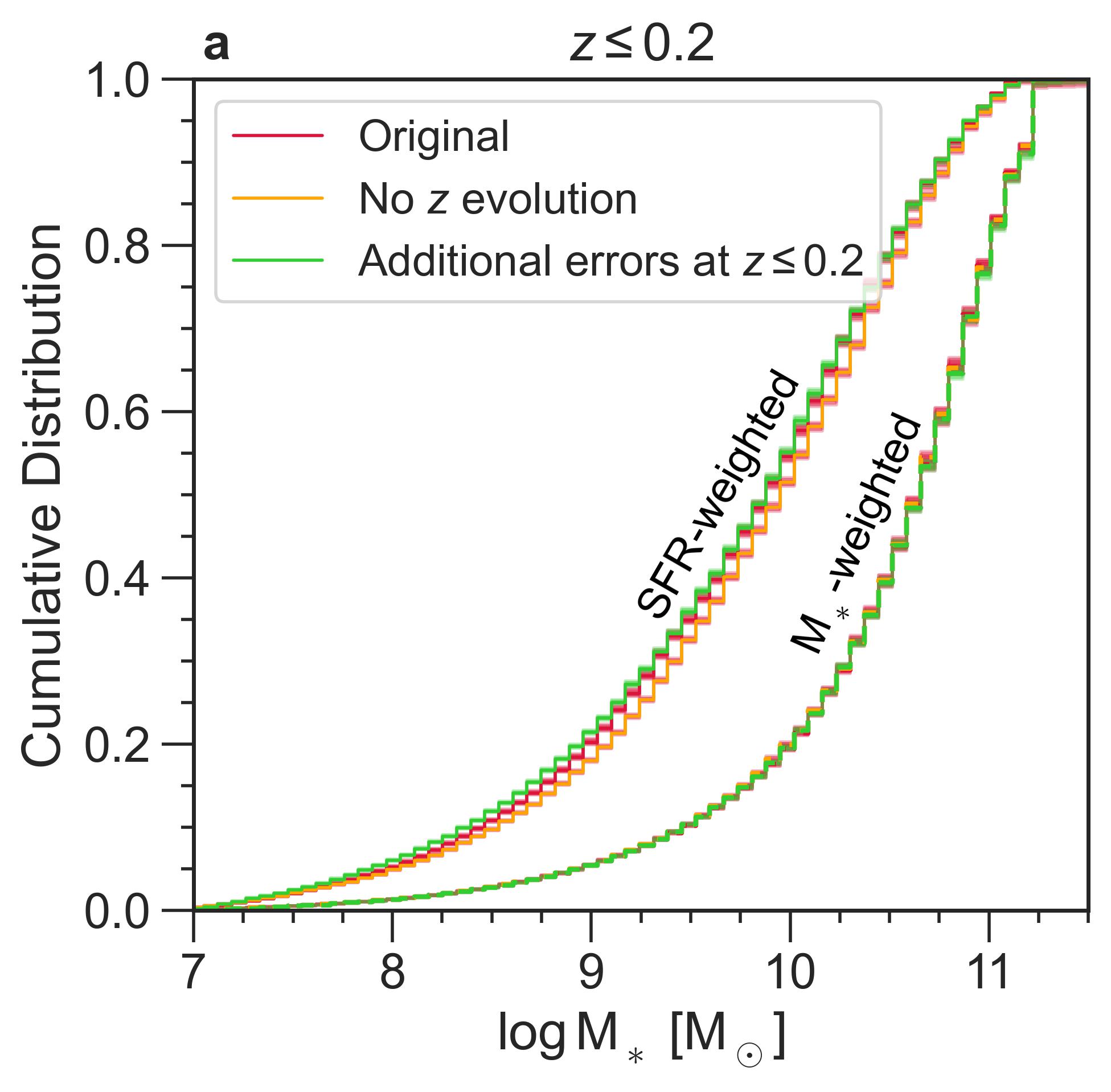}
\includegraphics[width=0.465\textwidth]{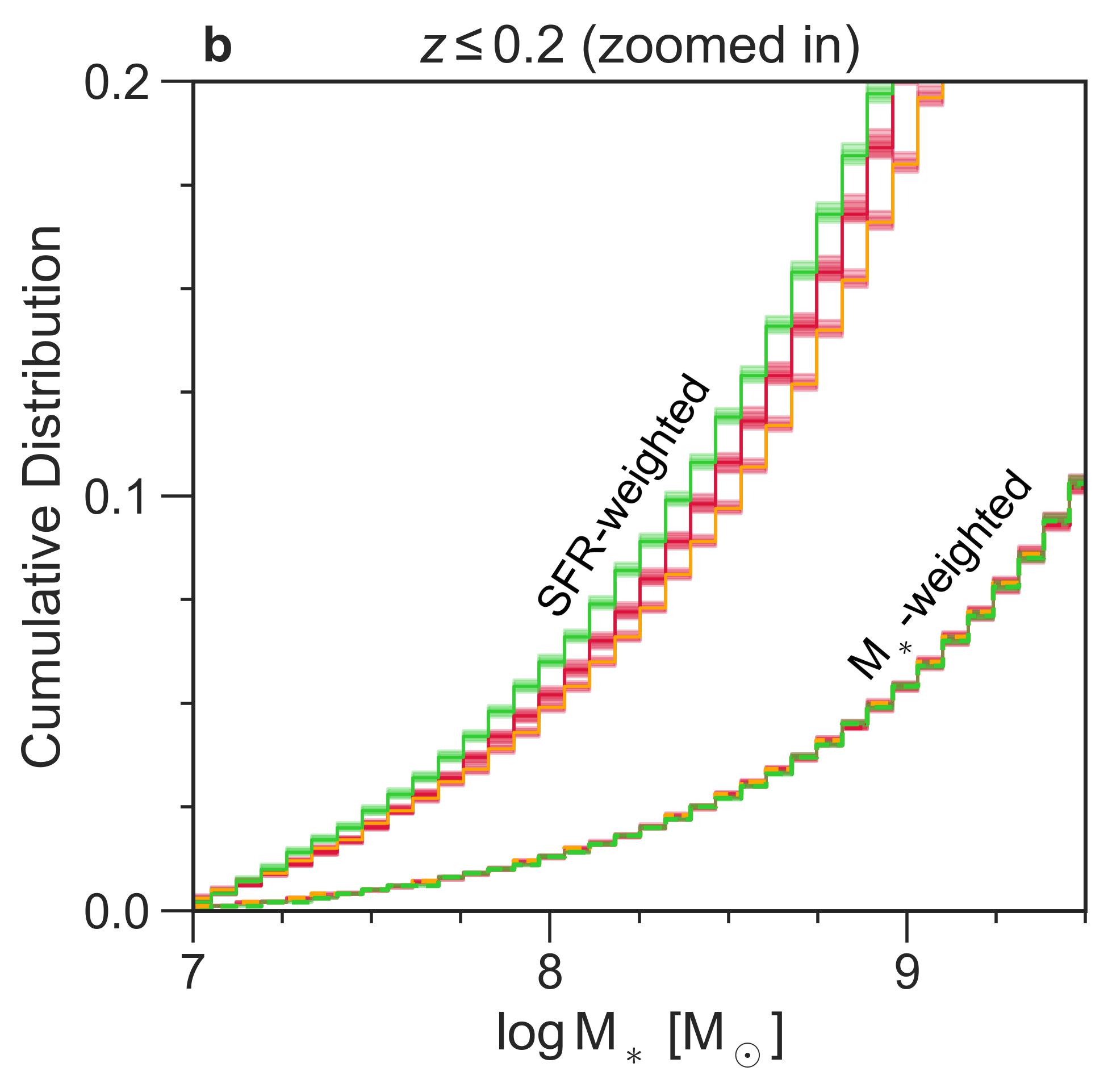}
\caption{{\supfigurelabel{fig:additional_uncertainties_no_z_evolution}} \ks{\textbf{Assessing the uncertainties on stellar mass and SFR-weighted galaxy stellar mass distribution from galaxy models.} Panel (b) is a zoomed in version of panel (a) on the low mass end for clarity. We test two extreme worst-case scenarios: (1) no redshift evolution to test extrapolation to low-redshift regime and (2) additional 10\% uncertainties on the power-law index of galaxy star-forming main sequence to test extrapolation to low-mass regime. The movement in mass-weighted distributions is negligible in both the scenarios. This is because (1) for no redshift evolution scenario, the difference in the mass function from redshift $z = 0.2$ to $z = 0$ is tiny and (2) for additional errors in galaxy main-sequence, only SFRs are impacted and stellar mass weighted stellar mass distribution remains the same. The movement in SFR-weighted distributions in both the scenarios is $\lesssim 0.1$~dex. Additional 10\% uncertainties on the low mass ends lead to marginally higher number of low mass galaxies with high SFRs, thus increasing the significance of our low-mass deficit. Despite of additional 10\% errors on the power-law index of the main sequence and assuming no redshift evolution, the shifts are small, implying that these uncertainties in galaxy models are small and the deficit of low-mass FRB host galaxies is robust.}}
\end{figure*}

\newpage

\setcounter{table}{0}
\renewcommand{\thetable}{\arabic{table}}
\renewcommand{\theHtable}{Supplement.\thetable}

\newcolumntype{C}{>{\footnotesize\raggedright\arraybackslash}c}

\newpage
\clearpage
\newpage
\newpage

\begin{table*}[ht!]
    \setlength{\tabcolsep}{4.2pt}
    \centering
    \captionsetup{labelformat=supplementarytablelabel}
    \caption{FRB host galaxy spectroscopy details.}
    
    \begin{tabular}{LLLLLLLLL}
        \toprule
        
        FRB & Facility & Instrument & Observation Date & Grating/Grism & Slit Width & Seeing & FWHM & Ref. \\
        
        & & & & & [arcsec] & [arcsec] & [\AA] & \\
        
        \hline 
        
        \frbmark & P-200 & DBSP & 2022 June 01 UT & R316/7500 & 1.5 & 1.20 & 8.24 & \protect\citemethods{2023arXiv230101000R}$^,$\citep{law2023deep} \\
        (Mark) & & & & & & & & \\

        \frbmizu & Keck:I & LRIS & 2023 Dec 17 UT & B400/3400, R400/8500 & 1.0 & 0.80 & 8.24 & This Work \\
        (Mizu) & & & & & & & & \\

        \frbzach & Keck:I & LRIS & 2022 May 26 UT & B400/3400, R400/8500 & 1.0 & 0.74 & 7.06 & \citep{law2023deep} \\
        (Zach) & & & & & & & & \\

        \frbjackie & Keck:I & LRIS & 2022 Oct 18 UT & B300/5000 & 1.0 & 0.95 & 12.48 & \citep{2023ApJ...950..175S}$^,$\citep{law2023deep} \\
        (Jackie) & & & & & & & & \\

        \frbleonidas & Keck:II & DEIMOS & 2023 May 14 UT & 600ZD, GG455, 6500~\AA & 1.0 & 0.91 & 4.78 & This Work \\
        (Leonidas) & & & & & & & & \\

        \frbelektra & Keck:I & LRIS & 2022 Oct 18 UT & B300/5000 & 1.0 & 0.95 & 8.64 & \citep{2023ApJ...950..175S}$^,$\citep{law2023deep} \\
        (Elektra) & & & & & & & & \\

        \frbtildis & Keck:II & DEIMOS & 2023 July 13 UT & 600ZD, GG455, 6500~\AA & 1.0 & 0.81 & 6.25 & This Work \\
        (Tildis) & & & & & & & & \\

        \frbetienne & P-200 & DBSP & 2022 Dec 17 UT & B600/4000, R316/7500 & 1.5 & 1.50 & 7.06 & \citep{law2023deep} \\
        (Etienne) & & & & & & & & \\

        \frbnina & P-200 & DBSP & 2023 Aug 26 UT & B600/4000, R316/7500 & 1.0 & 0.90 & 7.06 & This Work \\
        (Nina) & & & & & & & & \\

        \frbansel & Keck:I & LRIS & 2022 Oct 18 UT & B300/5000 & 1.0 & 0.67 & 8.57 & \citep{law2023deep} \\
        (Ansel) & & & & & & & & \\

        \frbalex & Keck:II & DEIMOS & 2023 July 13 UT & 600ZD, GG455, 6500~\AA & 1.0 & 1.15 & 8.24 & \citep{law2023deep} \\
        (Alex) & & & & & & & & \\

        \frbisha & P-200 & DBSP & 2023 Aug 26 UT & B600/4000, R316/7500 & 1.5 & 1.40 & 7.06 & This Work \\
        (Isha) & & & & & & & & \\

        \frbsquanto & Keck:I & LRIS & 2024 Feb 14 UT & B400/3400, R400/8500  & 1.0 & 1.10 & 7.06 & This Work \\
        
        (Squanto) & & & & & & & & \\
        
        \frbphineas & Keck:II & DEIMOS & 2023 May 14 UT & 600ZD, GG455, 6500~\AA & 1.0 & 0.68 & 8.24 & This Work \\
        (Phineas) & & & & & & & & \\

        \frbcharlotte & Keck:II & DEIMOS & 2023 Sept 9 UT & 600ZD, GG455, 6500~\AA & 1.0 & 0.70 & 11.19 & This Work \\
        (Charlotte) & & & & & & & & \\

        \frbjuan & Keck:I & LRIS & 2022 Oct 18 UT & B300/5000 & 1.0 & 0.72 & 13.45 & \citep{law2023deep} \\
        (Juan) & & & & & & & & \\

        \frboran & Keck:I & LRIS & 2022 July 21 UT & B400/3400, R400/8500 & 1.0 & 0.80 & 10.83 & \citep{law2023deep} \\
        (Oran) & & & & & & & & \\

        \frbmikayla & Keck:II & DEIMOS & 2023 Aug 14 UT & 600ZD, GG455, 6500~\AA & 1.0 & 0.77 & 9.36 & This Work \\
        (Mikayla) & & & & & & & & \\

        \frbfatima & P-200 & DBSP & 2023 Aug 26 UT & B600/4000, R316/7500 & 1.0 & 0.90 & 9.42 & This Work \\
        (Fatima) & & & & & & & & \\

        \frbishita & Keck:I & LRIS & 2022 Aug 2 UT & B400/3400, R400/8500 & 1.0 & 0.80 & 9.60 & This Work \\
        (Ishita) & & & & & & & & \\

        \frbgertrude & Keck:II & DEIMOS & 2023 Sept 9 UT & 600ZD, GG455, 6500~\AA & 1.0 & 1.12 & 9.42 & This Work \\
        (Gertrude) & & & & & & & & \\
         
        \hline
    \end{tabular}
\end{table*}

\newpage
\clearpage
\newpage
\newpage

\begin{table*}[ht!]
    \ContinuedFloat
    \setlength{\tabcolsep}{4.2pt}
    \centering
    \captionsetup{labelformat=supplementarytablelabel}
    \caption{FRB host galaxy spectroscopy details {\textit{(cont.)}}.}
    
    \begin{tabular}{LLLLLLLLL}
        \toprule
        
        FRB & Facility & Instrument & Observation Date & Grating/Grism & Slit Width & Seeing & FWHM & Ref. \\
        
        & & & & & [arcsec] & [arcsec] & [\AA] & \\
        
        \hline 

        \frberdos & Keck:I & LRIS & 2023 Dec 17 UT & B400/3400, R400/8500 & 1.0 & 0.80 & 12.11 & This Work \\
        (Erdos) & & & & & & & & \\
        
        \frbfen & Keck:II & DEIMOS & 2023 May 14 UT & 600ZD, GG455, 6500~\AA & 1.0 & 0.88 & 8.10 & This Work \\
        (Fen) & & & & & & & & \\

        \frbpingu & Keck:II & DEIMOS & 2023 Dec 13 UT & 600ZD, GG455, 6500~\AA & 1.0 & 1.15 & 5.70 & This Work \\
        (Pingu) & & & & & & & & \\

        \frbwhitney & Keck:I & LRIS & 2022 May 26 UT & B400/3400, R400/8500 & 1.0 & 0.80 & 8.69 & \citep{law2023deep} \\
        (Whitney) & & & & & & & & \\

        \frbferb & Keck:II & DEIMOS & 2023 June 17 UT & 600ZD, GG455, 6500~\AA & 1.0 & 0.82 & 8.01 & This Work \\
        (Ferb) & & & & & & & & \\

        \frbkoyaanisqatsi & Keck:II & DEIMOS & 2023 Apr 21 UT & 600ZD, OG550, 8150~\AA & 1.0 & 0.86 & 5.05 & This Work \\
        (Koy.) & & & & & & & & \\

        \frbnihari & Keck:I & LRIS & 2023 June 13 UT & B400/3400,R400/8500 & 1.0 & 0.80 & 8.12 & This Work \\
        (Nihari) & & & & & & & & \\

        20220418A & Keck:I & LRIS & 2022 July 21 UT & B400/3400,R400/8500 & 1.0 & 0.80 & 7.08 & \citep{law2023deep} \\
        (Quincy) & & & & & & & & \\

        20221029A & Keck:II & DEIMOS & 2023 Dec 10 UT & 830G, OG550, 8000 \AA & 0.8 & 0.84 & 6.01 & This Work \\
        (Mifanshan) & & & & & & & & \\
        
        \hline
    \end{tabular}
    \footnotesize{\raggedright{ N{\scriptsize{OTE}}: The details of instruments, configurations and facilities used for the spectroscopy of FRB hosts presented in this work. For DEIMOS, we note the grating, filter and central wavelengths used to obtain these observations.\par}}
    \label{table:host_spectroscopy_details}
\end{table*}

\newpage
\clearpage
\newpage
\newpage
    
\begin{table*}[ht!]
    \centering
    \captionsetup{labelformat=supplementarytablelabel}
    \caption{Log of host galaxies imaging and photometry.}
    \begin{tabular}{LLLLLLL}
        \toprule
        
        FRB & Facility & Instrument & Observation Date & Filter & M$_\mathrm{AB}$ & Reference \\ 

        & & & & & [mag] & \\ 
        
        \hline 

        \frbmark (Mark) & Pan-STARRS & & & $g$ & 13.63 $\pm$ 0.02 & \protect\citemethods{2023arXiv230101000R}$^,$\citep{2016arXiv161205560C}$^,$\citep{law2023deep} \\
        & Pan-STARRS & & & $r$ & 13.25 $\pm$ 0.01 & \protect\citemethods{2023arXiv230101000R}$^,$\citep{2016arXiv161205560C}$^,$\citep{law2023deep} \\
        & Pan-STARRS & & & $i$ & 13.00 $\pm$ 0.01 & \protect\citemethods{2023arXiv230101000R}$^,$\citep{2016arXiv161205560C}$^,$\citep{law2023deep} \\
        & Pan-STARRS & & & $z$ & 12.81 $\pm$ 0.01 & \protect\citemethods{2023arXiv230101000R}$^,$\citep{2016arXiv161205560C}$^,$\citep{law2023deep} \\
        & Pan-STARRS & & & $y$ & 12.45 $\pm$ 0.01 & \protect\citemethods{2023arXiv230101000R}$^,$\citep{2016arXiv161205560C}$^,$\citep{law2023deep} \\
        & SDSS & & & $u$ & 14.20 $\pm$ 0.14 & \citep{2015ApJS..219...12A}, This Work \\
        & SDSS & & & $g$ & 13.67 $\pm$ 0.02 & \citep{2015ApJS..219...12A}, This Work \\
        & SDSS & & & $r$ & 13.19 $\pm$ 0.01 & \citep{2015ApJS..219...12A}, This Work \\
        & SDSS & & & $i$ & 12.95 $\pm$ 0.01 & \citep{2015ApJS..219...12A}, This Work \\
        & SDSS & & & $z$ & 12.67 $\pm$ 0.02 & \citep{2015ApJS..219...12A}, This Work \\
        & 2MASS & & & $J$ & 12.53 $\pm$ 0.02 & \protect\citemethods{2023arXiv230101000R}$^,$\citep{2006AJ....131.1163S}$^,$\citep{law2023deep} \\
        & 2MASS & & & $H$ & 12.35 $\pm$ 0.02 & \protect\citemethods{2023arXiv230101000R}$^,$\citep{2006AJ....131.1163S}$^,$\citep{law2023deep} \\
        & 2MASS & & & $K_s$ & 12.55 $\pm$ 0.02 & \protect\citemethods{2023arXiv230101000R}$^,$\citep{2006AJ....131.1163S}$^,$\citep{law2023deep} \\
        & ALLWISE & & & $W1$ & 12.91 $\pm$ 0.03 & \protect\citemethods{2023arXiv230101000R}$^,$\citep{2014yCat.2328....0C}$^,$\citep{law2023deep} \\
        & ALLWISE & & & $W2$ & 13.42 $\pm$ 0.03 & \protect\citemethods{2023arXiv230101000R}$^,$\citep{2014yCat.2328....0C}$^,$\citep{law2023deep} \\
        & ALLWISE & & & $W3$ & 12.16 $\pm$ 0.03 & \protect\citemethods{2023arXiv230101000R}$^,$\citep{2014yCat.2328....0C}$^,$\citep{law2023deep} \\
        & ALLWISE & & & $W4$ & 12.00 $\pm$ 0.04 & \protect\citemethods{2023arXiv230101000R}$^,$\citep{2014yCat.2328....0C}$^,$\citep{law2023deep} \\
        
        \hline

        \frbmizu (Mizu) & Pan-STARRS & & & $g$ & 16.08 $\pm$ 0.01 & \citep{2016arXiv161205560C}, This Work \\
        & Pan-STARRS & & & $r$ & 15.39 $\pm$ 0.01 & \citep{2016arXiv161205560C}, This Work \\
        & Pan-STARRS & & & $i$ & 15.11 $\pm$ 0.00 & \citep{2016arXiv161205560C}, This Work \\
        & Pan-STARRS & & & $z$ & 14.86 $\pm$ 0.01 & \citep{2016arXiv161205560C}, This Work \\
        & Pan-STARRS & & & $y$ & 14.60 $\pm$ 0.01 & \citep{2016arXiv161205560C}, This Work \\
        & BASS & 90Prime & & $g$ & 16.18 $\pm$ 0.21 & \citep{2017PASP..129f4101Z}, This Work \\
        & BASS & 90Prime & & $r$ & 15.41 $\pm$ 0.21 & \citep{2017PASP..129f4101Z}, This Work \\
        & MzLS & MOSAIC-3 & & $z$ & 14.86 $\pm$ 0.20 & \citep{2019AJ....157..168D}, This Work \\
        & 2MASS & & & $J$ & 14.44 $\pm$ 0.04 & \citep{2006AJ....131.1163S}, This Work \\
        & 2MASS & & & $H$ & 14.35 $\pm$ 0.06 & \citep{2006AJ....131.1163S}, This Work \\
        & 2MASS & & & $K_s$ & 14.30 $\pm$ 0.05 & \citep{2006AJ....131.1163S}, This Work \\
        & ALLWISE & & & $W1$ & 15.43 $\pm$ 0.01 & \citep{2014yCat.2328....0C}, This Work \\
        & ALLWISE & & & $W2$ & 15.97 $\pm$ 0.01 & \citep{2014yCat.2328....0C}, This Work \\
        & ALLWISE & & & $W3$ & 14.70 $\pm$ 0.02 & \citep{2014yCat.2328....0C}, This Work \\
        & ALLWISE & & & $W4$ & 13.81 $\pm$ 0.07 & \citep{2014yCat.2328....0C}, This Work \\

        \hline

        \frbzach (Zach) & Pan-STARRS & & & $g$ & 16.43 $\pm$ 0.03 & \citep{2016arXiv161205560C}$^,$\citep{law2023deep} \\
        & Pan-STARRS & & & $r$ & 16.15 $\pm$ 0.01 & \citep{2016arXiv161205560C}$^,$\citep{law2023deep} \\
        & Pan-STARRS & & & $i$ & 15.96 $\pm$ 0.01 & \citep{2016arXiv161205560C}$^,$\citep{law2023deep} \\
        & Pan-STARRS & & & $z$ & 15.77 $\pm$ 0.01 & \citep{2016arXiv161205560C}$^,$\citep{law2023deep} \\
        & Pan-STARRS & & & $y$ & 15.67 $\pm$ 0.02 & \citep{2016arXiv161205560C}$^,$\citep{law2023deep} \\
        & 2MASS & & & $H$ & 15.45 $\pm$ 0.04 & \citep{2006AJ....131.1163S}$^,$\citep{law2023deep} \\
        & ALLWISE & & & $W1$ & 16.43 $\pm$ 0.01 & \citep{2014yCat.2328....0C}$^,$\citep{law2023deep} \\
        & ALLWISE & & & $W2$ & 17.10 $\pm$ 0.03 & \citep{2014yCat.2328....0C}$^,$\citep{law2023deep} \\
        & ALLWISE & & & $W3$ & 15.84 $\pm$ 0.11 & \citep{2014yCat.2328....0C}$^,$\citep{law2023deep} \\
        & ALLWISE & & & $W4$ & 15.14 $\pm$ 0.46 & \citep{2014yCat.2328....0C}$^,$\citep{law2023deep} \\
        
        \hline

        \frbjackie (Jackie) & GALEX & & & FUV & 21.05 $\pm$ 0.16 & \citep{2005ApJ...619L...1M}, This Work \\
        & GALEX & & & NUV & 19.93 $\pm$ 0.07 & \citep{2005ApJ...619L...1M}, This Work \\
        
        \hline
        
    \end{tabular}
    
    \label{table:imaging_log}
\end{table*}

\begin{table*}[ht!]
    \ContinuedFloat
    \centering
    \captionsetup{labelformat=supplementarytablelabel}
    \caption{Log of host galaxies imaging and photometry {\textit{(cont.)}}.}
    \begin{tabular}{LLLLLLL}
        \toprule
        
        FRB & Facility & Instrument & Observation Date & Filter & M$_\mathrm{AB}$ & Reference \\ 

        & & & & & [mag] & \\ 

        \hline 

        & Pan-STARRS & & & $g$ & 17.20 $\pm$ 0.02 & \citep{2016arXiv161205560C, 2023ApJ...950..175S}$^,$\citep{law2023deep} \\
        & Pan-STARRS & & & $r$ & 16.51 $\pm$ 0.01 & \citep{2016arXiv161205560C, 2023ApJ...950..175S}$^,$\citep{law2023deep} \\
        & Pan-STARRS & & & $i$ & 16.24 $\pm$ 0.01 & \citep{2016arXiv161205560C, 2023ApJ...950..175S}$^,$\citep{law2023deep} \\
        & Pan-STARRS & & & $z$ & 16.11 $\pm$ 0.01 & \citep{2016arXiv161205560C, 2023ApJ...950..175S}$^,$\citep{law2023deep} \\
        & Pan-STARRS & & & $y$ & 15.87 $\pm$ 0.02 & \citep{2016arXiv161205560C, 2023ApJ...950..175S}$^,$\citep{law2023deep} \\
        & BASS & 90Prime & & $g$ & 17.26 $\pm$ 0.21 & \citep{2017PASP..129f4101Z}, This Work \\
        & BASS & 90Prime & & $r$ & 16.51 $\pm$ 0.21 & \citep{2017PASP..129f4101Z}, This Work \\
        & MzLS & MOSAIC-3 & & $z$ & 16.05 $\pm$ 0.19 & \citep{2019AJ....157..168D}, This Work \\
        & 2MASS & & & $J$ & 15.67 $\pm$ 0.04 & \citep{2006AJ....131.1163S, 2023ApJ...950..175S}$^,$\citep{law2023deep} \\
        & 2MASS & & & $H$ & 15.68 $\pm$ 0.04 & \citep{2006AJ....131.1163S, 2023ApJ...950..175S}$^,$\citep{law2023deep} \\
        & 2MASS & & & $K_s$ & 15.80 $\pm$ 0.05 & \citep{2006AJ....131.1163S, 2023ApJ...950..175S}$^,$\citep{law2023deep} \\
        & ALLWISE & & & $W1$ & 16.48 $\pm$ 0.01 & \citep{2014yCat.2328....0C, 2023ApJ...950..175S}$^,$\citep{law2023deep} \\
        & ALLWISE & & & $W2$ & 17.08 $\pm$ 0.02 & \citep{2014yCat.2328....0C, 2023ApJ...950..175S}$^,$\citep{law2023deep} \\
        & ALLWISE & & & $W3$ & 16.55 $\pm$ 0.03 & \citep{2014yCat.2328....0C, 2023ApJ...950..175S}$^,$\citep{law2023deep} \\
        & ALLWISE & & & $W4$ & 15.84 $\pm$ 0.23 & \citep{2014yCat.2328....0C, 2023ApJ...950..175S}$^,$\citep{law2023deep} \\

        \hline

        \frbleonidas (Leonidas) & Pan-STARRS & & & $g$ & 19.37 $\pm$ 0.05 & \citep{2016arXiv161205560C}, This Work \\
        & Pan-STARRS & & & $r$ & 19.18 $\pm$ 0.03 & \citep{2016arXiv161205560C}, This Work \\
        & Pan-STARRS & & & $i$ & 18.85 $\pm$ 0.04 & \citep{2016arXiv161205560C}, This Work \\
        & Pan-STARRS & & & $z$ & 18.75 $\pm$ 0.05 & \citep{2016arXiv161205560C}, This Work \\
        & Pan-STARRS & & & $y$ & 18.73 $\pm$ 0.07 & \citep{2016arXiv161205560C}, This Work \\
        & BASS & 90Prime & & $g$ & 19.59 $\pm$ 0.20 & \citep{2017PASP..129f4101Z}, This Work \\
        & BASS & 90Prime & & $r$ & 19.10 $\pm$ 0.20 & \citep{2017PASP..129f4101Z}, This Work \\
        & MzLS & MOSAIC-3 & & $z$ & 18.67 $\pm$ 0.19 & \citep{2019AJ....157..168D}, This Work \\
        & 2MASS & & & $J$ & 18.49 $\pm$ 0.10 & \citep{2006AJ....131.1163S}, This Work \\
        & ALLWISE & & & $W1$ & 18.06 $\pm$ 0.03 & \citep{2014yCat.2328....0C}, This Work \\

        \hline

        \frbelektra (Elektra) & Pan-STARRS & & & $g$ & 20.55 $\pm$ 0.04 & \citep{2016arXiv161205560C} \\
        & Pan-STARRS & & & $r$ & 20.05 $\pm$ 0.04 & \citep{2016arXiv161205560C, 2023ApJ...950..175S}$^,$\citep{law2023deep} \\
        & Pan-STARRS & & & $i$ & 19.80 $\pm$ 0.03 & \citep{2016arXiv161205560C, 2023ApJ...950..175S}$^,$\citep{law2023deep} \\
        & Pan-STARRS & & & $z$ & 19.73 $\pm$ 0.02 & \citep{2016arXiv161205560C, 2023ApJ...950..175S}$^,$\citep{law2023deep} \\
        & BASS & 90Prime & & $g$ & 20.69 $\pm$ 0.20 & \citep{2017PASP..129f4101Z}, This Work \\
        & BASS & 90Prime & & $r$ & 20.20 $\pm$ 0.19 & \citep{2017PASP..129f4101Z}, This Work \\
        & MzLS & MOSAIC-3 & & $z$ & 19.70 $\pm$ 0.18 & \citep{2019AJ....157..168D}, This Work \\
        & ALLWISE & & & $W1$ & 18.88 $\pm$ 0.09 & \citep{2014yCat.2328....0C, 2023ApJ...950..175S}$^,$\citep{law2023deep} \\
        & ALLWISE & & & $W2$ & 19.41 $\pm$ 0.13 & \citep{2014yCat.2328....0C, 2023ApJ...950..175S}$^,$\citep{law2023deep} \\

        \hline

        \frbtildis (Tildis) & GALEX & & & FUV & 22.09 $\pm$ 0.19 & \citep{2005ApJ...619L...1M}, This Work \\
        & GALEX & & & NUV & 22.15 $\pm$ 0.19 & \citep{2005ApJ...619L...1M}, This Work \\
        & Pan-STARRS & & & $g$ & 20.70 $\pm$ 0.04 & \citep{2016arXiv161205560C}, This Work \\
        & Pan-STARRS & & & $r$ & 20.15 $\pm$ 0.04 & \citep{2016arXiv161205560C}, This Work \\
        & Pan-STARRS & & & $i$ & 19.88 $\pm$ 0.03 & \citep{2016arXiv161205560C}, This Work \\
        & Pan-STARRS & & & $z$ & 19.79 $\pm$ 0.02 & \citep{2016arXiv161205560C}, This Work \\
        & Pan-STARRS & & & $y$ & 19.49 $\pm$ 0.08 & \citep{2016arXiv161205560C}, This Work \\
        & BASS & 90Prime & & $g$ & 20.46 $\pm$ 0.20 & \citep{2017PASP..129f4101Z}, This Work \\
        & BASS & 90Prime & & $r$ & 19.94 $\pm$ 0.20 & \citep{2017PASP..129f4101Z}, This Work \\

        \hline
        
    \end{tabular}
    
\end{table*}

\begin{table*}[ht!]
    \ContinuedFloat
    \centering
    \captionsetup{labelformat=supplementarytablelabel}
    \caption{Log of host galaxies imaging and photometry {\textit{(cont.)}}.}
    \begin{tabular}{LLLLLLL}
        \toprule
        
        FRB & Facility & Instrument & Observation Date & Filter & M$_\mathrm{AB}$ & Reference \\ 

        & & & & & [mag] & \\ 
        
        \hline

        & MzLS & MOSAIC-3 & & $z$ & 19.53 $\pm$ 0.19 & \citep{2019AJ....157..168D}, This Work \\
        & ALLWISE & & & $W1$ & 20.14 $\pm$ 0.03 & \citep{2014yCat.2328....0C}, This Work \\
        & ALLWISE & & & $W2$ & 20.35 $\pm$ 0.05 & \citep{2014yCat.2328....0C}, This Work \\

        \hline

        \frbetienne (Etienne) & Pan-STARRS & & & $g$ & 19.59 $\pm$ 0.07 & \citep{2016arXiv161205560C}$^,$\citep{law2023deep} \\
        & Pan-STARRS & & & $g$ & 19.46 $\pm$ 0.03 & \citep{2016arXiv161205560C}$^,$\citep{law2023deep} \\
        & Pan-STARRS & & & $i$ & 18.96 $\pm$ 0.06 & \citep{2016arXiv161205560C}$^,$\citep{law2023deep} \\
        & Pan-STARRS & & & $z$ & 18.99 $\pm$ 0.05 & \citep{2016arXiv161205560C}$^,$\citep{law2023deep} \\
        & Pan-STARRS & & & $y$ & 18.54 $\pm$ 0.07 & \citep{2016arXiv161205560C}$^,$\citep{law2023deep} \\        
        & BASS & 90Prime & & $g$ & 19.65 $\pm$ 0.20 & \citep{2017PASP..129f4101Z} \\
        & BASS & 90Prime & & $r$ & 19.43 $\pm$ 0.20 & \citep{2017PASP..129f4101Z} \\
        & MzLS & MOSAIC-3 & & $z$ & 19.05 $\pm$ 0.19 & \citep{2019AJ....157..168D} \\
        & ALLWISE & & & $W1$ & 18.15 $\pm$ 0.03 & \citep{2014yCat.2328....0C} \\
        & ALLWISE & & & $W2$ & 18.52 $\pm$ 0.04 & \citep{2014yCat.2328....0C} \\
        & ALLWISE & & & $W3$ & 16.90 $\pm$ 0.06 & \citep{2014yCat.2328....0C} \\

        \hline

        \frbnina (Nina) & Pan-STARRS & & & $g$ & 19.08 $\pm$ 0.12 & \citep{2016arXiv161205560C}, This Work \\
         & Pan-STARRS & & & $r$ & 18.81 $\pm$ 0.04 & \citep{2016arXiv161205560C}, This Work \\
         & Pan-STARRS & & & $i$ & 18.40 $\pm$ 0.05 & \citep{2016arXiv161205560C}, This Work \\
         & Pan-STARRS & & & $z$ & 18.27 $\pm$ 0.02 & \citep{2016arXiv161205560C}, This Work \\
         & Pan-STARRS & & & $y$ & 17.98 $\pm$ 0.03 & \citep{2016arXiv161205560C}, This Work \\
         & 2MASS & & & $J$ & 17.16 $\pm$ 0.18 & \citep{2006AJ....131.1163S}, This Work \\
         & 2MASS & & & $K_s$ & 16.77 $\pm$ 0.17 & \citep{2006AJ....131.1163S}, This Work \\
         & ALLWISE & & & $W1$ & 16.91 $\pm$ 0.02 & \citep{2014yCat.2328....0C}, This Work \\
         & ALLWISE & & & $W2$ & 17.34 $\pm$ 0.05 & \citep{2014yCat.2328....0C}, This Work \\
         & ALLWISE & & & $W3$ & 17.22 $\pm$ 0.24 & \citep{2014yCat.2328....0C}, This Work \\
         & ALLWISE & & & $W4$ & 15.89 $\pm$ 0.25 & \citep{2014yCat.2328....0C}, This Work \\

        \hline

        \frbansel (Ansel) 
        & Pan-STARRS & & & $g$ & 21.36 $\pm$ 0.10 & \citep{2016arXiv161205560C}$^,$\citep{law2023deep} \\
        & Pan-STARRS & & & $r$ & 20.74 $\pm$ 0.16 & \citep{2016arXiv161205560C}$^,$\citep{law2023deep} \\
        & Pan-STARRS & & & $i$ & 20.34 $\pm$ 0.03 & \citep{2016arXiv161205560C}$^,$\citep{law2023deep} \\
        & Pan-STARRS & & & $z$ & 19.97 $\pm$ 0.07 & \citep{2016arXiv161205560C}$^,$\citep{law2023deep} \\
        & Pan-STARRS & & & $y$ & 19.56 $\pm$ 0.10 & \citep{2016arXiv161205560C}$^,$\citep{law2023deep} \\
        & ALLWISE & & & $W1$ & 18.80 $\pm$ 0.09 & \citep{2014yCat.2328....0C}$^,$\citep{law2023deep} \\
        & ALLWISE & & & $W2$ & 18.98 $\pm$ 0.10 & \citep{2014yCat.2328....0C}$^,$\citep{law2023deep} \\
        & ALLWISE & & & $W3$ & 17.72 $\pm$ 0.30 & \citep{2014yCat.2328....0C}$^,$\citep{law2023deep} \\

        \hline

        \frbalex (Alex) & Pan-STARRS & & & $g$ & 21.11 $\pm$ 0.13 & \citep{2016arXiv161205560C}$^,$\citep{law2023deep} \\
        & Pan-STARRS & & & $r$ & 20.45 $\pm$ 0.05 & \citep{2016arXiv161205560C}$^,$\citep{law2023deep} \\
        & Pan-STARRS & & & $i$ & 20.14 $\pm$ 0.09 & \citep{2016arXiv161205560C}$^,$\citep{law2023deep} \\
        & Pan-STARRS & & & $z$ & 19.82 $\pm$ 0.06 & \citep{2016arXiv161205560C}$^,$\citep{law2023deep} \\
        & Pan-STARRS & & & $y$ & 19.43 $\pm$ 0.09 & \citep{2016arXiv161205560C}$^,$\citep{law2023deep} \\
        & P-200 & WIRC & 2022 Aug 17 UT & $J$ & 19.50 $\pm$ 0.02 & \citep{law2023deep} \\
        & P-200 & WIRC & 2022 Aug 17 UT & $H$ & 18.82 $\pm$ 0.02 & \citep{law2023deep} \\

        \hline

        \frbisha (Isha) & Pan-STARRS & & & $g$ & 21.42 $\pm$ 0.18 & \citep{2016arXiv161205560C}, This Work \\
        & Pan-STARRS & & & $r$ & 21.38 $\pm$ 0.10 & \citep{2016arXiv161205560C}, This Work \\
        & Pan-STARRS & & & $i$ & 21.16 $\pm$ 0.13 & \citep{2016arXiv161205560C}, This Work \\
        & Pan-STARRS & & & $z$ & 20.99 $\pm$ 0.09 & \citep{2016arXiv161205560C}, This Work \\

        \hline
        
    \end{tabular}
\end{table*}

\begin{table*}[ht!]
    \ContinuedFloat
    \centering
    \captionsetup{labelformat=supplementarytablelabel}
    \caption{Log of host galaxies imaging and photometry {\textit{(cont.)}}.}
    \begin{tabular}{LLLLLLL}
        \toprule
        
        FRB & Facility & Instrument & Observation Date & Filter & M$_\mathrm{AB}$ & Reference \\ 

        & & & & & [mag] & \\ 
        
        \hline

        & P-200 & WIRC & 2022 Aug 17 UT & $H$ & 20.70 $\pm$ 0.18 & This Work \\
        & P-200 & WIRC & 2022 Aug 17 UT & $K_s$ & 20.65 $\pm$ 0.21 & This Work \\

        \hline

        \frbsquanto (Squanto) & Pan-STARRS & & & $g$ & 19.62 $\pm$ 0.05 & \citep{2016arXiv161205560C}, This Work \\
        & Pan-STARRS & & & $r$ & 18.49 $\pm$ 0.02 & \citep{2016arXiv161205560C}, This Work \\
        & Pan-STARRS & & & $i$ & 18.14 $\pm$ 0.01 & \citep{2016arXiv161205560C}, This Work \\
        & Pan-STARRS & & & $z$ & 17.89 $\pm$ 0.02 & \citep{2016arXiv161205560C}, This Work \\
        & Pan-STARRS & & & $y$ & 17.65 $\pm$ 0.04 & \citep{2016arXiv161205560C}, This Work \\
        & BASS & 90Prime & & $g$ & 19.56 $\pm$ 0.20 & \citep{2017PASP..129f4101Z}, This Work \\
        & BASS & 90Prime & & $r$ & 18.53 $\pm$ 0.20 & \citep{2017PASP..129f4101Z}, This Work \\
        & MzLS & MOSAIC-3 & & $z$ & 17.83 $\pm$ 0.19 & \citep{2019AJ....157..168D}, This Work \\
        & ALLWISE & & & $W1$ & 17.40 $\pm$ 0.03 & \citep{2014yCat.2328....0C}, This Work \\
        & ALLWISE & & & $W2$ & 17.56 $\pm$ 0.05 & \citep{2014yCat.2328....0C}, This Work \\
        & ALLWISE & & & $W3$ & 16.22 $\pm$ 0.06 & \citep{2014yCat.2328....0C}, This Work \\
        & ALLWISE & & & $W4$ & 15.37 $\pm$ 0.28 & \citep{2014yCat.2328....0C}, This Work \\

        \hline

        \frbphineas (Phineas) & GALEX & & & NUV & 22.88 $\pm$ 0.10 & \citep{2005ApJ...619L...1M}, This Work \\
        & Pan-STARRS & & & $g$ & 20.55 $\pm$ 0.04 & \citep{2016arXiv161205560C}, This Work \\
        & Pan-STARRS & & & $r$ & 19.41 $\pm$ 0.02 & \citep{2016arXiv161205560C}, This Work \\
        & Pan-STARRS & & & $i$ & 19.02 $\pm$ 0.01 & \citep{2016arXiv161205560C}, This Work \\
        & Pan-STARRS & & & $z$ & 18.84 $\pm$ 0.02 & \citep{2016arXiv161205560C}, This Work \\
        & Pan-STARRS & & & $y$ & 18.83 $\pm$ 0.05 & \citep{2016arXiv161205560C}, This Work \\
        & SDSS & & & $g$ & 20.67 $\pm$ 0.04 & \citep{2015ApJS..219...12A}, This Work \\
        & SDSS & & & $r$ & 19.56 $\pm$ 0.02 & \citep{2015ApJS..219...12A}, This Work \\
        & SDSS & & & $i$ & 19.08 $\pm$ 0.02 & \citep{2015ApJS..219...12A}, This Work \\
        & SDSS & & & $z$ & 18.75 $\pm$ 0.09 & \citep{2015ApJS..219...12A}, This Work \\        
        & BASS & 90Prime & & $g$ & 20.69 $\pm$ 0.20 & \citep{2017PASP..129f4101Z}, This Work \\
        & BASS & 90Prime & & $r$ & 19.47 $\pm$ 0.19 & \citep{2017PASP..129f4101Z}, This Work \\
        & MzLS & MOSAIC-3 & & $z$ & 18.71 $\pm$ 0.18 & \citep{2019AJ....157..168D}, This Work \\
        & ALLWISE & & & $W1$ & 18.36 $\pm$ 0.05 & \citep{2014yCat.2328....0C}, This Work \\
        & ALLWISE & & & $W2$ & 18.80 $\pm$ 0.08 & \citep{2014yCat.2328....0C}, This Work \\
        & ALLWISE & & & $W3$ & 17.57 $\pm$ 0.18 & \citep{2014yCat.2328....0C}, This Work \\

        \hline

        \frbcharlotte (Charlotte)
        & Pan-STARRS & & & $r$ & 20.54 $\pm$ 0.14  & \citep{2016arXiv161205560C}, This Work \\
        & Pan-STARRS & & & $i$ & 19.90 $\pm$ 0.07 & \citep{2016arXiv161205560C}, This Work \\
        & Pan-STARRS & & & $z$ & 19.48 $\pm$ 0.07 & \citep{2016arXiv161205560C}, This Work \\
        & Pan-STARRS & & & $y$ & 18.93 $\pm$ 0.12 & \citep{2016arXiv161205560C}, This Work \\
        & P-200 & WaSP & 2023 July 09 UT & $g^\prime$ & 21.47 $\pm$	0.05 & This Work \\
        & P-200 & WaSP & 2023 July 09 UT & $r^\prime$ & 20.54 $\pm$ 0.03 & This Work \\
        & P-200 & WaSP & 2023 July 09 UT & $i^\prime$ & 19.90 $\pm$ 0.03 & This Work \\
        & ALLWISE & & & $W1$ & 17.70 $\pm$ 0.04 & \citep{2014yCat.2328....0C}, This Work \\
        & ALLWISE & & & $W2$ & 18.02 $\pm$ 0.05 & \citep{2014yCat.2328....0C}, This Work \\
        & ALLWISE & & & $W3$ & 17.28 $\pm$ 0.18 & \citep{2014yCat.2328....0C}, This Work \\

        \hline

        \frbjuan (Juan) & Pan-STARRS & & & $g$ & 20.81 $\pm$ 0.02 & \citep{2016arXiv161205560C}$^,$\citep{law2023deep} \\
        & Pan-STARRS & & & $r$ & 19.67 $\pm$ 0.03 & \citep{2016arXiv161205560C}$^,$\citep{law2023deep} \\
        & Pan-STARRS & & & $i$ & 19.04 $\pm$ 0.02 & \citep{2016arXiv161205560C}$^,$\citep{law2023deep} \\
        & Pan-STARRS & & & $z$ & 18.84 $\pm$ 0.03 & \citep{2016arXiv161205560C}$^,$\citep{law2023deep} \\
        
        \hline
        
    \end{tabular}
    
\end{table*}

\begin{table*}[ht!]
    \ContinuedFloat
    \centering
    \captionsetup{labelformat=supplementarytablelabel}
    \caption{Log of host galaxies imaging and photometry {\textit{(cont.)}}.}
    \begin{tabular}{LLLLLLL}
        \toprule
        
        FRB & Facility & Instrument & Observation Date & Filter & M$_\mathrm{AB}$ & Reference \\ 

        & & & & & [mag] & \\ 

        \hline

        & Pan-STARRS & & & $y$ & 18.54 $\pm$ 0.02 & \citep{2016arXiv161205560C}$^,$\citep{law2023deep} \\
        & BASS & 90Prime & & $g$ & 21.13 $\pm$ 0.20 & \citep{2017PASP..129f4101Z}, This Work \\
        & BASS & 90Prime & & $r$ & 19.64 $\pm$ 0.19 & \citep{2017PASP..129f4101Z}, This Work \\
        & MzLS & MOSAIC-3 & & $z$ & 18.75 $\pm$ 0.19 & \citep{2019AJ....157..168D}, This Work \\
        & 2MASS & & & $J$ & 17.79 $\pm$ 0.11 & \citep{2006AJ....131.1163S}$^,$\citep{law2023deep} \\
        & 2MASS & & & $H$ & 17.83 $\pm$ 0.07 & \citep{2006AJ....131.1163S}$^,$\citep{law2023deep} \\
        & 2MASS & & & $K_s$ & 17.90 $\pm$ 0.09 & \citep{2006AJ....131.1163S}$^,$\citep{law2023deep} \\
        & ALLWISE & & & $W1$ & 17.83 $\pm$ 0.05 & \citep{2014yCat.2328....0C}$^,$\citep{law2023deep} \\
        & ALLWISE & & & $W2$ & 18.24 $\pm$ 0.05 & \citep{2014yCat.2328....0C}$^,$\citep{law2023deep} \\

        \hline

        \frboran (Oran) & P-200 & WaSP & 2023 July 09 UT & $r^\prime$ & 19.91 $\pm$ 0.02 & This Work \\
        & P-200 & WaSP & 2023 July 09 UT & $i^\prime$ & 19.55 $\pm$ 0.01 & This Work \\
        & P-200 & WaSP & 2023 July 09 UT & $y^\prime$ & 19.40 $\pm$ 0.03 & This Work \\
        & P-200 & WIRC & 2022 Aug 17 UT & $J$ & 18.77 $\pm$ 0.06 & This Work \\
        & P-200 & WIRC & 2022 Aug 17 UT & $H$ & 18.69 $\pm$ 0.04 & This Work \\
        & ALLWISE & & & $W1$ & 18.32 $\pm$ 0.05 & \citep{2014yCat.2328....0C}$^,$\citep{law2023deep} \\
        & ALLWISE & & & $W2$ & 18.28 $\pm$ 0.05 & \citep{2014yCat.2328....0C}$^,$\citep{law2023deep} \\

        \hline
        
        \frbmikayla (Mikayla) & P-200 & WaSP & 2023 July 9 UT & $r^\prime$ & 21.86 $\pm$ 0.01 & This Work \\
        & P-200 & WaSP & 2023 July 9 UT & $i^\prime$ & 21.74 $\pm$ 0.02 & This Work \\
        & P-200 & WaSP & 2023 July 9 UT & $z^\prime$ & 20.72 $\pm$ 0.03 & This Work \\
        & P-200 & WIRC & 2023 July 28 UT & $J$ & 20.42 $\pm$ 0.08 & \citep{2003SPIE.4841..451W}, This Work \\

        \hline

        \frbfatima (Fatima) & Pan-STARRS & & & $g$ & 21.18 $\pm$ 0.11 & \citep{2016arXiv161205560C}, This Work \\
        & Pan-STARRS & & & $r$ & 20.30 $\pm$ 0.03 & \citep{2016arXiv161205560C}, This Work \\
        & Pan-STARRS & & & $i$ & 19.93 $\pm$ 0.02 & \citep{2016arXiv161205560C}, This Work \\
        & Pan-STARRS & & & $z$ & 19.53 $\pm$ 0.05 & \citep{2016arXiv161205560C}, This Work \\
        & Pan-STARRS & & & $y$ & 19.20 $\pm$ 0.08 & \citep{2016arXiv161205560C}, This Work \\
        & BASS & 90Prime & & $g$ & 21.46 $\pm$ 0.20 & \citep{2017PASP..129f4101Z}, This Work \\
        & BASS & 90Prime & & $r$ & 19.97 $\pm$ 0.19 & \citep{2017PASP..129f4101Z}, This Work \\
        & MzLS & MOSAIC-3 & & $z$ & 19.51 $\pm$ 0.18 & \citep{2019AJ....157..168D}, This Work \\
        & P-200 & WIRC & 2023 July 31 UT & $J$ & 19.23 $\pm$ 0.03 & This Work \\
        & ALLWISE & & & $W1$ & 19.02 $\pm$ 0.07 & \citep{2014yCat.2328....0C}, This Work \\
        & ALLWISE & & & $W2$ & 19.70 $\pm$ 0.20 & \citep{2014yCat.2328....0C}, This Work \\

        \hline

        \frbishita (Ishita) & Pan-STARRS & & & $r$ & 20.73 $\pm$ 0.22 & \citep{2016arXiv161205560C}, This Work \\
        & Pan-STARRS & & & $i$ & 20.63 $\pm$ 0.14 & \citep{2016arXiv161205560C}, This Work \\
        & Pan-STARRS & & & $z$ & 20.96 $\pm$ 0.21 & \citep{2016arXiv161205560C}, This Work \\
        & P-200 & WIRC & 2022 Aug 16 UT & $J$ & 20.80 $\pm$ 0.08 & This Work \\
        
        \hline

        \frbgertrude (Gertrude) & Pan-STARRS & & & $r$ & 20.95 $\pm$ 0.19 & \citep{2016arXiv161205560C}, This Work \\
        & Pan-STARRS & & & $i$ & 20.92 $\pm$ 0.05 & \citep{2016arXiv161205560C}, This Work \\
        & Pan-STARRS & & & $z$ & 20.54 $\pm$ 0.06 & \citep{2016arXiv161205560C}, This Work \\
        & P-200 & WIRC & 2023 July 30 UT & $J$ & 20.13 $\pm$ 0.05 & This Work \\
        & ALLWISE & & & $W1$ & 18.71 $\pm$ 0.17 & \citep{2014yCat.2328....0C}, This Work \\
        & ALLWISE & & & $W2$ & 19.29 $\pm$ 0.21 & \citep{2014yCat.2328....0C}, This Work \\
        & ALLWISE & & & $W3$ & 17.72 $\pm$ 0.42 & \citep{2014yCat.2328....0C}, This Work \\

        \hline

        \frberdos (Erdos) & BASS & 90Prime & & $g$ & 24.60 $\pm$ 0.28 & \citep{2017PASP..129f4101Z}, This Work \\
        & BASS & 90Prime & & $r$ & 23.35 $\pm$ 0.20 & \citep{2017PASP..129f4101Z}, This Work \\

        \hline

        \end{tabular}

\end{table*}

\begin{table*}[ht!]
    \ContinuedFloat
    \centering
    \captionsetup{labelformat=supplementarytablelabel}
    \caption{Log of host galaxies imaging and photometry {\textit{(cont.)}}.}
    \begin{tabular}{LLLLLLL}
        \toprule
        
        FRB & Facility & Instrument & Observation Date & Filter & M$_\mathrm{AB}$ & Reference \\ 

        & & & & & [mag] & \\ 
        
        \hline 

        & MzLS & MOSAIC-3 & & $r$ & 23.09 $\pm$ 0.19 & \citep{2019AJ....157..168D}, This Work \\

        \hline

        \frbfen (Fen) & Pan-STARRS & & & $y$ & 22.37 $\pm$ 0.40 & \citep{2016arXiv161205560C}, This Work \\
        & P-200 & WaSP & 2023 July 9 UT & $g^\prime$ & 23.87 $\pm$ 0.05 & This Work \\
        & P-200 & WaSP & 2023 July 9 UT & $r^\prime$ & 23.00 $\pm$ 0.02 & This Work \\
        & P-200 & WaSP & 2023 July 9 UT & $i^\prime$ & 22.75 $\pm$ 0.02 & This Work \\
        & P-200 & WaSP & 2023 July 9 UT & $z^\prime$ & 22.20 $\pm$ 0.03 & This Work \\

        \hline

        \frbpingu (Pingu) & Pan-STARRS & & & $r$ & 21.46 $\pm$ 0.05 & \citep{2016arXiv161205560C}, This Work \\
        & Pan-STARRS & & & $i$ & 21.20 $\pm$ 0.18 & \citep{2016arXiv161205560C}, This Work \\
        & Pan-STARRS & & & $z$ & 20.41 $\pm$ 0.16 & \citep{2016arXiv161205560C}, This Work \\
        & BASS & 90Prime & & $g$ & 23.06 $\pm$ 0.22 & \citep{2017PASP..129f4101Z}, This Work \\
        & BASS & 90Prime & & $r$ & 21.36 $\pm$ 0.16 & \citep{2017PASP..129f4101Z}, This Work \\
        & MzLS & MOSAIC-3 & & $r$ & 20.50 $\pm$ 0.18 & \citep{2019AJ....157..168D}, This Work \\

        \hline

        \frbwhitney (Whitney) & GALEX & & & NUV & 22.51 $\pm$ 0.07 & \citep{2005ApJ...619L...1M}, This Work \\
        & Pan-STARRS & & & $r$ & 21.20 $\pm$ 0.08 & \citep{2016arXiv161205560C}$^,$\citep{law2023deep} \\
        & Pan-STARRS & & & $i$ & 20.72 $\pm$ 0.12 & \citep{2016arXiv161205560C}$^,$\citep{law2023deep} \\
        & Pan-STARRS & & & $z$ & 20.71 $\pm$ 0.01 & \citep{2016arXiv161205560C}$^,$\citep{law2023deep} \\
        & P-200 & WIRC & 2023 Feb 10 UT & $J$ & 20.22 $\pm$ 0.21 & \citep{law2023deep} \\
        & P-200 & WIRC & 2023 Feb 10 UT & $H$ & 20.29 $\pm$ 0.23 & \citep{law2023deep} \\
        & P-200 & WIRC & 2023 Feb 10 UT & $K_s$ & 20.41 $\pm$ 0.30 & \citep{law2023deep} \\
        & ALLWISE & & & $W1$ & 20.00 $\pm$ 0.23 & \citep{2014yCat.2328....0C}$^,$\citep{law2023deep} \\
        & ALLWISE & & & $W2$ & 20.48 $\pm$ 0.36 & \citep{2014yCat.2328....0C}$^,$\citep{law2023deep} \\
        
        \hline

        \frbferb (Ferb) & GALEX & & & NUV & 22.44 $\pm$ 0.06 & \citep{2005ApJ...619L...1M}, This Work \\
        & Pan-STARRS & & & $g$ & 21.98 $\pm$ 0.05 & \citep{2016arXiv161205560C}, This Work \\
        & Pan-STARRS & & & $r$ & 21.43 $\pm$ 0.20 & \citep{2016arXiv161205560C}, This Work \\
        & Pan-STARRS & & & $i$ & 21.23 $\pm$ 0.06 & \citep{2016arXiv161205560C}, This Work \\
        & Pan-STARRS & & & $z$ & 20.90 $\pm$ 0.09 & \citep{2016arXiv161205560C}, This Work \\
        & SDSS & & & $g$ & 22.07 $\pm$ 0.27 & \citep{2015ApJS..219...12A}, This Work \\
        & SDSS & & & $r$ & 21.31 $\pm$ 0.20 & \citep{2015ApJS..219...12A}, This Work \\
        & SDSS & & & $i$ & 21.17 $\pm$ 0.24 & \citep{2015ApJS..219...12A}, This Work \\
        & BASS & 90Prime & & $g$ & 22.01 $\pm$ 0.21 & \citep{2017PASP..129f4101Z}, This Work \\
        & BASS & 90Prime & & $r$ & 21.38 $\pm$ 0.20 & \citep{2017PASP..129f4101Z}, This Work \\
        & MzLS & MOSAIC-3 & & $z$ & 21.02 $\pm$ 0.22 & \citep{2019AJ....157..168D}, This Work \\

        \hline

        \frbkoyaanisqatsi (Koy) & BASS & 90Prime & & $g$ & 22.84 $\pm$ 0.20 & \citep{2017PASP..129f4101Z}, This Work \\
        & BASS & 90Prime & & $r$ & 22.84 $\pm$ 0.19 & \citep{2017PASP..129f4101Z}, This Work \\
        & MzLS & MOSAIC-3 & & $z$ & 22.55 $\pm$ 0.18 & \citep{2019AJ....157..168D}, This Work \\
        & P-200 & WIRC & 2023 Feb 2 UT & $J$ & 21.95 $\pm$ 0.27 & This Work \\
        & P-200 & WIRC & 2024 Jan 1 UT & $H$ & 21.53 $\pm$ 0.38 & This Work \\

        \hline 

        \frbnihari (Nihari)
        & Pan-STARRS & & & $r$ & 22.37 $\pm$ 0.13 & \citep{2016arXiv161205560C}, This Work \\
        & Pan-STARRS & & & $i$ & 22.28 $\pm$ 0.13 & \citep{2016arXiv161205560C}, This Work \\
        & Pan-STARRS & & & $z$ & 21.70 $\pm$ 0.14 & \citep{2016arXiv161205560C}, This Work \\
        & Pan-STARRS & & & $y$ & 21.51 $\pm$ 0.27 & \citep{2016arXiv161205560C}, This Work \\
        & SDSS & & & $u$ & 24.37 $\pm$ 1.09 & \citep{2015ApJS..219...12A}, This Work \\
        & SDSS & & & $g$ & 24.51 $\pm$ 0.57 & \citep{2015ApJS..219...12A}, This Work \\
        & SDSS & & & $r$ & 22.62 $\pm$ 0.18 & \citep{2015ApJS..219...12A}, This Work \\
        
        \hline

        \end{tabular}

\end{table*}

\newpage
\clearpage
\newpage
\newpage

\begin{table*}[ht!]
    \ContinuedFloat
    \centering
    \captionsetup{labelformat=supplementarytablelabel}
    \caption{Log of host galaxies imaging and photometry {\textit{(cont.)}}.}
    \begin{tabular}{LLLLLLL}
        \toprule
        
        FRB & Facility & Instrument & Observation Date & Filter & M$_\mathrm{AB}$ & Reference \\ 

        & & & & & [mag] & \\ 
        
        \hline 

        & SDSS & & & $i$ & 22.21 $\pm$ 0.16 & \citep{2015ApJS..219...12A}, This Work \\
        & SDSS & & & $z$ & 21.60 $\pm$ 0.38 & \citep{2015ApJS..219...12A}, This Work \\
        & BASS & 90Prime & & $g$ & 24.07 $\pm$ 0.21 & \citep{2017PASP..129f4101Z}, This Work \\
        & BASS & 90Prime & & $r$ & 22.61 $\pm$ 0.20 & \citep{2017PASP..129f4101Z}, This Work \\
        & MzLS & MOSAIC-3 & & $z$ & 21.71 $\pm$ 0.19 & \citep{2019AJ....157..168D}, This Work \\
        & P-200 & WIRC & 2023 April 29 UT & $K_s$ & 20.54 $\pm$ 0.09 & This Work \\
        & ALLWISE & & & $W1$ & 19.51 $\pm$ 0.09 & \citep{2014yCat.2328....0C}, This Work \\
        & ALLWISE & & & $W2$ & 20.19 $\pm$ 0.27 & \citep{2014yCat.2328....0C}, This Work \\

        \hline
        
        \frbquincy (Quincy) & GALEX & & & NUV & 22.58 $\pm$ 0.07 & \citep{2005ApJ...619L...1M}, This Work \\
        & Pan-STARRS & & & $r$ & 21.27 $\pm$ 0.18 & \citep{2016arXiv161205560C}$^,$\citep{law2023deep} \\
        & Pan-STARRS & & & $i$ & 21.29 $\pm$ 0.08 & \citep{2016arXiv161205560C}$^,$\citep{law2023deep} \\
        & Pan-STARRS & & & $z$ & 21.00 $\pm$ 0.21 & \citep{2016arXiv161205560C}$^,$\citep{law2023deep} \\
        & BASS & 90Prime & & $g$ & 22.18 $\pm$ 0.20 & \citep{2017PASP..129f4101Z}, This Work \\
        & BASS & 90Prime & & $r$ & 21.22 $\pm$ 0.19 & \citep{2017PASP..129f4101Z}, This Work \\
        & MzLS & MOSAIC-3 & & $z$ & 21.00 $\pm$ 0.18 & \citep{2019AJ....157..168D}, This Work \\        
        & P-200 & WIRC & 2022 Aug 16 UT & $J$ & 20.78 $\pm$ 0.10 & \citep{law2023deep} \\
        & ALLWISE & & & $W1$ & 20.69 $\pm$ 0.06 & \citep{2014yCat.2328....0C}$^,$\citep{law2023deep} \\
        
        \hline

        \frbmifanshan (Mifanshan) & BASS & 90Prime & & $g$ & 23.90 $\pm$ 0.22 & \citep{2017PASP..129f4101Z}, This Work \\
        & BASS & 90Prime & & $r$ & 23.17 $\pm$ 0.22 & \citep{2017PASP..129f4101Z}, This Work \\
        & MzLS & MOSAIC-3 & & $z$ & 22.03 $\pm$ 0.20 & \citep{2019AJ....157..168D}, This Work \\
        & P-200 & WIRC & 2023 Oct 20 UT & $J$ & 21.87 $\pm$ 0.26 & \citep{law2023deep}, This Work \\
        & P-200 & WIRC & 2024 Jan 1 UT & $H$ & 22.08 $\pm$ 0.31 & This Work \\
        
        \hline

        \end{tabular}

        \caption*{\footnotesize{N{\scriptsize{OTE}}: Photometry of all imaging observations used in this work. The listed AB magnitudes of the galaxy are corrected for the Milky Way galactic dust extinction~\protect\citemethods{2018JOSS....3..695M, 1999PASP..111...63F}.}
        }
    
\end{table*}

\newcolumntype{L}{>{\footnotesize\raggedright\arraybackslash}l}

\newpage
\clearpage
\newpage
\newpage

\begin{table*}[ht!]
    \centering
    \captionsetup{labelformat=supplementarytablelabel}
    \caption{Summary of Parameters Used in Spectral Energy Distribution (SED) Analysis.}
    \begin{tabular}{LLL}
        \toprule
        {Parameter} & {Definition} & {Prior or Value} \\
         \hline 

         \multicolumn{3}{l}{\footnotesize{Star Formation History}} \\
        
         SFH & Continuity SFH & 3 \\

         IMF & Kroupa IMF~\protect\citemethods{2001MNRAS.322..231K} & 2 \\
         
         $z_{\mathrm{red}}$ & 
         Spectroscopic redshift & $z_{\mathrm{spec}}+\mathcal{U}(-0.01, 0.01)$ \\
         
         $\log{\mathrm{M}_\mathrm{f}}$ & Total stellar mass formed & $\mathcal{U}(8, 12)$ \\
         
         $\log{\mathrm{Z}/\mathrm{Z}_\odot}$ & Metallicity (Follows mass-metallicity relation~\citep{2005MNRAS.362...41G}) & $\mathcal{U}(-2, 0.2)$ \\

         N$_{\mathrm{bins}}$ & Number of SFH bins & 7 \\

         $r_i$ & Ratios of SFRs in adjascent SFH bins & $\mathcal{T}(0, 3, 2)$ \\
         
         \hline

         \multicolumn{3}{l}{\footnotesize{Dust Attenuation}} \\

         A$_{\mathrm{V,~type}}$ & Kriek \& Conroy~\protect\citemethods{2013ApJ...775L..16K} dust attenuation curve & 4 \\

         $\Gamma_{\mathrm{dust}}$ & Power law modification of Calzetti et al.~\protect\citemethods{2000ApJ...533..682C} dust attenuation curve & $\mathcal{U}(-1, 0.4)$ \\

         $\tau_{\mathrm{old}}$ & Dust attenuation of old stellar light & $\mathcal{CN}(0.3, 1, 0, 4)$ \\
         $\tau_{\mathrm{young}}/\tau_{\mathrm{old}}$ & Describes dust attenuation of young stellar light & $\mathcal{CN}(1, 0.3, 0, 2)$ \\

         \hline

         \multicolumn{3}{l}{\footnotesize{Dust Emission}} \\

         u$_{\mathrm{dust}}$ & Add dust emission & True \\

         q$_{\mathrm{PAH}}$ & Describes grain size distribution of PAHs & $\mathcal{U}(0.5, 7)$ \\

         U$_{\mathrm{min}}$ & Minimum dust emission radiation field strength & 1 \\

         $\gamma_{\mathrm{dust}}$ &  Relative contribution of dust heated at and above U$_{\mathrm{min}}$ & 0.01 \\

         \hline

         \multicolumn{3}{l}{\footnotesize{AGN}} \\

         u$_{\mathrm{AGN}}$ & Add AGN-heated dust torus emission model~\protect\citemethods{2008ApJ...685..160N} & True \\

         $f_{\mathrm{AGN}}$ & Fraction of AGN luminosity relative to stellar bolometric luminosity & $\mathcal{LU}(10^{-5}, 3)$ \\

         $\tau_{\mathrm{AGN}}$ & Optical depth of clouds in the AGN dust torus model~\protect\citemethods{2008ApJ...685..160N} & $\mathcal{LU}(5, 150)$ \\

         \hline

         \multicolumn{3}{l}{\footnotesize{Nebular Emission}} \\

         u$_{\mathrm{nebular}}$ & Add nebular emission & True \\

         u$_{\mathrm{marginalize}}$ & Marginalize over observed emission lines & True \\

         U$_{\mathrm{neb}}$ & Nebular gas ionization parameter & $\mathcal{U}(-4, -1)$ \\

         $\log{\mathrm{Z}_{\mathrm{gas}}/{\mathrm{Z}_{\odot}}}$ & Gas-phase metallicity (Tied to the stellar metallicity) & $\log(\mathrm{Z}/\mathrm{Z}_{\odot})$ \\ 

         $\sigma_{\mathrm{smooth}}$ & Velocity dispersion [km/s] & $\mathcal{U}(50, 500)$ \\

         w$_{\mathrm{eline}}$ & Width of emission line amplitude prior & $\mathcal{U}(0.01, 100)$ \\

         $\sigma_{\mathrm{eline}}$ & Emission lines broadening &  $\mathcal{U}(30, 500)$ \\

         \hline
         
         \multicolumn{3}{l}{\footnotesize{Spectral and Photometric Calibration}} \\


         f$_{\mathrm{spec,~outlier}}$ & Fraction of outlier spectrum pixels & $\mathcal{U}(10^{-5}, 0.5)$ \\

         f$_{\mathrm{phot,~outlier}}$ & Fraction of outlier photometric bands & $\mathcal{U}(0, 0.5)$ \\

         j$_{\mathrm{spec}}$ & Spectral jitter parameter to inflate noise & $\mathcal{U}(1, 3)$ \\

         n$_{\sigma,~\mathrm{spec},~\mathrm{out.}}$ & Factor of inflation for errors determined by f$_{\mathrm{spec,~outlier}}$ & 50 \\

         n$_{\sigma,~\mathrm{phot},~\mathrm{outlier}}$ & Factor of inflation for errors determined by f$_{\mathrm{phot,~out.}}$ & 50 \\

         n$_{\mathrm{spec}}$ & Spectrum normalization & 1 \\

         u$_{\mathrm{FFT smooth}}$ & Use FFT spectral smoothing & True \\

         C & Chebyshev polynomial order for spectrum calibration & 12 \\

         u$_{\mathrm{regul}}$ & Regularization of spectral calibration vector coefficients & 0 \\
         
         \hline

         \end{tabular}

    \footnotesize{{\raggedright N{\scriptsize{OTE}}: $\mathcal{U}$ denotes a Uniform distribution, $\mathcal{CN}$ denotes a Clipped Normal distribution, $\mathcal{T}$ denotes a Student T-distribution and $\mathcal{LU}$ denotes a Log Uniform distribution.
    \par}}
    
    \label{table:sed_params}
\end{table*}

\newpage
\clearpage
\newpage
\newpage

\begin{table*}[ht!]
    \centering
    \captionsetup{labelformat=supplementarytablelabel}
    \caption{2D Two-sample KS-tests between background population and FRB host galaxies.}
    
     \begin{tabular}{LCCCCCC}
        \toprule
        Galaxy Type & All Types & HII Regions & LINER & Seyfert & HII + LINER & HII + Seyfert \\
        \hline

        $p( \text{[NII]}/\text{H}\alpha)$ & 0.101 & 0.023 & $\approx 10^{-18}$ & $\approx 10^{-11}$ & 0.022 & 0.161 \\ 

        $p( \text{[SII]}/\text{H}\alpha)$ & 0.076 & 0.224 & $\approx 10^{-15}$ & $\approx 10^{-15}$ & 0.250 & 0.085 \\

        $p( \text{[OI]}/\text{H}\alpha)$ & 0.063 & 0.081 & $\approx 10^{-17}$ & $\approx 10^{-13}$ & 0.116 & 0.067 \\
        
        \hline
    \end{tabular}

    \footnotesize{{\raggedright N{\scriptsize{OTE}}: P-values from the KS-tests comparing FRB host populations to different subsets of galaxy population in three different empirical optical emission line ratio spaces.
    \par}}
    \label{table:BPT_analysis}
\end{table*}

\begin{table*}[ht!]
    \centering
    \captionsetup{labelformat=supplementarytablelabel}
    \caption{KS-tests and test statistics between the hosts of different transient classes.}
    \begin{tabular}{LCCCCCC}
        \toprule
        Transient & $p_{\Delta \mathrm{R}}$ & $p_{\mathrm{M}_\ast}$ & $p_{\mathrm{SFR}}$ & $p_{\mathrm{sSFR}}$ & TS$_{\mathrm{M}_\ast}$ & TS$_{\mathrm{SFR}}$ \\
        \hline

        SLSNe & $10^{-4}$ & $10^{-14}$ & 0.001 & $10^{-10}$ & 0.136 & 0.054 \\

        lGRB & $10^{-8}$ & $10^{-11}$ & 0.105 & $10^{-7}$ & 0.128 & 0.061 \\

        CCSNe & 0.051 & 0.003 & 0.141 & 0.053 & 0.056 & 0.098 \\

        sGRB & 0.306 & 0.002 & 0.016 & 0.283 & 0.043 & 0.038 \\

        Type Ia & 0.223 & 0.058 & 0.415 & 0.003 & 0.056 & 0.067 \\

        ULX & 0.090 & $10^{-6}$ & 0.036 & 0.129 & 0.012 & 0.016 \\ 

        FRB & & & & & 0.017 & 0.024 \\

        \hline
    \end{tabular}

    \footnotesize{{\raggedright N{\scriptsize{OTE}}: P-values from the KS-tests and test statistics from comparing galactocentric offsets, stellar mass, SFR and sSFR of FRB hosts and different transient classes.
    \par}}
    \label{table:compare_with_other_transients}
\end{table*}

\end{document}